\pdfoutput=1
\documentclass[12pt]{article}
\usepackage{amsfonts,amssymb,amsmath,amsthm,dsfont,graphicx,hyperref,mathrsfs,euscript,enumitem,framed,bm, soul, bbm}
\usepackage[title, titletoc]{appendix}
\usepackage[doublespacing]{setspace}
\usepackage{ifthen,caption,subcaption,lscape}
\hypersetup{pdfborder = {0 0 0},colorlinks=true,linkcolor=blue,urlcolor=blue,citecolor=blue}
\usepackage{natbib}
\usepackage{comment}
\usepackage{anyfontsize}

\usepackage[top=1in,bottom=1in,left=1in,right=1in]{geometry}

\newtheorem{thm}{Theorem}

\newtheorem{assumption}{Assumption}

\numberwithin{equation}{section}

\theoremstyle{definition}
\newtheorem{remark_tmp}{Remark}

\newtheorem{example_tmp}{Example}
\newenvironment{example}
	{ \begin{example_tmp} 	}
	{ 
		\hfill{\LARGE$\lrcorner$}
		\end{example_tmp} 
	}
 
\allowdisplaybreaks[1]

\renewcommand*{\arraystretch}{.6}

	\DeclareMathOperator*{\argmin}{arg\,min}

\renewcommand{\P}{\mathbb{P}}
\newcommand{\E}{\mathbb{E}}

\newcommand{\I}{\mathds{1}}
\newcommand{\cval}{\mathfrak{c}}

\newcommand{\bb}{\mathbf{b}}

\newcommand{\bN}{\mathbf{N}}
\newcommand{\bQ}{\mathbf{Q}}
\newcommand{\bw}{\mathbf{w}}

\newcommand{\bbeta}{\boldsymbol{\beta}}
\newcommand{\btheta}{\boldsymbol{\theta}}
\newcommand{\bgamma}{\boldsymbol{\gamma}}

\newcommand{\bSigma}{\boldsymbol{\Sigma}}

\newcommand{\evalw}{\mathsf{w}}

\begin{document}
\fontsize{11.8pt}{14.5pt}\selectfont

{\singlespacing
\title{\vspace{-0.0in}Nonlinear Binscatter Methods\thanks{We thank Mart{\'i}n Almuzara, Isaiah Andrews, Ricardo Masini, Boris Shigida, Rocio Titiunik, and Rae Yu for helpful comments and discussions. Ignacio Lopez Gaffney provided excellent research assistance.  Cattaneo gratefully acknowledges financial support from the National Science Foundation through grants SES-1947805, SES-2019432, and SES-2241575, and the John Simon Guggenheim Memorial Foundation through a 2026 Guggenheim Fellowship. Feng gratefully acknowledges financial support from the National Natural Science Foundation of China (NSFC) through grants 72522001 and 72203122.
The views expressed in this paper are those of the authors and do not necessarily reflect the position of the Federal Reserve Bank of New York or the Federal Reserve System. Software and replication files are available at \url{https://github.com/nppackages/binsreg}.}
\medskip }
\author{Matias D. Cattaneo\thanks{Department of Economics, Princeton University.} \and
	    Richard K. Crump\thanks{Capital Markets, Federal Reserve Bank of New York.} \and
	    Max H. Farrell\thanks{Department of Economics, UC Santa Barbara.} \and 
	    Yingjie Feng\thanks{School of Economics and Management, Tsinghua University.}}
\maketitle
}

\begin{abstract}
Binscatters are a powerful tool for empirical work in the social, behavioral, and biomedical sciences. Available tools rely on least squares estimation of the conditional mean. We introduce novel binscatter methods based on nonlinear, possibly nonsmooth M-estimation, covering generalized linear, robust, and quantile regression models. We provide theoretical results and practical tools, including optimal bin selection, confidence bands, and statistical tests regarding functional form or shape restrictions. We demonstrate our methods by studying the relationship of income and (lack of) health insurance. We provide software for \texttt{Python}, \texttt{R}, and \texttt{Stata}. Our technical results may be of independent interest. 
\end{abstract}

\thispagestyle{empty}
\clearpage

\setcounter{page}{1}

\pagestyle{plain}

\section{Introduction}
    \label{sec:intro}

Data visualization is a crucial step in any statistical analysis. In the era of big data it has become increasingly important to have simple yet informative visual tools to guide, supplement, or in some cases even supplant, numerical statistical analyses. However, it is important to maintain statistical formality and rigor to ensure the validity of any conclusions based on the data. We seek to accomplish both of these goals---effective visualization couched in a formal framework---with binned scatter plot methods.

Often known simply as a \textit{binscatter}, the binned scatter plot has become a popular tool for visualization in large data sets, particularly in the social and behavioral sciences. The goal is to flexibly estimate, and visualize, features of the conditional distribution of a scalar outcome $y_i$, which may be discrete or continuous, given a covariate or treatment variable $x_i$, which is scalar and continuous, while possibly also controlling for a $d$-dimensional vector of additional factors $\bw_i$. For estimating the conditional mean, as in the traditional regression analysis, binning has a long history: so familiar is this approach that over $60$ years ago \cite{Tukey1961_Berkeley}, calling it a \textit{regressogram}, went so far as to claim that ``[a]ll statisticians who handle data know how to attack the simple case of this situation where $y$ and $x$ are both single real numbers'' (p. 682). Going on to describe the construction, Tukey writes: ``[t]he $x$-axis is to be divided into suitable intervals, the mean of all the $y$-values corresponding to $x$-values falling in each given interval is to be found, the results are then to be plotted, either as points, each located above the center of the corresponding $x$-interval, or better as horizontal bars, each extending over the corresponding $x$-interval'' (p. 682).

This construction (perhaps disappointingly often with plotted dots rather than horizontal lines) has gained popularity in statistics, economics, and data science. The prevalence of binscatter plots can be partly ascribed to its intuitive construction and compelling visualization properties: given only data on $x_i$ and $y_i$, the plot is a clean and interpretable depiction of the conditional mean. Moreover, several limitations of the classical scatter plot account for the rising use of binned scatter plots in modern analyses. First, in big data sets the classical scatter plot is too dense to be informative, particularly about general ``patterns'' in the data which are to be modeled in subsequent analyses. Second, somewhat conversely, in cases where privacy is a concern the scatter plot is not allowed, regardless of its informational use as a visualization tool. Third, classical scatter plots do not provide a well-defined way to control for other factors, a common goal in treatment effect estimation and causal inference. Finally, particularly relevant to our setting, a scatter plot is not useful when outcomes are discrete.  In contrast, a \emph{binned} scatter plot provides a simple, yet flexible way of visualizing features of the conditional distribution of a (possibly discrete) outcome variable given a continuous covariate (or treatment) of interest, while controlling for other important factors. 

Formally, a binscatter is grounded in the classical semilinear regression model. To date, however, binscatters have been available only to visualize (and estimate) conditional mean functions fitted using least squares. A common usage in this setting is comparing the nonparametric estimate to a linear fit, as a precursor to linear regression analysis. See \cite{Starr-Goldfarb_2020_SMJ} for a practical review and background references. In the least squares setting, a binscatter is formally an estimator of a semilinear model for the conditional mean, nonparametric in the covariate of interest and linear in the controls, where the nonparametric component is estimated by partitioned regression. \cite{Cattaneo-Crump-Farrell-Feng_2024_AER} used that framework to derive formal statistical properties of canonical binscatter, including correcting a common mistake in empirical practice when using controls, and provide asymptotically valid confidence bands and optimal tuning parameter selection.

The restriction to least squares semilinear regression to estimate the conditional mean has limited the applicability of binscatter methods. For one, important features of the data, such as spread or variability, cannot be visualized. Further, existing methods (and theory) can be misleading in settings where the outcome is discrete or otherse restricted. For example, in the empirical illustration we use throughout, we study uninsuredness rates using a fractional outcome model, most naturally fitted using quasi-likelihood methods based on the logistic link. Last but not least, binscatter methods for quantile regression analysis are currently lacking in the literature, despite  their usefulness for empirical work. 

This paper introduces and studies a broad class of binscatter M-estimation methods allowing for (i) a nonlinear and/or nonsmooth loss function and (ii) a nonlinear link function. Our results provide for the use of binned scatter plots for various visualization goals and different data types, particular leading cases being semiparametric conditional quantile regression and generalized partially linear models. We make several methodological and theoretical contributions: (i) we propose a feasible method for optimal tuning parameter selection to choose the appropriate number of bins; (ii) we provide (pointwise and) uniform inference procedures to construct confidence bands and hypothesis tests for parametric specifications and shape restrictions, and (iii) we develop group-wise comparisons for continuous treatment effects or for treatment effect heterogeneity. Developing these methods relies on novel technical work: allowing for a large class of binning methods, including random binning, we prove new uniform (in $x$) Bahadur representations and strong approximations, and thus uniform distribution theory, for the broad class of nonlinear semiparametric models considered. Obtaining these results for nonlinear, nonsmooth models, with data-dependent partitions and (a growing number of) additional controls, represents the main technical contributions of our paper, some of which may be of independent interest. For example, our results may be useful for first-step nonparametric estimation in larger inference tasks, which often involves nonlinear models.

Our nonlinear binscatter methods help restore, and in cases such as discrete outcomes or additional controls, surpass, the utility of the conventional scatter plot. We offer principled ways to visually assess patterns in the data, quantify uncertainty, and test hypotheses about the findings. Our results on quantile regression allow researchers to assess the spread of the conditional distribution, detect outliers or influential observations in the data, and study a larger class of treatment effects, formalizing and expanding common practices based on the classical scatter plot in small data, while controlling for additional factors. Our confidence bands properly quantify and communicate the uncertainty around the estimated function of interest, and can also be used to guide further analyses. We also develop formal uniform hypothesis testing procedures regarding those functions, to assess shape constraints and parametric specifications. Causal inference is an important application area of our uniform inference results: studying treatment effect heterogeneity for binary treatments or the dose response function for a continuous treatment without imposing a functional form (e.g., to evaluate important hypotheses such as monotonicity in the dosage).

The paper proceeds as follows. We next discuss the connections between our work and the existing literature. Section \ref{sec:setup} introduces binned scatter plots, defines the statistical model, and clarifies the parameters of interest. Section \ref{sec:theory} gives details on our theoretical contributions and states our most important technical results. Applied-minded readers can skip this section. Then, Sections \ref{sec:choice of J} and \ref{sec:inference} build on this theory to deliver tuning parameter selection and uniform inference, respectively. Section \ref{sec:conclusion} concludes.

Throughout the paper we illustrate our methods and results with a running empirical application using zip code-level data from the American Community Survey (ACS). The dependent variable, $y_i$, is the percentage of individuals without health insurance, and the independent variable of interest, $x_i$, is per capita income. These data are natural for our purposes because the presence of Medicaid, and other known economic/empirical regularities, tell us in advance how different features of the data should look, and so, rather than aim for novel empirical findings, we illustrate how our methods recover these known features.

The Supplemental Appendix (SA hereafter) contains additional technical and implementation details, all proofs, and further discussion of how our technical contributions improve on the literature. General-purpose software in \texttt{Python}, \texttt{R}, and \texttt{Stata} and replication files are available at \url{https://github.com/nppackages/binsreg}; see \cite{Cattaneo-Crump-Farrell-Feng_2025_Stata}.

\subsection{Related Literature}
    \label{sec:literature}

This paper contributes to several strands of the literature. First,  our work builds upon and extends existing binned scatter plot methods available for applied research. Binned scatter plots are particularly popular in applied microeconomics, but have become commonly used in a wide variety of disciplines. Examples of applications include research in climate studies \citep{Schlenker-Roberts2009_PNAS}, health economics \citep{Agarwal-etal2019_AER}, marketing \citep{Shapiro2020_MS}, urban economics \citep{Su2022_AEJPolicy}, and finance \citep{Gathergood-Olafsson2024_RFS}. \cite{Starr-Goldfarb_2020_SMJ} and \cite{Cattaneo-Crump-Farrell-Feng_2024_AER} give further examples. All these existing applications are linear binscatter methods used least squares loss. This motivates our work, as our main methodological contribution is to introduce nonlinear binscatter methods, constructed using a general, possibly nonsmooth semilinear M-estimation approach. As a result, we propose a broad array of new binscatter methods for generalized linear models (e.g., Logit or Probit), robust semiparametric regression (e.g., Huber), and quantile regression.

Second, our theoretical results contribute to the literature on series/sieve estimation in general, and partitioning-based methods in particular (i.e., piecewise polynomials and splines). See \cite{Gyorfi-etal_2002_book} for a textbook introduction, and \cite{Huang_2003_AoS}, \cite{Cattaneo-Farrell_2013_JoE}, \cite{Belloni-Chernozhukov-Chetverikov-Kato_2015_JoE}, \cite{Cattaneo-Farrell-Feng_2020_AoS}, as well as references therein, for prior convergence rates and distribution theory. These prior works studied uniform estimation and inference for linear piecewise polynomials and spline series regression without data-driven partitioning and without additional covariates, often imposing strong regularity conditions. Our primary technical contributions as compared to the recent literature are (i) allowing for general, possibly nonlinear and nonsmooth M-estimation, (ii) allowing for random partitions and hence random basis functions in the series estimator, (iii) controlling for a diverging number of other factors in a semilinear model, and (iv) obtaining novel strong approximations and uniform inference under weaker conditions than those previously available. A substrand of the series estimation literature studies quantile regression, the closest antecedent to our work being \cite{Belloni-Chernozhukov-Chetverikov-FernandezVal_2019_JoE}. Unlike that prior work, we consider general nonlinear, possibly nonsmooth, M-estimation problems and allow for random partitions and additional controls, and our technical results are obtained under weaker regularity conditions, which in particular permit the use of piecewise constant fitting necessary for a binned scatter plot. Notably, none of the results in \cite{Cattaneo-Crump-Farrell-Feng_2024_AER} are applicable to the large class of nonlinear binscatter estimators considered in this paper, because they only consider least squares semilinear binscatter models. Section \ref{sec:theory} below expands on how our theoretical techniques and results are novel relative to the extant literature. Even more detail is given throughout the SA.

Finally, our paper also contributes to the literature on data visualization, which has become an increasingly active field of study in recent years due to the rise of big data and machine learning methods. Our results speak directly to this literature, and in particular to the need for clear and explicit depictions of uncertainty, both in terms of variance and of estimation error \citep{Healy2018_book}. These are crucial in data visualization in science and research contexts as this ``builds trust and credibility'' \citep[p.189]{Schwabish2021_book}.

\section{Setup}
    \label{sec:setup}

A binned scatter plot is designed to provide a flexible, nonparametric estimate of a regression function. The construction and interpretation of a binned scatter plot is simple and intuitive, which drives their appeal for applied work. But as we will see, there are some subtleties when binned scatter plots are applied to nonlinear, nonsmooth models, especially with controls.

To describe the construction, it is helpful to first make precise the model and objects of interest. Our goal is to learn a regression function (which need not be the conditional mean) that features in the conditional distribution of a scalar outcome $y_i$, which may be discrete or continuous, given a covariate or treatment variable $x_i$, which is scalar and continuous, while possibly also controlling for a $d$-dimensional vector of additional factors $\bw_i$. In applications, the goal is to flexibly study the relationship of $y_i$ to $x_i$, but not necessarily to discover (or allow for) heterogeneity or nonlinearity in $\bw_i$. The set of controls $\bw_i$ often contains factor variables, and can be larger in dimension ($d$ may diverge with $n$ as made precise below). We assume the regression function depends on the scalar index $\theta_0(x_i, \bw_i) := \mu_0(x_i)+\bw_i'\bgamma_0$, for an unknown function $\mu_0$ and vector $\bgamma_0$, and is thus partially linear in nature.

The model is defined by the following structure, which determines how the scalar index $\theta_0(x_i, \bw_i)$ relates to the outcome $y_i$. Let $\theta$ be a generic value of the index. For a loss function $\rho(y;\eta(\theta))$ and an inverse link function $\eta(\theta)$, let
\begin{equation}
    \label{eq:model}
	(\mu_0(\cdot), \bgamma_0)=\argmin_{\mu\in\mathcal{M}, \bgamma\in\mathbb{R}^d}\; \E\big[\rho(y_i;\eta(\mu(x_i)+\bw_i'\bgamma))\big],
\end{equation}
where we assume the solution is unique, the model is correctly specified, and $\rho(y;\eta(\theta))$, $\eta(\theta)$, and the function class $\mathcal{M}$ obey typical boundedness and smoothness restrictions discussed in Section \ref{sec:theory}. For different choices of $\rho(\cdot)$ and $\eta(\cdot)$ this formulation covers a large class of problems including generalized linear models, robust regression, quantile regression, and nonlinear least squares. In the Appendix below we use the running empirical application to illustrate benefits of nonlinear modeling in the context of binscatter plots, which include robustness to outliers and corrupted data, valid predictions not exceeding the support of the outcome, and informative measures of spread distinct from uncertainty around the mean. Below are several leading examples that illustrate the breadth of possible applications of our framework. Textbook treatments of other nonlinear models and their use in economics are available in \citet[][Part IV]{Cameron-Trivedi2005_book} and \citet[][Part IV]{Wooldridge2010_book}.

\begin{example}[Least Squares Regression]
	\label{eg:least squares}
    Setting $\eta(\theta)=\theta$ and $\rho(y; \eta) = (y-\eta)^2$ recovers semiparametric least squares regression for partially linear models.
\end{example}

\begin{example}[Logistic Regression]
    \label{eg:logit}
    Assume that the binary outcome $y_i$, conditional on $x_i$ and $\bw_i$, is distributed Bernoulli with probability  $\eta(\mu(x_i)+\bw_i'\bgamma)$, where $\eta(\theta) = (1 + \exp(-\theta))^{-1}$, then $\rho(y;\eta) = -y \log(\eta) - (1-y)\log(1-\eta)$. 
\end{example}

\begin{example}[Quantile Regression]
	\label{eg:quantile}
    Set $\rho(y;\eta) = [\tau - \I(y < \eta)](y - \eta)$ with $\eta(\theta)=\theta$ for a user-specified quantile $\tau \in (0,1)$.
\end{example}

The key statistical challenge is (uniform in $x$) recovery of the function $\mu_0(x)$. Once accomplished, we can cover a wide variety of objects derived from \eqref{eq:model}. For concreteness we will focus on the following three objects, as they are of primary practical importance:
\begin{enumerate}[label=(\roman*)]
    \item the level of the regression function, $\vartheta_0(x,\evalw) = \eta(\mu_0(x)+\evalw'\bgamma_0) = \eta(\theta_0(x,\evalw))$,
    \item the nonparametric component itself (or its derivative), $\mu_0^{(v)}(x) = \frac{d^v}{d x^v} \mu_0(x)$, $v\geq 0$, and
    \item the marginal effect $\zeta_0(x,\evalw)=\frac{\partial}{\partial x}\eta(\mu_0(x)+\evalw'\bgamma_0) = \eta^{(1)}(\theta_0(x,\evalw))\mu_0^{(1)}(x)$,
\end{enumerate}
where $h^{(1)}(u) = \frac{\mathrm{d}}{\mathrm{d}u} h(u)$ denotes the derivative of a function with respect to its scalar argument and $\evalw$ is a user-chosen evaluation point for the additional controls. 

Most commonly, $\evalw$ is chosen as a summary statistic, such as the mean or the median or other quantile, or a pre-specified value such as the baseline category for discrete variables or fixed effects. Category-specific values or user-specified evaluation profiles may be useful for comparison across groups. We require that $\evalw$ be consistently estimable. The role of $\evalw$ in both plotting and inference introduces some important nuances that are discussed below.

Each of these parameters corresponds to different questions. The level, $\vartheta_0(x,\evalw)$, is directly useful for visualization of the relationship of $y_i$ to $x_i$ and is commonly used in causal inference. If the variable $x_i$ is a continuous treatment, our results yield a nonparametric estimate of the dose response function, while controlling for relevant factors $\bw_i$. A plot of $\vartheta_0(x,\evalw)$ shows this function for the subgroup defined by $\bw_i = \evalw$. We can also obtain separate dose response functions for different subgroups of the data to be used in multi-sample comparisons. On the other hand, if $x_i$ is a pre-treatment variable, the same multi-sample results provide an analysis of treatment effect heterogeneity for discrete (often binary) treatments, and our uniform inference allows for discovery of treatment effect heterogeneity. Finally, $\vartheta_0(x,\evalw)$ in the quantile regression case can be used to assess the spread of the conditional distribution (such as the interquartile range or quantiles close to zero and one) or robust measures of central tendency, as would be done with a classical scatter plot.

The nonparametric component is most often studied to assess functional form, generally against a parsimonious parametric specification to be considered for later analyses or for a shape restriction that is of substantive interest, such as monotonicity or convexity. Historically, a common use of binscatter was to visually (and informally) assess if $\mu_0(x)$ is well-approximated by a linear model, and if so, proceed under that specification. Our results provide rigor to such practice, and expand the idea to a much richer class of models and hypotheses. 

The marginal (or partial) effect $\zeta_0(x,\evalw)$ is a standard object in economic analysis of  nonlinear models. In binary choice models it is common to study how the probability of $y=1$ changes as a function of $x$. The marginal effect at the average, obtained by setting $\evalw$ to the sample mean, is a standard way to summarize nonlinear models by giving the effect for the ``average'' individual. For example, we show that the marginal effect of income changes sign in our application, indicating a changing response in the uninsuredness rate as a result of Medicaid. Comparing marginal effects across groups for heterogeneity analysis, in causal or noncausal settings, is a common goal in social science applications with nonlinear models. Note that we obtain the marginal effect at the average, but not the average marginal effect. This distinction is not present for linear binscatters \citep{Cattaneo-Crump-Farrell-Feng_2024_AER}, but for nonlinear models, these differ in both implementation and interpretation (see \citet[][\S 5.2.4]{Cameron-Trivedi2005_book}). It is conceivable that our results can be used as a first step, to obtain $\zeta_0(x,\evalw=\bw_i)$, prior to integrating over the distribution of $\bw_i$ for the average marginal effect.

\subsection{Estimation}
    \label{sec:estimation}

Given an i.i.d.\! sample $(y_i, x_i, \bw_i')'$, $i=1, \ldots, n$, the binscatter estimator is constructed by solving the empirical analogue of \eqref{eq:model} using a partitioning-based approximation to the unknown function $\mu_0(x)$. This nonparametric approximation requires two choices: the partitioning of the support of $x_i$ and the estimation within each bin. 

To fix ideas, it is useful to begin with the simplest case where local constant fitting is used and $\bw_i$ is absent. First, the support of $x_i$ is divided into $J < n$ disjoint bins. $J$ is the main tuning parameter for this nonparametric estimation problem, and its choice is crucial both visually and statistically. To describe the estimator, we will take $J < n$ as given at present, and return to its choice in Section \ref{sec:choice of J} below.

Coupled with a choice of $J$ is a method to divide the support. A theoretical innovation of our work is that the bin breakpoints themselves can be data-dependent, provided the $y_i$'s are not used. This is distinct from a data-driven choice of $J$. The partition is denoted by $\widehat{\Delta} = \{\widehat{\mathcal{B}}_1, \widehat{\mathcal{B}}_2, \dots, \widehat{\mathcal{B}}_J\}$, with the bins and breakpoints denoted by
\[\widehat{\mathcal{B}}_j =
  \begin{cases}
    \big[\widehat{\tau}_{j-1}, \widehat{\tau}_{j}\big)  & \qquad \text{if } j=1,2,\dots,J-1,\\
    \big[\widehat{\tau}_{J-1}, \widehat{\tau}_{J}\big]  & \qquad \text{if } j=J,\\
  \end{cases}
\]
and the union of the bins equal to the support of $x_i$.
The breakpoints $\{\widehat{\tau}_{j}\}_{j=1}^J$ result from a user-chosen, possibly data-driven, partitioning method. Our theoretical results cover any random partition that is independent of the outcomes $y_i$'s (given $x_i$'s and $\bw_i$'s), and  ``quasi-uniform'', which intuitively requires the bins to be sufficiently similar. The formal condition is stated in the next section. The simplest approach is evenly-spaced breakpoint locations ($\widehat{\tau}_{j}=x_\mathtt{min} + j(x_\mathtt{max}-x_\mathtt{min})/J$). The most popular choice, however, is to use the empirical quantiles of $x_i$, setting $\widehat{\tau}_{j}=\widehat{F}^{-1}(j/J)$ with $\widehat{F}(u)=\frac{1}{n}\sum_{i=1}^n \I(x_i \leq u)$ the standard empirical cumulative distribution function, and $\widehat{F}^{-1}$ its generalized inverse. In the SA we show that both of these satisfy our generic assumptions. Other possible methods include certain adaptive regression trees and related partitioning methods, such as those with the so-called ``X-property'' or via sample splitting; see \cite{Devroye-etal2013_book}, \cite{zhang2010recursive}, and references therein. For concreteness, we will use quantile spacing for empirical analysis throughout the paper.

Given a partition $\widehat{\Delta}$, the binscatter estimate is formed by fitting the sample analogue of \eqref{eq:model} within each bin, using only an intercept. In the simple case of least squares regression, this is identical to computing the sample average of $y_i$ for observations in each bin, exactly as \cite{Tukey1961_Berkeley} described, yielding a piecewise constant approximation to the unknown conditional expectation. The same method is followed for all other models. For example, in the case of binary data or fractional outcomes, a logistic regression of $y_i$ on a constant is fit for each bin. For the median, or any other quantile, one simply computes the empirical quantile of $y_i$ using observations only within the bin.

\begin{figure}[!ht]
    \caption{\small{\bf Illustration of Nonlinear Binned Scatter Plots.} This figure illustrates the construction of a nonlinear binned scatter plot using the ACS data. All analyses are performed at the zip code tabulation area level for the United States (excluding Puerto Rico). The dependent variable is the percentage of individuals without health insurance and the independent variable of interest is per capita income.}
    \label{fig:intro}
    \begin{subfigure}[t]{0.5\columnwidth}
        \centering	
        \caption{\it Raw Scatter Plot}
        \includegraphics[trim={0cm 0cm 0cm 1.5cm}, clip, scale=.45]{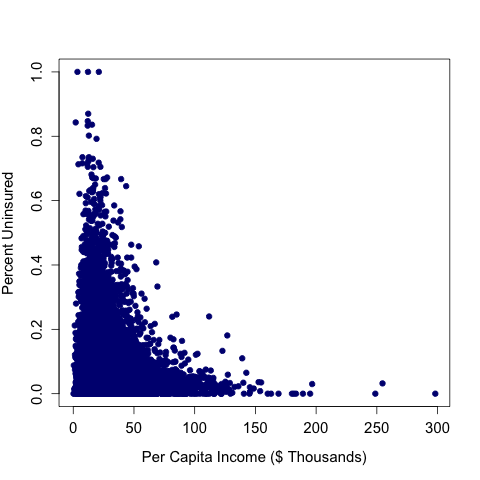}
    \end{subfigure}\hfill
    \begin{subfigure}[t]{0.5\columnwidth}
        \centering	
        \caption{\it Bin Cutoffs and Binscatter}
        \includegraphics[trim={0cm 0cm 0cm 1.5cm}, clip, scale=.45]{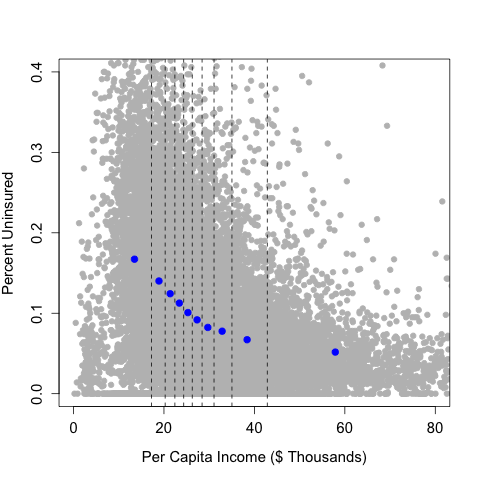}
    \end{subfigure}\\[.5 ex]
    \begin{subfigure}[t]{0.5\columnwidth}
        \centering	
        \caption{\it Ad Hoc Binscatter Plot}
        \includegraphics[trim={0cm 0cm 0cm 1.5cm}, clip, scale=.45]{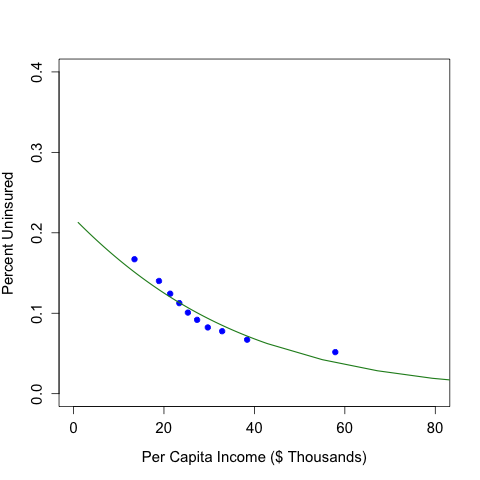}
    \end{subfigure}\hfill
    \begin{subfigure}[t]{0.5\columnwidth}
        \centering	
        \caption{\it IMSE-Optimal Binscatter}
        \includegraphics[trim={0cm 0cm 0cm 1.5cm}, clip, scale=.45]{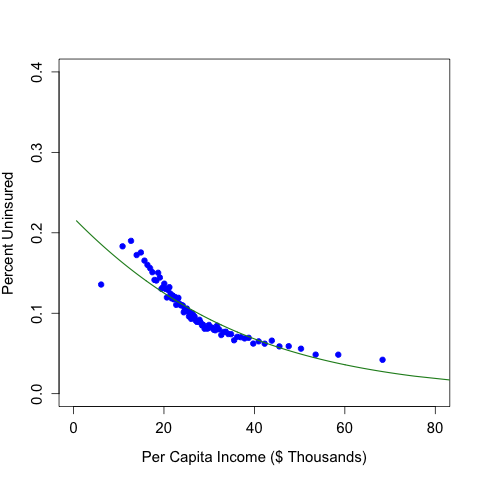}
    \end{subfigure}
\end{figure}

Formally, we define the basis functions $\widehat{\bb}_0(x) = [\I_{\widehat{\mathcal{B}}_1}(x),\I_{\widehat{\mathcal{B}}_2}(x),\cdots,\I_{\widehat{\mathcal{B}}_J}(x)]'$, consisting of indicators for each bin. We then obtain
\begin{equation}
    \label{eq:simple binscatter}
	\widehat{\mu}(x) = \widehat{\bb}_0(x)'\widehat{\bbeta}, \qquad
    \widehat{\bbeta} = \argmin_{\bbeta\in\mathbb{R}^J} \sum_{i=1}^{n} \rho\Big(y_i; \;\eta\big(\widehat{\bb}_0(x_i)'\bbeta\big)\Big).
\end{equation}
To help fix ideas, Figure \ref{fig:intro} illustrates the basics of how a binscatter is constructed. The data are obtained from the ACS using the 5-year survey estimates beginning in 2013 and ending in 2017 (available from the Census Bureau website). All analyses are performed at the zip code tabulation area level for the United States (excluding Puerto Rico). The dependent variable, $y_i$, is the percentage of individuals without health insurance, and the independent variable of interest, $x_i$, is per capita income. The fractional nature of the outcome motivates the use of logistic quasi-maximum likelihood for estimation and inference \citep{Papke-Wooldridge1996_JAE}.\footnote{
For piecewise constant modeling of the conditional mean without controls, the fitted values from QMLE approaches will be identical to those from least squares, as both return the sample means in each bin. Since valid inference (\S \ref{sec:inference} ahead) requires bias correction, the equivalence is for fitted values only. Further, this equivalence does not apply to marginal effects, quantiles, or other estimands, nor to when other basis functions are used, nor to when controls are added. The Appendix provides further discussion and examples of the benefits of nonlinear modeling, including showing that in these data, linear binscatter plots may lie outside $y_i \in [0,1]$.\label{foot:linear}
}
Figure \ref{fig:intro}(a) shows the classical scatter plot of the raw data. This data set has about 32,000 observations, far from the millions commonly encountered, and already this plot fails to be useful for assessing the functional form: the visualization is dominated by a dense cloud of data with a few outlying observations. Figure \ref{fig:intro}(b) shows a binned scatter plot being constructed, with the raw data in the background. The dots are the fitted values of applying \eqref{eq:simple binscatter} following Example \ref{eg:logit}, i.e., we show  $\widehat{\vartheta}(x) =\eta(\widehat{\mu}(x))$. Figure \ref{fig:intro}(c) isolates the binscatter and overlays a linear fit (i.e., a global logistic quasi-likelihood with $\mu_0(x)$ assumed linear in $x$). The linear approximation to $\mu_0(x)$ appears satisfactory at first, but this is only because the nonparametric estimate is oversmoothed. We know this to be the case both \emph{statistically}, as 10 bins is far fewer than the IMSE-optimal value (81 in this case; see Section \ref{sec:choice of J}), and \emph{economically}, because the existence of Medicaid means fewer uninsured people at the lowest incomes. Figure \ref{fig:intro}(d) presents the estimate using the optimal number of bins and recovers the correct shape, which is missed in the informal analysis using an ad hoc choice of $J$. Figure \ref{fig:intro}(d) also shows the inadequacy of parametric modeling in this case, and the need for the flexibility provided by nonlinear binscatter. As a preview, Table \ref{table:testing} below shows that formal tests reject polynomial parametric specifications and reject the hypothesis that the uninsurance rate is monotonically decreasing with per capita income.

Graphs like Figures \ref{fig:intro}(c) and (d) have a long tradition in statistics and data science, and have recently become ubiquitous in applied microeconomics. Visually assessing functional forms is the typical use. Importantly, in this case the visualization shows an estimate of $\vartheta_0(x,\evalw)$, not $\mu_0(x)$ directly. Further, although the binned scatter plot invites the viewer to ``connect the dots'' smoothly, the actual estimator is piecewise constant, which generally gives a less appealing visualization but underpins any formal analysis.

We expand on \eqref{eq:simple binscatter} in two ways: adding the covariates $\bw_i$ and enriching the set of allowable basis functions. The covariates can be directly incorporated into the loss, exactly as they are in \eqref{eq:model}. Moreover, the additively separable and linear nature of the controls makes this generalization straightforward empirically. Importantly, the presence of controls invalidates bin-by-bin estimation, as the coefficients $\bgamma_0$ are global parameters. The SA discusses different approaches to estimating $\mu_0$ and $\bgamma_0$.

Next, we replace the piecewise constant approximation based on  $\widehat{\bb}_0(x)$ with an order-$p$ polynomial approximation in each bin that is $(s-1)$-times continuously differentiable at the breakpoints of the bins, with the convention that $s=0$ corresponds to discontinuous fits and $s=1$ is continuous but nondifferentiable. This change to the basis gives additional flexibility that is crucial for bias reduction and derivative estimation, the latter being instrumental for studying shape restrictions and specification testing. 
By construction, $p \geq s \geq 0$, and derivative estimation requires that the derivative of interest, $v \geq 0$, is no larger than $p$. The general basis is defined as $\widehat{\bb}_{p,s}(x):=\widehat{\mathbf{T}}_s[\widehat{\bb}_0(x)\otimes (1,x,\dots,x^p)']$, where $\otimes$ denotes the Kronecker product and $\widehat{\mathbf{T}}_s$ is a $[(p+1)J-(J-1)s]\times (p+1)J$ transformation matrix ensuring that the $(s-1)$-th derivative of the estimate is continuous ($\widehat{\mathbf{T}}_s$  depends on the random partition and its exact form is given in Section SA-5.2 of \cite{Cattaneo-Crump-Farrell-Feng_2024_AER}).


The class of bases we consider includes both piecewise polynomials as well as standard splines. We draw the distinction only to highlight that we do not necessarily enforce smoothness at the breakpoints. Textbook treatments often separate the two with the same distinction \citep{Gyorfi-etal_2002_book}. But all the bases are locally-supported and partition-based, and thus similar, compared to, for example, global polynomials, which are ruled out. When $s=0$, $\widehat{\mathbf{T}}_s$ is the identity matrix, and the fit is a piecewise polynomial of degree $p$ (i.e.\! a separate fit within each bin). The standard binscatter uses piecewise constant fit, shown as lines or as dots, corresponding to $s=p=0$. Another popular choice are cubic B-splines, obtained by setting $s=p=3$. On account of its popularity and to simplify notation, we will assume throughout the paper that $s=p$ and use the notation $\widehat{\bb}_p(x) := \widehat{\bb}_{p,p}(x)$. The SA treats the general case of $s \leq p$. Finally, for technical reasons, $\widehat{\bb}_p(x)$ and associated coefficients are understood to be represented in the standardized coordinates defined in SA-2. This is an invertible reparameterization of the basis defined above and leaves fitted functions and variance formulas unchanged, and thus we omit the details here.

The generalized, covariate-adjusted binscatter can now be defined. We first solve\footnote{It is possible to add nonstochastic, per-observation weights to the loss function \eqref{eq:binscatter-QMLE}, as might be natural in aggregate data such as our running example. Under appropriate conditions on these weights, our theory will continue to hold. We do not make this explicit to save notation, but our software implementation allows certain types of weights, depending on the platform's native implementation. See \url{https://github.com/nppackages/binsreg} for platform-specific documentation.}
\begin{equation}
    \label{eq:binscatter-QMLE}	\begin{bmatrix}\;\widehat{\bbeta}\;\\\;\widehat{\bgamma}\;\end{bmatrix}
        = \argmin_{\bbeta,\bgamma} \sum_{i=1}^{n} \rho\Big(y_i; \;\eta\big(\widehat{\bb}_p(x_i)'\bbeta+\bw_{i}'\bgamma\big)\Big).
\end{equation}
 Next, using \eqref{eq:binscatter-QMLE} the estimators of the three functions of interest are:
\begin{equation}
    \label{eq:estimator-level}
    \widehat{\vartheta}_p(x,\widehat\evalw) = \eta(\widehat{\theta}_p(x,\widehat\evalw)),
\end{equation}
\begin{equation}
    \label{eq:estimator-mu}
    \widehat{\mu}_p^{(v)}(x) = \widehat{\bb}_p^{(v)}(x)'\widehat{\bbeta},
\end{equation}
\begin{equation}
    \label{eq:estimators-marginal}
    \widehat{\zeta}_p(x,\widehat\evalw) = \eta^{(1)}(\widehat{\theta}_p(x,\widehat\evalw))\widehat{\mu}_p^{(1)}(x),
\end{equation}
respectively, where $\widehat{\theta}_p(x,\widehat\evalw) = \widehat{\mu}_p(x)+\widehat\evalw'\widehat{\bgamma}$ is the plug-in estimator of the true index $\theta_0(x,\evalw)$, and $\widehat\evalw$ (non-random or generated based on $\{\bw_i\}_{i=1}^n$) is a consistent estimator for the desired evaluation point $\evalw$. We will often make the polynomial order $p$ explicit, as this is needed for clarity when constructing confidence bands and hypothesis tests in Section \ref{sec:inference}; dependence on $\widehat{\Delta}$ and choice of $J$ is suppressed, but also important. 

Nonlinear binscatter methods can be constructed for both fixed $J < \infty$ and large $J \to \infty$ as $n\to\infty$, naturally leading to different interpretations. The (true, nonparametric) functions of interest defined at the beginning of this section cannot be recovered when $J$ is fixed, but coarsened versions thereof will be, and these objects can have an interesting interpretation: if the partition ``settles'' as $n\to\infty$ to some fixed $\Delta_0$ with associated fixed basis $\bb_p(x)$ (see Assumption \ref{asmpt:IMSE} below for a precise definition), then the probability limit of the fixed-$J$ binscatter is the solution to \eqref{eq:model} where the function class $\mathcal{M}$ is restricted to be $\mathcal{M}=\{ \mu(x)= \bb_p(x)'\bbeta : \bbeta\in\mathbb{R}^K, K=\dim(\bb_p(x))\}$. This is most natural with $p=0$ and quantile-spacing, because the binscatter shows average outcomes across quantiles of a continuous covariate. For $p=0$ and $J=100$, for example, the results allow for comparison of $y_i$ across percentiles of $x_i$ (possibly controlling for $\bw_i$), which is standard for $x_i$ variables such as test scores or measures of wealth. All our estimation and inference results remain valid when $J$ is fixed, provided the target parameter is adjusted accordingly and $d$ is treated as fixed (keeping focus on the relationship of $y_i$ and $x_i$ being the more difficult statistical problem). See \cite{Cattaneo-Crump-Farrell-Feng_2024_AER,Cattaneo-Crump-Farrell-Feng_2025_Stata} for further discussion of fixed-$J$ binscatter methods. Here we will consider only the case $J \to \infty$ as $n\to\infty$ to streamline the presentation and theory.

\section{Theory}
    \label{sec:theory}

This section presents two main novel technical results: a uniform Bahadur representation (Theorem \ref{thm:bahadur}) and a feasible strong approximation (Theorem \ref{thm:strong approximation}). The methodological results in subsequent sections---tuning parameter selection, confidence bands, and hypothesis testing---are built from these results. To conserve space and notation, in this section we only show the results for $\widehat{\vartheta}_p(x,\widehat\evalw)$. The analogous results for $\widehat{\mu}_p^{(v)}(x)$ and $\widehat{\zeta}_p(x,\widehat\evalw)$ are deferred to the SA (specific references below) and are conceptually similar. The SA also gives other important technical results and additional discussion of how our theory improves on the existing literature. Note that because the results in this section are technical in nature and serve as building blocks for the methodology, this section may be skipped by the applied-minded reader and we do not revisit the running empirical application here.

First we state the assumptions required on the data generating process and statistical model. These are mostly standard and not restrictive, but involve some notation and technicalities, particularly to allow for $d = \dim(\bw_i) \to \infty$. To reduce clutter the dependence on $n$ is left implicit (from e.g.\! $J$, $d$, $\evalw$, and so forth). Unless explicitly stated otherwise, all constants and all upper, lower, moment, and Lipschitz bounds in the assumptions below are uniform in $n$. The class of data generating processes is restricted by the following.

\begin{assumption}[Data Generating Process]\label{asmpt:dgp} \
\begin{enumerate}[label=(\roman*)]
    \item $\{(y_{i}, x_{i}, \bw_i'): 1\leq i\leq n\}$ are i.i.d. random vectors supported on $\mathcal{Y} \times \mathcal{X} \times \mathcal{W}$ and satisfy the correctly specified model \eqref{eq:model}, where $(\mu_0(\cdot), \bgamma_0)$ is the unique solution, and $\sup_n \|\bgamma_0\| < \infty$, where $\|\cdot\|$ is the Euclidean norm. $\mathcal{X}$ is a compact interval, $\mathcal{W}$ is a compact set, and $\sup_n\sup_{\bw\in\mathcal W}\|\bw\|_\infty<\infty$, where $\|\cdot\|_\infty$ is the sup norm.
    
    \item The marginal distribution of $x_{i}$ has a Lipschitz continuous (Lebesgue) density bounded away from zero on $\mathcal{X}$.
    
    \item If the conditional distribution of $y_i$ given $(x_i,\bw_i')$ has a (conditional) density with respect to Lebesgue measure, this density is uniformly bounded over its support and $\mathcal{X} \times \mathcal{W}$.
\end{enumerate}
\end{assumption}

This assumption restricts attention to cross-sectional data and bounded covariates with minimal and standard regularity imposed on the underlying joint distribution. Requiring $x_i$ to be continuously distributed is natural given the visualization and estimation goals, but our results can also be applied to discrete $x_i$ by taking each mass point as its own bin to conduct simultaneous estimation and inference over those support points.

The next assumption gives regularity conditions on the class of statistical models.

\begin{assumption}[Statistical Model]\label{asmpt:model} \
\begin{enumerate}[label=(\roman*)]
    \item For some $\delta>0$, there exists a compact interval $\mathcal I$ contained in the interior of the domain of $\eta(\cdot)$ such that
    $[\theta_0(x,\bw)-\delta,\theta_0(x,\bw)+\delta]\subseteq\mathcal I$
    uniformly over $n$, $x\in\mathcal X$, and $\bw\in\mathcal W$, where neither $\delta$ nor $\mathcal{I}$ depend on $n$. $\rho(y;\eta)$ is absolutely continuous with respect to $\eta$ on an open set containing $\eta(\mathcal I)$ and admits a derivative $\psi(y,\eta) := \psi^\dagger(y-\eta)\psi^\ddagger(\eta)$ almost everywhere on this set. $\psi^\ddagger(\cdot)$ is continuously differentiable and strictly positive or negative on this set. If the conditional distribution of $y_i$ given $(x_i,\bw_i')$ does not have a Lebesgue density, then $\psi^\dagger(\cdot)$ is Lipschitz continuous, otherwise it is piecewise Lipschitz with finitely many discontinuity points.

    \item $\rho(y;\eta(\theta))$ is convex with respect to $\theta$ and $\eta(\cdot)$ is strictly monotonic and thrice continuously differentiable.

    \item $\Psi(x,\bw;\eta):=\E[\psi(y_i,\eta)|x_i=x,\bw_i=\bw]$ is twice continuously differentiable in $\eta$ on $\eta(\mathcal I)$ with derivatives bounded uniformly in $n$ and $\mathcal{X} \times \mathcal{W} \times \eta(\mathcal I)$. $\Psi(x_i,\bw_i;\eta(\theta_0(x_i, \bw_i))) = 0$. For $\Upsilon(x, \bw) := \frac{\partial}{\partial \eta} \Psi(x,\bw;\eta)|_{\eta = \eta(\theta_0(x,\bw))}$, $\Upsilon(x, \bw)\eta^{(1)}(\theta_0(x,\bw))^2$ is bounded away from zero uniformly over $\mathcal{X} \times \mathcal{W}$ and $\E[\Upsilon(x_i, \bw_i)\eta^{(1)}(\theta_0(x_i,\bw_i))^2|x_i=x]$ is Lipschitz continuous on $\mathcal{X}$. $\E[|\psi(y_i,\eta(\theta_0(x_i, \bw_i)))|^\nu|x_i=x, \bw_i=\bw]$ is uniformly bounded over $\mathcal{X} \times \mathcal{W}$ for some $\nu>2$. 

    \item $\sigma^2(x,\bw):=\E[\psi(y_i,\eta(\theta_0(x_i, \bw_i)))^2|x_i=x,\bw_i=\bw]$ is bounded away from zero uniformly in $\mathcal{X} \times \mathcal{W}$. $\E[\eta^{(1)}(\theta_0(x_i, \bw_i))^2\sigma^2(x_i,\bw_i)|x_i=x]$ is Lipschitz continuous. 

    \item $\mu_0(\cdot)$ is $\varsigma$-times continuously differentiable for some $\varsigma\geq p+1$ with derivatives bounded uniformly in $n$ and $\mathcal{X}$.
\end{enumerate}
\end{assumption}

This assumption imposes regularity conditions on the statistical model in \eqref{eq:model}, particularly on the loss function and resulting parameters of interest. The complexity of part (i) reflects the breadth of the class of models and parameters we cover. When $y_i$ is continuous the loss function can have points of nondifferentiability, but for discrete outcomes the loss must be smoother. To illustrate, consider first Example \ref{eg:quantile} in Section \ref{sec:setup}: the loss function for quantile regression is continuous but not differentiable everywhere, which is covered by our assumptions with $\psi(y,\eta) = \I(y-\eta<0)-\tau$, where $\psi^\dagger(y-\eta)=\I(y-\eta<0)-\tau$ exhibits a discontinuity point, and $\psi^\ddagger(\eta)=1$ is smooth. 
Alternatively, for logistic regression (Example \ref{eg:logit}), $y_i\in\{0,1\}$ and we have $\psi(y,\eta) = -(y-\eta) [\eta(1-\eta)]^{-1}$, which exactly matches the required structure of $\psi^\dagger(y-\eta)\psi^\ddagger(\eta)$. Both functions are clearly as smooth as required and the definition of $\eta(\theta)$ ensures that $\psi^\ddagger > 0$. The rest of the assumption gives standard moment and boundedness conditions to ensure that the parameters and their estimators are well-defined; those regularity conditions are also satisfied in all examples of interest. 

Finally, the nonparametric object $\mu_0(\cdot)$ is assumed to be smooth, as is standard in the nonparametric inference literature. Our theory assumes sufficient smoothness based on a fixed $p$, stemming from the viewpoint that practitioners select a polynomial degree for their estimator and then conduct estimation and inference based on that choice. This is particularly true for binscatter applications, where piecewise constant estimators are de facto standard, but also gives rise to the ubiquity of (e.g.) cubic splines for regression or local linear methods in regression discontinuity designs. Taking $p$ as given, we assume sufficient regularity for feasible tuning parameter selection and inference ($\varsigma\geq p+1$). In contrast, if one views $\varsigma$ as known and fixed, $p$ can be chosen to be equal or larger than the smoothness assumed on $\mu_0(\cdot)$, which generally yields the fastest convergence rate by speeding up the bias rate to the smoothness-imposed limit, while only inflating the variance constant. The implausibility of knowing the exact smoothness aside, this comes at the expense of feasible tuning parameter selection. \cite{Calonico-Cattaneo-Farrell_2022_Bernoulli} give more discussion on assumed versus exploited smoothness and how this relates to inference in practice. In typical applied research, assumed smoothness is not binding and does not dictate the choice of estimator.

\subsection{Bahadur Representation}
    \label{sec:Bahadur}

Here we state our novel Bahadur representation. This result is essentially a stochastic linearization of the estimator uniformly in $x \in \mathcal{X}$. In addition to the assumptions stated above, we require the following high-level conditions on the estimation procedure. These conditions ensure that the partitioning scheme is sufficiently regular and that both the coefficients and the evaluation point for the control variables can be estimated sufficiently well. 

\begin{assumption}[High-Level Estimation Conditions]
    \label{asmpt:estimation} \
    \begin{enumerate}[label=(\roman*)]	
        \item The partition $\widehat\Delta$ is independent of $\{y_i\}_{i=1}^n$ given $\{x_i,\bw_i\}_{i=1}^n$ and, with probability approaching $1$, $\max_{1\leq j \leq J} |\widehat{\tau}_{j} - \widehat{\tau}_{j-1} | \leq C  \min_{1\leq j \leq J} |\widehat{\tau}_{j} - \widehat{\tau}_{j-1} |$, for an absolute constant $C>0$. 
        
		\item $\|\widehat{\bgamma}-\bgamma_0\|=O_\P(r_\gamma)$ for a deterministic sequence $r_\gamma$ satisfying $\sqrt{d}\,r_\gamma=o\big(\sqrt{J/n}+J^{-p-1}\big)$. Moreover, $\evalw=\evalw_n\in\mathcal W$, $\|\widehat\evalw-\evalw\|=o_\P(1)$ and $\P[\widehat\evalw\in\mathcal W]\to1$.
    \end{enumerate}
\end{assumption}

The requirement that the partition intervals are not too dissimilar in length is satisfied for evenly spaced partitioning, trivially, and is shown to hold for quantile spacing in the SA (Lemma SA-5.2). For other data-driven methods this condition must be checked. This assumed property of the random binning structure is often called quasi-uniformity \citep{Huang_2003_AoS,Cattaneo-Farrell-Feng_2020_AoS}, and is important for controlling the approximation bias and, when combined with the assumptions on the density of $x_i$, for ensuring that each bin contains sufficient data to control the variance. Much more discussion is available in the SA. 

Part (ii) requires that the coefficient vector $\bgamma_0$ can be estimated sufficiently accurately and the desired evaluation point of $\bw_i$ can be estimated consistently. The latter is not restrictive in general. Any summary statistic, such as the mean, minimum, maximum, or a specific quantile, is feasible for $\evalw$. Prespecified values, such as categories for discrete variables, require no estimation and trivially satisfy this condition. Combining the two, which might serve for category-specific binscatters, is also allowed. For estimation of $\bgamma_0$ a rate is required. For fixed (or very slowly-diverging) $d$ this is not generally restrictive, as the nonparametric estimation of $\mu_0(x)$ is the most statistically difficult part of the problem. However, when $d \to \infty$, estimation becomes more difficult for the coefficients and the stated convergence requirement interacts with choices of $J$ and $p$ and $\nu$ in Assumption \ref{asmpt:model}. We do not cover the case where $\bw_i$ contains individual-specific intercepts as fixed effects, such as in panel data, which would require $d$ proportional to $n$: in all cases $d = o(n)$. The most popular implementation of local constant estimation ($p=0$) with an IMSE-optimal choice of $J = J_{\mathtt{IMSE}} \asymp n^{1/3}$ (see Theorem \ref{thm:IMSE} below) is allowed if $\nu > 6$ and $d = o(n^{1/6} / \sqrt{\log n})$. For the remainder of the main text we maintain the high-level condition above for notational simplicity, as this covers all cases of fixed and diverging dimension. Section SA-3.9 gives complete details, including verification of the assumption under primitive conditions and discussion of different rate restrictions.

We can now give our first theoretical result: a uniform (in $x$) Bahadur representation for $\widehat{\vartheta}_p(x,\widehat\evalw)$, as defined in \eqref{eq:estimator-level}.
\begin{thm}[Bahadur Representation]
    \label{thm:bahadur}
		Let $\ell_n:=\log n+\sqrt{d\log n}$. Suppose that Assumptions \ref{asmpt:dgp}, \ref{asmpt:model}, and \ref{asmpt:estimation} hold, and
		\[
		    \frac{d+\log n}{J}
		    +\frac{J^{\frac{\nu}{\nu-2}}\log n}{n}
		    +\frac{J^{\frac{2\nu}{\nu-1}}(\log n)^{\frac{\nu}{\nu-1}}}{n}
		    =o(1),
		    \qquad
		    \frac{J\ell_n^2}{n^{1-1/\nu}}
		    \lesssim
		    \sqrt{\frac{J\log n}{n}}+J^{-p-1}.
		\]
		Then,
	    \[\sup_{x\in\mathcal{X}} \big| \widehat{\vartheta}_p(x,\widehat\evalw) - \vartheta_0(x,\evalw) - \widehat{\mathsf{L}}_p(x,\evalw)\big| \lesssim_\P r_n+\|\widehat\evalw-\evalw\|,\]
    where the infeasible Gram matrix $\bar{\bQ}_p := n^{-1}\sum_{i=1}^{n} \widehat{\bb}_{p}(x_i)\widehat{\bb}_{p}(x_i)'\Upsilon(x_i, \bw_i) \eta^{(1)}(\theta_0(x_i, \bw_i))^2$, 
    \begin{align*}
        \widehat{\mathsf{L}}_p(x,\evalw) := - \eta^{(1)}(\theta_0(x,\evalw)) \widehat{\bb}_{p}(x)'\bar{\bQ}_p^{-1} \frac{1}{n}\sum_{i=1}^n\widehat{\bb}_{p}(x_i)\eta^{(1)}(\theta_0(x_i, \bw_i))\psi(y_i, \eta(\theta_0(x_i,\bw_i))),
    \end{align*}
	    and
	    \[
	    r_n:=\left(\frac{J\log n}{n}\right)^{3/4}\ell_n
	    +J^{-\frac{p+1}{2}}\sqrt{\frac Jn}\,\ell_n
	    +\frac{J\ell_n^2}{n}
	    +J^{-p-1}
	    +\sqrt d\,r_\gamma.
	    \]
\end{thm}

This result is one fundamental building block for the strong approximation in the next section. Further, this directly yields important consequences including the mean squared error expansion used to choose $J$ and the asymptotic variance formula for inference. This will be key in our results and is discussed below. The analogous Bahadur representations for $\widehat{\mu}_p^{(v)}(x)$ and $\widehat{\zeta}_p(x,\widehat\evalw)$, under the same assumptions, are given in Theorem SA-3.1. The ``linear'' term is slightly different to account for the different structure of the three estimands and the remainder rate for derivative estimation is slower. 

Theorem \ref{thm:bahadur} improves on the econometric literature in several ways. Here we give a short summary, but further information is available in the SA (Remark SA-3.3), wherein we also discuss the contributions of several preliminary lemmas, which may be of separate interest in the series estimation literature. First, this result substantially generalizes the least squares results in \cite{Cattaneo-Crump-Farrell-Feng_2024_AER}, under essentially the same rate restrictions, with an error of approximation that is optimal up to $\log(n)$ terms.

Second, compared to other work in the series estimation literature, our results allow for random partitioning schemes and do not require that the partition be ``convergent'' in any sense, which allows for general partitioning schemes. It is not possible to obtain the uniform results of Theorem \ref{thm:bahadur} (or the strong approximations below) by expanding around a nonrandom limit. Thus $\widehat{\bb}_{p}^{(v)}(x)$ is left as random in the Bahadur representation, including in the matrix $\bar{\bQ}_p$, which are infeasible but not population objects.

Third, we require only weak rate restrictions. Specifically, when $d$ is fixed, we require $J^{\frac{8}{3}}/n=o(1)$ up to $\log n$ terms when $\nu\geq 4$ (see Assumption \ref{asmpt:model}), thus accommodating IMSE-optimal (Theorem \ref{thm:IMSE} ahead) piecewise constant binscatter estimators which are ruled out by prior work. In fact, for piecewise polynomials ($s=0$), we can show that the Bahadur representation still holds under $J/n=o(1)$ up to $\log n$ terms when a subexponential moment restriction holds for the ``score'' $\psi(y_i, \eta_i)$, which is analogous to the results of \citet{Kong-Linton-Xia_2010_ET} for kernel-based estimators. \citet{Belloni-Chernozhukov-Chetverikov-FernandezVal_2019_JoE} considers series-based quantile regression, and Theorem 2 and Corollary 2 therein can be used to establish a Bahadur representation and uniform convergence of the resulting estimators under (the substantially stronger) $J^4/n^{1-\varepsilon}=o(1)$ for some $\varepsilon>0$. However, \citet{Kong-Linton-Xia_2010_ET} handle weakly dependent data, which we do not, and \cite{Belloni-Chernozhukov-Chetverikov-FernandezVal_2019_JoE} obtain a result uniform in both $x\in\mathcal{X}$ and in the quantile index (over a compact set), whereas our result is for a specific quantile. Our theory could be extended in both directions.

We use several novel theoretical techniques to achieve these improvements. 
First, unlike \cite{Cattaneo-Crump-Farrell-Feng_2024_AER} where the least-squares estimators have explicit closed-form expressions, our proof derives the linear representation directly from the M-estimation loss and its scores, allowing for nonlinear links and possibly nonsmooth loss functions. Second, relative to prior general sieve theory, we exploit the locally supported, sparse binscatter basis and the exponential off-diagonal decay of the inverse Gram matrix to localize each coefficient perturbation used in the objective comparison to a small neighborhood; this accommodates random partitions and avoids the additional full-dimensional complexity penalty from treating all $J$ basis coefficients as generic regressors, thereby yielding weaker rate restrictions.

\subsection{Strong Approximation}
    \label{sec:strong}

With the Bahadur representation in place, we can develop tools for inference. Our main result is a strong approximation for the (Studentized) $t$-statistic process for each of the three estimators, allowing us to obtain a feasible asymptotic distributional approximation. Again we give the details only for $\widehat{\vartheta}_p(x,\widehat\evalw)$ and defer the others to the SA. 

The variance is an immediate consequence of the expansion in Theorem \ref{thm:bahadur}, and is made feasible by replacing unknown objects by their estimators. The form of the variance is reminiscent of its parametric counterpart (e.g., for generalized linear models), but estimation is more complicated. Herein we give a general high-level condition that the infeasible Gram matrix can be estimated sufficiently well, complementing the conditions of Assumption \ref{asmpt:estimation}, which can justify several alternatives commonly used in practice. These are discussed in Section SA-4.1. See Section SA-4.1 for examples of $\widehat{\Upsilon}(x_i, \bw_i)$ for different models.
\begin{assumption}[High-Level Variance Estimation Condition]
    \label{asmpt:Gram}
    For the infeasible Gram matrix $\bar{\bQ}_p$ defined in Theorem \ref{thm:bahadur}, there is an estimator $\widehat{\Upsilon}(x_i, \bw_i)$ such that 
    $\|\bar{\bQ}_p - \widehat{\bQ}_p \|_o = O_\P\big( J^{-p-1}+\big(\frac{J\log n}{n^{1-2/\nu}}\big)^{1/2}\big)$, where $\widehat{\bQ}_p := n^{-1}\sum_{i=1}^{n} \widehat{\bb}_{p}(x_i)\widehat{\bb}_{p}(x_i)'\widehat{\Upsilon}(x_i, \bw_i) \eta^{(1)}(\widehat{\theta}_p(x_i,\bw_i))^2$ and $\|\cdot \|_o$ is the operator norm.
\end{assumption}

With the Gram matrix estimator in place, we define the $t$-statistic as follows, for a fixed $p$ as a process over $x \in \mathcal{X}$.
\begin{equation}
    \label{eqn:t stat}
    T_{\vartheta,p}(x)=
    \frac{\widehat{\vartheta}_p(x,\widehat\evalw)-\vartheta_0(x,\evalw)}{\sqrt{\widehat{\Omega}_{\vartheta,p}(x)/n}}
    , \qquad
    \widehat\Omega_{\vartheta,p}(x) := \eta^{(1)}(\widehat{\theta}_p(x,\widehat\evalw))^2 \widehat{\bb}_{p}(x)'\widehat{\bQ}_p^{-1}\widehat{\bSigma}_p\widehat{\bQ}_p^{-1}\widehat{\bb}_{p}(x),
\end{equation}
where $\widehat{\bSigma}_p := n^{-1}\sum_{i=1}^n \widehat{\bb}_{p}(x_i)\widehat{\bb}_{p}(x_i)' \psi(y_i, \eta(\widehat{\theta}_p(x_i, \bw_i)) )^2\eta^{(1)}(\widehat{\theta}_p(x_i, \bw_i))^2$ and $\widehat{\bQ}_p$ is defined in Assumption \ref{asmpt:Gram}. The $t$-statistics $T_{\mu^{(v)},p}(x)$ and $T_{\zeta,p}(x)$, for $\mu^{(v)}_0(x)$ and $\zeta_0(x,\evalw)$ respectively, are entirely analogous, and are defined in Section SA-3.3.

Our inference results will follow from the next key theorem.
\begin{thm}[Strong Approximation]
    \label{thm:strong approximation}
		Suppose that the conditions of Theorem \ref{thm:bahadur} and Assumption \ref{asmpt:Gram} hold, and let $(a_n: n\geq 1)$ be a positive deterministic sequence bounded away from zero such that
	    \[
	        \frac{J(\log n)^2}{n^{1-\frac{2}{\nu}}}
	        +\Big(\frac{J(\log n)^3\ell_n^4}{n}\Big)^{1/2}
	        +nJ^{-2p-3}
	        +\frac{\ell_n^2}{J^{p+1}}
	        +\frac{nd r_\gamma^2}{J}
	        =o(a_n^{-2})  ,
	    \]
	    and $\|\widehat\evalw-\evalw\|=o_\P(a_n^{-1}\sqrt{J/n})$. Then, on a properly enriched probability space, there exists a $(J+p)$-dimensional standard Normal random vector $\bN$ such that for any $\xi>0$,
	\[
    	\P\Big(\sup_{x\in\mathcal{X}}|T_{\vartheta,p}(x) -
            \bar{Z}_{\vartheta,p}(x)|>\xi a_n^{-1}\Big)=o(1), \qquad
    	\bar{Z}_{\vartheta,p}(x)=\frac{\widehat{\bb}_{p}(x)' \eta^{(1)}(\theta_0(x,\evalw)) \bar{\bQ}_p^{-1} \bar{\bSigma}_p^{1/2}}{\sqrt{\bar\Omega_{\vartheta,p}(x)}}\bN,
	\]
    where $\bar{\bSigma}_p$ and $\bar\Omega_{\vartheta,p}(x)$ are shown in 
    Section SA-3. On a further enriched space, there exists a conformable standard Normal vector $\bN^\star$, independent of $\{(y_i, x_i, \bw_i')'\}_{i=1}^n$ and $\widehat{\Delta}$. In the following display, $\bar{Z}_{\vartheta,p}(\cdot)$ denotes a conditionally equal-in-distribution copy driven by the same $\bN^\star$. Then, for any $\xi>0$,
     \begin{align*}
	&\P\Big(\sup_{x\in\mathcal{X}}|\bar{Z}_{\vartheta,p}(x) -  \widehat{Z}_{\vartheta,p}(x) |>\xi a_n^{-1}\Big|\{(y_i, x_i, \bw_i')'\}_{i=1}^n,\widehat{\Delta}\Big)=o_\P(1),\\
    &\qquad
    \widehat{Z}_{\vartheta,p}(x) = \frac{\widehat{\bb}_{p}(x)' \eta^{(1)}(\widehat{\theta}_p(x,\widehat\evalw))  \widehat{\bQ}_p^{-1}\widehat{\bSigma}_p^{1/2}}{\sqrt{\widehat{\Omega}_{\vartheta,p}(x)}}\bN^\star.
\end{align*}
\end{thm}
The approximating process, $\bar{Z}_{\vartheta,p}(\cdot)$, is a Gaussian process conditional on $\{x_i,\bw_i\}_{i=1}^n$ and $\widehat{\Delta}$ by construction, and the elements of $\bar{\bSigma}_p$, $\bar\Omega_{\vartheta,p}(x)$ and $\widehat{\bb}_p(x)$ reflect this conditioning. This process is infeasible but the second result shows that all the unknown quantities in $\bar{Z}_{\vartheta,p}(\cdot)$ can be replaced by their sample analogues to obtain a feasible approximation. Theorems SA-3.5 and SA-3.6 give the corresponding results for $T_{\mu^{(v)},p}(x)$ and $T_{\zeta,p}(x)$, under the same assumptions. Pointwise inference results are also given in the SA for completeness.

Compared to the existing literature on uniform nonparametric, Theorem \ref{thm:strong approximation} gives several important improvements. It inherits all the novelty discussed at the close of Section \ref{sec:Bahadur} above, which is not duplicated here. The closest antecedents are \cite{Belloni-Chernozhukov-Chetverikov-FernandezVal_2019_JoE} and \cite{Cattaneo-Crump-Farrell-Feng_2024_AER}. Again, Theorem \ref{thm:strong approximation} substantially generalizes the least squares results in \cite{Cattaneo-Crump-Farrell-Feng_2024_AER}, under essentially the same rate restrictions, with an error of approximation that is optimal up to $\log(n)$ terms. 

\cite{Belloni-Chernozhukov-Chetverikov-FernandezVal_2019_JoE} considers strong approximations of more general series estimators, but restricts attention to quantile regression (though again, uniform in the quantiles). We focus on partitioning-based methods, but consider a broader class of models. We have weaker assumptions than would be found by applying their results to binscatter estimators. Their Theorem 11 can be applied to construct strong approximation of the $t$-statistic process based on pivotal coupling that achieves the approximation rate $a_n=n^{-\varepsilon'}$ under $J^4/n^{1-\varepsilon}=o(1)$ for some constants $\varepsilon, \varepsilon'>0$, whereas their Theorem 12 can be used to construct strong approximation based on Gaussian processes under $J^{5}/n^{1-\varepsilon}=o(1)$. In contrast, when $d$ is fixed, for $a_n=\sqrt{\log n}$ and $\nu\geq 4$, we require $J^{\frac{8}{3}}/n=o(1)$, up to $\log n$ terms. Prior work has obtained such sharp results, under weak conditions, only for the supremum of the process \citep{Chernozhukov-Chetverikov-Kato_2014a_AoS}, whereas our theory yields an approximation for the entire $t$-statistic process.
The coupling technique employed here is the same as in \cite{Cattaneo-Crump-Farrell-Feng_2024_AER}, once the leading linear term is obtained from our Bahadur representation. Specifically, conditional on the covariates and the random partition, and  after ordering observations by $x_i$, 
we couple scalar partial sums with Gaussian partial sums and exploit the local binscatter structure, rather than directly approximating an increasing-dimensional coefficient vector as in generic series methods. This approach accommodates random partitions and delivers weaker rate restrictions than prior general series methods.

\section{Tuning Parameter Selection}
    \label{sec:choice of J}

We can use our theoretical results from the previous section to directly inform implementation. Our first task is selecting the number of bins, $J$, which determines both the visual and statistical properties of the estimator. Consistent estimation and valid inference is possible for a range of diverging sequences of $J$, but this does not provide sufficiently precise guidance for implementation. Thus, our first methodological result is a Nagar-type integrated mean squared error expansion that enables an optimal choice of $J$ in empirical applications. 

We need one further assumption to characterize the leading terms of the expansion. Intuitively, we require that the random partition $\widehat\Delta$ ``converges'' to a fixed one which obeys the same restrictions as in Assumption \ref{asmpt:estimation}. This assumption is not needed for any other results.
\begin{assumption}
    \label{asmpt:IMSE}
    There exists a non-random partition 
        $\Delta_0=\{\mathcal{B}_1, \cdots, \mathcal{B}_J\}$ with $\mathcal{B}_j=[\tau_{j-1}, \tau_j)$ for $j\leq J-1$ and $\mathcal{B}_J=[\tau_{J-1}, \tau_J]$ such that
        $\max_{1\leq j \leq J} |\tau_{j} - \tau_{j-1} | \leq C  \min_{1\leq j \leq J} |\tau_{j} - \tau_{j-1} |$ for an absolute constant $C>0$ and $\max_{0\leq j\leq J}|\hat{\tau}_j-\tau_j| = o_\P( J^{-1})$. For $p>0$, also assume $\max_{2\leq j\leq J}\left|(\tau_j-\tau_{j-1})-(\tau_{j-1}-\tau_{j-2})\right|=o(J^{-1})$.
\end{assumption}
This condition trivially holds for well-behaved nonrandom partitions, but also holds for the leading case of quantile-spacing, since sample quantiles converge to population counterparts. In more general cases this condition could fail. However, all our other results remain valid, and furthermore, even if this condition fails a rule-of-thumb choice of $J$ is available, which has the optimal rate but suboptimal constants. See Section SA-4.2. 

Our IMSE result for $\widehat{\vartheta}_p(x,\widehat\evalw)$ is given by the following. The corresponding results for $\widehat{\mu}_p^{(v)}(x)$ and $\widehat{\zeta}_p(x,\widehat\evalw)$ are stated in Theorem SA-3.4.
\begin{thm}
    \label{thm:IMSE}
	    Set $\omega(x)$ to be a continuous weighting function over $\mathcal{X}$ bounded away from zero. Suppose that the conditions of Theorem \ref{thm:bahadur} and Assumption \ref{asmpt:IMSE} hold, and $\varsigma \geq (p+2)$. Assume $\|\widehat\evalw-\evalw\|=o_\P(\sqrt{J/n}+J^{-p-1})$, $J(\log n)^3\ell_n^4 = o(n)$, and $\ell_n^2 = o(J)$. Then
    \[
	   \int_{\mathcal{X}}\Big(\widehat{\vartheta}_p(x,\widehat\evalw) - \vartheta_0(x,\evalw) \Big)^2 \omega(x)dx
	=\mathtt{AISE}_{\vartheta}+o_\P\Big(\frac{J}{n}+J^{-2(p+1)}\Big)
	\]
 	where, for nonrandom terms $\mathscr{V}_n(p)$ and $\mathscr{B}_n(p)$ shown in 
    Theorem SA-3.4 that are bounded and nonzero in general, $\mathtt{AISE}_{\vartheta}$ obeys
	\[
       \E[\mathtt{AISE}_{\vartheta}|\{(x_i,\bw_i')'\}_{i=1}^n,\widehat\Delta]=
        \frac{J}{n}\mathscr{V}_n(p)+J^{-2(p+1)}\mathscr{B}_n(p)
        +o_\P\Big(\frac{J}{n}+J^{-2(p+1)}\Big.
    \]
\end{thm}
This result is stated in terms of $J$, as it is the tuning parameter, but the rates and constants depend on the fixed polynomial order $p$ (recall that $p=s$, see Section \ref{sec:estimation}). An $L_2$ convergence rate immediately follows from this result. An $L_\infty$ convergence rate is available in Corollary SA-3.1. This theorem, and the $L_2$ and $L_\infty$ rates, are new to the literature, even in the case of non-random partitioning and without covariate adjustment, for nonlinear series estimators and binscatter methods. The SA contains detailed discussions of our contributions.

The practical consequence of Theorem \ref{thm:IMSE} is that we can balance the (squared) bias and variance to obtain an IMSE-optimal choice of $J$, which is given by
\begin{equation}
    \label{eqn:J IMSE}
    J_{\mathtt{IMSE}}(p) := \bigg(\frac{2(p+1) \mathscr{B}_n(p)} {\mathscr{V}_n(p)}\bigg)^{\frac{1}{2p+3}}  n^{\frac{1}{2p+3}}.
\end{equation}
Implementing binscatter with this $J$ is optimal in the sense of providing the IMSE-optimal estimate of the unknown function $\vartheta_0(\cdot,\evalw)$. In the next section we discuss the use of $J_{\mathtt{IMSE}}(p)$ for inference, where, as is typical, a bias correction must be applied. A feasible version of $J_{\mathtt{IMSE}}(p)$ is described in Section SA-4 and implemented in the {\tt binsreg} package \citep{Cattaneo-Crump-Farrell-Feng_2025_Stata}. Section SA-4.3 discusses binned scatter plots with a fixed choice of $J$, which can be visually appealing and interpretable in some applications.

\begin{figure}[!ht]
    \caption{\small{\bf Conditional Quantiles.} This figure illustrates the use of quantile regression to visualize the spread in the ACS data (see Example \ref{eg:quantile}). As the link function is the identity, panel (a) shows estimates of $\vartheta_0(x,\evalw)$ for $\tau = 0.1, 0.5, 0.9$, while panel (b) shows the same quantiles including additional covariates controlling for demographics listed in footnote \ref{foot:controls}.}
    \label{fig:quantile}
    \begin{subfigure}[t]{0.5\columnwidth}
    \centering	
    \caption{\it Quantile Regression w/o Controls}
    \includegraphics[trim={0cm 0cm 0cm 1.5cm}, clip, scale=.45]{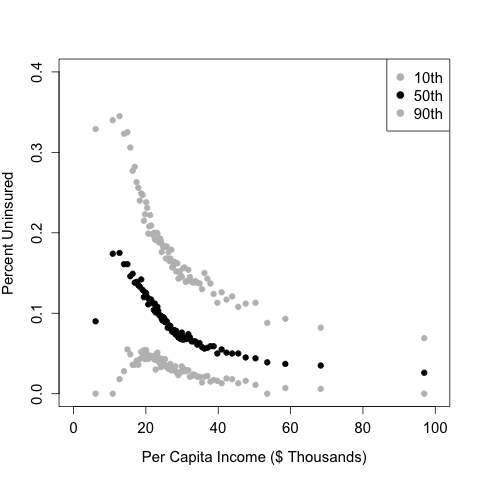}
    \end{subfigure}\hfill
    \begin{subfigure}[t]{0.5\columnwidth}
    \centering	
    \caption{\it Quantile Regression w/ Controls}
    \includegraphics[trim={0cm 0cm 0cm 1.5cm}, clip, scale=.45]{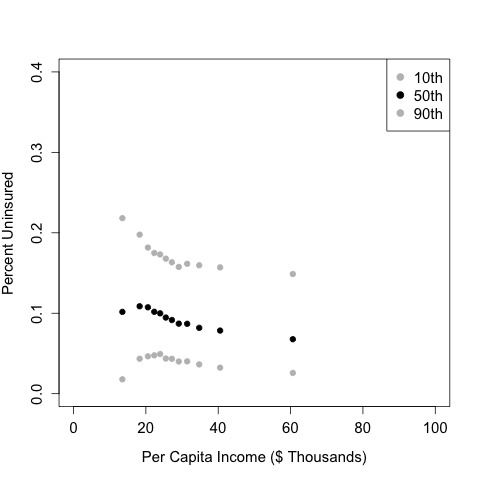}
    \end{subfigure}
\end{figure}

Figure \ref{fig:quantile} demonstrates the use of the optimal $J$ for quantile estimation (see Figure \ref{fig:intro}(d) for the mean). Figure \ref{fig:quantile}(a) shows the quantiles of uninsuredness conditional on per capita income, while Figure \ref{fig:quantile}(a) adds demographic controls following the linear structure of Equation \eqref{eq:model}.\footnote{
Beginning with \cite{Hall-Mishkin1982_Ecma}, it has become commonplace in the literature on dynamics of income and consumption to (linearly) residualize the outcome of interest with respect to a set of demographic controls prior to analyzing nonlinear dynamics using quantile methods, to focus on the unpredictable aspects of the outcome. For a recent example see \cite{Arellano-Blundell-Bonhomme2017_Ecma}. The analogue in our setting would be to apply the simple binscatter of Equation \eqref{eq:simple binscatter}, with quantile loss, using residualized $\widehat{y}_i$ and $\widehat{x}_i$. While such an approach may be sensible in time series applications, in our cross-sectional context this practice is unlikely to be appropriate and is difficult to interpret in general. See \cite{Cattaneo-Crump-Farrell-Feng_2024_AER} for discussion and examples. The partially linear structure of $\theta_0(x_i, \bw_i) = \mu_0(x_i)+\bw_i'\bgamma_0$ naturally has the interpretation of ``controlling for $\bw_i$'' which obviates any need for preprocessing of the data in our setting.
}
Quantile regression can be used to visualize the spread of the conditional distribution. Observe that Figure \ref{fig:quantile} restores the visualization of the dispersion in the data that is present in Figure \ref{fig:intro}(a) but hidden by the averaging in Figure \ref{fig:intro}(c). This (conditional) dispersion is distinct from uncertainty around the mean estimate, shown in Figure \ref{fig:bands} ahead and further in the Appendix. Figure \ref{fig:quantile}(a) shows that there is much larger spread in the fraction uninsured in lower income areas, but Figure \ref{fig:quantile}(b) shows that this gap narrows when controlling for demographics ($\widehat\evalw$ is set to the sample mean).\footnote{
    We add nine controls: 
    percentage of residents with a high school degree, 
    percentage of residents with a bachelor’s degree, 
    median resident age, 
    the local unemployment rate,
    mean household size,
    percentage of residents without internet access,
    percentage of residents in the armed forces,
    percentage of residents speaking only English,
    and 
    percentage of residents aged 65 and older.\label{foot:controls}
}
Although our empirical exercise primarily serves to illustrate our new methodology and visualization, uncovering this spread at low incomes may have policy implications that warrant further study. Although Medicaid is designed to reach low-income individuals, the spread identified by the quantile curves in Figure \ref{fig:quantile} shows that there is substantial inequality that may be important for policy makers. For example, this could arise from barriers to access that differ across regions (states administer separate Medicaid programs) or across demographics, or inequality in the provision of healthcare facilities. These inequalities could further be crucial in designing targeted subsidies to better approach universal coverage \citep{Finkelstein-Hendren-Shepard2019_AER}. See \cite{Koenker2017_ARE} for a comprehensive discussion of quantile models and their empirical relevance.

\section{Uniform Inference}
    \label{sec:inference}

We now turn to uniform inference for the three functions defined in Section \ref{sec:setup}. Uniform inference is required to make statistical statements about the functions $\vartheta_0(x,\evalw)$, $\mu_0^{(v)}(x)$, and $\zeta_0(x,\evalw)$, rather than about their values at a specific point $x$. Pointwise inference methods (e.g. confidence intervals) will not suffice for our main applications of interest, including treatment effect heterogeneity and continuous treatment effects, as well as shape restrictions and functional forms. For completeness, pointwise inference results are given in the SA and implemented in the {\tt binsreg} package, but omitted here to save space.

A key element of our uniform inference results---from a practical point of view---is the pairing of a feasible tuning parameter choice with valid inference. To give the most accurate estimate, and therefore also visualization, of $\vartheta_0(x,\evalw)$ we prefer to use $J_{\mathtt{IMSE}}(p)$ of \eqref{eqn:J IMSE}. However, as is typical for nonparametric problems, the (I)MSE-optimal tuning parameter choice delivers invalid inference, as it fails to eliminate a first-order bias. We therefore use robust bias correction to ensure that $J_{\mathtt{IMSE}}(p)$ remains a valid choice across all uses, delivering optimal estimation and valid inference. With this eye toward practicality, we state the result below at the IMSE-optimal rate $J\asymp n^{1/(2p+3)}$, which includes $J_{\mathtt{IMSE}}(p)$ in the nondegenerate case under Assumption \ref{asmpt:IMSE}, while the SA gives the more general result under mild rate restrictions on $J$.

Bias correction involves estimating, and removing, the leading smoothing bias term, and is made ``robust'' by correcting the standard errors to account for the additional sampling variability that has been introduced. Robust bias correction has theoretically superior higher-order inference properties \citep{Calonico-Cattaneo-Farrell_2018_JASA,Calonico-Cattaneo-Farrell_2022_Bernoulli}, performs well in simulations, and has been empirically validated in specific contexts \citep{Hyytinen-Tukiainen-etal2018_QE,Magalhaes-Tukiainen-etal2025_PA}. Robust bias correction is operationalized in the present context by (i) selecting a degree $p$ and creating a partition $\widehat{\Delta}$ based on $J_{\mathtt{IMSE}}(p)$ to form the optimal point estimate of $\widehat{\vartheta}_p(x,\widehat\evalw)$ and then (ii) conducting inference using $T_{\vartheta,p+1}(x)$ (or its feasible analogue $\widehat{Z}_{\vartheta,p+1}(x)$), i.e. the statistic formed using a higher-degree polynomial but the partitioning scheme based on $p$ in (i): $\widehat{\Delta} = \widehat{\Delta}(J_{\mathtt{IMSE}}(p))$. Any higher-degree polynomial may be used, but $p+1$ is simple and robust. \cite{Cattaneo-Farrell-Feng_2020_AoS} give further discussion of robust bias correction in the context of partition regression, including alternative strategies.

\subsection{Confidence Bands}

Our first uniform inference result delivers confidence bands for the functions $\vartheta_0(x,\evalw)$, $\mu_0^{(v)}(x)$, and $\zeta_0(x,\evalw)$. Confidence bands are similar in spirit to the more familiar concept of confidence intervals, but instead cover the entire function (uniformly over $x \in \mathcal{X}$) with a prespecified probability. Confidence bands are the appropriate tool for visualizing the uncertainty around the estimated function. The size of the band also changes to reflect the presence of heteroskedasticity in the data. These  bands can be used directly to identify interesting or important features of the function, for example, regions where it is statistically indistinguishable from zero or from a constant function. Bands are also useful for assessing the functional form or shape, such as regions of linearity or monotonicity, and therefore visually complement the formal hypothesis tests we introduce below. 

The confidence bands are built from Theorem \ref{thm:strong approximation}, coupled with robust bias correction, as discussed above. Confidence bands are defined as the area between an upper and lower bounding function. Recall that we employ robust bias correction, so that $\widehat{\vartheta}_{p+1}(x,\widehat\evalw)$ is the bias-corrected version of $\widehat{\vartheta}_p(x,\widehat\evalw)$, and is thus the ``center'' of the confidence band, and using $\widehat{\Omega}_{\vartheta,p+1}(x)$ in the standard error accounts for the additional variability. The robust bias-corrected confidence band for $\vartheta_0(x,\evalw)$ is given by
\begin{equation}
    \label{eqn:cb}
    \widehat{I}_{\vartheta,p+1}(x)=\Big[\widehat{\vartheta}_{p+1}(x,\widehat\evalw) \pm  \cval_{\vartheta} \sqrt{\widehat{\Omega}_{\vartheta,p+1}(x)/n}\Big], 
\end{equation}
where the critical value is determined by
\begin{equation}
    \label{eqn:cval}
    \cval_{\vartheta} = \inf\left\{c\in\mathbb{R}_+:\P\left[ \left. \sup_{x\in\mathcal{X}}|\widehat{Z}_{\vartheta,p+1}(x)|\leq c \; \right| \{(y_i, x_i, \bw_i')'\}_{i=1}^n,\widehat{\Delta} \right]\geq 1-\alpha\right\}.
\end{equation}
The validity of this confidence band follows from Theorem \ref{thm:strong approximation}, and inherits the novelty discussed above, which allows us to approximate the distribution of the supremum of $T_{\vartheta,p+1}(x)$ by applying the same functional to $\widehat{Z}_{\vartheta,p+1}(x)$. This yields the following result. Here we assume directly that $J$ has the IMSE-optimal rate. Theorem SA-3.8 states a more general result, valid for a range of $J$, as well as inference results for $\mu_0^{(v)}(x)$ and $\zeta_0(x,\evalw)$.
\begin{thm}
    \label{thm:CB}
	    Set $J \asymp n^{1/(2p+3)}$. Suppose that the conditions of Theorem \ref{thm:strong approximation} hold, with $p+1$ in place of $p$ and $a_n=\sqrt{\log J}$. Then, for $\widehat{I}_{\vartheta,p+1}(x)$ defined in \eqref{eqn:cb},
    \[\P\Big[\vartheta_0(x,\evalw) \in \widehat{I}_{\vartheta,p+1}(x),
      \text{ for all }x\in\mathcal{X}\Big]=1-\alpha+o(1).
	\]
\end{thm}

This result establishes valid confidence bands for generalized, covariate-adjusted binscatters. We can use this result to visually assess uncertainty about the form and shape of the regression function. One can visually ``test'' hypotheses of interest, though formal testing (Section \ref{sec:testing}) is recommended. Plotting both the estimate $\widehat{\vartheta}_{p}(x,\widehat\evalw)$ and the band $\widehat{I}_{\vartheta,p+1}(x)$ is advisable in applications because doing so presents both the IMSE-optimal point estimate and a valid measure of uncertainty (and one that uses the same bins). 

In nonlinear models, particularly in social sciences, partial effects are often the preferred way of summarizing the relationship (causal or not) of $x_i$ to $y_i$, controlling for $\bw_i$. Here,  this corresponds to the estimate of $\zeta_0(x,\evalw)$ and its associated confidence band. When $x_i$ is a treatment variable, $\zeta_0(x,\evalw)$ captures the effect of increasing the treatment dosage, and the band can help identify regions of $\mathcal{X}$ with the largest effects, or any other noteworthy shape. 

\begin{figure}[!ht]
    \caption{\small{\bf Confidence Bands.} This figure illustrates confidence bands in a nonlinear binned scatter plot using the ACS data. Panel (a) shows the point estimate (dots) and robust bias corrected confidence band (shaded region) for the conditional mean function with no controls, i.e., $\vartheta_0(x) = \eta(\mu_0(x))$, while panel (b) shows the corresponding point estimate and confidence band for the marginal effect, $\zeta_0(x)$. Shaded regions denote 95\% confidence bands and are based on 50,000 random draws.}
    \label{fig:bands}
    \begin{subfigure}[t]{0.5\columnwidth}
    \centering	
    \caption{\it Conditional Mean}
    \includegraphics[trim={0cm 0cm 0cm 0cm}, clip, scale=.45]{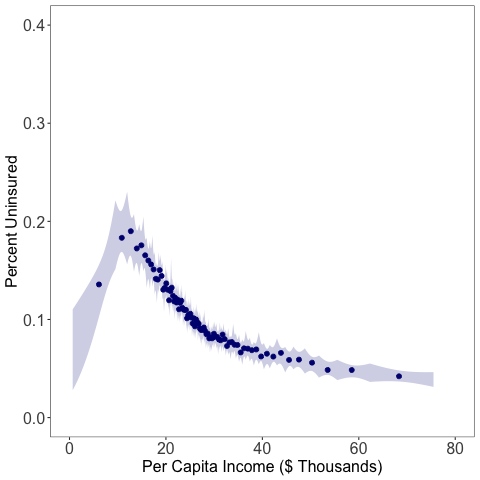}
    \end{subfigure}\hfill
    \begin{subfigure}[t]{0.5\columnwidth}
    \centering	
    \caption{\it Marginal Effect}
    \includegraphics[trim={0cm 0cm 0cm 0cm}, clip, scale=.45]{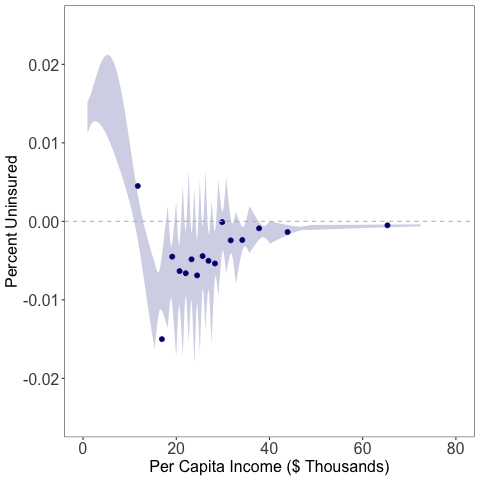}
    \end{subfigure}
\end{figure}

Figure \ref{fig:bands} shows examples of confidence bands using our running empirical application. The confidence band in Figure \ref{fig:bands}(a) displays the uncertainty surrounding the estimate first shown in Figure \ref{fig:intro}(d). The presence of the Medicaid program is clearly delineated by the shape of the band at lower income levels. From the band we can immediately conclude that the relationship is nonmonotonic. This is further emphasized in Figure \ref{fig:bands}(b), showing the marginal effect. Using the bands, we can reject the null hypothesis of monotonicity as the band lies completely on either side of zero at low and high income levels.

There are two features of our confidence bands that warrant mention. First, the user-selected point of evaluation $\evalw$ can impact the shape, placement, and size, of the band. One might expect that since the additional controls are modeled as additively linear, the evaluation point $\evalw$ (and the coefficient $\bgamma_0$) should not impact conclusions about the nonparametric relationship between $y$ and $x$. But this intuition overlooks the fact that the function $\mu_0(x)$ is only defined relative to how $\bw_i$ is coded. For example, if $\bw_i$ contains a binary variable indicating groups of different sizes, then the uncertainty will be different between the two groups. For $\vartheta_0(x,\evalw)$ and $\zeta_0(x,\evalw)$, the shape of the confidence bands may change with the choice of $\evalw$. This impacts all aspects of inference, both visual and the formal tests below. This is not particular to our method; it is always present in analyses of models like \eqref{eq:model}.

Second, the bias correction may result in the point estimate lying outside the confidence band. This is formally correct but can be visually unappealing. Figure \ref{fig:bands}(b) shows an example of this phenomenon. This occurs in areas of high bias, and may arise in any application of bias correction. Being based on a polynomial of order $p+1$, the confidence band removes more bias than does the point estimate. This is not necessarily a failing: the point estimate remains IMSE-optimal and inference remains valid, and, importantly, both are based on theory-founded partitioning choices. Adding bias correction to the point estimate itself would render it no longer optimal, and further, would require additional ad-hoc tuning parameter choices, potentially spoiling the formal guarantees of the current approach.

\subsection{Hypothesis Testing}
    \label{sec:testing}

We also provide formal hypothesis tests for substantive questions including functional form or shape restrictions for $\vartheta_0(x,\evalw)$, $\mu_0(x)$, and $\zeta_0(x,\evalw)$. Our discussion here is brief. Full details are given in the SA. 

A leading case is testing a parametric form for $\mu_0(x)$, where under the null there exists some finite-dimensional parameter $\btheta$ such that $\mu_0(x) = m(x;\btheta)$, uniformly in $x \in \mathcal{X}$ (we can also test any derivative of $\mu_0(x)$). The two-sided testing problem is
\begin{alignat*}{2}
    &\dot{\mathsf{H}}_0^{\mu}:\quad &&\sup_{x\in\mathcal{X}} \Big|\mu_0(x) - m(x;\btheta)\Big|=0, \quad \text{ for some } \btheta, \qquad vs.\\
    &\dot{\mathsf{H}}_\text{A}^{\mu}: &&\sup_{x\in\mathcal{X}} \Big|\mu_0(x) - m(x;\btheta)\Big|>0, \quad \text{ for all } \btheta.
\end{alignat*}
This test formalizes the notion of visual inspection in plots like Figure \ref{fig:intro}(c) and (d), beyond what is already done by adding a confidence band. We test this hypothesis using the statistic
\[
    \dot{T}_{\mu,p+1}(x)
    :=\frac{\widehat{\mu}_{p+1}(x)-m(x;\widetilde{\btheta})}
    {\sqrt{\widehat{\Omega}_{\mu,p+1}(x)/n}},
\]
where $\widetilde{\btheta}$ and $\widetilde{\bgamma}$ are estimators of $\btheta$ and $\bgamma_0$ under $\dot{\mathsf{H}}_0^{\mu}$, where $\theta_0(x, \evalw) = m(x;\btheta)+\evalw'\bgamma_0$. Conditions on the estimators are mild, and indeed, satisfied when obtaining $\widetilde{\btheta}$ and $\widetilde{\bgamma}$ by fitting the hypothesized parametric model. For example, in testing a parametric model for $\mu_0(x)$, $\widetilde{\btheta}$ must obey $\sup_{x\in\mathcal X}|\mu_0(x)-m(x;\widetilde{\btheta})| = o_\P(\sqrt{J / (n \log n)})$. Section SA-3.7 gives complete details. Theorem \ref{thm:strong approximation} again provides the tools to obtain the appropriate critical value. Theorem SA-3.9 gives the formal result showing size control and consistency of the test, as well as the corresponding tests for $\vartheta_0(x, \evalw)$ and $\zeta_0(x, \evalw)$, which require further, but similarly mild, regularity conditions. Last, we also provide for testing shape restrictions, see Theorem SA-3.10, which are conceptually similar but are generally one-sided testing problems. A leading example would be testing monotonicity of $\vartheta_0(x,\evalw)$ or $\zeta_0(x,\evalw)$.

Table \ref{table:testing} shows several testing examples using the $L_\infty$ norm. Consider first the left column of results. We test against the linear specification, formalizing the visual comparison in Figure \ref{fig:intro}(d). Here $\widetilde{\btheta}$ and $\widetilde{\bgamma}$ are obtained by fitting this linear index specification parametrically. We also test against a cubic (in $x$) logistic quasi-likelihood model for added flexibility, again obtaining $\widetilde{\btheta}$ and $\widetilde{\bgamma}$ from this parametric model. Both parametric specifications are rejected, and moreover, also rejected when including other controls (listed in footnote \ref{foot:controls}). This highlights the need for nonparametric modeling in this application. Finally we test the substantive null hypothesis that the uninsuredness rate is monotonically decreasing with income. This null is also strongly rejected due to the existence of Medicaid. This motivates the right column of results, where we repeat the analysis after restricting the sample to zip codes with per capita income above 138\% of the 2013--2017 average federal poverty line for a single-person household (\$16,248). This is the cutoff for expanded Medicaid eligibility based only on income. When we restrict to this sample, which diminishes the influence of the Medicaid program, we fail to reject the null hypothesis of a monotonic decline, but still reject the parametric specifications. This aligns with the need for flexible estimation and matches the conclusions we draw from the shape of the confidence bands shown in Figure \ref{fig:bands}.


\begin{table}[ht]
	\renewcommand{\arraystretch}{1.12}
	\begin{center}
		\caption{{\bf Specification and Shape Testing} 
		\label{table:testing}}
		\vspace{-.1in}
    \resizebox{\textwidth}{!}{
    \begin{tabular}{lrrrcrrr}
        \hline\hline\\[-2.25ex]
        & \multicolumn{3}{c}{\bfseries Full Sample} & &\multicolumn{3}{c}{\bfseries Above Income Cutoff}  \\
        \cline{2-4} \cline{6-8}\\[-2.25ex]
        & Test Stat. & $p$-value & {$\hat{J}_{\texttt{IMSE}}$} & & Test Stat. & $p$-value & {$\hat{J}_{\texttt{IMSE}}$} \\
        \hline
        {\bfseries Test of Linear Fit}\\
        ~~No Covariates & 22.319 & 0.000 & 81 & & 21.344 & 0.000 & 40 \\
        ~~Covariates, $\widehat\evalw = \bar{\bw}$ & 7.176 & 0.000 & 12 & & 4.265 & 0.000 & 8 \\
        \hline
        {\bfseries Test of Cubic Fit}\\
        ~~No Covariates & 9.961 & 0.000 & 81 & & 75.426 & 0.000 & 40 \\
        ~~Covariates, $\widehat\evalw = \bar{\bw}$ & 9.014 & 0.000 & 12 & & 3.506 & 0.001 & 8 \\
        \hline
        {\bfseries Test of Monotonic Decline}\\
        ~~No Covariates & 7.155 & 0.000 & 16 & & 1.095 & 0.944 & 10 \\
        ~~Covariates, $\widehat\evalw = \bar{\bw}$ & 5.929 & 0.000 & 11 & & 0.803 & 0.973 & 9 \\
        \hline\hline
    \end{tabular}
    }
    \end{center}
	\footnotesize\textit{Notes}. This table reports the test statistics and associated p-values along with the IMSE-optimal choice of $J$ from hypothesis tests of parametric specifications and shape restrictions using the ACS data. The first and second panels report test results for the null hypotheses of linear and cubic (in $x$) logistic quasi-likelihood models, respectively, while the third panel reports results for the null hypothesis of a monotonic decline in the level (i.e., negative derivative). All tests are performed with and without control variables (controls are listed in footnote \ref{foot:controls}). The left panel (``Full Sample'') reports results for the full sample whereas the right panel (``Above Income Cutoff'') restricts to the sample of zip codes with per capita income above \$16,248. All $p$-values are based on 50,000 random draws.
\end{table}

\subsection{Multi-Sample Comparisons}
    \label{sec:multi}

Our results extend to comparisons between different samples, or groups, within the data. This is a common goal in program evaluation and causal inference settings. With discrete (e.g., binary) treatments, the groups are defined by treatment arms and the differences define heterogeneous (in $x_i$) effects. In the continuous case, the grouping is the dimension of heterogeneity and $x_i$ is the treatment. Our results extend naturally to this setting. For a grouping indicator $t_i = 0, \ldots, L$, we replace the scalar index in the model \eqref{eq:model} with $\theta_0(x_i, \bw_i, t_i) := \sum_{t=0}^L \mathbbm{1}\{t_i = t\}  \theta_{0,t}(x_i, \bw_i)$, where each $\theta_{0,t}(x_i, \bw_i) = \mu_{0,t}(x_i)+\bw_i'\bgamma_{0,t}$. The level and marginal effect can then be defined groupwise, as $\vartheta_{0,t}(x,\evalw) = \eta(\theta_{0,t}(x,\evalw))$ and $\zeta_{0,t}(x,\evalw) = \eta^{(1)}(\theta_{0,t}(x,\evalw))\mu_{0,t}^{(1)}(x)$ for some evaluation point $\evalw$ of control variables.

For example, in a randomized experiment $\vartheta_{0,1}(x,\evalw) - \vartheta_{0,0}(x,\evalw)$ is the conditional average treatment effect (CATE) function, and the binscatter naturally captures treatment effect heterogeneity along the $x_i$ dimension holding fixed $\bw_i = \evalw$. The rate of change in this heterogeneity is $\zeta_{0,1}(x,\evalw) - \zeta_{0,0}(x,\evalw)$. Our methods can be used to formally test the null hypothesis that $\vartheta_{0,1}(x,\evalw) = \vartheta_{0,0}(x,\evalw)$ for all $x\in\mathcal{X}$, which captures the idea of no (heterogeneous) treatment effect. As a second example, our theory can be used to quantify uncertainty for the largest heterogeneous treatment effect: $\widehat{x}^\star = \underset{x\in\mathcal{X}}{\arg\sup}\; \big|\widehat\vartheta_{0,1}(x,\evalw) - \widehat\vartheta_{0,0}(x,\evalw)\big|$.
These and many other problems of interest in applied microeconometrics concern the uniform discrepancy of two or more binscatter function estimators, which can be analyzed using our strong approximation and related theoretical results in the supplemental appendix. We do not provide further details here to conserve space, but our software implements several multi-sample estimation, uncertainty quantification, and hypothesis testing procedures.

\begin{figure}[!ht]
    \caption{\small{\bf Two Sample Comparison.} This figure uses the same ACS data to compare areas in low density states (blue) and high density states (orange). Low density states are defined as those with average population per square mile below 100. Panels (a) and (b) show the point estimate (squares or dots) and robust bias corrected confidence band (shaded region) for each group, first without control variables and then with controls added (listed in footnote \ref{foot:controls}). Panels (c) and (d) show the estimated difference (evaluated using the binning of the low density states) and the associated confidence bands. Shaded regions denote 95\% confidence bands and are based on 50,000 random draws.}
    \label{fig:multi-sample}
    \begin{subfigure}[t]{0.5\columnwidth}
        \centering	
        \caption{\it Conditional means w/o controls}
        \includegraphics[trim={0cm 0cm 0cm 0cm}, clip, scale=.45]{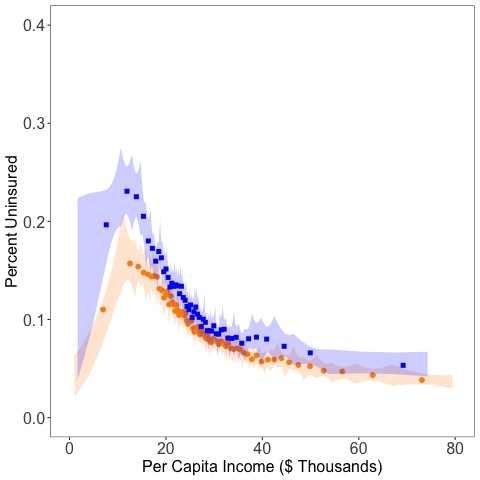}
    \end{subfigure}\hfill
    \begin{subfigure}[t]{0.5\columnwidth}
        \centering	
        \caption{\it Conditional means w/ controls}
        \includegraphics[trim={0cm 0cm 0cm 0cm}, clip, scale=.45]{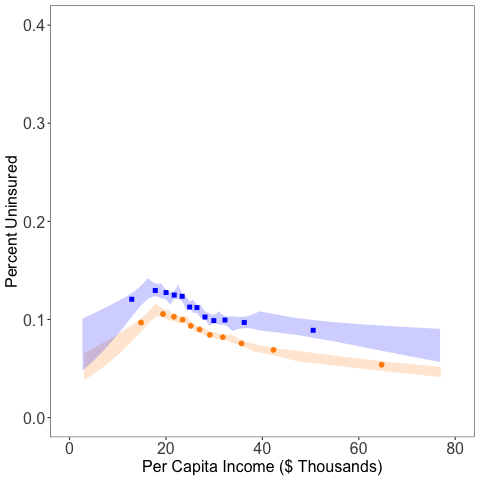}
    \end{subfigure}\\[.5 ex]
    \begin{subfigure}[t]{0.5\columnwidth}
        \centering	
        \caption{\it Difference of means w/o controls}
        \includegraphics[trim={0cm 0cm 0cm 0cm}, clip, scale=.45]{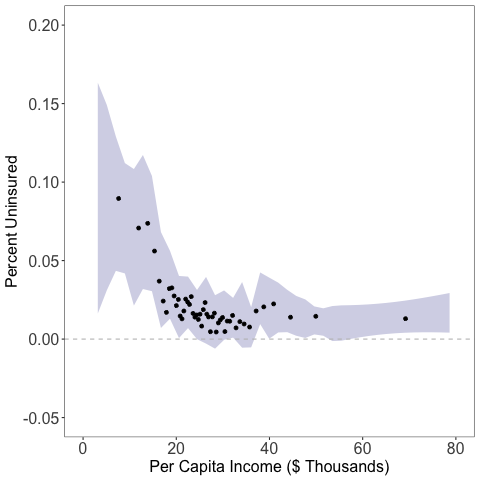}
    \end{subfigure}\hfill
    \begin{subfigure}[t]{0.5\columnwidth}
        \centering	
        \caption{\it Difference of means w/ controls}
        \includegraphics[trim={0cm 0cm 0cm 0cm}, clip, scale=.45]{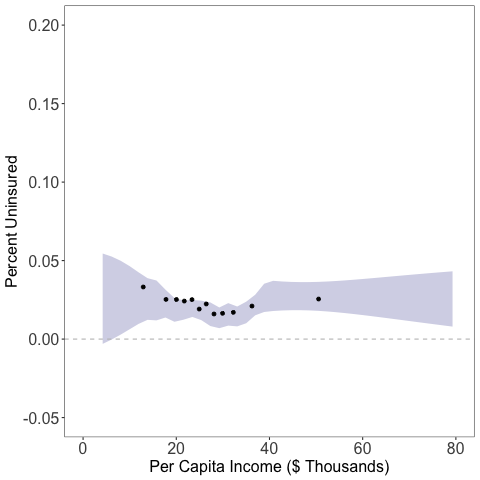}
    \end{subfigure}\\[.5 ex]
\end{figure}

Figure \ref{fig:multi-sample} shows an example of this type of analysis. We divide states into two groups based on their population density, with low and high density states as those with population densities below or above 100 people per square mile, respectively. Density is defined as the average population per square mile, and the data is available from the Census Bureau. Panel (a) shows $\widehat{\vartheta}_{t}(x)$ for each group, i.e. without controls, while panel (b) adds controls and shows $\widehat{\vartheta}_{t}(x,\widehat\evalw)$, with $\widehat\evalw$ set to the sample mean. Note that bin locations differ between the groups because the bias-variance tradeoff is data-dependent. The point estimates show higher uninsured rates in zip codes in low population density states as compared to high density states. Without controls, there is generally overlap in the confidence bands except for very low incomes. In contrast, when covariates are added, there is a much clearer delineation between the two groups at all but the lowest of income levels. This is made clear in panels (c) and (d), which plot the point estimate of the difference (the CATE) and the associated confidence band, enforcing common binning between the groups. The null hypothesis that $\vartheta_{0,1}(x,\evalw) = \vartheta_{0,0}(x,\evalw)$ for all $x\in\mathcal{X}$ is rejected in both cases, with test statistics of 7.719 and 8.936, respectively, and negligible p-values. Multi-sample comparisons share the same sensitivity to the chosen evaluation point as discussed above. These issues are unavoidable; researchers must be mindful when implementing the tests and interpreting the results. 


\section{Conclusion} 
    \label{sec:conclusion}

With the rise of large data sets, new visualization tools, such as binned scatter plots, have emerged and gained in popularity. This paper has thoroughly studied binned scatter plots in nonlinear, nonsmooth regression models. Our main contributions are to propose novel nonlinear binscatter methods, together with IMSE-optimal tuning parameter selection and uniform inference methods, including valid confidence bands and functional testing. Our companion \texttt{binsreg} software package makes these tools available for applications. 

One avenue for future work would be to generalize the analysis beyond a scalar covariate of interest. For example, in two dimensions such an approach would produce ``heat maps'' which are the bivariate extension of binned scatter plots. Extending our results to that case would be a valuable addition to the practitioner's toolkit.

\section*{References} 
\begingroup
\renewcommand{\section}[2]{}	
\bibliography{CCFF_2026_RESTAT--bib.bib}

@article{Koenker2017_ARE,
  title={Quantile regression: 40 years on},
  author={Koenker, Roger},
  journal={Annual review of economics},
  volume={9},
  pages={155--176},
  year={2017},
  publisher={Annual Reviews}
}

@article{Shapiro2020_MS,
  title={Advertising in health insurance markets},
  author={Shapiro, Bradley T},
  journal={Marketing Science},
  volume={39},
  number={3},
  pages={587--611},
  year={2020},
  publisher={INFORMS}
}

@article{Gathergood-Olafsson2024_RFS,
  title={The coholding puzzle: New evidence from transaction-level data},
  author={Gathergood, John and Olafsson, Arna},
  journal={The Review of Financial Studies},
  year={in press},
  publisher={Oxford University Press}
}

@article{Schlenker-Roberts2009_PNAS,
  title={Nonlinear temperature effects indicate severe damages to US crop yields under climate change},
  author={Schlenker, Wolfram and Roberts, Michael J},
  journal={Proceedings of the National Academy of sciences},
  volume={106},
  number={37},
  pages={15594--15598},
  year={2009},
  publisher={National Acad Sciences}
}

@article{Su2022_AEJPolicy,
  title={The rising value of time and the origin of urban gentrification},
  author={Su, Yichen},
  journal={American Economic Journal: Economic Policy},
  volume={14},
  number={1},
  pages={402--439},
  year={2022},
  publisher={American Economic Association 2014 Broadway, Suite 305, Nashville, TN 37203-2425}
}

@article{Agarwal-etal2019_AER,
  title={Market failure in kidney exchange},
  author={Agarwal, Nikhil and Ashlagi, Itai and Azevedo, Eduardo and Featherstone, Clayton R and Karaduman, {\"O}mer},
  journal={American Economic Review},
  volume={109},
  number={11},
  pages={4026--4070},
  year={2019},
}

@article{Magalhaes-Tukiainen-etal2025_PA,
	title={When Can We Trust Regression Discontinuity Design Estimates from Close Elections? Evidence from Experimental Benchmarks},
	author={De Magalh{\~a}es, Leandro and Hangartner, Dominik and Hirvonen, Salomo and Meril{\"a}inen, Jaakko and Ruiz, Nelson and Tukiainen, Janne},
	journal={Political Analysis},
	volume={33},
	number={3},
	pages={258--265},
	year={2025},
	publisher={Cambridge University Press}
}

@article{Hall-Mishkin1982_Ecma,
	author = {Robert E. Hall and Frederic S. Mishkin},
	title = {The Sensitivity of Consumption to Transitory Income: Estimates from Panel Data on Households},
	journal = {Econometrica},
	volume = {50},
	number = {2},
	pages = {461-482},
	year = {1982}
}

@article{Finkelstein-Hendren-Shepard2019_AER,
    Author = {Finkelstein, Amy and Hendren, Nathaniel and Shepard, Mark},
    Title = {Subsidizing Health Insurance for Low-Income Adults: Evidence from Massachusetts},
    Journal = {American Economic Review},
    Volume = {109},
    Number = {4},
    Year = {2019},
    Month = {April},
    Pages = {1530–67}
}

@article{Arellano-Blundell-Bonhomme2017_Ecma,
	title={Earnings and consumption dynamics: a nonlinear panel data framework},
	author={Arellano, Manuel and Blundell, Richard and Bonhomme, Stephane},
	journal={Econometrica},
	volume={85},
	number={3},
	pages={693--734},
	year={2017},
	publisher={Wiley Online Library}
}

@Book{Bhatia_2013_Book,
	Title                    = {Matrix Analysis},
	Author                   = {Bhatia, Rajendra},
	Publisher                = {Springer},
	Year                     = {2013}
}

@article{Belloni-Chernozhukov-Chetverikov-Kato_2015_JoE,
  title={Some New Asymptotic Theory for Least Squares Series: Pointwise and Uniform Results},
  author={Belloni, Alexandre and Chernozhukov, Victor and Chetverikov, Denis and Kato, Kengo},
  journal={Journal of Econometrics},
  volume={186},
  number={2},
  pages={345--366},
  year={2015}
}

@Article{Belloni-Chernozhukov-Chetverikov-FernandezVal_2019_JoE,
	title={Conditional Quantile Processes based on Series or Many Regressors},
	author={Belloni, Alexandre and Chernozhukov, Victor and Chetverikov, Denis and Fernandez-Val, Ivan},
	journal={Journal of Econometrics}, volume={213}, number={1}, pages={4--29}, year={2019}
}

@Article{Calonico-Cattaneo-Farrell_2018_JASA,
	Author = {Calonico, Sebastian and Matias D. Cattaneo and Max H. Farrell},
	Title = {On the Effect of Bias Estimation on Coverage Accuracy in Nonparametric Inference},
	Journal = {Journal of the American Statistical Association}, Number = {522}, Pages = {767-779}, Volume = {113}, Year = {2018}
}

@Article{Calonico-Cattaneo-Farrell_2022_Bernoulli,
	Author = {Calonico, Sebastian and Matias D. Cattaneo and Max H. Farrell},
	Title = {Coverage Error Optimal Confidence Intervals for Local Polynomial Regression},
	Journal = {Bernoulli}, Volume = {28}, Number = {4}, Pages = {2998--3022}, Year = {2022}
}

@book{Cameron-Trivedi2005_book,
  title={Microeconometrics: methods and applications},
  author={Cameron, A Colin and Trivedi, Pravin K},
  year={2005},
  publisher={Cambridge university press}
}

@article{Cattaneo-Crump-Farrell-Feng_2024_AER,
	author  = {Cattaneo, Matias D. and Richard K. Crump and Max H. Farrell and Yingjie Feng},
	title   = {On Binscatter},
	journal = {American Economic Review}, volume={114}, number={5}, pages={1488-1514}, year={2024}
}

@Article{Cattaneo-Crump-Farrell-Feng_2025_Stata,
	author  = {Matias D. Cattaneo and Richard K. Crump and Max H. Farrell and Yingjie Feng},
	title   = {Binscatter Regressions},
	journal = {Stata Journal},
    volume={25}, number={1}, pages={3--50}, year={2025}
}

@article{Cattaneo-Farrell-Feng_2020_AoS,
  title={Large Sample Properties of Partitioning-Based Series Estimators},
  author={Cattaneo, Matias D and Farrell, Max H and Feng, Yingjie},
  journal={Annals of Statistics},
  volume={48},
  number={3},
  pages={1718--1741},
  year={2020},
  publisher={Institute of Mathematical Statistics}
}

@book{Devroye-etal2013_book,
  title={A Probabilistic Theory of Pattern Recognition},
  author={Devroye, Luc and Gy{\"o}rfi, L{\'a}szl{\'o} and Lugosi, G{\'a}bor},
  volume={31},
  year={2013},
  publisher={Springer Science \& Business Media}
}

@Book{Gyorfi-etal_2002_book,
	Title                    = {A Distribution-Free Theory of Nonparametric Regression},
	Author                   = {L{\'a}szl{\'o} Gy{\"o}rfi and Michael Kohler and Adam Krzy{\.z}ak and Harro Walk},
	Publisher                = {Springer-Verlag},
	Year                     = {2002}
}

@book{Healy2018_book,
  title={Data Visualization: A Practical Introduction},
  author={Healy, Kieran},
  year={2018},
  publisher={Princeton University Press}
}

@Article{Huang_2003_AoS,
	Title                    = {Local Asymptotics for Polynomial Spline Regression},
	Author                   = {Huang, Jianhua Z},
	Journal                  = {Annals of Statistics},
	Number                   = {5},
	Pages                    = {1600--1635},
	Volume                   = {31},
	Year                     = {2003}
}

@article{Hyytinen-Tukiainen-etal2018_QE,
  title={When Does Regression Discontinuity Design Work? Evidence from Random Election Outcomes},
  author={Hyytinen, Ari and Meril{\"a}inen, Jaakko and Saarimaa, Tuukka and Toivanen, Otto and Tukiainen, Janne},
  journal={Quantitative Economics},
  volume={9},
  number={2},
  pages={1019--1051},
  year={2018}
}

@book{Koenker_2005_book,
  title={Quantile Regression},
  author={Koenker, Roger},
  volume={38},
  year={2005},
  publisher={Cambridge University Press}
}

@Article{Kong-Linton-Xia_2010_ET,
	Author={Kong, Efang and Linton, Oliver and Xia, Yingcun},
	Title={Uniform Bahadur Representation for Local Polynomial Estimates of M-Regression and Its Application to the Additive Model},
	Journal={Econometric Theory}, Volume={26}, Number={5}, Pages={1529--1564}, Year={2010}
}

@article{Papke-Wooldridge1996_JAE,
  title={Econometric Methods for Fractional Response Variables with an Application to 401(k) Plan Participation Rates},
  author={Papke, Leslie E and Wooldridge, Jeffrey M},
  journal={Journal of Applied Econometrics},
  volume={11},
  number={6},
  pages={619--632},
  year={1996}
}

@Article{Starr-Goldfarb_2020_SMJ,
	Title                    = {Binned Scatterplots: A Simple Tool to Make Research Easier and Better},
	Author                   = {Starr, Evan and Goldfarb, Brent},
	Journal                  = {Strategic Management Journal},
	Number                   = {12},
	Pages                    = {2261--2274},
	Volume                   = {41},
	Year                     = {2020}
}

@book{Schwabish2021_book,
  title={Better Data Visualizations: A Guide for Scholars, Researchers, and Wonks},
  author={Schwabish, Jonathan},
  year={2021},
  publisher={Columbia University Press}
}

@InProceedings{Tukey1961_Berkeley,
  Title                    = {Curves As Parameters, and Touch Estimation},
  Author                   = {John W. Tukey},
  Booktitle                = {Fourth Berkeley Symposium on Mathematical Statistics and Probability},
  Year                     = {1961},
  Editor                   = {Jerzy Neyman},
  Pages                    = {681-694},
  Volume                   = {1}
}

@Book{vandevarrt-Wellner_1996_book,
	title={Weak Convergence and Empirical Processes: With Application to Statistics},
	author={van der {V}aart, A.W. and Wellner, Jon},
	year={1996},
	publisher={Springer}
}

@book{zhang2010recursive,
  title={Recursive Partitioning and Applications},
  author={Zhang, Heping and Singer, Burton H},
  year={2010},
  publisher={Springer Science \& Business Media}
}

@article{Cattaneo-Farrell_2013_JoE,
  title={Optimal Convergence Rates, Bahadur Representation, and Asymptotic Normality of Partitioning Estimators},
  author={Cattaneo, Matias D and Farrell, Max H},
  journal={Journal of Econometrics},
  volume={174},
  number={2},
  pages={127--143},
  year={2013},
  publisher={Elsevier}
}

@article{Cattaneo-Feng-Underwood_2024_JASA,
	author  = {Cattaneo, Matias D. and Yingjie Feng and William G. Underwood},
	title   = {Uniform Inference for Kernel Density Estimators with Dyadic Data},
	journal = {Journal of the American Statistical Association}, volume={119}, number={524}, pages={2695--2708}, year={2024},
}

@Article{Chernozhukov-Chetverikov-Kato_2014a_AoS,
	Title                    = {Gaussian Approximation of Suprema of Empirical Processes},
	Author                   = {Victor Chernozhukov and Denis Chetverikov and Kengo Kato},
	Journal                  = {Annals of Statistics},
	Number                   = {4},
	Pages                    = {1564--1597},
	Volume                   = {42},
	Year                     = {2014}
}

@Article{Chernozhukov-Chetverikov-Kato_2014b_AoS,
	Title                    = {Anti-Concentration and Honest Adaptive Confidence Bands},
	Author                   = {Victor Chernozhukov and Denis Chetverikov and Kengo Kato},
	Journal                  = {Annals of Statistics},
	Number                   = {5},
	Pages                    = {1787--1818},
	Volume                   = {42},
	Year                     = {2014}
}

@article{Chernozhukov-Imbens-Newey_2007_JoE,
	Title={Instrumental Variable Estimation of Nonseparable Models},
	Author={Chernozhukov, Victor and Imbens, Guido W and Newey, Whitney K},
	Journal={Journal of Econometrics},
	Volume={139},
	Number={1},
	Pages={4--14},
	Year={2007},
	Publisher={Elsevier}
}

@article{Demko_1977_SIAM,
	title={Inverses of Band Matrices and Local Convergence of Spline Projections},
	author={Demko, Stephen},
	journal={SIAM Journal on Numerical Analysis},
	volume={14},
	number={4},
	pages={616--619},
	year={1977},
	publisher={SIAM}
}

@Book{Gine-Nickl_2016_book,
	title={Mathematical Foundations of Infinite-Dimensional Statistical Models},
	author={Gin{\'e}, Evarist and Nickl, Richard},
	volume={40},
	year={2016},
	publisher={Cambridge University Press}
}

@Article{Sakhanenko_1991_SAM,
	title={On the Accuracy of Normal Approximation in the Invariance Principle},
	author={Sakhanenko, AI},
	journal={Siberian Advances in Mathematics},
	volume={1},
	pages={58--91},
	year={1991}
}

@Book{Schumaker_2007_book,
	title={Spline Functions: Basic Theory},
	author={Schumaker, Larry},
	year={2007},
	publisher={Cambridge University Press}
}

@Book{Wooldridge2010_book,
  Title                    = {Econometric Analysis of Cross Section and Panel Data},
  Author                   = {Jeffrey M. Wooldridge},
  Publisher                = {MIT Press},
  Year                     = {2010},
  Address                  = {Cambridge},
  Edition                  = {2}
}
\bibliographystyle{jasa}
\endgroup

\begin{appendix}
\numberwithin{figure}{section}

\section{Appendix: Additional Empirical Illustrations}

Here we show three additional empirical analyses designed to illustrate the added value of nonlinear and/or nonsmooth modeling in applied contexts. We are grateful to the editor for suggesting these analyses. The discussion and figures below build on several issues raised in the main paper. We stress that these analyses are designed to illustrate methodological issues and not to find novel empirical conclusions.

\textbf{Robustness.} The goal is to estimate a measure of central tendency, similar to Figure \ref{fig:intro}. Control variables are not used. A linear binscatter (i.e., least squares loss estimation of the conditional mean) may be suitable (footnote \ref{foot:linear}) but outliers or corrupted data can render it unreliable. This is common in practice and is of particular concern in large data sets. To demonstrate the value of our results in this context, we intentionally corrupt the original data in a controlled way to show how the median binscatter is more robust than least squares, using the unaltered data as a benchmark. Our empirical application is well-suited for this because economic facts and regularities dictate the shape of the conditional mean: the existence of Medicaid implies lower uninsuredness for the poorest areas and cost of insurance implies a decreasing function above certain income levels (see Table \ref{table:testing}). 

\begin{figure}[!ht]
    \caption{\small{\bf Robustness to Corrupted Data or Outliers.} This figure illustrates the robustness of median regression compared to least squares estimation as a measure of centrality. Panels (a) and (b) show, respectively, corruption of data at low and high income levels. In both cases, we corrupt 0.1\% of the data into upper outliers.}
    \label{appxfig:robust}
    \begin{subfigure}[t]{0.5\columnwidth}
    \centering	
    \caption{\it Corruption of Low-Income Data}
    \includegraphics[trim={0cm 0cm 0cm 1.5cm}, clip, scale=.45]{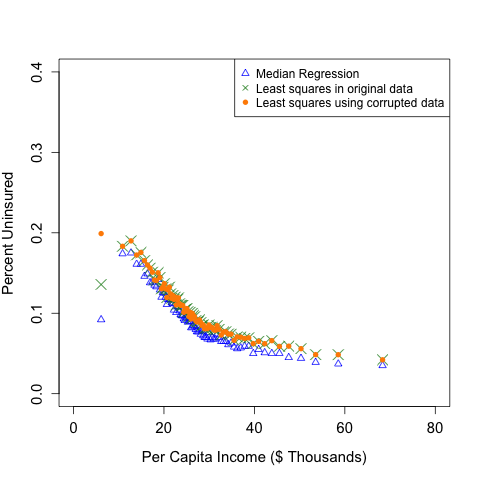}
    \end{subfigure}\hfill
    \begin{subfigure}[t]{0.5\columnwidth}
    \centering	
    \caption{\it Corruption of High-Income Data}
    \includegraphics[trim={0cm 0cm 0cm 1.5cm}, clip, scale=.45]{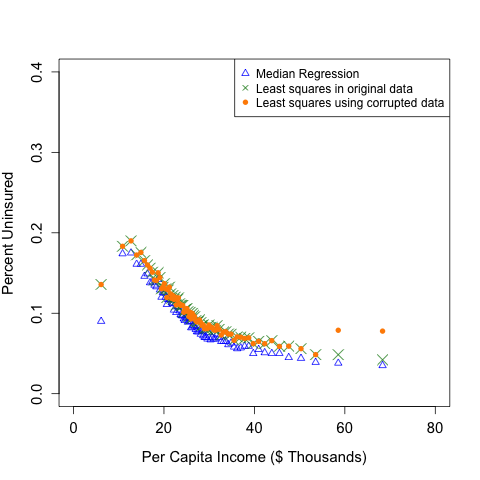}
    \end{subfigure}
\end{figure}

    Figure \ref{appxfig:robust} shows the results of corrupting 32 observations (one tenth of one percent of the data) at low income (Panel (a)) and high income (Panel (b)). This small corruption at low income levels destroys the lower uninsuredness due to Medicaid when fitting the conditional mean using least squares loss, but not when using the conditional median. The same is true at high income levels: the curve is no longer downward sloping when a linear binscatter is applied to the corrupted data. 
    In sum, this analysis shows that the advantages of robust regression methods, long used in parametric modeling, carry over to binscatter contexts.

\begin{figure}[!ht]
    \caption{\small{\bf Predictions Outside the Unit Interval.} This figure shows that estimating the conditional mean using least squares loss, Panel (a) can lead to fitted values outside the unit interval but logitistic QMLE, Panel (b), will not. Changing the evaluation point $\evalw$ results in a shift of the fitted binscatter and least squares does not respect the bounded nature of the outcome. Each light blue line corresponds to a different $\evalw$, cycling through all possible combinations of realized minima and maxima for the controls. The black line shows the binscatter at the sample mean of $\bw_i$.}
    \label{appxfig:zeroone}
    \begin{subfigure}[t]{0.5\columnwidth}
    \centering	
    \caption{\it Linear Least Squares}
    \includegraphics[trim={0cm 0cm 0cm 1.5cm}, clip, scale=.45]{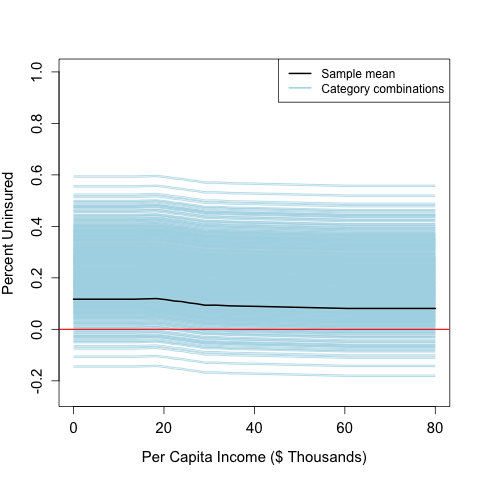}
    \end{subfigure}\hfill
    \begin{subfigure}[t]{0.5\columnwidth}
    \centering	
    \caption{\it Logistic QMLE}
    \includegraphics[trim={0cm 0cm 0cm 1.5cm}, clip, scale=.45]{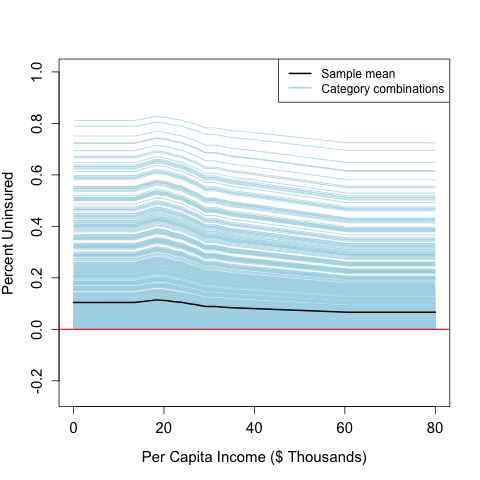}
    \end{subfigure}
\end{figure}

\textbf{Predictions outside the support.} A different form of robustness is illustrated by nonlinear modeling of the conditional mean with controls. Here we apply \eqref{eq:binscatter-QMLE} using both least squares estimation (Example \ref{eg:least squares}) and logistic QMLE (Example \ref{eg:logit}) with the additional controls of footnote \ref{foot:controls}. Following the discussion in footnote \ref{foot:linear}, due to the presence of controls these two do not result in identical binscatters as was the case in Figure \ref{fig:intro}. However, our focus in this analysis is on the evaluation point $\evalw$ and how this shifts the binscatter. For least squares, this can easily result in predictions outside the unit interval, despite $y_i$ being a fractional outcome. This is similar to the perils of linear probability modeling.

    To examine this sensitivity, we create a binscatter plot for each combination of the minimum and maximum values of the coordinates of $\bw_i$.\footnote{Some of the manufactured $\evalw$ may lie outside the support $\mathcal{W}$ as required by Assumption \ref{asmpt:dgp}, but as this is for illustration we abstract from this detail.} This would correspond, for example, to setting different baseline categories for fixed effects. With nine controls this yields 512 binscatter plots. Figure \ref{appxfig:zeroone} shows the results of this exercise, with Panel (a) containing the least squares fits and Panel (b) the logistic QMLE. The light blue curves show each of the 512 fits, in each panel. Here we connect the underlying dots for visual purposes, so that the reader can see each fit individually, though this would not be standard practice. The thicker black line shows, for concreteness, setting $\widehat\evalw$ to the sample mean. 

    The overall shape is not too dissimilar from the two methods, for the sample mean, although the curves are flattened visually due to the expanded $y$-axis. The important feature is the number of least squares (linear) binscatters which are below zero. Loosely speaking, these curves largely represent small-household areas, high percentages of armed forces, high percentages of residents aged 65 and above, and high percentages of English-only speakers. With aggregate data the identities of these is difficult to interpret, but with individual level data these curves would represent distinct ``types'' as defined by $\bw_i$. Due to the modeling assumptions, the logistic QMLE approach does not suffer from this problem. 

\textbf{\bf Visualizing Uncertainty and Target Parameter.} Often the object of interest is defined by a specific loss function. This occurs naturally when visualizing uncertainty. Studying the interquartile range as a measure of dispersion is a typical example. As discussed briefly in the main text, quantile binscatters can be used to evaluate the spread in the conditional distribution more flexibly and richly than reporting single values, just as examining a classic scatter plot can better inform spread than the summary statistics of the 25th and 75th percentiles alone. This is one motivation for the quantile binscatters (Example \ref{eg:quantile} and Figure \ref{fig:quantile}). (Quantiles also have policy-relevant reasons for study on their own, as discussed in the text.) 

    We emphasize that in these cases the target parameter is itself a measure of the spread in the data. The goal is not to visualize the uncertainty of an estimator of, for example, the conditional mean. Although both depend on the variability in the underlying data, they represent different questions. The estimation uncertainty around the mean is relevant for testing hypotheses and making scientific decisions, particularly in causal inference contexts such as identifying most-impacted groups or heterogeneous effects. See Sections \ref{sec:testing} and \ref{sec:multi}.
    
\begin{figure}[!ht]
    \caption{\small{\bf Visualizing Uncertainty.} This figure illustrates how quantile regression can be used to visualize uncertainty differently than confidence bands, capturing the spread instead of estimation uncertainty. The grey dots show the 10th and 90th percentiles, the dark blue dots the conditional mean estimate, and the shaded region its associated confidence band.}
    \label{appxfig:uncertainty}
    \begin{subfigure}[t]{0.5\columnwidth}
    \centering	
    \caption{\it Bands vs. Quantiles w/o Controls}
    \includegraphics[trim={0cm 0cm 0cm 0cm}, clip, scale=.4]{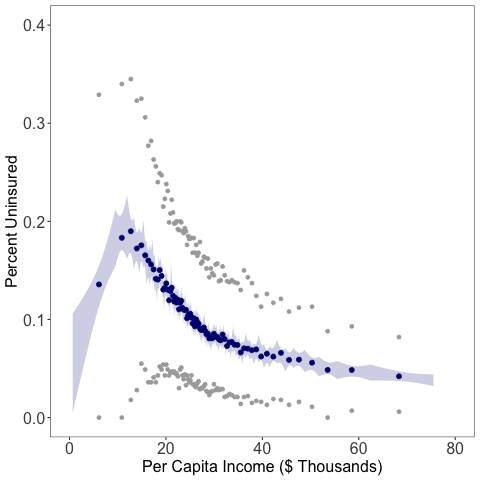}
    \end{subfigure}\hfill
    \begin{subfigure}[t]{0.5\columnwidth}
    \centering	
    \caption{\it Bands vs. Quantiles w/ Controls}
    \includegraphics[trim={0cm 0cm 0cm 0cm}, clip, scale=.4]{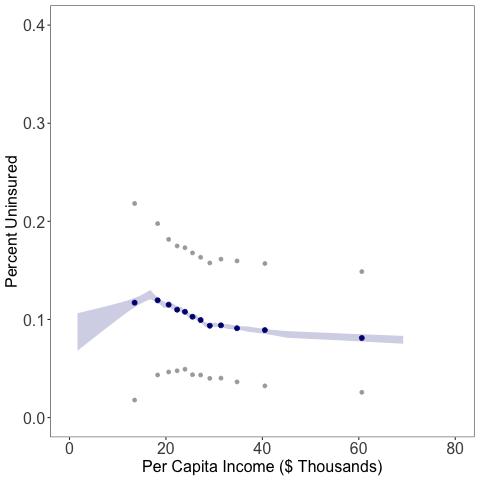}
    \end{subfigure}
\end{figure}

    Figure \ref{appxfig:uncertainty} overlays the 10th and 90th percentiles (grey dots) with an estimate of the conditional mean (blue dots) and its 80\% confidence band (shaded region). Panel (a) has no controls; Panel (b) adds the nine controls listed in footnote \ref{foot:controls}. We have little uncertainty regarding the conditional mean, hence the narrow bands. However, there is considerable uncertainty in the outcome $y_i$ given $x_i$ (and $\bw_i$ in Panel (b)).

\end{appendix}

\end{document}


\title{\vspace{-0.0in}Nonlinear Binscatter Methods\\
	Supplemental Appendix\bigskip }
\author{Matias D. Cattaneo\thanks{Department of Economics, Princeton University.} \and
	    Richard K. Crump\thanks{Capital Markets, Federal Reserve Bank of New York.} \and
	    Max H. Farrell\thanks{Department of Economics, UC Santa Barbara.} \and 
	    Yingjie Feng\thanks{School of Economics and Management, Tsinghua University.}}
\date{February 12, 2026}
\maketitle

\begin{abstract}
    This supplement collects all technical proofs for more general theoretical results than those reported in the main paper. Several of our new theoretical results for nonlinear partitioning-based series estimation may be of independent interest. More details on methodological aspects of nonlinear binscatter are also provided. Companion general-purpose software and replication files are available at \url{https://github.com/nppackages/binsreg}.
\end{abstract}
\thispagestyle{empty}

\pagestyle{plain}

\clearpage
\doublespacing
\setcounter{page}{1}

{\onehalfspacing
\setcounter{tocdepth}{2}
\addtocontents{toc}{\protect\thispagestyle{plain}}
\enlargethispage{\baselineskip}
\tableofcontents
}
\clearpage

\section{Introduction}\label{SA section: Intro}

This supplemental appendix is a comprehensive collection of all our new theoretical results for nonlinear binscatter estimators with semi-linear covariate-adjustment and random partitioning. Many of our results contribute to the broader literature on nonparametric estimation and inference, particularly when using series estimators, and are thus of independent interest outside of binned scatter plots. To help place our results in the literature, we include a remark labelled ``Improvements over literature'' at the end of each technical subsection that discusses in detail the technical improvements presented in that subsection and gives related references.

Here we give a brief summary of this appendix, including pointers to some of the major new results. We proceed as follows. The next subsection lists notation used throughout; further notation is defined throughout Section \ref{SA section: Setup} and at the outset of Section \ref{SA section: general loss fn}. Then, Section \ref{SA section: Setup} describes the setup for nonlinear binscatter methods, including the statistical model, parameters of interest, and assumptions, as well as the (random) partitioning and estimation details.
Specifically, Assumption \ref{SA Assumption DGP} imposes some basic conditions on the data generating process.
Assumption \ref{SA Assumption GL} imposes some technical conditions that characterize and restricts the statistical model of interest.
The loss function specified there is general enough to cover many practically important examples such as mean regression, quantile regression, logit/probit estimation, and Huber regression. 
Assumption \ref{SA Assumption HLE} imposes some mild high-level conditions on the estimation procedure. 
Assumption \ref{SA Assumption RP} summarizes the key conditions on the partitioning scheme used in our theory. We allow for a large class of random partitions. Importantly, the ``convergence'' of the random partition (Assumption \ref{SA Assumption RP}(ii)) is not necessary for most of our main theoretical results, thereby allowing for flexible data-driven partitioning methods, including certain recursive adaptive partitioning methods: see \citet{Devroye-etal2013_book}, \citet{zhang2010recursive}, and references therein.

Section \ref{SA section: prelim lemmas} presents some preliminary technical lemmas for analyzing nonlinear binscatter (and thus also partitioning-based estimators more broadly). New results include precise non-asymptotic concentration results related to Gram matrices (Lemma \ref{SA lem: Gram, GL}), asymptotic variances (Lemmas \ref{SA lem: asymp variance, GL} and \ref{SA lem: uniform converge var part, GL}),
approximation errors (Lemma \ref{SA lem: proj approx error, GL}), and uniform convergence (Lemma \ref{SA lem: uniform consistency, GL}). Sharp control of these objects is a crucial ingredient for obtaining results under weak conditions, as below.

Section \ref{SA section: Bahadur representation} presents a tight uniform (in $x$) Bahadur representation for nonlinear binscatter (Theorem \ref{SA thm: Bahadur representation, GL}). This is our first main result. We allow for random partitions and much weaker rate restrictions (on the tuning parameter $J$) than previously imposed in the literature, in addition to additional controls. The data-dependent partitioning means our series estimator uses random basis functions, and this is entirely new. {In terms of tuning parameter rate restrictions,
when the number of additional control variables is fixed, previous results required $J^4/n\to 0$ (up to $\log(n)$ terms) or something stricter, while our main moment-driven restrictions are $\frac{J^{\frac{\nu}{\nu-2}}\log n}{n}=o(1)$ and $\frac{J^{\frac{2\nu}{\nu-1}}(\log n)^{\frac{\nu}{\nu-1}}}{n}=o(1)$, with $\nu>2$ denoting the number of finite moments of the ``score''; in particular, when $\nu\geq3$, the second restriction is the binding polynomial rate and may be substantially weaker than $J^4/n\to0$.}
{Furthermore, our theory can allow the number of covariates to grow with the sample size,
if the control coefficient estimates converge faster than the nonparametric components.
We also provide primitive sufficient conditions to verify the required fast convergence rates (Lemma \ref{SA lem: growing d gamma rate}).
Note that our class of models is broader than prior work in other aspects as well.}
Importantly, our results can now allow for piecewise constant binscatters, i.e., with degree $p=0$, which is excluded by prior results in the literature (i.e., for previous technical results there was no sequence $J \to\infty$ such that the bias and variance are simultaneously controlled). In addition, employing our novel uniform Bahadur representation, we can establish the uniform convergence rates of nonlinear binscatter (Corollary \ref{SA coro: uniform convergence, GL}) and variance estimators (Theorem \ref{SA thm: meat matrix, GL}) under similarly weak restrictions.

Section \ref{SA sec: pointwise inference} studies the pointwise distributional approximation for nonlinear binscatter estimators. These results are omitted from the main paper to save space, but are standard properties of interest in the nonparametrics literature and thus are included for completeness. The main result is Theorem \ref{SA thm: pointwise normality, GL}, which establishes pointwise asymptotic normality for our point estimators, again allowing for random (and possibly ``non-convergent'') partitions, and under mild rate restrictions similar to those for the (uniform) Bahadur representation.

Section \ref{SA section: IMSE, GL} presents a new Nagar-type approximate IMSE expansion for nonlinear binscatter estimators with semi-linear covariate-adjustment and random partitions (Theorem \ref{SA thm: IMSE, GL}), which has no antecedent in the literature. Our results can be used to design data-driven procedures for selecting IMSE-optimal choices of tuning parameters for nonlinear binscatter. Again, these results are novel in their breadth, the weakness of the assumptions, and the conditions on the partitioning. Here we do require an extra assumption on the partitioning in order to characterize the leading terms in the expansion: intuitively, the random partitioning must ``settle'' to a population partition so that the leading constants of the expansion can be expressed. For example, sample quantiles converge to population quantiles, so this assumption is satisfied.

Uniform inference is dealt with in the next several sections of this appendix. First, Section \ref{SA section: uniform inference} establishes a uniform (in $x$) distributional approximation for nonlinear binscatter estimators. The two main results, which are combined into one in the main text (Theorem 2), are the (conditional) strong approximation in Theorem \ref{SA thm: strong approximation, GL} and the feasible implementation thereof in Theorem \ref{SA thm: plug-in approx, GL}. Again, we allow for a large class of random partitions, a broad class of (possibly) nonlinear and nonsmooth models, and additional controls. Here the partitions do not need to be ``convergent'' in any sense. These results are obtained under weak assumptions, including in particular mild rate restrictions, as in the case of the uniform Bahadur representation, all of which improve on the literature in various directions as explained in the text below. Finally, Theorem \ref{SA thm: sup approx} shows a distributional approximation for the suprema of the $t$-statistic processes in the case of the convergent partition (as in the previous paragraph).

Sections \ref{SA section: confidence band}--\ref{SA section: test shape restriction} employ the strong approximation results 
to study uniform inference for various parameters of interest in the specific context of nonlinear binscatter. These results rely on, and inherit the novelty of, Theorems \ref{SA thm: strong approximation, GL} and \ref{SA thm: plug-in approx, GL}. New results include valid uniform confidence bands (Theorem \ref{SA thm: CB}), consistent hypothesis tests about parametric specification (Theorem \ref{SA thm: testing specification}) and tests for shape restrictions (Theorem \ref{SA thm: testing shape restriction}). All these results explicitly account for the possibly random partitioning scheme and semi-linear covariate-adjustment with random evaluation points. 

Section \ref{SA section: implementation} discusses implementation details for nonlinear binscatter, including standard error computation, feasible data-driven number of bins selector, and choices of polynomial orders given the number of bins. For a more explicit treatment of the package \texttt{binsreg} per se, see \cite{Cattaneo-Crump-Farrell-Feng_2025_Stata} and \url{https://github.com/nppackages/binsreg}.

Finally, Section \ref{SA section: proof} contains the proofs for all the technical results in Section \ref{SA section: general loss fn}.

\subsection{Notation}

See \citet{vandevarrt-Wellner_1996_book}, \citet{Bhatia_2013_book}, \citet{Gine-Nickl_2016_book}, and references therein, for background definitions.

\textit{\textbf{Matrices and Norms}}. For (column) vectors, $\|\cdot\|$ denotes the Euclidean norm, $\|\cdot\|_1$ denotes the $L_1$ norm, $\|\cdot\|_\infty$ denotes the sup-norm, and $\|\cdot\|_0$ denotes the number of nonzeros. For matrices, $\|\cdot\|$ is the operator matrix norm induced by the $L_2$ norm, and $\|\cdot\|_\infty$ is the matrix norm induced by the supremum norm, i.e., the maximum absolute row sum of a matrix. For a square matrix $\mathbf{A}$, $\lambda_{\max}(\mathbf{A})$ and $\lambda_{\min}(\mathbf{A})$ are the maximum and minimum eigenvalues of $\mathbf{A}$, respectively. $[\mathbf{A}]_{ij}$ denotes the $(i,j)$th entry of a generic matrix $\mathbf{A}$. We will use $\mathcal{S}^{L}$ to denote the unit sphere in $\mathbb{R}^L$, i.e., $\|\mathbf{a}\|=1$ for any $\mathbf{a}\in \mathcal{S}^L$. For a real-valued function $g(\cdot)$ defined on a measure space $\mathcal{Z}$, let $\|g\|_{\mathbb{Q}, 2}:=(\int_\mathcal{Z} |g|^2\d{\mathbb{Q}})^{1/2}$ be its $L_2$-norm with respect to the measure $\mathbb{Q}$. In addition, let $\|g\|_\infty=\sup_{z\in\mathcal{Z}}|g(z)|$ be $L_\infty$-norm of $g(\cdot)$, and
if $g$ is a univariate function, let $g^{(v)}(z)=\d{}^v g(z)/\d{z}^v$ be the $v$th derivative for $v\geq 0$.

\textit{\textbf{Asymptotics}}. For sequences of numbers or random variables, we use $l_n\lesssim m_n$ to denote that $\limsup_n|l_n/m_n|$ is finite, $l_n\lesssim_\P m_n$ or $l_n=O_\P(m_n)$ to denote $\limsup_{\varepsilon\to\infty}\limsup_n\P[|l_n/m_n|\geq\varepsilon]=0$, $l_n=o(m_n)$ implies $l_n/m_n\to 0$, and $l_n=o_\P(m_n)$ implies that $l_n/m_n\to_\P 0$, where $\to_\P$ denotes convergence in probability. 
Accordingly, we write $l_n\gtrsim m_n$ if $m_n\lesssim l_n$, and $l_n\gtrsim_\P m_n$ if $m_n\lesssim_\P l_n$. 
$l_n\asymp m_n$ implies that $l_n\lesssim m_n$ and $m_n\lesssim l_n$.

\textit{\textbf{Empirical Process}}. We employ standard empirical process notation: $\E_n[g(\bv_i)]=\frac{1}{n}\sum_{i=1}^ng(\bv_i)$, and $\G_n[g(\bv_i)]=\frac{1}{\sqrt{n}}\sum_{i=1}^n(g(\bv_i)-\E[g(\bv_i)])$ for a sequence of random variables $\{\bv_i\}_{i=1}^n$. In addition, we employ the notion of covering number extensively in the proofs. Specifically, given a measurable space $(A, \mathcal{A})$ and a suitably measurable class of functions $\mathcal{G}$ mapping $A$ to $\mathbb{R}$ equipped with a measurable envelope function $\bar{G}(z)\geq \sup_{g\in\mathcal{G}}|g(z)|$, 
the \textit{covering number} $N(\mathcal{G}, L_2(\mathbb{Q}), \varepsilon)$ is the minimal number of $L_2(\mathbb{Q})$-balls of radius $\varepsilon$ needed to cover $\mathcal{G}$ for a measure $\mathbb{Q}$. The covering number of $\mathcal{G}$ relative to the envelope is denoted as $N(\mathcal{G}, L_2(\mathbb{Q}), \varepsilon\|\bar{G}\|_{\mathbb{Q},2})$.

\textit{\textbf{Other}}. 
$\lceil z \rceil$ outputs the smallest integer no less than $z$ and $a\wedge b=\min\{a,b\}$.
``w.p.a. $1$'' means ``with probability approaching one''.

\section{Setup}\label{SA section: Setup}

{For each $n$, suppose that $\{(y_i,x_i, \bw_i'):1\leq i\leq n\}$ is i.i.d. across $i$, where $y_i\in\mathcal{Y}$ is a scalar response variable, $x_i\in\mathcal{X}$ is a scalar covariate, and $\bw_i\in\mathcal{W}_n\subset\mathbb R^{d_\gamma}$ is a vector of additional control variables, where $d_\gamma$ may increase with $n$. Population objects may depend on $n$, with this dependence suppressed. To simplify notation, we write $\mathcal W=\mathcal W_n$ below. Let $\bD= [(y_i, x_i, \bw_i')' : i=1,2,\dots,n]$.}

For a loss function $\rho(\cdot; \cdot)$ and a strictly monotonic transformation function $\eta(\cdot)$, define 
\begin{equation}\label{SA eq: parameter}
	(\mu_0(\cdot), \bgamma_0)=\argmin_{\mu\in\mathcal{M}, \bgamma\in\mathbb{R}^{d_\gamma}}\; \E\Big[\rho\Big(y_i;\eta(\mu(x_i)+\bw_i'\bgamma)\Big)\Big],
\end{equation}
where $\mathcal{M}$ is a space of functions satisfying certain smoothness conditions to be specified later. 

Equation \eqref{SA eq: parameter} assumes that the minimizers $\mu_0(\cdot)$ and $ \bgamma_0$ are unique. For example, this holds when the index $\theta_0(x,\bw)=\mu_0(x)+\bw'\bgamma_0$ is the unique solution to the population optimization criterion, almost surely, and
\[
c_w\bI_{d_\gamma}
\leq
\E\left[
\{\bw_i-\E[\bw_i\mid x_i]\}
\{\bw_i-\E[\bw_i\mid x_i]\}'
\mid x_i=x
\right]
\leq
C_w\bI_{d_\gamma},
\]
for some $0<c_w<C_w<\infty$, where
the intercept is
absorbed into $\mu_0(\cdot)$, and therefore $\bw_i$ does not contain a
constant.
These conditions rule out any nonzero
vector $\mathbf{a}$ such that $\bw_i'\mathbf{a}$ is a function of $x_i$ almost surely and,
therefore, identify $\mu_0(\cdot)$ and $\bgamma_0$ separately.

This setup is general. For example, consider $\bgamma_0=\mathbf{0}$. If $\rho(\cdot;\cdot)$ is a squared loss and $\eta(\cdot)$ is the identity function, $\mu_0(x)$ is the conditional expectation of $y_i$ given $x_i=x$. Let $\I(\cdot)$ denote the indicator function. If $\rho(y;\eta)=(q-\I(y<\eta))(y-\eta)$ for some $0<q<1$ and $\eta(\cdot)$ is an identity function, then $\mu_0(x)$ is the $q$th conditional quantile of $y_i$ given $x_i=x$. Introducing a transformation function $\eta(\cdot)$ is useful. For instance, it may accommodate logistic regression for binary responses. When $\bgamma_0\neq \mathbf{0}$, the parametric and the nonparametric components are additively separable, and thus \eqref{SA eq: parameter} becomes a generalized partially linear model.

Binscatter estimators are typically constructed based on a (possibly random) partition of the support of the covariate $x_i$.
Specifically, the relevant support of $x_i$ is partitioned into $J$ disjoint intervals, leading to the partitioning scheme $\widehat{\Delta} = \{\widehat{\mathcal{B}}_1, \widehat{\mathcal{B}}_2, \dots, \widehat{\mathcal{B}}_J\}$, where
\[
\widehat{\mathcal{B}}_j = \begin{cases}
[\widehat{\tau}_{j-1}, \widehat{\tau}_j)& \qquad \text{if } j=1, \cdots, J-1\\
[\widehat{\tau}_{J-1}, \widehat{\tau}_J]& \qquad \text{if } j=J
\end{cases},
\]
One popular choice in binscatter applications is the quantile-based partition:
$\widehat\tau_0=\inf\mathcal X$, $\widehat\tau_J=\sup\mathcal X$, and
$\widehat\tau_j=\widehat{F}_X^{-1}(j/J)$ for $j=1,\ldots,J-1$, with $\widehat{F}_X(u)=n^{-1}\sum_{i=1}^n\I(x_i\leq u)$ the empirical cumulative distribution function and $\widehat{F}_X^{-1}$ its generalized inverse.
Our theory is general enough to cover other partitioning schemes satisfying certain regularity conditions specified below. 
An innovation herein is accounting for the additional randomness from the partition $\widehat\Delta$.
The number of bins $J$ plays the role of the tuning parameter for the binscatter method, and is assumed to diverge: $J\to\infty$ as $n\to\infty$ throughout the supplement, unless explicitly stated otherwise.

The piecewise polynomial basis of degree $p$, for some choice of $p=0,1,2,\dots$, is defined as
\[\Big[\begin{array}{cccc}
\I_{\widehat{\mathcal{B}}_1}(x)&
\I_{\widehat{\mathcal{B}}_2}(x)&
\cdots&
\I_{\widehat{\mathcal{B}}_J}(x)
\end{array} \Big]' \otimes
\Big[\begin{array}{cccc}
1& x& \cdots& x^p
\end{array} \Big]',
\]
where $\I_\mathcal{A}(x)=\I(x\in \mathcal{A})$ and $\otimes$ is the Kronecker product operator. For convenience of later analysis, we use $\widehat{\bb}_{p,0}(x)$ to denote a \textit{standardized rotated} basis, the $j$th element of which is given by
\[
\sqrt{J}\times\I_{\widehat{\mathcal{B}}_{\bar{j}}}(x)\times\Big(\frac{x-\widehat{\tau}_{\bar{j}-1}}{\hat{h}_{\bar{j}}} \Big)^{j-1-(\bar{j}-1)(p+1)},
\quad j=1, \cdots, (p+1)J,
\]
where $\bar{j}=\lceil j/(p+1)\rceil$, $\lceil\cdot\rceil$ is the ceiling operator, 
and $\hat{h}_{\bar{j}}=\widehat\tau_{\bar{j}}-\widehat\tau_{\bar{j}-1}$.
Thus, each local polynomial is centered at the start of each bin and scaled by the length of the bin. $\sqrt{J}$ is an additional scaling factor which  helps simplify some expressions of our results. 
The standardized rotated basis $\widehat{\bb}_{p,0}(x)$ is equivalent to the original piecewise polynomial basis in the sense that they represent the same (linear) function space.

To impose the restriction that the estimated function is $(s-1)$-times continuously differentiable for $1\leq s \leq p$, we introduce 
the following basis 
\[
\widehat{\bb}_{p,s}(x)=\Big(\widehat{b}_{p,s,1}(x),\ldots, \widehat{b}_{p,s,K_{p,s}}(x)\Big)'
=\widehat{\mathbf{T}}_s\widehat{\bb}_{p,0}(x),
\qquad K_{p,s}=(p+1)J-s(J-1),
\] 
where
$\widehat{\mathbf{T}}_s:=\widehat{\bT}_s(\widehat{\Delta})$ is a $K_{p,s}\times(p+1)J$ matrix depending on $\widehat{\Delta}$, which transforms a piecewise polynomial basis into a smoothed binscatter basis. 
Some useful properties of $\widehat\bT_s$ are given in Lemma \ref{SA lem: spline transform} in Section \ref{SA section: proof},  
and the explicit representation of $\widehat\bT_s$ is available in the proof of Lemma SA-3.2 in \citet*{Cattaneo-Crump-Farrell-Feng_2024_AER}. 
When $s=0$, we let $\widehat{\mathbf{T}}_0=\mathbf{I}_{(p+1)J}$, the identity matrix of dimension $(p+1)J$. When $s=p$, $\widehat{\bb}_{p,s}(x)$ is the well-known $B$-spline basis of order $p+1$ with simple knots, 
which is $(p-1)$-times continuously differentiable. When $0<s<p$, they can be defined similarly as $B$-splines with knots of certain multiplicities. See Definition 4.1 in Section 4 of \cite{Schumaker_2007_book} for more details about spline functions.  
We require $s\leq p$, since if $s=p+1$, $\widehat{\bb}_{p,s}(x)$ reduces to a global polynomial basis of degree $p$.

Given a choice of basis, we consider the following generalized binscatter estimator:
\begin{equation}\label{SA eq:binscatter-QMLE}
    \widehat{\mu}^{(v)}_{p,s}(x) := \widehat{\bb}_{p,s}^{(v)}(x)'\widehat{\bbeta}, \qquad
    \begin{bmatrix}\;\widehat{\bbeta}\;\\\;\widehat{\bgamma}\;\end{bmatrix}
	= \argmin_{\bbeta,\bgamma} \sum_{i=1}^{n} \rho\Big(y_i; \;\eta\big(\widehat{\bb}_{p,s}(x_i)'\bbeta+\bw_{i}'\bgamma\big)\Big),
\end{equation}
where $\widehat{\bb}_{p,s}^{(v)}(x)=\frac{\d{}^v}{\d{x}^v}\widehat{\bb}_{p,s}(x)$ for some $v\in\mathbb{Z}_+$ such that $v\leq p$.
At interior partition knots, derivatives of the basis and resulting estimators are interpreted as right derivatives, while at the endpoints of $\mathcal X$ they are interpreted as inward derivatives; using left derivatives at interior knots instead leaves all results unchanged.
This estimator can be written as:
\begin{equation}\label{SA eq:binscatter-s-order, general}
	\widehat{\mu}^{(v)}_{p,s}(x) = \widehat{\bb}_{p,s}^{(v)}(x)'\widehat{\bbeta}, \quad
	\widehat{\bbeta} :=\widehat{\bbeta}(\widehat{\bgamma}):= \argmin_{\bbeta\in\mathbb{R}^{K_{p,s}}} \sum_{i=1}^{n} \rho\Big(y_i; \;\eta(\widehat{\bb}_{p,s}(x_i)'\bbeta+\bw_{i}'\widehat{\bgamma})\Big).
\end{equation}
{Throughout, each displayed optimization problem is understood to admit a finite solution w.p.a. $1$, with any measurable solution used when it is not unique.}
We take the representation \eqref{SA eq:binscatter-s-order, general} as the starting point for theoretical analysis, 
which allows us to be more general and agnostic about the estimation of $\bgamma_0$, and also simplifies some of the proofs. 
More specifically, our theory requires only a sufficiently fast convergence rate of $\widehat{\bgamma}$ (see Assumption \ref{SA Assumption HLE} below), 
which, in nonlinear estimation models, can be justified in different ways, e.g., joint estimation, backfitting, profiling, and split-sampling, among other possibilities. 
For the special case of the joint estimator \eqref{SA eq:binscatter-QMLE}, Lemma \ref{SA lem: growing d gamma rate} below 
verifies the fast convergence of $\widehat{\bgamma}$ under intuitive primitive conditions, which allows $d_\gamma$ to possibly grow.
Our software implementation \citep*{Cattaneo-Crump-Farrell-Feng_2025_Stata} relies on joint estimation, as done by the default base estimation packages in \texttt{Python}, \texttt{R}, and \texttt{Stata}.

In this supplement, we focus on estimation and inference of the following three parameters:
\begin{enumerate}[label=(\roman*)]
    \item the nonparametric component $\mu_0^{(v)}(x)$ for any $v\geq 0$, 
    \item the level function 
    $\vartheta_0(x,\evalw)=\eta(\mu_0(x)+\evalw'\bgamma_0)$, and
    \item the marginal effect $\zeta_0(x,\evalw)=\frac{\partial}{\partial x}\eta(\mu_0(x)+\evalw'\bgamma_0)$,
\end{enumerate}
where {$\evalw=\evalw_n\in\mathcal W$} is a user-chosen evaluation point of the control variables, and thus these parameters are viewed as functions of $x$ only in our theory.
Nevertheless, all our results are readily applied to other linear or nonlinear transformations of $\mu_0(x)$, such as the higher-order derivatives $\frac{\partial^v}{\partial x^v}\eta(\mu_0(x)+\evalw'\bgamma_0)$.
Given the binscatter estimates $\widehat{\mu}_{p,s}(x)$ and $\widehat\bgamma$ in \eqref{SA eq:binscatter-QMLE}, the estimators of the three parameters defined above are given by 
\begin{equation*}
    \widehat{\mu}_{p,s}^{(v)}(x), \quad
    \widehat{\vartheta}_{p,s}(x,\widehat\evalw) = \eta(\widehat{\mu}_{p,s}(x)+\widehat\evalw'\widehat{\bgamma}), \quad
    \text{and}\quad
    \widehat{\zeta}_{p,s}(x,\widehat\evalw) = \eta^{(1)}(\widehat{\mu}_{p,s}(x)
    +\widehat\evalw'\widehat{\bgamma})\widehat{\mu}_{p,s}^{(1)}(x)
\end{equation*}
respectively, for some consistent estimate $\widehat\evalw$ (non-random or generated based on 
$\{\bw_i\}_{i=1}^n$) of the evaluation point $\evalw$.
As a reminder, we need to require $p\geq v$ to get $\widehat\mu_{p,s}^{(v)}(x)$, 
$p\geq 0$ to get $\widehat\vartheta_{p,s}(x,\widehat\evalw)$, and $p\geq 1$ to get $\widehat\zeta_{p,s}(x,\widehat\evalw)$.

Recall that in the main text we always set $s=p$ and omit the dependence of estimators on $s$. Thus, 
$\widehat{\mu}_p^{(v)}(x)=\widehat{\mu}_{p,p}^{(v)}(x)$,
$\widehat{\vartheta}_p(x,\widehat\evalw)=
\widehat{\vartheta}_{p,p}(x,\widehat\evalw)$, and
$\widehat{\zeta}_p(x,\widehat\evalw)=
\widehat{\zeta}_{p,p}(x,\widehat\evalw)$.
In this supplement, however, all our results hold for a general choice of the degree and the smoothness of the basis. For ease of notation, the subscripts $p$ and $s$ of the above estimators are dropped  hereafter:
\[
\widehat{\mu}^{(v)}(x):=\widehat{\mu}_{p,s}^{(v)}(x), \quad
\widehat{\vartheta}(x,\widehat\evalw):=\widehat{\vartheta}_{p,s}(x,\widehat\evalw),\quad \text{and}\quad
\widehat{\zeta}(x,\widehat\evalw):=\widehat{\zeta}_{p,s}(x,\widehat\evalw).
\]

\begin{remark}[Smoothness and Bias Correction]
    This supplemental appendix presents \textit{all} results under general choices of the number of bins $J$, the degree of the basis $p$, and the smoothness of the basis $s$. By contrast, for simplicity, the main paper employs the basis with the maximum smoothness, i.e. choosing $s=p$, and considers the special case in which $J$ is taken to be the IMSE-optimal choice for a fixed $p$ (see Theorem \ref{SA thm: IMSE, GL}), and inference is conducted based on the binscatter basis of degree $(p+1)$. Such a choice of $J$ guarantees that the smoothing bias of the binscatter estimator is negligible in inference under mild conditions and thus can be viewed as a bias correction strategy.
\end{remark}


{Throughout, $n$-dependent quantities (e.g., $\bw_i$, $\bgamma_0$, and $\widehat{\bb}_{p,s}$)
are written without an $n$ subscript for notational simplicity,
whereas generic constants and upper and lower bounds are independent of $n$, unless explicitly stated otherwise.}
We first assume the following basic conditions on the data generating process.

\begin{assumDGP}[Data Generating Process]\label{SA Assumption DGP} \
\begin{enumerate}[label=(\roman*)]
    \item {For each $n$, $\{(y_{i}, x_{i}, \bw_i'): 1\leq i\leq n\}$ are i.i.d. across $i$, satisfy \eqref{SA eq: parameter}, and are supported on $\mathcal{Y} \times \mathcal{X} \times \mathcal{W}$, where $\mathcal{X}$ is a compact interval, $\mathcal{W}\subset\mathbb R^{d_\gamma}$ is compact for each $n$, and $\sup_n\sup_{\bw\in\mathcal W}\|\bw\|_\infty<\infty$.}
    
    \item $F_X(x):=\P[x_i\leq x]$ has a Lipschitz continuous (Lebesgue) density $f_X(x)$ bounded away from zero on $\mathcal{X}$.
    
    \item Let $F_{Y|XW}(y|x,\bw):=\P[y_i\leq y|x_i=x,\bw_i=\bw]$. If this conditional distribution is absolutely continuous with respect to Lebesgue measure, its conditional density $f_{Y|XW}(y|x,\bw)$ is uniformly bounded over its support and $x\in\mathcal{X}$, $\bw\in\mathcal{W}$.
\end{enumerate}
\end{assumDGP}

Next, we impose several technical conditions related to the  statistical model of interest.
\begin{assumGL}[Statistical Model]\label{SA Assumption GL} \
\begin{enumerate}[label=(\roman*)]

    \item For some $\delta>0$, there exists a compact interval $\mathcal I$ not depending on $n$ and contained in the interior of the domain of $\eta$ such that
    $[\theta_0(x,\bw)-\delta,\theta_0(x,\bw)+\delta]\subseteq\mathcal I$
    uniformly over $n$, $x\in\mathcal X$, and $\bw\in\mathcal W$. $\rho(y;\cdot)$ is absolutely continuous on an open set containing $\eta(\mathcal I)$ and admits a derivative $\psi(y,\eta) := \psi^\dagger(y-\eta)\psi^\ddagger(\eta)$ almost everywhere on this set. $\psi^\ddagger(\cdot)$ is continuously differentiable and strictly positive or negative on this set. $\psi^\dagger(\cdot)$ is Lipschitz continuous if $F_{Y|XW}(y|x_i,\bw_i)$ does not have a Lebesgue density, or piecewise Lipschitz with finitely many discontinuity points otherwise.
    
    \item $\rho(y;\eta(\theta))$ is convex with respect to $\theta$. $\eta(\cdot)$ is strictly monotonic and three-times continuously differentiable.

    \item $\E[\psi(y_i,\eta(\mu_0(x_i)+\bw_i'\bgamma_0))|x_i,\bw_i]=0$. For $\sigma^2(x,\bw):=\E[\psi(y_i,\eta(\mu_0(x_i)+\bw_i'\bgamma_0))^2|x_i=x,\bw_i=\bw]$, $\inf_{x\in\mathcal{X},\bw\in\mathcal{W}}\sigma^2(x,\bw)\gtrsim 1$. $\E[\eta^{(1)}(\mu_0(x_i)+\bw_i'\bgamma_0)^2\sigma^2(x_i,\bw_i)|x_i=x]$ is Lipschitz continuous on $\mathcal{X}$, and $\sup_{x\in\mathcal{X},\bw\in\mathcal{W}}\E[|\psi(y_i,\eta(\mu_0(x_i)+\bw_i'\bgamma_0))|^\nu|x_i=x, \bw_i=\bw] \lesssim 1$ for some $\nu>2$.
    Let $\Psi(x,\bw;\tau):=\E[\psi(y_i,\tau)|x_i=x,\bw_i=\bw]$. For every $n$ and every $(x,\bw)\in\mathcal X\times\mathcal W$, the map $\tau\mapsto\Psi(x,\bw;\tau)$ is twice continuously differentiable on $\eta(\mathcal I)$. For $j\in\{1,2\}$, define $\Psi_j(x,\bw;\tau):=\frac{\partial^j}{\partial\tau^j}\Psi(x,\bw;\tau)$ and assume
    \[
        \sup_{n\geq1}\sup_{(x,\bw,\tau)\in\mathcal X\times\mathcal W\times\eta(\mathcal I)}|\Psi_j(x,\bw;\tau)|\lesssim 1, \qquad j\in\{1,2\}.
    \]
            
    \item $\inf_{x\in\mathcal{X},\bw\in\mathcal{W}}\varkappa(x,\bw)\gtrsim 1$ and $\E[\varkappa(x_i,\bw_i)|x_i=x]$ is Lipschitz continuous on $\mathcal{X}$ where  
    $\varkappa(x,\bw):=\Psi_1(x,\bw;\eta(\mu_0(x)+\bw'\bgamma_0))(\eta^{(1)}(\mu_0(x)+\bw'\bgamma_0))^2$.
    
    \item $\mu_0(\cdot)$ is $\varsigma$-times continuously differentiable for some $\varsigma\geq p+1$, and 
	$$\sup_{n\geq 1}\sup_{x\in\mathcal X}|\mu_0^{(p+1)}(x)|\lesssim 1.$$
\end{enumerate}
\end{assumGL}

Our next assumption imposes mild high-level conditions on 
the estimator $\widehat{\bgamma}$ of the coefficient vector $\bgamma_0$, the estimator $\widehat\evalw$ of the evaluation point $\evalw$ for control variables, and 
the estimator of the function $\Psi_1$ defined previously in Assumption \ref{SA Assumption GL}(iv).

\begin{assumHLE}[High-Level Estimation Conditions]\label{SA Assumption HLE} \
\begin{enumerate}[label=(\roman*)]	
    \item {$\|\widehat{\bgamma}-\bgamma_0\|\lesssim_\P \mathfrak{r}_\gamma$ for a deterministic sequence $\mathfrak{r}_\gamma$ satisfying $\sqrt{d_\gamma}\mathfrak{r}_\gamma=o(\sqrt{J/n}+J^{-p-1})$. Moreover, $\|\widehat\evalw-\evalw\|=o_\P(1)$, $\P[\widehat\evalw\in\mathcal W]\to1$, and $\sup_n\|\bgamma_0\|<\infty$.}
   
    \item For some estimator $\widehat{\Psi}_1$ of $\Psi_1$,  
	$\|\E_n[\widehat{\bb}_{p,s}(x_i)\widehat{\bb}_{p,s}(x_i)'
    (\widehat{\varkappa}(x_i, \bw_i)-\varkappa(x_i,\bw_i))]\|\lesssim_\P J^{-p-1}+\big(\frac{J\log n}{n^{1-2/\nu}}\big)^{1/2}$ where $\widehat{\varkappa}(x_i,\bw_i)=\widehat{\Psi}_1(x_i, \bw_i;\eta(\widehat{\mu}(x_i)+\bw_i'\widehat{\bgamma}))(\eta^{(1)}(\widehat{\mu}(x_i)+\bw_i'\widehat{\bgamma}))^2$.
\end{enumerate}
\end{assumHLE}

Note that $\Upsilon(x,\bw)=\Psi_1(x, \bw;\eta(\mu_0(x)+\bw'\bgamma_0))$ in the main paper to streamline the presentation. {Part (i) is a high-level condition on the convergence of $\widehat{\bgamma}$ and $\widehat\evalw$. The factor $\sqrt{d_\gamma}$ converts the Euclidean coefficient rate into a uniform fitted-index rate under the coordinatewise support bound in Assumption \ref{SA Assumption DGP}(i).}
Part (ii) is a high-level condition that ensures we have a valid feasible estimator of the Gram matrix ($\bar\bQ$ or $\bQ_0$ defined at the outset of Section \ref{SA section: general loss fn} below). 
{Note that the convergence rate of $\eta^{(1)}(\widehat{\mu}(x_i)+\bw_i'\widehat{\bgamma})$ can be deduced from Corollary \ref{SA coro: uniform convergence, GL} below, which requires only part (i) of this assumption.} Thus, part (ii) can be largely viewed as a restriction on $\widehat{\Psi}_1$ only. Note that $\widehat{\Psi}_1$ does not have to be consistent for $\Psi_1$ in any sense;
it suffices that the estimator $\E_n[\widehat{\bb}_{p,s}(x_i)\widehat{\bb}_{p,s}(x_i)'
\widehat{\varkappa}(x_i, \bw_i)]$ based on $\widehat{\Psi}_1$ as a whole is consistent.
See Section \ref{SA section: implementation} for several examples of the estimator $\widehat{\Psi}_1$.

\subsection{Partitions}\label{SA section: Partitions}

We need some regularity conditions on the partitioning scheme, which can be verified on a case-by-case basis. We first define a family of ``quasi-uniform'' partitions for some absolute constant $C>0$:
\begin{equation} \label{SA eq: quasi-uniform}
	\Pi_C = \Big\{ \Delta:
	\frac{\max_{1\leq j \leq J} h_j(\Delta)}
	{\min_{1\leq j \leq J} h_j(\Delta)}\leq C \Big\},
\end{equation}
where $h_j(\Delta)$ denotes the length of the $j$th bin in the partition $\Delta$. Roughly speaking, \eqref{SA eq: quasi-uniform} says that the bins in any $\Delta\in\Pi_C$ do not differ too much in length. Also, let
$\bX = [x_1, \ldots, x_n]'$, $\bW=[\bw_1,\cdots, \bw_n]'$ and $\bY=[y_1, \cdots, y_n]'$.

\begin{assumRP}[Random Partition]\label{SA Assumption RP} \leavevmode
    \begin{enumerate}[label=(\roman*)]
        \item $\widehat\Delta\indep \bY |(\bX, \bW)$ and $\widehat\Delta\in\Pi_C$ w.p.a. $1$ for some absolute constant $C>0$.
        \item There exists a non-random partition of $\mathcal{X}$,
        $\Delta_0=\{\mathcal{B}_1, \cdots, \mathcal{B}_J\}$ with $\mathcal{B}_j=[\tau_{j-1}, \tau_j)$ for $j\leq J-1$ and $\mathcal{B}_J=[\tau_{J-1}, \tau_J]$ such that
    $\frac{\max_{1\leq j\leq J}h_j}{\min_{1\leq j\leq J}h_j}\leq c_{\tt QU}$ for some absolute constant $c_{\tt QU}>0$, and   
    $|\hat{\tau}_0-\tau_0|\lesssim_\P \mathfrak{r}_{\tt RP}$ and
    $\max_{1\leq j\leq J}|\hat{h}_j-h_j|\lesssim_\P J^{-1}\mathfrak{r}_{\tt RP}$ for $\mathfrak{r}_{\tt RP}=o(1)$.
    \end{enumerate}
\end{assumRP}
Part (i) is the key condition for our main results and will be imposed throughout. First, it requires that the possibly random partition $\widehat\Delta$ be independent of the outcome $\bY$ given the covariates $(\bX,\bW)$. 
This conditional independence assumption is trivially satisfied if $\widehat\Delta$ is deterministic (e.g., equally-spaced partition) or depends on $\bX$ and $\bW$ only (e.g., quantile-spaced partition based on $\bX$). It also holds if a sample splitting scheme is used: a subsample (including the information about the outcome) is used for constructing the partition, and the other is employed to construct the binscatter estimator (so that $\widehat\Delta$ is independent of the data $(\bX, \bW, \bY)$). Second, $\widehat\Delta$ is required to be ``quasi-uniform'' with large probability. It is trivially true for equally-spaced partitions and can be verified for quantile-spaced partitions under the mild conditions on the covariate density imposed before
(see Lemma \ref{SA lem: quantile partition}). However, this condition may be too restrictive for other modern machine-learning-based partitioning methods, in which case some additional regularization may be necessary to recover the quasi-uniformity property.

Part (ii) requires that the random partition $\widehat\Delta$ ``stabilizes'' to a fixed one in large samples. This is true if the partition is deterministic or generated by sample quantiles (since sample quantiles converge to population quantiles), but more generally, it is not always possible. Fortunately, this ``convergence'' requirement is not necessary for most of our main results (except Theorem \ref{SA thm: IMSE, GL} and Theorem \ref{SA thm: sup approx}). Thus, we will always make it clear if part (ii) of Assumption \ref{SA Assumption RP} is imposed.

Given the random partition $\widehat{\Delta}$, $\E_{\widehat{\Delta}}[\cdot]$ denotes unconditional population expectation with $\widehat{\Delta}$ held fixed (this is not a conditional expectation given $\widehat{\Delta}$).
We use $\mathscr{F}_{XW\Delta}$ to denote the $\sigma$-field generated by
$(\bX,\bW,\widehat\Delta)$.
To further simplify notation, 
let $\hat{h}_j=\hat{\tau}_j-\hat{\tau}_{j-1}$ be the width of the $j$th bin $\widehat{\mathcal{B}}_j$, and when the ``limiting'' partition $\Delta_0=\{\mathcal{B}_1, \cdots, \mathcal{B}_J\}$ is defined (Assumption \ref{SA Assumption RP}(ii) holds),  
let $h_j$ be the width of $\mathcal{B}_j$.
Analogously to $\widehat{\bb}_{p,s}(x)$, $\bb_{p,s}(x)$ denotes the binscatter basis of degree $p$ that is  $(s-1)$-times continuously differentiable and is constructed based on the \textit{nonrandom} partition $\Delta_0$. We sometimes write $\bb_{p,s}(x;\Delta)=(b_{p,s,1}(x;\Delta), \ldots, b_{p,s,K_{p,s}}(x;\Delta))'$ to emphasize a binscatter basis is constructed based on a particular partition $\Delta$. Therefore, $\widehat{\bb}_{p,s}(x)=\bb_{p,s}(x;\widehat{\Delta})$ and $\bb_{p,s}(x)=\bb_{p,s}(x;\Delta_0)$. 
Accordingly, we use $\bT_s$ to denote the transformation matrix based on the non-random partition $\Delta_0$ (which transforms $\bb_{p,0}(x)$ to $\bb_{p,s}(x)$).


\section{Main Results} \label{SA section: general loss fn}

We introduce the following quantities that will be extensively used throughout the supplement: 
\begin{align*}
	&\eta_i=\eta(\mu_0(x_i)+\bw_i'\bgamma_0),\qquad \widehat{\eta}_i=\eta(\widehat{\mu}(x_i)+\bw_i'\widehat{\bgamma}),\\ 
 &\eta_{i,1}=\eta^{(1)}(\mu_0(x_i)+\bw_i'\bgamma_0),\qquad \widehat{\eta}_{i,1}=\eta^{(1)}(\widehat{\mu}(x_i)+\bw_i'\widehat{\bgamma}),\\ 
 &\eta_{0,1}(x,\evalw)=\eta^{(1)}(\mu_0(x)+\evalw'\bgamma_0), \qquad
 \widehat\eta_{0,1}(x,\widehat\evalw)=\eta^{(1)}(\widehat\mu(x)+\widehat\evalw'\widehat\bgamma),\\
 &\widehat{\mu}(x_i)=\widehat{\bb}_{p,s}(x_i)'\widehat{\bbeta}, \qquad
	\epsilon_i=y_i-\eta_i, \qquad
	\widehat{\epsilon}_{i}=y_i-\widehat{\eta}_i,\\
	&\widehat{\bQ}_{p,s}:=\widehat{\bQ}_{p,s}(\widehat{\Delta}):=
	\E_n[\widehat{\bb}_{p,s}(x_i)\widehat{\bb}_{p,s}(x_i)'\widehat{\Psi}_1(x_i, \bw_i;\widehat{\eta}_i)
	\widehat{\eta}_{i,1}^2],\\	
	&\bar{\bQ}_{p,s}:=\bar{\bQ}_{p,s}(\widehat{\Delta}):=
	\E_n[\widehat{\bb}_{p,s}(x_i)\widehat{\bb}_{p,s}(x_i)'\Psi_1(x_i, \bw_i;\eta_i)
	\eta_{i,1}^2],\\
	&\bQ_{0,p,s}:=\bQ_{p,s}(\Delta_0):=\E[\bb_{p,s}(x_i)\bb_{p,s}(x_i)'\Psi_1(x_i,\bw_i;\eta_i)\eta_{i,1}^2], \\
	&\widehat{\bSigma}_{p,s}:=\widehat{\bSigma}_{p,s}(\widehat{\Delta}):=
	\E_n[\widehat{\bb}_{p,s}(x_i)\widehat{\bb}_{p,s}(x_i)'
	\psi(y_i, \widehat{\eta}_i)^2\widehat{\eta}_{i,1}^2], \\
	&\bar{\bSigma}_{p,s}:=\bar{\bSigma}_{p,s}(\widehat{\Delta}):=
	\E_n\Big[\widehat{\bb}_{p,s}(x_i)\widehat{\bb}_{p,s}(x_i)'
	\sigma^2(x_i,\bw_i)\eta_{i,1}^2\Big], \\
	&\bSigma_{0,p,s}:=\bSigma_{p,s}(\Delta_0):=\E\Big[\bb_{p,s}(x_i)\bb_{p,s}(x_i)'
	\psi(y_i,\eta_i)^2\eta_{i,1}^2\Big], \\
	&\widehat{\Omega}_{\mu^{(v)},p,s}(x):=\widehat{\Omega}_{\mu^{(v)},p,s}(x;\widehat{\Delta}):=\widehat{\bb}_{p,s}^{(v)}(x)'\widehat{\bQ}_{p,s}^{-1}\widehat{\bSigma}_{p,s}\widehat{\bQ}_{p,s}^{-1}\widehat{\bb}_{p,s}^{(v)}(x),\\
	&\bar{\Omega}_{\mu^{(v)},p,s}(x):=\bar{\Omega}_{\mu^{(v)},p,s}(x;\widehat{\Delta}):=
	\widehat{\bb}_{p,s}^{(v)}(x)'\bar{\bQ}_{p,s}^{-1}\bar{\bSigma}_{p,s}\bar{\bQ}_{p,s}^{-1}\widehat{\bb}_{p,s}^{(v)}(x), \\ 
 &\Omega_{\mu^{(v)},p,s}(x):=\Omega_{\mu^{(v)},p,s}(x;\widehat{\Delta}):=\widehat{\bb}_{p,0}^{(v)}(x)'\bT_s'\bQ_{0,p,s}^{-1}\bSigma_{0,p,s}\bQ_{0,p,s}^{-1}\bT_s\widehat{\bb}_{p,0}^{(v)}(x),\\
    &\widehat\Omega_{\vartheta,p,s}(x):=\widehat\Omega_{\vartheta,p,s}(x;\widehat\Delta):=
    [\eta^{(1)}(\widehat{\mu}(x)+\widehat\evalw'\widehat\bgamma)]^2
    \widehat{\bb}_{p,s}(x)'\widehat\bQ_{p,s}^{-1}\widehat\bSigma_{p,s}
    \widehat\bQ_{p,s}^{-1}\widehat{\bb}_{p,s}(x),\\
    &\bar\Omega_{\vartheta,p,s}(x):=\bar\Omega_{\vartheta,p,s}(x;\widehat\Delta):=
    [\eta^{(1)}(\mu_0(x)+\evalw'\bgamma_0)]^2
    \widehat{\bb}_{p,s}(x)'\bar\bQ_{p,s}^{-1}\bar\bSigma_{p,s}\bar\bQ_{p,s}^{-1}\widehat{\bb}_{p,s}(x),\\
    &\Omega_{\vartheta,p,s}(x):=\Omega_{\vartheta,p,s}(x;\widehat\Delta):=
    [\eta^{(1)}(\mu_0(x)+\evalw'\bgamma_0)]^2
    \widehat{\bb}_{p,0}(x)'\bT_s'\bQ_{0,p,s}^{-1}\bSigma_{0,p,s}\bQ_{0,p,s}^{-1}\bT_s\widehat{\bb}_{p,0}(x),\\
    &\widehat\Omega_{\zeta,p,s}(x):=\widehat\Omega_{\zeta,p,s}(x;\widehat\Delta):=[\eta^{(1)}(\widehat{\mu}(x)+\widehat\evalw'\widehat\bgamma)]^2\widehat{\bb}^{(1)}_{p,s}(x)'\widehat\bQ_{p,s}^{-1}\widehat\bSigma_{p,s}\widehat\bQ_{p,s}^{-1}\widehat{\bb}^{(1)}_{p,s}(x),\\
    &\bar\Omega_{\zeta,p,s}(x):=\bar\Omega_{\zeta,p,s}(x;\widehat\Delta):=
    [\eta^{(1)}(\mu_0(x)+\evalw'\bgamma_0)]^2
    \widehat{\bb}^{(1)}_{p,s}(x)'\bar\bQ_{p,s}^{-1}\bar\bSigma_{p,s}\bar\bQ_{p,s}^{-1}\widehat{\bb}^{(1)}_{p,s}(x),\quad \text{and}\\
    &\Omega_{\zeta,p,s}(x):=\Omega_{\zeta,p,s}(x;\widehat\Delta):=
    [\eta^{(1)}(\mu_0(x)+\evalw'\bgamma_0)]^2
    \widehat{\bb}^{(1)}_{p,0}(x)'\bT_s'\bQ_{0,p,s}^{-1}\bSigma_{0,p,s}\bQ_{0,p,s}^{-1}\bT_s\widehat{\bb}^{(1)}_{p,0}(x).
\end{align*}

Recall that in the main text we always set $s=p$ and omit the dependence on $s$ whenever there is no confusion. Thus,
\begin{align*}
    &\widehat{\bQ}_p=\widehat{\bQ}_{p,p},\quad
    \bar{\bQ}_p=\bar{\bQ}_{p,p},\quad
    \bQ_{0,p}=\bQ_{0,p,p},\\
    &\widehat{\bSigma}_p=\widehat{\bSigma}_{p,p},\quad
    \bar{\bSigma}_p=\bar{\bSigma}_{p,p},\quad
    \bSigma_{0,p}=\bSigma_{0,p,p},\\
    &\widehat{\Omega}_{\mu^{(v)},p}(x)=
    \widehat{\Omega}_{\mu^{(v)},p,p}(x),\quad
    \bar{\Omega}_{\mu^{(v)},p}(x)=
    \bar{\Omega}_{\mu^{(v)},p,p}(x),\quad
    \Omega_{\mu^{(v)},p}(x)=
    \Omega_{\mu^{(v)},p,p}(x),\\
    &\widehat{\Omega}_{\vartheta,p}(x)=
    \widehat{\Omega}_{\vartheta,p,p}(x),\quad
    \bar{\Omega}_{\vartheta,p}(x)=
    \bar{\Omega}_{\vartheta,p,p}(x),\quad
    \Omega_{\vartheta,p}(x)=
    \Omega_{\vartheta,p,p}(x),\\
    &\widehat{\Omega}_{\zeta,p}(x)=
    \widehat{\Omega}_{\zeta,p,p}(x),\quad
    \bar{\Omega}_{\zeta,p}(x)=
    \bar{\Omega}_{\zeta,p,p}(x), \quad\text{and}\quad
    \Omega_{\zeta,p}(x)=
    \Omega_{\zeta,p,p}(x).
\end{align*}

In this supplement, however, all our results hold for a general choice of the degree and the smoothness of the basis. For ease of notation, the subscripts $p$ and $s$ of the above quantities are dropped hereafter:
\begin{align*}
    &\widehat{\bQ}=\widehat{\bQ}_{p,s},\quad
    \bar{\bQ}=\bar{\bQ}_{p,s},\quad
    \bQ_{0}=\bQ_{0,p,s},\\
    &\widehat{\bSigma}=\widehat{\bSigma}_{p,s},\quad
    \bar{\bSigma}=\bar{\bSigma}_{p,s},\quad
    \bSigma_{0}=\bSigma_{0,p,s},\\
    &\widehat{\Omega}_{\mu^{(v)}}(x)=
    \widehat{\Omega}_{\mu^{(v)},p,s}(x),\quad
    \bar{\Omega}_{\mu^{(v)}}(x)=
    \bar{\Omega}_{\mu^{(v)},p,s}(x),\quad
    \Omega_{\mu^{(v)}}(x)=
    \Omega_{\mu^{(v)},p,s}(x),\\
    &\widehat{\Omega}_{\vartheta}(x)=
    \widehat{\Omega}_{\vartheta,p,s}(x),\quad
    \bar{\Omega}_{\vartheta}(x)=
    \bar{\Omega}_{\vartheta,p,s}(x),\quad
    \Omega_{\vartheta}(x)=
    \Omega_{\vartheta,p,s}(x),\\
    &\widehat{\Omega}_{\zeta}(x)=
    \widehat{\Omega}_{\zeta,p,s}(x),\quad
    \bar{\Omega}_{\zeta}(x)=
    \bar{\Omega}_{\zeta,p,s}(x), \quad\text{and}\quad
    \Omega_{\zeta}(x)=
    \Omega_{\zeta,p,s}(x).
\end{align*}

In addition, given the family $\Pi_C$ of the quasi-uniform partitions defined in \eqref{SA eq: quasi-uniform}, for any $\Delta\in \Pi_C$, we let $\bbeta_0(\Delta)\in\mathbb{R}^{K_{p,s}}$ be any vector such that for every
$v\leq p$,
\begin{equation*}
\sup_{x\in\mathcal{X}}\Big|\mu_0^{(v)}(x)-\bb_{p,s}^{(v)}(x;\Delta)'\bbeta_0(\Delta)\Big|
\lesssim J^{-p-1+v}.
\end{equation*}
Let $r_{0,v}(x;\Delta)=\mu_0^{(v)}(x)-\bb_{p,s}^{(v)}(x;\Delta)'\bbeta_0(\Delta)$ denote the corresponding approximation error. 
Accordingly, given the random partition $\widehat\Delta$, we let $\widehat\bbeta_0:=\bbeta_0(\widehat\Delta)$, 
and $\widehat{r}_{0,v}(x)=\mu_0^{(v)}(x)-\widehat\bb_{p,s}^{(v)}(x)'\widehat\bbeta_0$ denote the corresponding approximation error.
The existence of such vectors is guaranteed by Assumptions \ref{SA Assumption DGP} and \ref{SA Assumption GL}(v), and is verified in Lemma \ref{SA lem: uniform approx rate} in Section \ref{SA section: proof}.

\subsection{Preliminary Lemmas}\label{SA section: prelim lemmas}

\begin{lem}[Gram] \label{SA lem: Gram, GL}
	{Suppose that Assumptions \ref{SA Assumption DGP}, \ref{SA Assumption GL}, and \ref{SA Assumption RP}(i) hold.}
	If $\frac{J\log J}{n}=o(1)$, then
	\[
 	1\lesssim \lambda_{\min}(\bar{\bQ})\leq\lambda_{\max}(\bar{\bQ})\lesssim 1,\quad
		\big|[\bar{\bQ}^{-1}]_{ij}\big|\lesssim \varrho^{|i-j|} \quad \text{w.p.a. } 1,\quad
		\text{and}\quad \|\bar{\bQ}^{-1}\|_\infty\lesssim_\P 1,
        \]
        where $\varrho\in(0,1)$ is some absolute constant.
	
        If, in addition, Assumption \ref{SA Assumption RP}(ii) holds. Then,
        \begin{eqnarray*}
            &1\lesssim \lambda_{\min}(\bQ_0)\leq \lambda_{\max}(\bQ_0)\lesssim 1,\\
		&\|\bar{\bQ}-\bQ_0\|
		\lesssim_\P \Big(\frac{J\log J}{n}\Big)^{1/2}+\mathfrak{r}_{\tt RP},
		\quad\text{and}\quad
		\|\bar{\bQ}^{-1}-\bQ_0^{-1}\|_\infty\lesssim_\P \Big(\frac{J\log J}{n}\Big)^{1/2}+\mathfrak{r}_{\tt RP}.
	\end{eqnarray*}
\end{lem}

The next lemma shows that the limiting variance is bounded from above and below.

\begin{lem}[Asymptotic Variance] \label{SA lem: asymp variance, GL}
	{Suppose that Assumptions \ref{SA Assumption DGP}, \ref{SA Assumption GL}, and \ref{SA Assumption RP}(i) hold.} If $\frac{J\log J}{n}=o(1)$, then w.p.a. $1$,
      \begin{eqnarray*}
      &J^{1+2v}\lesssim\inf_{x\in\mathcal{X}}\bar{\Omega}_{\mu^{(v)}}(x)
		\leq\sup_{x\in\mathcal{X}}\bar{\Omega}_{\mu^{(v)}}(x)
		\lesssim J^{1+2v},\\
      &J\lesssim \inf_{x\in\mathcal{X}}\bar{\Omega}_{\vartheta}(x)
		\leq\sup_{x\in\mathcal{X}}\bar{\Omega}_{\vartheta}(x)
		\lesssim J,\\
      &J^3\lesssim \inf_{x\in\mathcal{X}}\bar{\Omega}_{\zeta}(x)
		\leq\sup_{x\in\mathcal{X}}\bar{\Omega}_{\zeta}(x)
		\lesssim J^3.
      \end{eqnarray*}	
      If, in addition, Assumption \ref{SA Assumption RP}(ii) holds, then w.p.a. $1$,
 \begin{eqnarray*}
		&J^{1+2v}\lesssim\inf_{x\in\mathcal{X}}\Omega_{\mu^{(v)}}(x)
		\leq\sup_{x\in\mathcal{X}}\Omega_{\mu^{(v)}}(x)
		\lesssim J^{1+2v},\\
		&J\lesssim\inf_{x\in\mathcal{X}}\Omega_{\vartheta}(x)
		\leq\sup_{x\in\mathcal{X}}\Omega_{\vartheta}(x)
		\lesssim J,\\
		&J^3\lesssim\inf_{x\in\mathcal{X}}\Omega_{\zeta}(x)
		\leq\sup_{x\in\mathcal{X}}\Omega_{\zeta}(x)
		\lesssim J^3.
	\end{eqnarray*}
\end{lem}

The next lemma gives a bound on the variance component of the nonlinear binscatter estimator.

\begin{lem}[Uniform Convergence: Variance] \label{SA lem: uniform converge var part, GL}
	{Suppose that Assumptions \ref{SA Assumption DGP}, \ref{SA Assumption GL}, and \ref{SA Assumption RP}(i) hold.} If $\frac{J^{\frac{\nu}{\nu-2}}\log J}{n}=o(1)$, then
	\[
	\sup_{x\in\mathcal{X}}\Big|\widehat{\bb}_{p,s}^{(v)}(x)'\bar{\bQ}^{-1}\E_n[\widehat{\bb}_{p,s}(x_i)\eta_{i,1}\psi(y_i,\eta_i)]\Big|\lesssim_\P J^v\Big(\frac{J\log J}{n}\Big)^{1/2}.
	\]
\end{lem}

\begin{lem}[Projection of Approximation Error] \label{SA lem: proj approx error, GL}
	{Suppose that Assumptions \ref{SA Assumption DGP}, \ref{SA Assumption GL}, and \ref{SA Assumption RP}(i) hold.} If $\frac{J^{\frac{\nu}{\nu-2}}\log J}{n}=o(1)$, then
	\[
	\begin{split}
		&\sup_{x\in\mathcal{X}}
		\Big|\widehat{\bb}_{p,s}^{(v)}(x)'\bar{\bQ}^{-1}
		\E_n\Big[\widehat{\bb}_{p,s}(x_{i})\Big(\eta_{i,1}\psi(y_i,\eta_i)-
		\eta^{(1)}(\widehat{\bb}_{p,s}(x_i)'\widehat{\bbeta}_0+\bw_i'\bgamma_0)
		\psi(y_i,\eta(\widehat{\bb}_{p,s}(x_i)'\widehat{\bbeta}_0+\bw_i'\bgamma_0))\Big)\Big]\Big|\\
		\lesssim_\P&\; J^{-p-1+v}+J^{\frac{2v-p-1}{2}}\Big(\frac{J\log J}{n}\Big)^{1/2}+\frac{J^{1+v}\log J}{n}.
	\end{split}
	\]
\end{lem}

\begin{lem}[Uniform Consistency]\label{SA lem: uniform consistency, GL}
	{Suppose that Assumptions \ref{SA Assumption DGP}, \ref{SA Assumption GL}, \ref{SA Assumption HLE}(i), and \ref{SA Assumption RP}(i) hold.} If
	{$\frac{J(J+d_\gamma)\log J}{n^{1-1/\nu}}=o(1)$}, then
	\[
	\|\widehat{\bbeta}-\widehat{\bbeta}_0\|_\infty=o_\P(J^{-1/2})\quad \text{and }\;
	\sup_{x\in\mathcal{X}}\,|\widehat{\mu}(x)-\mu_0(x)|=o_\P(1).
	\]
\end{lem}

\begin{remark}[Side rate conditions]
      {When $\nu\to\infty$, the rate restriction in Lemma \ref{SA lem: uniform consistency, GL} tends to $J(J+d_\gamma)\log J/n=o(1)$,
	  which reduces to $J^2\log J/n=o(1)$ when $d_\gamma=o(J)$. 
	  We conjecture this rate restriction is stronger than needed. 
	  For piecewise polynomials (i.e., $s=0$) and fixed $d_\gamma$, the weaker condition $\frac{J^{\frac{\nu}{\nu-1}}(\log J)^{\frac{\nu}{\nu-1}}}{n}=o(1)$ suffices to establish uniform consistency, and is redundant in our main theorems in view of the condition $\frac{J^{\frac{\nu}{\nu-2}}(\log n)^{\frac{\nu}{\nu-2}}}{n}=o(1)$ imposed below.}
\end{remark}

Our result holds without imposing any smoothness restrictions on the estimation space. Specifically, the estimation procedure \eqref{SA eq:binscatter-s-order, general} searches for solutions in $\mathbb{R}^{K_{p,s}}$, leading to an estimation space $\{\widehat{\bb}_{p,s}(x)'\bbeta: \bbeta\in\mathbb{R}^{K_{p,s}}\}$. In contrast, many studies of series (or sieve) methods restrict the functions in the estimation space to satisfy certain smoothness conditions, e.g., Lipschitz continuity, to derive the uniform consistency. See, for example, \cite*{Chernozhukov-Imbens-Newey_2007_JoE} and references therein.

\begin{remark}[Improvements over literature]
    Most of the results in this subsection are new to the literature, even in the case of non-random partitioning and without covariate-adjustments, because they take advantage of the specific binscatter structure (i.e., locally bounded series basis). The closest antecedent in the literature is  \citet*{Belloni-Chernozhukov-Chetverikov-FernandezVal_2019_JoE}, while it focuses on series-based quantile regression only. Furthermore, relative to prior work, our results
    allow for random partitioning schemes, formally taking into account both the potential randomness of the partition and the semi-linear regression estimation structure.
    Importantly, we highlight the key conditions on the possibly random partition 
    (Assumptions \ref{SA Assumption RP}(i) and \ref{SA Assumption RP}(ii)) used to derive various properties of the Gram matrix, asymptotic variance and other quantities.
\end{remark}

\subsection{Bahadur Representation}\label{SA section: Bahadur representation}

{Define the dimension-adjusted entropy factor and Bahadur remainder
\[
\ell_{\gamma,n}:=\log n+\sqrt{d_\gamma\log n},\qquad
\mathfrak r_{\mathrm{BR},n}:=\left(\frac{J\log n}{n}\right)^{3/4}\ell_{\gamma,n}+J^{-\frac{p+1}{2}}\sqrt{\frac Jn}\ell_{\gamma,n}+\frac{J\ell_{\gamma,n}^2}{n}+J^{-p-1}+\sqrt{d_\gamma}\mathfrak r_\gamma.
\]
The factor $\sqrt{d_\gamma}\mathfrak r_\gamma$ controls the fitted-index error under coordinatewise bounded controls,
while $\ell_{\gamma,n}$ accounts for the empirical-process complexity of estimating $d_\gamma$ coefficients.
For fixed $d_\gamma$, $\ell_{\gamma,n}\asymp\log n$, and $J\ell_{\gamma,n}^2/n$ is absorbed by the first term under the imposed rate conditions.
The second rate condition in the theorem below is an additional heavy-tail restriction used to keep this concise remainder.}

\begin{thm}[Bahadur Representation]\label{SA thm: Bahadur representation, GL}
 {Suppose that Assumptions \ref{SA Assumption DGP}, \ref{SA Assumption GL}, \ref{SA Assumption HLE}(i), and \ref{SA Assumption RP}(i) hold, $d_\gamma=o(J)$,
 \[ \frac{J^{\frac{\nu}{\nu-2}}\log n}{n}+\frac{J\log n\,\ell_{\gamma,n}^{4/3}}{n}+\frac{J^{\frac{2\nu}{\nu-1}}(\log n)^{\frac{\nu}{\nu-1}}}{n}+\frac{\log n}{J}=o(1),\qquad \frac{J\ell_{\gamma,n}^2}{n^{1-1/\nu}}\lesssim\sqrt{\frac{J\log n}{n}}+J^{-p-1}. \]
 Then,}
 \begin{enumerate}[label=(\roman*)]
     \item $\widehat\mu^{(v)}(x)$ satisfies that
     \[
	\sup_{x\in\mathcal{X}}\Big|\widehat{\mu}^{(v)}(x)-\mu_0^{(v)}(x)+\widehat{\bb}_{p,s}^{(v)}(x)'\bar{\bQ}^{-1}\E_n[\widehat{\bb}_{p,s}(x_i)\eta_{i,1}\psi(y_i,\eta_i)]\Big|\lesssim_\P J^v\mathfrak r_{\mathrm{BR},n}.
	\]

    \item
       $\widehat\vartheta(x,\widehat\evalw)$ satisfies that
       \begin{align*}
       &\sup_{x\in\mathcal{X}}\Big|\widehat\vartheta(x,\widehat\evalw)-\vartheta_0(x,\evalw)+
       \eta^{(1)}(\mu_0(x)+\evalw'\bgamma_0)\widehat{\bb}_{p,s}(x)'\bar{\bQ}^{-1}\E_n[\widehat\bb_{p,s}(x_i)\eta_{i,1}\psi(y_i,\eta_i)]\Big|\\
       &{\lesssim_\P \mathfrak r_{\mathrm{BR},n}+\|\widehat\evalw-\evalw\|.}
    \end{align*}

      \item $\widehat\zeta(x,\widehat\evalw)$ satisfies that
    \begin{align*}
     &\sup_{x\in\mathcal{X}}\Big|\widehat\zeta(x,\widehat\evalw)-\zeta_0(x,\evalw)+\eta^{(1)}(\mu_0(x)+\evalw'\bgamma_0)\widehat{\bb}_{p,s}^{(1)}(x)'\bar{\bQ}^{-1}\E_n[\widehat\bb_{p,s}(x_i)\eta_{i,1}\psi(y_i,\eta_i)]
       \Big|\\[.5em]
       &{\lesssim_\P \Big(\frac{J\log n}{n}\Big)^{1/2}+J\mathfrak r_{\mathrm{BR},n}}
	+\|\widehat\evalw-\evalw\|\Big(1+J\Big(\frac{J\log n}{n}\Big)^{1/2}\Big).
       \end{align*}
 \end{enumerate}

\end{thm}

The following corollary is an immediate result of Lemma \ref{SA lem: uniform converge var part, GL} and Theorem \ref{SA thm: Bahadur representation, GL}, and hence its proof is omitted.
\begin{coro}[Uniform Convergence]\label{SA coro: uniform convergence, GL}
	Suppose that the conditions of Theorem \ref{SA thm: Bahadur representation, GL} hold and {$\frac{J\log n\,\ell_{\gamma,n}^4}{n}+\frac{\ell_{\gamma,n}^2}{J\log n}\lesssim 1$}. Then
    \[
    \sup_{x\in\mathcal{X}}|\widehat{\mu}^{(v)}(x)-\mu_0^{(v)}(x)|
    \lesssim_\P J^{v}\Big(\Big(\frac{J\log n}{n}\Big)^{1/2}+J^{-p-1}\Big).
    \]
    If, in addition, $\|\widehat\evalw-\evalw\|\lesssim_\P
    \Big(\frac{J\log n}{n}\Big)^{1/2}+J^{-p-1}$, then
    \begin{align*}
    &\sup_{x\in\mathcal{X}}|\widehat{\vartheta}(x,\widehat\evalw)-\vartheta_0(x,\evalw)|
	\lesssim_\P \Big(\frac{J\log n}{n}\Big)^{1/2}+J^{-p-1}\quad \text{and}\\
    & \sup_{x\in\mathcal{X}}|\widehat{\zeta}(x,\widehat\evalw)-\zeta_0(x,\evalw)|
	\lesssim_\P J\Big(\Big(\frac{J\log n}{n}\Big)^{1/2}+J^{-p-1}\Big).
 \end{align*}
\end{coro}

The next theorem shows that the proposed variance estimator is consistent.
\begin{thm}[Variance Estimate] \label{SA thm: meat matrix, GL}
{
	Suppose that the conditions of Theorem \ref{SA thm: Bahadur representation, GL} and Assumption \ref{SA Assumption HLE}(ii) hold. If
	$\frac{J^{\frac{\nu}{\nu-2}}(\log n)^{\frac{\nu}{\nu-2}}}{n}+\frac{J^{\frac{2\nu}{\nu-1}}(\log n)^{\frac{\nu}{\nu-1}}}{n}+\frac{J\log n\,\ell_{\gamma,n}^4}{n}+\frac{\ell_{\gamma,n}^2}{J\log n}=o(1)$, then
       \begin{align*}
        &\Big\|\widehat{\bSigma}-\bar\bSigma\Big\|
	\lesssim_\P J^{-p-1}+\Big(\frac{J\log n}{n^{1-\frac{2}{\nu}}}\Big)^{1/2},\\
        &\sup_{x\in\mathcal{X}}\Big|\widehat{\Omega}_{\mu^{(v)}}(x)-\bar\Omega_{\mu^{(v)}}(x)\Big|\lesssim_\P J^{1+2v}\Big(J^{-p-1}+\Big(\frac{J\log n}{n^{1-\frac{2}{\nu}}}\Big)^{1/2}\Big),\\
        &\sup_{x\in\mathcal{X}}\Big|\widehat{\Omega}_{\vartheta}(x)-\bar\Omega_{\vartheta}(x)\Big|\lesssim_\P J\Big(J^{-p-1}+\Big(\frac{J\log n}{n^{1-\frac{2}{\nu}}}\Big)^{1/2}+\|\widehat\evalw-\evalw\|\Big),\quad\text{and}\\
        &\sup_{x\in\mathcal{X}}\Big|\widehat{\Omega}_\zeta(x)-\bar\Omega_\zeta(x)\Big|\lesssim_\P J^3\Big(J^{-p-1}+\Big(\frac{J\log n}{n^{1-\frac{2}{\nu}}}\Big)^{1/2}+\|\widehat\evalw-\evalw\|\Big).
       \end{align*}
       If Assumption \ref{SA Assumption RP}(ii) also holds, then
        \begin{align*}
            &\Big\|\widehat{\bSigma}-\bSigma_0\Big\|
	    \lesssim_\P J^{-p-1}+\Big(\frac{J\log n}{n^{1-\frac{2}{\nu}}}\Big)^{1/2}+\mathfrak{r}_{\tt RP},\\
            &\sup_{x\in\mathcal{X}}\Big|\widehat{\Omega}_{\mu^{(v)}}(x)-\Omega_{\mu^{(v)}}(x)\Big|\lesssim_\P J^{1+2v}\Big(J^{-p-1}+\Big(\frac{J\log n}{n^{1-\frac{2}{\nu}}}\Big)^{1/2}+\mathfrak{r}_{\tt RP}\Big),\\
            &\sup_{x\in\mathcal{X}}\Big|\widehat{\Omega}_{\vartheta}(x)-\Omega_{\vartheta}(x)\Big|\lesssim_\P J\Big(J^{-p-1}+\Big(\frac{J\log n}{n^{1-\frac{2}{\nu}}}\Big)^{1/2}+\mathfrak{r}_{\tt RP}+\|\widehat\evalw-\evalw\|\Big),\quad \text{and}\\
            &\sup_{x\in\mathcal{X}}\Big|\widehat{\Omega}_\zeta(x)-\Omega_\zeta(x)\Big|\lesssim_\P J^3\Big(J^{-p-1}+\Big(\frac{J\log n}{n^{1-\frac{2}{\nu}}}\Big)^{1/2}+\mathfrak{r}_{\tt RP}+\|\widehat\evalw-\evalw\|\Big).
        \end{align*}
}
\end{thm}
\begin{remark}[Improvements over literature]
    Theorem \ref{SA thm: Bahadur representation, GL} and Corollary \ref{SA coro: uniform convergence, GL} construct the Bahadur representation and uniform convergence of nonlinear binscatter-based M-estimators, which improve upon prior results in the literature in at least two aspects.
    First, our results allow for random partitioning schemes, and the key condition imposed on the partition is Assumption \ref{SA Assumption RP}(i), i.e., the conditional independence between the partition and the outcome and the quasi-uniformity of the partition.
    The ``convergence'' of the random partition (Assumption \ref{SA Assumption RP}(ii)) is not required, which implies that our results can accommodate more complex partitioning schemes other than evenly-spaced or empirical-quantile-spaced partitions.

    Second, our results are established under weaker rate restrictions. {When $d_\gamma$ is fixed}, we require $J^{\frac{8}{3}}/n=o(1)$ up to $\log n$ terms when $\nu\geq 4$, thus accommodating IMSE-optimal piecewise constant binscatter estimators. In fact, for piecewise polynomials ($s=0$), we can show that the Bahadur representation still holds under $J/n=o(1)$ up to $\log n$ terms when a subexponential moment restriction holds for the ``score'' $\psi(y_i, \eta_i)$, which is analogous to the result for kernel-based estimators in the literature \citep[see, e.g.,][]{Kong-Linton-Xia_2010_ET}. For series estimators, similar results were established for particular choices of loss functions under more stringent conditions in the literature. For example, \citet*{Belloni-Chernozhukov-Chetverikov-FernandezVal_2019_JoE} considers series-based quantile regression, and Theorem 2 and Corollary 2 therein can be used to establish a Bahadur representation and uniform convergence of the resulting estimators under $J^4/n^{1-\varepsilon}=o(1)$ for some $\varepsilon>0$.

    When $d_\gamma$ may grow, the comparison is less direct. 
	We focus on the binscatter component and use a profiling-type argument, which requires the estimation error of $\widehat{\bgamma}$ to be negligible relative to that of the nonparametric component 
	and therefore imposes additional restrictions on $d_\gamma$. 
	By contrast, no separate negligibility restriction is needed in \cite{Belloni-Chernozhukov-Chetverikov-FernandezVal_2019_JoE},
	where $(\widehat{\bb}_{p,s}(x)',\bw')'$ is treated as a single generic increasing-dimensional regressor vector. See Remark \ref{SA rem: fast gamma rate} for details.
	Now, assuming that the required fast convergence of $\widehat{\bgamma}$ holds, when $\nu=4$, the rate restrictions imposed directly in our Bahadur representation reduce to $J^{8/3}/n=o(1)$ and $Jd_\gamma^2/\sqrt n\lesssim1$, ignoring logarithmic factors. More generally, these restrictions are $J^{2\nu/(\nu-1)}/n=o(1)$ and $Jd_\gamma^2/n^{1-2/\nu}\lesssim1$, and approach $J^2/n=o(1)$ and $Jd_\gamma^2/n\lesssim1$ as $\nu\to\infty$. 
	They still improve upon those obtained in \cite{Belloni-Chernozhukov-Chetverikov-FernandezVal_2019_JoE} by treating $(\widehat{\bb}_{p,s}(x)',\bw')'$ as a generic increasing-dimensional regressor vector.

	Note that the results in \cite{Belloni-Chernozhukov-Chetverikov-FernandezVal_2019_JoE} are slightly stronger than our Theorem \ref{SA thm: Bahadur representation, GL} in the sense that the expansion holds uniformly over both the evaluation point $x\in\mathcal{X}$ and the desired quantiles $u\in\mathcal{U}$ for a compact set of quantile indices $\mathcal{U}\subset(0,1)$. Our results regarding Bahadur representation can be extended to achieve the same level of uniformity. In general, the parameter of interest \eqref{SA eq: parameter} and the estimator \eqref{SA eq:binscatter-QMLE} are defined for each particular choice of the loss function within a function class $\mathcal{F}$. For the class of check functions used in quantile regression or other function classes with low complexity, it can be shown that the Bahadur representation still holds uniformly over the evaluation point $x\in\mathcal{X}$ and the loss function $\rho\in\mathcal{F}$ under rate restrictions similar to those in Theorem \ref{SA thm: Bahadur representation, GL}, thereby providing an improvement over the literature.
\end{remark}

\subsection{Pointwise Inference}\label{SA sec: pointwise inference}
Starting from this section,
we consider statistical inference on $\mu_0^{(v)}(x)$, $\vartheta_0(x,\evalw)$ and $\zeta_0(x,\evalw)$ based on the following Studentized $t$-statistics:
\begin{align*}
    &T_{\mu^{(v)},p}(x)=\frac{\widehat{\mu}^{(v)}(x)-\mu_0^{(v)}(x)}{\sqrt{\widehat{\Omega}_{\mu^{(v)}}(x)/n}},\\
    &T_{\vartheta,p}(x)=
    \frac{\widehat{\vartheta}(x,\widehat\evalw)-\vartheta_0(x,\evalw)}{\sqrt{\widehat{\Omega}_{\vartheta}(x)/n}}\quad \text{and}\\
    &T_{\zeta,p}(x)=\frac{\widehat{\zeta}(x,\widehat\evalw)-\zeta_0(x,\evalw)}{\sqrt{\widehat{\Omega}_{\zeta}(x)/n}}.
\end{align*}

The next theorem shows the pointwise asymptotic normality of the binscatter estimators.

\begin{thm}[Pointwise Asymptotic Distribution] \label{SA thm: pointwise normality, GL}
	{Suppose that the conditions of Theorem \ref{SA thm: Bahadur representation, GL} and Assumption \ref{SA Assumption HLE}(ii) hold},
	{$\sup_{x\in\mathcal{X},\,\bw\in\mathcal{W}}\E[|\psi(y_i,\eta_i)|^\nu\mid x_i=x,\bw_i=\bw]\lesssim 1$ for some $\nu\geq 3$}, and
	{$\frac{J^{\frac{\nu}{\nu-2}}(\log n)^{\frac{\nu}{\nu-2}}}{n}+\frac{J^{\frac{2\nu}{\nu-1}}(\log n)^{\frac{\nu}{\nu-1}}}{n}+\frac{J(\log n)^3\ell_{\gamma,n}^4}{n}+\frac{\ell_{\gamma,n}^2}{J^{p+1}}+nJ^{-2p-3}=o(1)$}.
	Then the following conclusions hold:
\begin{enumerate}[label=(\roman*)]
    \item For $\widehat\mu^{(v)}(x)$,
         \[
	\sup_{u\in\mathbb{R}}\Big|\P(T_{\mu^{(v)},p}(x)\leq u)-\Phi(u)\Big|=o(1),\quad
	\text{for each } x\in\mathcal{X}.
	\]

    \item  For $\widehat\vartheta(x,\widehat\evalw)$,
        if, in addition, $\|\widehat\evalw-\evalw\|=o_\P(\sqrt{J/n})$, then
	\[
	\sup_{u\in\mathbb{R}}\Big|\P(T_{\vartheta,p}(x)\leq u)-\Phi(u)\Big|=o(1)\quad
	\text{for each } x\in\mathcal{X}.
        \]

	\item For $\widehat\zeta(x,\widehat\evalw)$,
        if, in addition,
        \[
        \|\widehat\evalw-\evalw\|
        =o_\P\left(\sqrt{J^3/n} \wedge \sqrt{1/\log n}\right),
        \]
        then
	\[
         \sup_{u\in\mathbb{R}}\Big|\P(T_{\zeta,p}(x)\leq u)-\Phi(u)\Big|=o(1)\quad
	\text{for each } x\in\mathcal{X}.
	\]
 \end{enumerate}
\end{thm}

\begin{remark}[Improvements over literature]
    The result in this subsection is new to the literature, even in the case of non-random partitioning and without covariate adjustments, because it takes advantage of the specific binscatter structure (i.e., locally bounded series basis). The closest antecedent in the literature is  \citet{Belloni-Chernozhukov-Chetverikov-FernandezVal_2019_JoE}, which focuses on series-based quantile regression only. Furthermore, relative to prior work, our results allow for more general partitioning schemes, formally take into account the potential randomness of the partition, and account for the semi-linear regression estimation structure. The key condition imposed on the partition for pointwise inference is Assumption \ref{SA Assumption RP}(i), and the ``convergence'' of the random partition is not required.
\end{remark}

\subsection{Integrated Mean Squared Error}\label{SA section: IMSE, GL}
In this section we give a Nagar-type approximate IMSE expansion for each of the three estimators $\widehat\mu^{(v)}(x)$, $\widehat\vartheta(x,\widehat\evalw)$ and $\widehat\zeta(x,\widehat\evalw)$, with explicit characterization of the leading constants. Define
\begin{align*}
	r_{0,v}^\star(x)
		=\frac{\mu_0^{(p+1)}(x)}{(p+1-v)!}h_x^{p+1-v}\mathscr{E}_{p+1-v}
		\Big(\frac{x-\tau_x^{\mathtt{L}}}{h_x}\Big).
\end{align*}
Here $\mathscr{E}_m(\cdot)$ is the $m$th Bernoulli polynomial for each $m\in\mathbb{Z}_+$, $\tau_x^{\mathtt{L}}$ is the start of the interval in the non-random partition $\Delta_0$ containing $x$, and $h_x$ denotes its length.

\begin{thm}[IMSE] \label{SA thm: IMSE, GL}
    Suppose that the conditions of Theorem \ref{SA thm: Bahadur representation, GL} and Assumption \ref{SA Assumption RP}(ii) hold,
	and $\mu_0(\cdot)$ is $(p+2)$-times continuously differentiable with $\sup_{n\geq 1}\sup_{x\in\mathcal X}|\mu_0^{(p+2)}(x)|\lesssim 1$. 
	If $s>0$, assume in addition that $\max_{2\leq j\leq J}|h_j-h_{j-1}|=o(J^{-1})$. Let $\omega(x)$ be a continuous weighting function over $\mathcal{X}$ bounded away from zero. Also, assume that {$\frac{J^{\frac{\nu}{\nu-2}}\log n}{n}+\frac{J^{\frac{2\nu}{\nu-1}}(\log n)^{\frac{\nu}{\nu-1}}}{n}+\frac{J(\log n)^3\ell_{\gamma,n}^4}{n}+\frac{\ell_{\gamma,n}^2}{J}=o(1)$}.
 \begin{enumerate}[label=(\roman*)]
     \item For $\widehat{\mu}^{(v)}(x)$,
     \[
	\int_{\mathcal{X}}\Big(\widehat{\mu}^{(v)}(x)-\mu_0^{(v)}(x)\Big)^2\omega(x)\d{x}
	=\mathtt{AISE}_{\mu^{(v)}}+o_\P\Big(\frac{J^{1+2v}}{n}+J^{-2(p+1-v)}\Big)
	\]
	where
	\[\begin{split}
	&\E[\mathtt{AISE}_{\mu^{(v)}}|\bX, \bW,\widehat\Delta]=\frac{J^{1+2v}}{n}\mathscr{V}_n(p,s,v)+J^{-2(p+1-v)}\mathscr{B}_n(p,s,v)
	+o_\P\Big(\frac{J^{1+2v}}{n}+J^{-2(p+1-v)}\Big),\\
	&\mathscr{V}_n(p,s,v):=J^{-(1+2v)}\tr\Big(
		\bQ_0^{-1}\bSigma_0\bQ_0^{-1}
		\int_{\mathcal{X}}\bb_{p,s}^{(v)}(x)\bb_{p,s}^{(v)}(x)'\omega(x)\d{x}\Big)
		\asymp 1, \\
	&\mathscr{B}_n(p,s,v):=J^{2p+2-2v}\int_{\mathcal{X}}
	\Big(r^\star_{0,v}(x)-\bb_{p,s}^{(v)}(x)'\bQ_0^{-1}\E[\bb_{p,s}(x_i)\varkappa(x_i,\bw_i)r^\star_{0,0}(x_i)]\Big)^2\omega(x)\d{x}\lesssim 1.
	\end{split}
	\]

        \item For $\widehat\vartheta(x,\widehat\evalw)$,
         if $\|\widehat\evalw-\evalw\|=o_\P(\sqrt{J/n}+J^{-p-1})$, then
        \[
	\int_{\mathcal{X}}\Big(\widehat{\vartheta}(x,\widehat\evalw)-\vartheta_0(x,\evalw)\Big)^2\omega(x)\d{x}
	=\mathtt{AISE}_{\vartheta}+o_\P\Big(\frac{J}{n}+J^{-2(p+1)}\Big)
	\]
	where
	\[\begin{split}
	&\E[\mathtt{AISE}_{\vartheta}|\bX, \bW,\widehat\Delta]=\frac{J}{n}\mathscr{V}_n(p,s)+J^{-2(p+1)}\mathscr{B}_n(p,s)
	+o_\P\Big(\frac{J}{n}+J^{-2(p+1)}\Big),\\
	&\mathscr{V}_n(p,s):=J^{-1}\tr\Big(
		\bQ_0^{-1}\bSigma_0\bQ_0^{-1}
		\int_{\mathcal{X}}\eta_{0,1}(x,\evalw)^2
  \bb_{p,s}(x)\bb_{p,s}(x)'\omega(x)\d{x}\Big)
		\asymp 1, \\
	&\mathscr{B}_n(p,s):=J^{2p+2}\int_{\mathcal{X}}
	\Big[\eta_{0,1}(x,\evalw)
        \Big(r^\star_{0,0}(x)-\bb_{p,s}(x)'\bQ_0^{-1}\E[\bb_{p,s}(x_i)\varkappa(x_i,\bw_i)r^\star_{0,0}(x_i)]\Big)\Big]^2\omega(x)\d{x}\lesssim 1.
	\end{split}
	\]

        \item For $\widehat\zeta(x,\widehat\evalw)$,
         if
         \[
         \|\widehat\evalw-\evalw\|
         =o_\P\left(
         \frac{\sqrt{J^3/n+J^{-2p}}}
         {1+\sqrt{J^3\log n/n}}
         \right),
         \]
         then
    \[
	\int_{\mathcal{X}}\Big(\widehat{\zeta}(x,\widehat\evalw)-\zeta_0(x,\evalw)\Big)^2\omega(x)\d{x}
	=\mathtt{AISE}_\zeta+o_\P\Big(\frac{J^{3}}{n}+J^{-2p}\Big)
	\]
	where
	\[\begin{split}
	&\E[\mathtt{AISE}_\zeta|\bX, \bW,\widehat\Delta]=\frac{J^{3}}{n}\mathscr{V}_n(p,s)
            +J^{-2p}\mathscr{B}_n(p,s)
	+o_\P\Big(\frac{J^{3}}{n}+J^{-2p}\Big),\\
	&\mathscr{V}_n(p,s):=J^{-3}\tr\Big(
		\bQ_0^{-1}\bSigma_0\bQ_0^{-1}
		\int_{\mathcal{X}}\eta_{0,1}(x,\evalw)^2\bb_{p,s}^{(1)}(x)\bb_{p,s}^{(1)}(x)'\omega(x)\d{x}\Big)
		\asymp 1, \\
	&\mathscr{B}_n(p,s):=J^{2p}\int_{\mathcal{X}}
	\Big[\eta_{0,1}(x,\evalw)
 \Big(r^\star_{0,1}(x)-\bb_{p,s}^{(1)}(x)'\bQ_0^{-1}\E[\bb_{p,s}(x_i)\varkappa(x_i,\bw_i)r^\star_{0,0}(x_i)]\Big)\Big]^2\omega(x)\d{x}\lesssim 1.
	\end{split}
	\]
 \end{enumerate}
\end{thm}

\begin{remark}[Smoothness for the IMSE Expansion]
The bound on $\mu_0^{(p+1)}$ in Assumption \ref{SA Assumption GL}(v) suffices to give an approximation error rate. 
However, for IMSE expansion, uniform equicontinuity of $\mu_0^{(p+1)}$ over $n$ is needed to guarantee the remainder is of smlaler order; 
the additional assumption that $\mu_0^{(p+2)}$ is uniformly bounded in Theorem \ref{SA thm: IMSE, GL} is a simpler sufficient condition. 
For fixed $\mu_0$, continuity of $\mu_0^{(p+1)}$ suffices.
\end{remark}

In general, $\mathscr{B}_n(p,s,v)\gtrsim 1$ (see Remark SA-3.7 in \cite{Cattaneo-Crump-Farrell-Feng_2024_AER}), and thus the above theorem implies that the (approximate) IMSE-optimal number of bins satisfies that $J_{\mathtt{AIMSE}}\asymp n^{\frac{1}{2p+3}}$. Relying on the IMSE expansion in Theorem \ref{SA thm: IMSE, GL}, one may design a data-driven procedure to select the IMSE-optimal number of bins for nonlinear binscatter-based M-estimators.

\begin{remark}[Leading Bias under Quantile-Spaced Partitions]
Suppose that $\Delta_0$ is the population quantile-spaced partition, so that $\tau_j=F_X^{-1}(j/J)$. Assumption \ref{SA Assumption DGP}(ii) implies that $\max_{2\leq j\leq J}|h_j-h_{j-1}|=O(J^{-2})$, so the additional condition in Theorem \ref{SA thm: IMSE, GL} is satisfied. By Assumption \ref{SA Assumption DGP}(ii), uniformly over $x\in\mathcal X$,
\[
h_x=\frac{1}{Jf_X(x)}\{1+o(1)\}.
\]
Consequently,
\[
r_{0,v}^\star(x)
=\frac{J^{-p-1+v}\mu_0^{(p+1)}(x)}{(p+1-v)!f_X(x)^{p+1-v}}
\mathscr E_{p+1-v}\left(\frac{x-\tau_x^{\mathtt L}}{h_x}\right)
+o(J^{-p-1+v})
\]
uniformly over $x\in\mathcal X$. Thus, the familiar density-based leading bias is recovered as a special case, while the definition above also applies to other limiting partitions covered by Theorem \ref{SA thm: IMSE, GL}. The variance expression in Theorem \ref{SA thm: IMSE, GL} already accommodates quantile-spaced partitions through the corresponding population basis and matrices.
\end{remark}

\begin{remark}[Improvements over literature]
	The results in this subsection are new to the literature, even in the case of non-random partitioning and without covariate-adjustments, for both nonlinear series estimators and binscatter (piecewise polynomials and splines) nonlinear series estimators in particular. Furthermore, our results allow for random partitioning schemes, formally take into account the potential randomness of the partition, and account for the semi-linear regression estimation structure. We highlight the key conditions imposed on the partition (Assumption \ref{SA Assumption RP}) for the approximate IMSE expansion.
    The ``convergence'' of the random partition (Assumption \ref{SA Assumption RP}(ii)) is needed to derive the non-random variance and bias constants $\mathscr{V}_n(p,s)$ and $\mathscr{B}_n(p,s)$. When $s>0$, Theorem \ref{SA thm: IMSE, GL} also requires adjacent bin widths in $\Delta_0$ to be asymptotically equal.
\end{remark}

\subsection{Uniform Inference}\label{SA section: uniform inference}
Let $(a_n: n\geq 1)$ be a sequence of positive non-vanishing constants.
We will first show that the (feasible) Studentized $t$-statistic processes
$T_{\mu^{(v)},p}(\cdot)$, $T_{\vartheta,p}(\cdot)$ and $T_{\zeta,p}(\cdot)$ can be approximated by Gaussian processes in a proper sense at certain rate.

\begin{thm}[Strong Approximation] \label{SA thm: strong approximation, GL}
	{Suppose that the conditions of Theorem \ref{SA thm: Bahadur representation, GL} and Assumption \ref{SA Assumption HLE}(ii) hold},
	{\[
	\begin{aligned}
	&\frac{J(\log n)^2}{n^{1-\frac{2}{\nu}}}+\Big(\frac{J(\log n)^3\ell_{\gamma,n}^4}{n}\Big)^{1/2}+nJ^{-2p-3}+\frac{\ell_{\gamma,n}^2}{J^{p+1}}+nJ^{-1}d_\gamma\mathfrak{r}_\gamma^2=o(a_n^{-2}),\\
	&\frac{J^{\frac{2\nu}{\nu-1}}(\log n)^{\frac{\nu}{\nu-1}}}{n}=o(1).
	\end{aligned}
	\]}
        Then the following conclusions hold:

       \begin{enumerate}[label=(\roman*)]
	\item  On a properly enriched probability space, there exists some $K_{p,s}$-dimensional standard normal random vector $\bN_{K_{p,s}}$ such that for any $\xi>0$,
	\[
	\P\Big(\sup_{x\in\mathcal{X}}|T_{\mu^{(v)},p}(x)-
        \bar{Z}_{\mu^{(v)},p}(x)|>\xi a_n^{-1}\Big)=o(1), \quad
	\bar{Z}_{\mu^{(v)},p}(x)=\frac{\widehat{\bb}_{p,0}^{(v)}(x)'\widehat{\bT}_s'\bar\bQ^{-1}\bar\bSigma^{1/2}}{\sqrt{\bar\Omega_{\mu^{(v)}}(x)}}\bN_{K_{p,s}}.
	\]
       If Assumption \ref{SA Assumption RP}(ii) also holds with $\mathfrak{r}_{\tt RP}=o(a_n^{-1}(\log n)^{-1/2})$, then
	\[
	\P\Big(\sup_{x\in\mathcal{X}}|T_{\mu^{(v)},p}(x)-Z_{\mu^{(v)},p}(x)|>\xi a_n^{-1}\Big)=o(1), \quad
	Z_{\mu^{(v)},p}(x)=\frac{\widehat{\bb}_{p,0}^{(v)}(x)'\bT_s'\bQ_0^{-1}\bSigma_0^{1/2}}{\sqrt{\Omega_{\mu^{(v)}}(x)}}\bN_{K_{p,s}}.
	\]

      \item  If $\|\widehat\evalw-\evalw\|=o_\P(a_n^{-1}\sqrt{J/n})$, then
      on a properly enriched probability space
      there exists some $K_{p,s}$-dimensional standard normal random vector $\bN_{K_{p,s}}$ such that for any $\xi>0$,
	\[
         \P\bigg(\sup_{x\in\mathcal{X}}|T_{\vartheta,p}(x)-
	\bar{Z}_{\vartheta,p}(x)|>\xi a_n^{-1}\bigg)=o(1),\quad
          \bar{Z}_{\vartheta,p}(x)=\frac{\widehat{\bb}_{p,0}(x)'\widehat\bT_s'\eta_{0,1}(x,\evalw)
          \bar\bQ^{-1}}{\sqrt{\bar\Omega_\vartheta(x)}}\bar\bSigma^{1/2}\bN_{K_{p,s}}.
	\]
       If Assumption \ref{SA Assumption RP}(ii) also holds with $\mathfrak{r}_{\tt RP}=o(a_n^{-1}(\log n)^{-1/2})$, then
	\[
         \P\bigg(\sup_{x\in\mathcal{X}}|T_{\vartheta,p}(x)-
	Z_{\vartheta,p}(x)|>\xi a_n^{-1}\bigg)=o(1),\quad
          Z_{\vartheta,p}(x)=\frac{\widehat{\bb}_{p,0}(x)'\bT_s'\eta_{0,1}(x,\evalw)\bQ_0^{-1}}{\sqrt{\Omega_\vartheta(x)}}\bSigma_0^{1/2}\bN_{K_{p,s}}.
	\]

       \item If
       \[
	       \|\widehat\evalw-\evalw\|
	       =o_\P\left(
	       \frac{a_n^{-1}}{\sqrt{n/J^3}+\sqrt{\log n}}
	       \right),
	       \]
       then on a properly enriched probability space there exists some $K_{p,s}$-dimensional standard normal random vector $\bN_{K_{p,s}}$ such that for any $\xi>0$,
       \[
       \P\bigg(\sup_{x\in\mathcal{X}}|T_{\zeta,p}(x)-
       \bar{Z}_{\zeta,p}(x)|>\xi a_n^{-1}\bigg)=o(1),\quad
       \bar{Z}_{\zeta,p}(x)=\frac{\widehat{\bb}_{p,0}^{(1)}(x)'\widehat\bT_s'\eta_{0,1}(x,\evalw)\bar\bQ^{-1}}{\sqrt{\bar\Omega_{\zeta}(x)}}\bar\bSigma^{1/2}\bN_{K_{p,s}}.
       \]
       If Assumption \ref{SA Assumption RP}(ii) also holds with $\mathfrak{r}_{\tt RP}=o(a_n^{-1}(\log n)^{-1/2})$, then
            \[
       \P\bigg(\sup_{x\in\mathcal{X}}|T_{\zeta,p}(x)-
       Z_{\zeta,p}(x)|>\xi a_n^{-1}\bigg)=o(1),\quad
       Z_{\zeta,p}(x)=\frac{\widehat{\bb}_{p,0}^{(1)}(x)'\bT_s'\eta_{0,1}(x,\evalw)\bQ_0^{-1}}{\sqrt{\Omega_{\zeta}(x)}}\bSigma_0^{1/2}\bN_{K_{p,s}}.
       \]
      \end{enumerate}

\end{thm}

The approximating processes $\bar{Z}_{\mu^{(v)},p}(\cdot)$, $\bar{Z}_{\vartheta,p}(\cdot)$ and $\bar{Z}_{\zeta,p}(\cdot)$ are Gaussian processes conditional on $\bX$, $\bW$ and $\widehat{\Delta}$, and
$Z_{\mu^{(v)},p}(\cdot)$, $Z_{\vartheta,p}(\cdot)$ and $Z_{\zeta,p}(\cdot)$ are Gaussian processes conditional on $\widehat{\Delta}$ by construction. In practice, one can replace all unknowns in $\bar{Z}_{\mu^{(v)},p}(\cdot)$, $\bar{Z}_{\vartheta, p}(\cdot)$ and $\bar{Z}_{\zeta, p}(\cdot)$ (or $Z_{\mu^{(v)},p}(\cdot)$, $Z_{\vartheta, p}(\cdot)$ and $Z_{\zeta, p}(\cdot)$) by their sample analogues, and then construct the following feasible (conditional) Gaussian processes:
\begin{align*}
&\widehat{Z}_{\mu^{(v)},p}(x)=\frac{\widehat{\bb}_{p,0}^{(v)}(x)'\widehat{\bT}_s'\widehat{\bQ}^{-1}\widehat{\bSigma}^{1/2}}{\sqrt{\widehat{\Omega}_{\mu^{(v)}}(x)}}\bN_{K_{p,s}}^\star=
\frac{\widehat{\bb}_{p,s}^{(v)}(x)'\widehat{\bQ}^{-1}\widehat{\bSigma}^{1/2}}{\sqrt{\widehat{\Omega}_{\mu^{(v)}}(x)}}\bN_{K_{p,s}}^\star,\\
&\widehat{Z}_{\vartheta,p}(x)=\frac{\widehat{\bb}_{p,0}(x)'\widehat{\bT}_s'\widehat\eta_{0,1}(x,\widehat\evalw)\widehat{\bQ}^{-1}\widehat{\bSigma}^{1/2}}{\sqrt{\widehat{\Omega}_\vartheta(x)}}\bN_{K_{p,s}}^\star=
\frac{\widehat{\bb}_{p,s}(x)'\widehat\eta_{0,1}(x,\widehat\evalw)\widehat{\bQ}^{-1}\widehat{\bSigma}^{1/2}}{\sqrt{\widehat{\Omega}_\vartheta(x)}}\bN_{K_{p,s}}^\star,\\
&\widehat{Z}_{\zeta,p}(x)=\frac{\widehat{\bb}_{p,0}^{(1)}(x)'
\widehat{\bT}_s'\widehat\eta_{0,1}(x,\widehat\evalw)\widehat{\bQ}^{-1}\widehat{\bSigma}^{1/2}}{\sqrt{\widehat{\Omega}_\zeta(x)}}\bN_{K_{p,s}}^\star=
\frac{\widehat{\bb}_{p,s}^{(1)}(x)'\widehat\eta_{0,1}(x,\widehat\evalw)\widehat{\bQ}^{-1}\widehat{\bSigma}^{1/2}}{\sqrt{\widehat{\Omega}_\zeta(x)}}\bN_{K_{p,s}}^\star,
\end{align*}
where $\bN_{K_{p,s}}^\star$ denotes a $K_{p,s}$-dimensional standard normal vector independent of the data $\bD$ and the partition $\widehat{\Delta}$.

For ease of presentation, we will always require a fast convergence rate of $\widehat\evalw$ hereafter: $\|\widehat\evalw-\evalw\|=o_\P(a_n^{-1}\sqrt{J/n})$.
Nevertheless, note that as shown in Theorem \ref{SA thm: strong approximation, GL}, such a rate restriction on $\widehat\evalw$ can be different for inference of $\vartheta_0(x,\evalw)$ and $\zeta_0(x,\evalw)$ and are unnecessary for inference of $\mu_0^{(v)}(x)$.
Also, in the following theorem, the infeasible processes $\bar Z$ are understood as conditionally equal-in-distribution copies driven by the same external Gaussian vector $\bN_{K_{p,s}}^\star$ as their feasible counterparts, on a sufficiently rich probability space.

\begin{thm}[Plug-in Approximation] \label{SA thm: plug-in approx, GL}
        {Suppose that the conditions of Theorem \ref{SA thm: Bahadur representation, GL} and Assumption \ref{SA Assumption HLE}(ii) hold},
	\begin{eqnarray*}
	&{\frac{J(\log n)^2}{n^{1-\frac{2}{\nu}}}
	+\Big(\frac{J(\log n)^3\ell_{\gamma,n}^4}{n}\Big)^{1/2}
	+nJ^{-2p-3}
	+\frac{\ell_{\gamma,n}^2}{J^{p+1}}+nJ^{-1}d_\gamma\mathfrak{r}_\gamma^2}
	=o(a_n^{-2}),\\
        &\frac{J^{\frac{2\nu}{\nu-1}}(\log n)^{\frac{\nu}{\nu-1}}}{n}=o(1), \quad\text{and}\quad \|\widehat\evalw-\evalw\|=o_\P(a_n^{-1}\sqrt{J/n}).
	\end{eqnarray*}
 Then on a properly enriched probability space, there exists a $K_{p,s}$-dimensional standard normal random vector $\bN_{K_{p,s}}^\star$ independent of $\bD$ and $\widehat{\Delta}$ such that for any $\xi>0$,
 \begin{enumerate}[label=(\roman*)]
	\item
        $
	\P\Big(\sup_{x\in\mathcal{X}}|\widehat{Z}_{\mu^{(v)},p}(x)-\bar{Z}_{\mu^{(v)},p}(x)|>\xi a_n^{-1}\Big|\bD,\widehat{\Delta}\Big)=o_\P(1),
	$

 \item
        $
	\P\Big(\sup_{x\in\mathcal{X}}|\widehat{Z}_{\vartheta,p}(x)-\bar{Z}_{\vartheta,p}(x)|>\xi         a_n^{-1}\Big|\bD,\widehat{\Delta}\Big)=o_\P(1),
	$
    \item
        $
	\P\Big(\sup_{x\in\mathcal{X}}|\widehat{Z}_{\zeta,p}(x)-\bar{Z}_{\zeta,p}(x)|>\xi                 a_n^{-1}\Big|\bD,\widehat{\Delta}\Big)=o_\P(1).
	$
  \end{enumerate}
  If Assumption \ref{SA Assumption RP}(ii) also holds with $\mathfrak{r}_{\tt RP}=o(a_n^{-1}(\log n)^{-1/2})$, then
   \begin{enumerate}[label=(\roman*)]\setcounter{enumi}{3}
	\item
        $
	\P\Big(\sup_{x\in\mathcal{X}}|\widehat{Z}_{\mu^{(v)},p}(x)-Z_{\mu^{(v)},p}(x)|>\xi a_n^{-1}\Big|\bD,\widehat{\Delta}\Big)=o_\P(1),
	$

 \item
        $
	\P\Big(\sup_{x\in\mathcal{X}}|\widehat{Z}_{\vartheta,p}(x)-Z_{\vartheta,p}(x)|>\xi         a_n^{-1}\Big|\bD,\widehat{\Delta}\Big)=o_\P(1),
	$
    \item
        $
	\P\Big(\sup_{x\in\mathcal{X}}|\widehat{Z}_{\zeta,p}(x)-Z_{\zeta,p}(x)|>\xi                 a_n^{-1}\Big|\bD,\widehat{\Delta}\Big)=o_\P(1).
	$
  \end{enumerate}
\end{thm}

\begin{remark}[Improvements over literature]
	Theorems \ref{SA thm: strong approximation, GL} and \ref{SA thm: plug-in approx, GL} provide empirical researchers with powerful tools for uniform inference based on binscatter methods.
	Importantly, we allow for random partitioning schemes, formally take into account the potential randomness of the partition, and construct a novel strong approximation of nonlinear binscatter-based M-estimators under mild rate restrictions.
	{For fixed $d_\gamma$}, $a_n=\sqrt{\log n}$ and $\nu\geq 4$, we require $J^{\frac{8}{3}}/n=o(1)$, up to $\log n$ terms. In the literature, similar results were only available in some special cases under stringent rate restrictions.
	For instance,  \cite{Belloni-Chernozhukov-Chetverikov-FernandezVal_2019_JoE} considers strong approximations of more general series-based quantile regression estimators. For the binscatter basis considered in this paper, their Theorem 11 can be applied to construct strong approximation of the $t$-statistic process based on pivotal coupling that achieves the approximation rate $a_n^{-1}=n^{-\varepsilon'}$ under $J^4/n^{1-\varepsilon}=o(1)$ for some constants $\varepsilon, \varepsilon'>0$, whereas their Theorem 12 can be used to construct strong approximation based on Gaussian processes under  $J^{5}/n^{1-\varepsilon}=o(1)$. It should be noted that their notion of strong approximation is stronger than ours in the sense that it holds uniformly over both the evaluation point $x\in\mathcal{X}$ and the desired quantile $u\in\mathcal{U}$ for a compact set of quantile indices $\mathcal{U}\subset(0,1)$. On the other hand, our methods allow for other loss functions (e.g., Huber regression), a large class of random partitions, and semi-linear covariate adjustment, leading to new results that were previously unavailable in the literature.
\end{remark}

Theorems \ref{SA thm: strong approximation, GL} and \ref{SA thm: plug-in approx, GL} offer a way to approximate the distribution of the \textit{whole} $t$-statistic process based on $\widehat\mu^{(v)}(\cdot)$, $\widehat\vartheta(\cdot, \widehat\evalw)$ or  $\widehat{\zeta}(\cdot,\widehat\evalw)$. 
A direct application of these results is the distributional approximation to the suprema of these $t$-statistic processes.
In subsequent results we take $a_n=\sqrt{\log J}$.

\begin{thm}[Supremum Approximation] \label{SA thm: sup approx}
	{Suppose that the conditions of Theorem \ref{SA thm: Bahadur representation, GL} and Assumptions \ref{SA Assumption HLE}(ii) and \ref{SA Assumption RP}(ii) hold},
	\begin{eqnarray*}
	&{\frac{J(\log n)^2}{n^{1-\frac{2}{\nu}}}
	+\Big(\frac{J(\log n)^3\ell_{\gamma,n}^4}{n}\Big)^{1/2}
	+nJ^{-2p-3}+\frac{\ell_{\gamma,n}^2}{J^{p+1}}
	+nJ^{-1}d_\gamma\mathfrak{r}_\gamma^2}
	=o((\log J)^{-1}),\\
        &\frac{J^{\frac{2\nu}{\nu-1}}(\log n)^{\frac{\nu}{\nu-1}}}{n}=o(1),\quad
        \|\widehat\evalw-\evalw\|=o_\P\Big(\sqrt{\frac{J}{n\log J}}\Big),
        \quad\text{and} \quad
        \mathfrak{r}_{\tt RP}=o\Big(\frac{1}{\sqrt{\log n\,\log J}}\Big).
        \end{eqnarray*}
        Then,
	\begin{align*}
        &\sup_{u\in\mathbb{R}}\Big|\P\Big(\sup_{x\in\mathcal{X}}
	|T_{\mu^{(v)},p}(x)|\leq u\Big)-
	\P\Big(\sup_{x\in\mathcal{X}}|\widehat{Z}_{\mu^{(v)},p}(x)|\leq u\Big|\bD,\widehat{\Delta}\Big)\Big|=o_\P(1),\\
        &\sup_{u\in\mathbb{R}}\Big|\P\Big(\sup_{x\in\mathcal{X}}
	|T_{\vartheta,p}(x)|\leq u\Big)-
	\P\Big(\sup_{x\in\mathcal{X}}|\widehat{Z}_{\vartheta,p}(x)|\leq u\Big|\bD,\widehat{\Delta}\Big)\Big|=o_\P(1),\quad \text{and}\\
        &\sup_{u\in\mathbb{R}}\Big|\P\Big(\sup_{x\in\mathcal{X}}
	|T_{\zeta,p}(x)|\leq u\Big)-
	\P\Big(\sup_{x\in\mathcal{X}}|\widehat{Z}_{\zeta,p}(x)|\leq u\Big|\bD,\widehat{\Delta}\Big)\Big|=o_\P(1).
	\end{align*}
\end{thm}

\subsection{Confidence Bands}\label{SA section: confidence band}

Let
\begin{align*}
&\widehat{I}_{\mu^{(v)},p}(x)=\Big[\widehat{\mu}^{(v)}(x)\pm \cval_{\mu^{(v)}}\sqrt{\widehat{\Omega}_{\mu^{(v)}}(x)/n}\Big],\\
&\widehat{I}_{\vartheta,p}(x,\evalw)=\Big[\widehat{\vartheta}(x,\widehat\evalw)\pm \cval_\vartheta\sqrt{\widehat{\Omega}_{\vartheta}(x)/n}\Big]\quad \text{and}\\
&\widehat{I}_{\zeta,p}(x,\evalw)=\Big[\widehat{\zeta}(x,\widehat\evalw)\pm \cval_\zeta\sqrt{\widehat{\Omega}_{\zeta}(x)/n}\Big]
\end{align*}
be confidence bands for $\mu_0^{(v)}(\cdot)$, $\vartheta_0(\cdot,\evalw)$ and $\zeta_0(\cdot,\evalw)$ respectively, where
$\cval_{\mu^{(v)}}$, $\cval_\vartheta$ and $\cval_\zeta$ are corresponding critical values to be specified. Recall that $\evalw$ here is taken as a fixed evaluation point for the control variables, and these bands are constructed based on a certain choice of $J$ and the $p$th-order binscatter basis. Using the previous results, we have the following theorem.

\begin{thm}\label{SA thm: CB}
	{Suppose that the conditions of Theorem \ref{SA thm: Bahadur representation, GL} and Assumption \ref{SA Assumption HLE}(ii) hold},
	\begin{eqnarray*}
	&{\frac{J(\log n)^2}{n^{1-\frac{2}{\nu}}}
	+\Big(\frac{J(\log n)^3\ell_{\gamma,n}^4}{n}\Big)^{1/2}
	+nJ^{-2p-3}+\frac{\ell_{\gamma,n}^2}{J^{p+1}}
	+nJ^{-1}d_\gamma\mathfrak{r}_\gamma^2}
	=o((\log J)^{-1}),\\
        &\frac{J^{\frac{2\nu}{\nu-1}}(\log n)^{\frac{\nu}{\nu-1}}}{n}=o(1),\quad\text{and}\quad\|\widehat\evalw-\evalw\|=o_\P\Big(\sqrt{\frac{J}{n\log J}}\Big).
        \end{eqnarray*}
    \begin{enumerate}[label=(\roman*)]
        \item  If $\cval_{\mu^{(v)}}=\inf\Big\{c\in\mathbb{R}_+:\P[\sup_{x\in\mathcal{X}}|\widehat{Z}_{\mu^{(v)},p}(x)|\leq c \;|\bD,\widehat{\Delta}]\geq 1-\alpha\Big\}$, then
	\[
	\P\Big[\mu_0^{(v)}(x)\in\widehat{I}_{\mu^{(v)},p}(x),\text{ for all }x\in\mathcal{X}\Big]=1-\alpha+o(1).
	\]

        \item If $\cval_\vartheta=\inf\Big\{c\in\mathbb{R}_+:\P[\sup_{x\in\mathcal{X}}|\widehat{Z}_{\vartheta,p}(x)|\leq c \;|\bD,\widehat{\Delta}]\geq 1-\alpha\Big\}$, then
	\[
	\P\Big[\vartheta_0(x,\evalw)\in\widehat{I}_{\vartheta,p}(x,\evalw),\text{ for all }x\in\mathcal{X}\Big]=1-\alpha+o(1).
        \]
        \item If $\cval_\zeta=\inf\Big\{c\in\mathbb{R}_+:\P[\sup_{x\in\mathcal{X}}|\widehat{Z}_{\zeta,p}(x)|\leq c \;|\bD,\widehat{\Delta}]\geq 1-\alpha\Big\}$, then
	\[
	\P\Big[\zeta_0(x,\evalw)\in\widehat{I}_{\zeta,p}(x,\evalw),\text{ for all }x\in\mathcal{X}\Big]=1-\alpha+o(1).
 \]
    \end{enumerate}
\end{thm}

\begin{remark}
	\setlength{\emergencystretch}{2em}
	The above results construct valid uniform confidence bands for nonlinear binscatter-based M-estimators under mild rate restrictions. Specifically, {when $d_\gamma$ is fixed and} $\nu\geq 4$, we require $J^{\frac{8}{3}}/n=o(1)$, up to $\log n$ terms. In contrast, \cite{Belloni-Chernozhukov-Chetverikov-FernandezVal_2019_JoE} considers more general series-based quantile regression estimators, and Theorem 15 therein can be used to construct confidence bands for binscatter estimators via various resampling methods under $J^4/n^{1-\varepsilon}=o(1)$ for some $\varepsilon>0$. Furthermore, our results allow for random partitioning schemes, formally accounting for their randomness and generic structure. The key condition imposed on the partition for the validity of confidence bands is Assumption \ref{SA Assumption RP}(i), but the ``convergence'' of the random partition (Assumption \ref{SA Assumption RP}(ii)) is not necessary.
\end{remark}

\subsection{Parametric Specification Tests}\label{SA section: test parametric spec}
We can test parametric specifications of and shape restrictions on
$\mu_0^{(v)}(x)$, $\vartheta_0(x,\evalw)$ and $\zeta_0(x,\evalw)$.
{Before we present the results, some clarifications are needed.
We will use $\bar{\bgamma}\in\mathbb R^{d_\gamma}$ to denote the (non-random) benchmark coefficient
corresponding to $\widehat{\bgamma}$. $\bar{\bgamma}$ may vary with $n$, although this dependence is suppressed.
Under the coordinatewise support bound on $\bw$, the contribution of estimating the benchmark index is bounded by
\[
\sqrt{d_\gamma}\|\widetilde{\bgamma}-\bar{\bgamma}\|
+\|\widehat\evalw-\evalw\|\|\bar{\bgamma}\|,
\]
so the displayed function-level convergence conditions below can, for example,
be verified by requiring this quantity, together with the contribution from estimating $\bar{\btheta}$,
to have the corresponding stated order.
}

We first consider parametric specification tests.
We introduce the following tests:
\begin{alignat*}{2}
&\dot{\mathsf{H}}_0^{\mu^{(v)}}:\quad &&\sup_{x\in\mathcal{X}} \Big|\mu_0^{(v)}(x) - m^{(v)}(x;\btheta)\Big|=0, \quad \text{ for some } \btheta, \qquad vs.\\
&\dot{\mathsf{H}}_\text{A}^{\mu^{(v)}}: &&\sup_{x\in\mathcal{X}} \Big|\mu_0^{(v)}(x) - m^{(v)}(x;\btheta)\Big|>0, \quad \text{ for all } \btheta.
\end{alignat*}
where $m(x;\btheta)$ is some known function depending on some finite dimensional parameter $\btheta$.
This testing problem can be viewed as a two-sided test where the equality between two functions holds \textit{uniformly} over $x\in\mathcal{X}$. In this case, we introduce $\widetilde{\btheta}$ and $\widetilde{\bgamma}$ as  consistent estimators of $\btheta$ and $\bgamma_0$ under $\dot{\mathsf{H}}_0^{\mu^{(v)}}$.
Then we rely on the following test statistic:
\[
\dot{T}_{\mu^{(v)},p}(x)
:=\frac{\widehat{\mu}^{(v)}(x)-m^{(v)}(x;\widetilde{\btheta})}
{\sqrt{\widehat{\Omega}_{\mu^{(v)}}(x)/n}}.
\]
The null hypothesis is rejected if $\sup_{x\in\mathcal{X}}|\dot{T}_{\mu^{(v)},p}(x)|>\cval_{\mu^{(v)}}$ for some critical value $\cval_{\mu^{(v)}}$.

Similarly, to test the specification of $\vartheta_0(x,\evalw)$, we introduce
\begin{alignat*}{2}
&\dot{\mathsf{H}}_0^\vartheta:\quad &&\sup_{x\in\mathcal{X}} \Big|\vartheta_0(x,\evalw) -
M(x,\evalw;\btheta,\bgamma_0)\Big|=0, \quad \text{ for some } \btheta, \qquad vs.\\
&\dot{\mathsf{H}}_\text{A}^\vartheta: &&\sup_{x\in\mathcal{X}} \Big|\vartheta_0(x,\evalw) - M(x,\evalw;\btheta,\bgamma_0)\Big|>0, \quad \text{ for all } \btheta.
\end{alignat*}
where $M(x,\evalw;\btheta,\bgamma_0)=\eta(m(x;\btheta)+\evalw'\bgamma_0)$.
We rely on the following test statistic:
\[
\dot{T}_{\vartheta,p}(x)
:=\frac{\widehat{\vartheta}(x,\widehat\evalw)-M(x,\widehat\evalw;\widetilde{\btheta},\widetilde\bgamma)}
{\sqrt{\widehat{\Omega}_\vartheta(x)/n}}.
\]
The null hypothesis is rejected if $\sup_{x\in\mathcal{X}}|\dot{T}_{\vartheta,p}(x)|>\cval_\vartheta$ for some critical value $\cval_\vartheta$.

To test the specification of $\zeta_0(x,\evalw)$, we introduce
\begin{alignat*}{2}
&\dot{\mathsf{H}}_0^\zeta:\quad &&\sup_{x\in\mathcal{X}} \Big|\zeta_0(x,\evalw) -
M^{(1)}(x,\evalw;\btheta,\bgamma_0)\Big|=0, \quad \text{ for some } \btheta, \qquad vs.\\
&\dot{\mathsf{H}}_\text{A}^\zeta: &&\sup_{x\in\mathcal{X}} \Big|\zeta_0(x,\evalw) -
M^{(1)}(x,\evalw;\btheta,\bgamma_0)\Big|>0, \quad \text{ for all } \btheta.
\end{alignat*}
where $M^{(1)}(x,\evalw;\btheta,\bgamma_0):=\eta^{(1)}(m(x;\btheta)+\evalw'\bgamma_0)m^{(1)}(x;\btheta)$.
We rely on the following test statistic:
\[
\dot{T}_{\zeta,p}(x)
:=\frac{\widehat{\zeta}(x,\widehat\evalw)-M^{(1)}(x,\widehat\evalw;\widetilde{\btheta},\widetilde\bgamma)}
{\sqrt{\widehat{\Omega}_\zeta(x)/n}}.
\]
The null hypothesis is rejected if $\sup_{x\in\mathcal{X}}|\dot{T}_{\zeta,p}(x)|>\cval_\zeta$ for some critical value $\cval_\zeta$.

\begin{thm}[Specification Tests] \label{SA thm: testing specification}
	{Suppose that the conditions in Theorem \ref{SA thm: CB} hold.}
	\begin{enumerate}[label=(\roman*)]
		\item
		Let $\cval_{\mu^{(v)}}=\inf\{c\in\mathbb{R}_+: \P[\sup_{x\in\mathcal{X}}|\widehat{Z}_{\mu^{(v)},p}(x)|\leq c |\bD,\widehat{\Delta}]\geq 1-\alpha \}$.

		Under $\dot{\mathsf{H}}_0^{\mu^{(v)}}$, if
		$\sup_{x\in\mathcal{X}} |\mu_0^{(v)}(x)-m^{(v)}
		(x;\widetilde{\btheta})|=o_\P\Big(\sqrt{\frac{J^{1+2v}}{n\log J}}\Big)$, then
		$$
		\lim_{n\to\infty}\P\Big[\sup_{x\in\mathcal{X}}|\dot{T}_{\mu^{(v)},p}(x)|>
		\cval_{\mu^{(v)}}\Big]=\alpha.
		$$

		Under $\dot{\mathsf{H}}_{\text{A}}^{\mu^{(v)}}$, if there exist some fixed $\bar{\btheta}$
            such that $\sup_{x\in\mathcal{X}} |m^{(v)}(x;\widetilde{\btheta})-m^{(v)}(x;\bar{\btheta})|=o_\P(1)$,
            \[
            \liminf_{n\to\infty}\sup_{x\in\mathcal X}
            |\mu_0^{(v)}(x)-m^{(v)}(x;\bar{\btheta})|>0,
            \]
            and $J^v\Big(\frac{J\log J}{n}\Big)^{1/2}=o(1)$, then
		\[
		\lim_{n\to\infty}\P\Big[\sup_{x\in\mathcal{X}}
		|\dot{T}_{\mu^{(v)},p}(x)|>\cval_{\mu^{(v)}}\Big]=1.
		\]

        \item 	Let $\cval_\vartheta=\inf\{c\in\mathbb{R}_+: \P[\sup_{x\in\mathcal{X}}|\widehat{Z}_{\vartheta,p}(x)|\leq c |\bD,\widehat{\Delta}]\geq 1-\alpha \}$.

		Under $\dot{\mathsf{H}}_0^\vartheta$, if
		$\sup_{x\in\mathcal{X}} |\vartheta_0(x,\evalw)-
        M(x,\widehat\evalw;\widetilde{\btheta},\widetilde{\bgamma})|=o_\P\Big(\sqrt{\frac{J}{n\log J}}\Big)$, then
		$$
		\lim_{n\to\infty}\P\Big[\sup_{x\in\mathcal{X}}|\dot{T}_{\vartheta,p}(x)|>
		\cval_\vartheta\Big]=\alpha.
		$$

		Under $\dot{\mathsf{H}}_{\text{A}}^\vartheta$, if there exist some fixed $\bar{\btheta}$ and {a (non-random) benchmark $\bar{\bgamma}\in\mathbb R^{d_\gamma}$} such that {$\sup_n\|\bar{\bgamma}\|<\infty$ and the benchmark index $m(x;\bar{\btheta})+\bw'\bar{\bgamma}$ is uniformly contained in a fixed compact subset of the interior of the domain of $\eta$,} $\sup_{x\in\mathcal{X}} |M(x,\widehat\evalw;\widetilde{\btheta},\widetilde{\bgamma})-M(x,\evalw;\bar{\btheta}, \bar{\bgamma})|=o_\P(1)$, {
		\[
		\liminf_{n\to\infty}\sup_{x\in\mathcal X}
		|\vartheta_0(x,\evalw)-M(x,\evalw;\bar{\btheta},\bar{\bgamma})|>0,
		\]}
		and $\Big(\frac{J\log J}{n}\Big)^{1/2}=o(1)$, then
		\[
		\lim_{n\to\infty}\P\Big[\sup_{x\in\mathcal{X}}
		|\dot{T}_{\vartheta,p}(x)|>\cval_\vartheta\Big]=1.
		\]

        \item 		Let $\cval_\zeta=\inf\{c\in\mathbb{R}_+: \P[\sup_{x\in\mathcal{X}}|\widehat{Z}_{\zeta,p}(x)|\leq c |\bD,\widehat{\Delta}]\geq 1-\alpha \}$.

		Under $\dot{\mathsf{H}}_0^\zeta$, if
		$\sup_{x\in\mathcal{X}} |\zeta_0(x,\evalw)-M^{(1)}
		(x,\widehat\evalw;\widetilde{\btheta},\widetilde{\bgamma})|=o_\P\Big(\sqrt{\frac{J^3}{n\log J}}\Big)$, then
		$$
		\lim_{n\to\infty}\P\Big[\sup_{x\in\mathcal{X}}|\dot{T}_{\zeta,p}(x)|>
		\cval_\zeta\Big]=\alpha.
		$$

		Under $\dot{\mathsf{H}}_{\text{A}}^\zeta$, if there exist some fixed $\bar{\btheta}$ and {a (non-random) benchmark $\bar{\bgamma}\in\mathbb R^{d_\gamma}$} such that {$\sup_n\|\bar{\bgamma}\|<\infty$ and the benchmark index $m(x;\bar{\btheta})+\bw'\bar{\bgamma}$ is uniformly contained in a fixed compact subset of the interior of the domain of $\eta$,} $\sup_{x\in\mathcal{X}} |M^{(1)}(x,\widehat\evalw;\widetilde{\btheta},\widetilde{\bgamma})-M^{(1)}(x,\evalw;\bar{\btheta}, \bar{\bgamma})|=o_\P(1)$, {
		\[
		\liminf_{n\to\infty}\sup_{x\in\mathcal X}
		|\zeta_0(x,\evalw)-M^{(1)}(x,\evalw;\bar{\btheta},\bar{\bgamma})|>0,
		\]}
		and $J\Big(\frac{J\log J}{n}\Big)^{1/2}=o(1)$, then
		\[
		\lim_{n\to\infty}\P\Big[\sup_{x\in\mathcal{X}}
		|\dot{T}_{\zeta,p}(x)|>\cval_\zeta\Big]=1.
		\]
  \end{enumerate}
\end{thm}

\subsection{Shape Restriction Tests}\label{SA section: test shape restriction}

The third application of our results is to test certain shape restrictions on $\mu_0^{(v)}(x)$, $\vartheta_0(x,\evalw)$ and $\zeta_0(x,\evalw)$.
To be specific, consider the following problem:
\[\begin{split}
	&\ddot{\mathsf{H}}_0^{\mu^{(v)}}:\;
	\sup_{x\in\mathcal{X}}\, (\mu_0^{(v)}(x)-m^{(v)}(x; \bar{\btheta}))\leq 0 \text{ for certain } \bar\btheta \quad \text{v.s.}\\
	&\ddot{\mathsf{H}}_\text{A}^{\mu^{(v)}}:\;
	\sup_{x\in\mathcal{X}}\, (\mu_0^{(v)}(x) -
	m^{(v)}(x;\bar{\btheta}))>0 \text{ for } \bar\btheta.
\end{split}\]

This testing problem can be viewed as a one-sided test where the inequality holds \textit{uniformly} over $x\in\mathcal{X}$. Importantly, it should be noted that under both $\ddot{\mathsf{H}}_0^{\mu^{(v)}}$ and  $\ddot{\mathsf{H}}_\text{A}^{\mu^{(v)}}$, we fix $\bar{\btheta}$. In such a case, we introduce
$\widetilde{\btheta}$ as a consistent estimator of $\bar{\btheta}$ under both $\ddot{\mathsf{H}}_0^{\mu^{(v)}}$ and $\ddot{\mathsf{H}}_\text{A}^{\mu^{(v)}}$. Then we will rely on the following test statistic:
\[
\ddot{T}_{\mu^{(v)},p}(x):=
\frac{\widehat{\mu}^{(v)}(x)-m^{(v)}(x;\widetilde{\btheta})}
{\sqrt{\widehat{\Omega}_{\mu^{(v)}}(x)/n}}.
\]
The null hypothesis is rejected if $\sup_{x\in\mathcal{X}} \ddot{T}_{\mu^{(v)},p}(x)>\cval_{\mu^{(v)}}$ for some critical value $\cval_{\mu^{(v)}}$.

For the following transformed targets, we fix $\bar{\btheta}$ and {use the same benchmark coefficient $\bar{\bgamma}$} under the null and alternative, and introduce $\widetilde{\btheta}$ and $\widetilde{\bgamma}$ as consistent estimators of $\bar{\btheta}$ and $\bar{\bgamma}$.
Similarly, define the test for the shape of $\vartheta_0(x,\evalw)$:
\[\begin{split}
	&\ddot{\mathsf{H}}_0^\vartheta:\;
	\sup_{x\in\mathcal{X}}\, (\vartheta_0(x,\evalw)-M(x,\evalw; \bar{\btheta},\bar{\bgamma}))\leq 0 \text{ for certain } \bar\btheta \text{ and } \bar{\bgamma} \quad \text{v.s.}\\
	&\ddot{\mathsf{H}}_\text{A}^\vartheta:\;
	\sup_{x\in\mathcal{X}}\, (\vartheta_0(x,\evalw) -
	M(x,\evalw;\bar{\btheta},\bar{\bgamma}))>0 \text{ for } \bar\btheta \text{ and } \bar{\bgamma}.
\end{split}\]
We will rely on the following test statistic:
\[
\ddot{T}_{\vartheta,p}(x):=
\frac{\widehat{\vartheta}(x,\widehat\evalw)-M(x,\widehat\evalw;\widetilde{\btheta},\widetilde{\bgamma})}
{\sqrt{\widehat{\Omega}_\vartheta(x)/n}}.
\]
The null hypothesis is rejected if $\sup_{x\in\mathcal{X}} \ddot{T}_{\vartheta,p}(x)>\cval_\vartheta$ for some critical value $\cval_\vartheta$.

Also, define the test for the shape of $\zeta_0(x,\evalw)$:
\[\begin{split}
	&\ddot{\mathsf{H}}_0^\zeta:\;
	\sup_{x\in\mathcal{X}}\, (\zeta_0(x,\evalw)-M^{(1)}(x,\evalw; \bar{\btheta},\bar{\bgamma}))\leq 0 \text{ for certain } \bar\btheta \text{ and } \bar{\bgamma} \quad \text{v.s.}\\
	&\ddot{\mathsf{H}}_\text{A}^\zeta:\;
	\sup_{x\in\mathcal{X}}\, (\zeta_0(x,\evalw) -
	M^{(1)}(x,\evalw;\bar{\btheta},\bar{\bgamma}))>0 \text{ for } \bar\btheta \text{ and } \bar{\bgamma}.
\end{split}\]
We will rely on the following test statistic:
\[
\ddot{T}_{\zeta,p}(x):=
\frac{\widehat{\zeta}(x,\widehat\evalw)-M^{(1)}(x,\widehat\evalw;\widetilde{\btheta},\widetilde{\bgamma})}
{\sqrt{\widehat{\Omega}_\zeta(x)/n}}.
\]
The null hypothesis is rejected if $\sup_{x\in\mathcal{X}} \ddot{T}_{\zeta,p}(x)>\cval_\zeta$ for some critical value $\cval_\zeta$.

The following theorem characterizes the size and power of such tests.

\begin{thm}[Shape Restriction Tests] \label{SA thm: testing shape restriction}
	{Suppose that the conditions in Theorem \ref{SA thm: CB} hold.}
 \begin{enumerate}[label=(\roman*)]
        \item 	Assume 	$\sup_{x\in\mathcal{X}} |m^{(v)}(x;\widetilde{\btheta})-
	m^{(v)}(x;\bar{\btheta})|=
	o_\P\Big(\sqrt{\frac{J^{1+2v}}{n\log J}}\Big)$.

        Let $\cval_{\mu^{(v)}}=\inf\{c\in\mathbb{R}_+: \P[\sup_{x\in\mathcal{X}}\widehat{Z}_{\mu^{(v)},p}(x)\leq c
		|\bD,\widehat{\Delta}]\geq 1-\alpha \}$.

		Under $\ddot{\mathsf{H}}_0^{\mu^{(v)}}$,
		\[
		\lim_{n\to\infty}\P\Big[\sup_{x\in\mathcal{X}}
		\ddot{T}_{\mu^{(v)},p}(x)>\cval_{\mu^{(v)}}\Big]\leq \alpha.
		\]

		Under $\ddot{\mathsf{H}}_\text{A}^{\mu^{(v)}}$, if
		\[
		\liminf_{n\to\infty}\sup_{x\in\mathcal X}
		\{\mu_0^{(v)}(x)-m^{(v)}(x;\bar{\btheta})\}>0,
		\]
		and $J^v\Big(\frac{J\log J}{n}\Big)^{1/2}=o(1)$,
		\[
		\lim_{n\to\infty}\P\Big[\sup_{x\in\mathcal{X}}
		\ddot{T}_{\mu^{(v)},p}(x)>\cval_{\mu^{(v)}}\Big]=1.
		\]

        \item 	{Assume that the (non-random) benchmark $\bar{\bgamma}\in\mathbb R^{d_\gamma}$ satisfies $\sup_n\|\bar{\bgamma}\|<\infty$ and that the benchmark index $m(x;\bar{\btheta})+\bw'\bar{\bgamma}$ is uniformly contained in a fixed compact subset of the interior of the domain of $\eta$.} Assume 	$\sup_{x\in\mathcal{X}} |M(x,\widehat\evalw;\widetilde{\btheta},\widetilde{\bgamma})-
	M(x,\evalw;\bar{\btheta},\bar{\bgamma})|=
	o_\P\Big(\sqrt{\frac{J}{n\log J}}\Big)$.
        Let $\cval_\vartheta=\inf\{c\in\mathbb{R}_+: \P[\sup_{x\in\mathcal{X}}\widehat{Z}_{\vartheta,p}(x)\leq c
		|\bD,\widehat{\Delta}]\geq 1-\alpha \}$.

		Under $\ddot{\mathsf{H}}_0^\vartheta$,
		\[
		\lim_{n\to\infty}\P\Big[\sup_{x\in\mathcal{X}}
		\ddot{T}_{\vartheta,p}(x)>\cval_\vartheta\Big]\leq \alpha.
		\]

		Under $\ddot{\mathsf{H}}_\text{A}^\vartheta$, if {
		\[
		\liminf_{n\to\infty}\sup_{x\in\mathcal X}
		\{\vartheta_0(x,\evalw)-M(x,\evalw;\bar{\btheta},\bar{\bgamma})\}>0,
		\]}
		and $\Big(\frac{J\log J}{n}\Big)^{1/2}=o(1)$,
		\[
		\lim_{n\to\infty}\P\Big[\sup_{x\in\mathcal{X}}
		\ddot{T}_{\vartheta,p}(x)>\cval_\vartheta\Big]=1.
		\]

        \item {Assume that the (non-random) benchmark $\bar{\bgamma}\in\mathbb R^{d_\gamma}$ satisfies $\sup_n\|\bar{\bgamma}\|<\infty$ and that the benchmark index $m(x;\bar{\btheta})+\bw'\bar{\bgamma}$ is uniformly contained in a fixed compact subset of the interior of the domain of $\eta$.} Assume 	$\sup_{x\in\mathcal{X}} |M^{(1)}(x,\widehat\evalw;\widetilde{\btheta},\widetilde{\bgamma})-
	M^{(1)}(x,\evalw;\bar{\btheta},\bar{\bgamma})|=
	o_\P\Big(\sqrt{\frac{J^3}{n\log J}}\Big)$.
        Let $\cval_\zeta=\inf\{c\in\mathbb{R}_+: \P[\sup_{x\in\mathcal{X}}\widehat{Z}_{\zeta,p}(x)\leq c
		|\bD,\widehat{\Delta}]\geq 1-\alpha \}$.

		Under $\ddot{\mathsf{H}}_0^\zeta$,
		\[
		\lim_{n\to\infty}\P\Big[\sup_{x\in\mathcal{X}}
		\ddot{T}_{\zeta,p}(x)>\cval_\zeta\Big]\leq \alpha.
		\]

		Under $\ddot{\mathsf{H}}_\text{A}^\zeta$, if {
		\[
		\liminf_{n\to\infty}\sup_{x\in\mathcal X}
		\{\zeta_0(x,\evalw)-M^{(1)}(x,\evalw;\bar{\btheta},\bar{\bgamma})\}>0,
		\]}
		and $J\Big(\frac{J\log J}{n}\Big)^{1/2}=o(1)$,
		\[
		\lim_{n\to\infty}\P\Big[\sup_{x\in\mathcal{X}}
		\ddot{T}_{\zeta,p}(x)>\cval_\zeta\Big]=1.
		\]
 \end{enumerate}
\end{thm}

\begin{remark}[Improvements over literature]
	The results in Sections \ref{SA section: confidence band}--\ref{SA section: test shape restriction} are new to the literature, even in the case of non-random partitioning and without covariate-adjustments, because they take advantage of the specific binscatter structure (i.e., locally bounded series basis). Furthermore, relative to prior work, our results allow for a large class of random partitioning schemes, formally take into account the potential randomness of the partition, account for the generalized semi-linear structure, and consider an array of possibly nonlinear estimation and inference problems. In particular, the approach taken in Theorems \ref{SA thm: strong approximation, GL} and \ref{SA thm: sup approx} to establish strong approximation and related distributional approximations for nonlinear binscatter statistics may be of independent interest. The key condition imposed on the partition for uniform inference (confidence bands and hypothesis testing)
    is Assumption \ref{SA Assumption RP}(i), while ``convergence'' of the random partition (Assumption \ref{SA Assumption RP}(ii)) is not required.
\end{remark}

\subsection{Convergence Rate of Covariate Coefficients}
In this section, we establish the convergence rate of $\widehat{\bgamma}$ under primitive conditions.
Set
$\widehat{\bb}_i=\widehat{\bb}_{p,s}(x_i)$,
$\varpi_i=\mu_0(x_i)+\bw_i'\bgamma_0$, and
{
\[
\begin{aligned}
	\varkappa_i&=\varkappa(x_i,\bw_i)
	=\Psi_1(x_i,\bw_i;\eta_i)\eta_{i,1}^2,
	&\qquad \bar{\bQ}&=\E_n[\varkappa_i\widehat{\bb}_i\widehat{\bb}_i'],\\
	\bpi_w(x_i)&=\E_n[\bw_i\widehat{\bb}_i'\varkappa_i]
	\bar{\bQ}^{-1}\widehat{\bb}_i,\qquad
	&\qquad \widetilde{\bw}_i&=\bw_i-\bpi_w(x_i),\\
	\widehat{\bG}_w&=\E_n[\varkappa_i
	\widetilde{\bw}_i\widetilde{\bw}_i'],
	&\qquad \bpi_{w,0}(x)&=\frac{\E[\varkappa_i\bw_i\mid x_i=x]}
	{\E[\varkappa_i\mid x_i=x]},\\
	\bpi_{w,J}(x)
	&=\E_{\widehat\Delta}[\varkappa_i\bw_i\widehat{\bb}_i']
	\E_{\widehat\Delta}[\varkappa_i\widehat{\bb}_i\widehat{\bb}_i']^{-1}
	\widehat{\bb}_{p,s}(x).
\end{aligned}
\]
}
Define the directional projection error by
\begin{equation}\label{SA eq: growing d projection error}
	\mathfrak a_{\pi,J}
	=\sup_{\|\balpha\|=1}
	\left\{\E_{\widehat\Delta}\left[
	\E[\varkappa_i\mid x_i]
	\{\balpha'(\bpi_{w,0}(x_i)-\bpi_{w,J}(x_i))\}^2
	\right]\right\}^{1/2},
\end{equation}
and set
{
\[
\begin{split}
	\btheta_0
	&=\widehat{\bbeta}_0+\bar{\bQ}^{-1}
	\E_n[\widehat{\bb}_i\bw_i'\varkappa_i]\bgamma_0,
	\qquad u_i(\btheta)=\widehat{\bb}_i'(\btheta-\btheta_0),\\
	\mathfrak r_{\theta,n}
	&=\sqrt{J/n}+J^{-p-1},\\
	\mathfrak r_{\gamma,n}
	&=\frac{\sqrt{\overline{\lambda}_{w,n}d_\gamma/n}
	+J^{-p-1}\mathfrak a_{\pi,J}
	+\sqrt{\overline{\lambda}_{w,n}}\mathfrak r_{\theta,n}^2
	{{}+\sqrt{\overline{\lambda}_{w,n}J\mathfrak r_{\theta,n}\log n/n}
	+J\sqrt{d_\gamma}\log n/n}}
	{\underline{\lambda}_{w,n}}.
\end{split}
\]
}

\begin{lem}[Covariate Coefficients Rate]\label{SA lem: growing d gamma rate}
{Suppose that $\widehat{\bgamma}$ is obtained from the joint estimator
in \eqref{SA eq:binscatter-QMLE}. Suppose that Assumptions \ref{SA Assumption DGP}, \ref{SA Assumption GL},
and \ref{SA Assumption RP}(i) hold uniformly}
over the triangular array with $\bw_i\in\mathbb R^{d_\gamma}$ and
$\mathcal W\subset\mathbb R^{d_\gamma}$. Suppose also that
$d_\gamma=o(J)$.
Assume the following additional conditions:
\begin{enumerate}[label=(\roman*)]
	\item There exist deterministic sequences
	$0<\underline{\lambda}_{w,n}\leq\overline{\lambda}_{w,n}<\infty$
	such that, w.p.a. $1$,
	\[
		\underline{\lambda}_{w,n}
		\leq\lambda_{\min}(\widehat{\bG}_w)
		\leq\lambda_{\max}(\widehat{\bG}_w)
		\leq\overline{\lambda}_{w,n}.
	\]

	\item The sample projection satisfies
	\[
		\left\|
		\E_n[\varkappa_i\widetilde{\bw}_i\widehat r_{0,0}(x_i)]
		-\E_{\widehat\Delta}\left[
		\varkappa_i\{\bw_i-\bpi_{w,J}(x_i)\}\widehat r_{0,0}(x_i)
		\right]
		\right\|
		\lesssim_\P
		J^{-p-1}\sqrt{\overline{\lambda}_{w,n}d_\gamma/n}.
	\]

	\item For every fixed $C<\infty$, uniformly over $\|\balpha\|=1$ and
	$\btheta$ such that
	$\E_n[u_i(\btheta)^2]\leq C\mathfrak r_{\theta,n}^2$ and
	$\max_i|u_i(\btheta)|\leq C\sqrt J\mathfrak r_{\theta,n}$,
	\[
		\E_n[(\balpha'\widetilde{\bw}_i)^2u_i(\btheta)^2]
		\lesssim_\P\overline{\lambda}_{w,n}
		\E_n[u_i(\btheta)^2].
	\]
	Moreover,
	\[
	\mathfrak r_{\theta,n}
	\left(\sqrt J+
	\sqrt{\frac{d_\gamma}{\underline{\lambda}_{w,n}}}\right)
	=o_\P(1),
	\qquad
	\frac{\overline{\lambda}_{w,n}}{\underline{\lambda}_{w,n}}
	\left\{\sqrt J\mathfrak r_{\theta,n}
	+\sqrt{d_\gamma}\mathfrak r_{\gamma,n}\right\}=o_\P(1).
	\]

	\item The following rate restrictions hold:
	{
	\[
	\begin{split}
	&\frac{J}{n^{1-1/\nu}}
	\left(\sqrt J+
	\sqrt{\frac{d_\gamma}{\underline{\lambda}_{w,n}}}\right)
	=o_\P(\mathfrak r_{\theta,n}),\qquad
	\frac{d_\gamma^{3/2}}{n^{1-1/\nu}}
	=o_\P(\underline{\lambda}_{w,n}\mathfrak r_{\gamma,n}),\\
	&\frac{J^{3/2}\mathfrak r_{\theta,n}}{n}
	\sqrt{d_\gamma}
	\left\{n^{1/\nu}+\sqrt J\mathfrak r_{\theta,n}
	+\sqrt{d_\gamma}\mathfrak r_{\gamma,n}\right\}
	=o_\P(\underline{\lambda}_{w,n}\mathfrak r_{\gamma,n}).
	\end{split}
	\]
	}
\end{enumerate}
{Then $\|\widehat{\bgamma}-\bgamma_0\|
=O_\P(\mathfrak r_{\gamma,n})$.}
{If, in addition, there exists a deterministic sequence $\mathfrak r_\gamma$ such that
\[ \mathfrak r_{\gamma,n}=O_\P(\mathfrak r_\gamma),\qquad \sqrt{d_\gamma}\mathfrak r_\gamma=o\!\left(\sqrt{J/n}+J^{-p-1}\right), \]
then the rate condition on $\widehat{\bgamma}$ in Assumption \ref{SA Assumption HLE}(i) holds.}
\end{lem}

\begin{remark}[Fast convergence of covariate coefficients]
\label{SA rem: fast gamma rate}
The last conclusion of Lemma \ref{SA lem: growing d gamma rate} verifies the fast-rate condition in Assumption \ref{SA Assumption HLE}(i) whenever $\mathfrak r_{\gamma,n}=O_\P(\mathfrak r_\gamma)$ and $\sqrt{d_\gamma}\mathfrak r_\gamma=o(\sqrt{J/n}+J^{-p-1})$. As a simple benchmark, suppose that $\nu=4$, the Gram matrix for the residualized controls is well conditioned, $\mathfrak r_{\theta,n}\asymp\sqrt{J/n}$, and the projection and quadratic terms in $\mathfrak r_{\gamma,n}$ are no larger than $\sqrt{d_\gamma/n}$. Then
\[
	\mathfrak r_{\gamma,n}\asymp
	\sqrt{d_\gamma/n}+(J/n)^{3/4}\sqrt{\log n}
	+J\sqrt{d_\gamma}\log n/n.
\]
Comparing $\sqrt{d_\gamma}\mathfrak r_{\gamma,n}$ with $\sqrt{J/n}$, the first term in the display requires $d_\gamma=o(\sqrt J)$, while each of the other two terms requires $d_\gamma^2J(\log n)^2/n=o(1)$. 
Because our analysis allows $\bw$ to be generic, sharper conditions may be available when $\bw$ has additional structure, such as sparsity or local support.
\end{remark}

\begin{remark}[Verification of Conditions in Lemma \ref{SA lem: growing d gamma rate}]
\label{SA rem: growing d primitive covariance}
We can verify conditions imposed in Lemma \ref{SA lem: growing d gamma rate} under more primitive conditions.

\textit{Condition (i).}
Suppose that there exist deterministic sequences
$0<c_n\leq C_n<\infty$ such that the following primitive
covariance condition holds uniformly over $x\in\mathcal X$:
\begin{equation}\label{SA eq: growing d primitive covariance}
	c_n\bI_{d_\gamma}
	\leq\E\left[(\bw_i-\E[\bw_i\mid x_i])
	(\bw_i-\E[\bw_i\mid x_i])'\mid x_i=x\right]
	\leq C_n\bI_{d_\gamma}.
\end{equation}
Under \eqref{SA eq: growing d primitive covariance}, since
$\varkappa_i$ is bounded and bounded away from zero and $\bpi_{w,0}(x)$ is the
weighted conditional mean, the conditional second-moment matrix of
$\varkappa_i^{1/2}\{\bw_i-\bpi_{w,0}(x_i)\}$ has eigenvalues bounded
below and above by constant multiples of $c_n$ and $C_n$, respectively,
uniformly over $x$. Suppose, in addition, that
$\mathfrak a_{\pi,J}^2\lesssim C_n$. Coordinatewise boundedness of
$\bw_i$, local support of the binscatter basis, and the bounded row-sum norm of
the inverse weighted Gram matrix imply the required projection stability.
If the uniform within-bin matrix-concentration error is
$o_\P(c_n)$, then condition (i) holds with
$\underline{\lambda}_{w,n}\asymp c_n$ and
$\overline{\lambda}_{w,n}\asymp C_n$. Under coordinatewise boundedness, a
sufficient concentration restriction is
{$C_n d_\gamma J\log(Jd_\gamma)/(n c_n^2)=o(1)$.}
When $c_n\asymp C_n\asymp1$, this reduces to
$d_\gamma J\log(Jd_\gamma)/n=o(1)$.

\textit{Condition (ii).}
The difference in condition (ii) equals
\[
	(\E_n-\E_{\widehat\Delta})\left[\varkappa_i\{\bw_i-\bpi_{w,J}(x_i)\}\widehat r_{0,0}(x_i)\right]-\E_n\left[\varkappa_i\{\bpi_w(x_i)-\bpi_{w,J}(x_i)\}\widehat r_{0,0}(x_i)\right].
\]
The projection normal equations imply
\[
	\begin{split}
	\E_n[\varkappa_i\{\bpi_w(x_i)-\bpi_{w,J}(x_i)\}
	\widehat r_{0,0}(x_i)]
	={}&(\E_n-\E_{\widehat\Delta})
	[\varkappa_i\{\bw_i-\bpi_{w,J}(x_i)\}\widehat{\bb}_i']
	\bar{\bQ}^{-1}\\
	&\times\E_n[\varkappa_i\widehat{\bb}_i\widehat r_{0,0}(x_i)].
	\end{split}
\]
Decomposing $\bar{\bQ}^{-1}$ around
$\{\E_{\widehat\Delta}[\varkappa_i\widehat{\bb}_i
\widehat{\bb}_i']\}^{-1}$ and applying the uniform approximation bound, vector
concentration, and projection stability bounds the resulting leading terms by
$J^{-p-1}\sqrt{\overline{\lambda}_{w,n}d_\gamma/n}$. The remaining term is
\[
	(\E_n-\E_{\widehat\Delta})[\varkappa_i\{\bw_i-\bpi_{w,J}(x_i)\}\widehat{\bb}_i']\times\left(\bar{\bQ}^{-1}-\{\E_{\widehat\Delta}[\varkappa_i\widehat{\bb}_i\widehat{\bb}_i']\}^{-1}\right)\E_n[\varkappa_i\widehat{\bb}_i\widehat r_{0,0}(x_i)],
\]
which is bounded by $J^{-p-1}J\sqrt{d_\gamma}/n$ and is of smaller order
provided $J^2/(nC_n)=o(1)$. This condition reduces to $J^2/n=o(1)$ in the
well-conditioned case. For a data-dependent partition, the concentration is
applied uniformly over the admissible partition class.

\textit{Condition (iii).}
{Under \eqref{SA eq: growing d primitive covariance}, conditioning
on $x_i$ bounds the conditional second moment of the weighted conditional-mean
residual $\bw_i-\bpi_{w,0}(x_i)$ uniformly by $C_n$. Smoothness and projection
stability control the approximation and sample projection errors
$\bpi_{w,0}(x_i)-\bpi_{w,J}(x_i)$ and
$\bpi_{w,J}(x_i)-\bpi_w(x_i)$, respectively. Iterated expectations and the
same uniform concentration arguments therefore give, uniformly over
$\|\balpha\|=1$ and the stated neighborhood of $\btheta$,
\[
	\E_n[(\balpha'\widetilde{\bw}_i)^2u_i(\btheta)^2]
	\lesssim_\P C_n\E_n[u_i(\btheta)^2],
\]
which is the first bound in condition (iii).} Moreover,
\[
	\mathfrak r_{\gamma,n}\asymp
	\frac{\sqrt{C_nd_\gamma/n}
	+J^{-p-1}\mathfrak a_{\pi,J}
	+\sqrt{C_n}\mathfrak r_{\theta,n}^2
	{{}+\sqrt{C_nJ\mathfrak r_{\theta,n}\log n/n}
	+J\sqrt{d_\gamma}\log n/n}}{c_n}.
\]
The last two terms account for crossing the finitely many
discontinuities of a piecewise-Lipschitz score and may be omitted when
$\psi^\dagger(\cdot)$ is globally Lipschitz.
Thus, for general $c_n$ and $C_n$, the remaining restrictions in
condition (iii) require
\[
	\mathfrak r_{\theta,n}\left(\sqrt J+\sqrt{d_\gamma/c_n}\right)=o_\P(1),
	\qquad
	\frac{C_n}{c_n}\left\{\sqrt J\mathfrak r_{\theta,n}
	+\sqrt{d_\gamma}\mathfrak r_{\gamma,n}\right\}=o_\P(1).
\]
In the well-conditioned case
$c_n\asymp C_n\asymp1$, compact support and projection stability also give
$\mathfrak a_{\pi,J}\lesssim_\P\sqrt{d_\gamma}$. Therefore,
\[
	\sqrt J\mathfrak r_{\theta,n}=o(1),
	\qquad
	\sqrt{d_\gamma}\mathfrak r_{\gamma,n}
	\lesssim_\P
	\frac{d_\gamma}{\sqrt n}+d_\gamma J^{-p-1}
	+\sqrt{d_\gamma}\mathfrak r_{\theta,n}^2
	{{}+\sqrt{d_\gamma J\mathfrak r_{\theta,n}\log n/n}
	+Jd_\gamma\log n/n}=o(1),
\]
where the last relations follow from
{$J^2\log n/n=o(1)$} and $d_\gamma=o(J)$.
These relations imply both rate restrictions in condition (iii)
in this well-conditioned case.

\textit{Condition (iv).}
For general $c_n$ and $C_n$, condition (iv) retains the
corresponding factors displayed in the lemma. In the well-conditioned case
$c_n\asymp C_n\asymp1$, it reduces to
\[
	\frac{J^{3/2}}{n^{1-1/\nu}}=o(\mathfrak r_{\theta,n}),
	\qquad
	\frac{d_\gamma^{3/2}}{n^{1-1/\nu}}
	=o(\mathfrak r_{\gamma,n}),
	\qquad
	\frac{J^{3/2}\sqrt{d_\gamma}\mathfrak r_{\theta,n}}
	{n^{1-1/\nu}}=o(\mathfrak r_{\gamma,n}).
\]
For finite $\nu$, all three relations follow, for example,
if {$J^{2\nu/(\nu-2)}/n=o(1)$,}
which also controls the corresponding $d_\gamma$ terms because
$d_\gamma=o(J)$. If the score is uniformly bounded, condition (iv) follows
directly from $J^2/n=o(1)$ and $d_\gamma=o(J)$.

\end{remark}

\begin{remark}[Projection approximation]
The directional approximation error $\mathfrak a_{\pi,J}$ is needed only to
bound the population component of
$\E_n[\varkappa_i\widetilde{\bw}_i\widehat r_{0,0}(x_i)]$. A primitive
sufficient condition is, for some $s_\pi>0$ and a deterministic
sequence $C_{\pi,n}<\infty$,
\begin{equation}\label{SA eq: growing d primitive projection approximation}
	\max_{1\leq k\leq d_\gamma}\sup_{x\in\mathcal X}
	|\pi_{w,0,k}(x)-\pi_{w,J,k}(x)|
	\lesssim C_{\pi,n}J^{-s_\pi}.
\end{equation}
Condition \eqref{SA eq: growing d primitive projection approximation}
and boundedness of $\E[\varkappa_i\mid x_i]$ give
{$\mathfrak a_{\pi,J}
\lesssim\sqrt{d_\gamma}C_{\pi,n}J^{-s_\pi}$,}
so the corresponding contribution to $\mathfrak r_{\gamma,n}$ is
$\sqrt{d_\gamma}C_{\pi,n}J^{-p-1-s_\pi}/c_n$. In particular, the
condition $\mathfrak a_{\pi,J}^2\lesssim C_n$ in the preceding remark follows
if $d_\gamma C_{\pi,n}^2J^{-2s_\pi}\lesssim C_n$. The displayed
$L_\infty$ approximation follows from coordinatewise smoothness of
$\bpi_{w,0}(x)$ with bound $C_{\pi,n}$, sparsity of the binscatter basis, and
stability of the weighted projection.

In general, it is not enough to approximate $\E[\bw_i\mid x_i=x]$, because
\[
	\bpi_{w,0}(x)-\E[\bw_i\mid x_i=x]
	=\frac{\E[\varkappa_i\{\bw_i-\E[\bw_i\mid x_i]\}\mid x_i=x]}
	{\E[\varkappa_i\mid x_i=x]}.
\]
An $L_\infty$ approximation condition on the ordinary conditional mean is
sufficient if the numerator is zero, as when $\varkappa_i$ depends only on
$x_i$.
\end{remark}

\section{Implementation Details}
\label{SA section: implementation}

\subsection{Standard Error Computation}\label{SA sec: standard error}
In Section \ref{SA section: general loss fn}, we have given the variance formulas
$\widehat\Omega_{\mu^{(v)}}(x)$, $\widehat\Omega_\vartheta(x)$ and $\widehat\Omega_\zeta(x)$ that can be used to obtain the standard errors of $\widehat{\mu}^{(v)}(x)$, $\widehat{\vartheta}(x,\widehat\evalw)$ and $\widehat{\zeta}(x,\widehat\evalw)$.
Recall that the formula for the estimator $\widehat{\bSigma}$ of $\bSigma_0$ is
$$
\widehat{\bSigma}=
\E_n\Big[\widehat{\bb}_{p,s}(x_i)\widehat{\bb}_{p,s}(x_i)'
\psi(y_i, \widehat{\eta}_i)^2\eta^{(1)}(\widehat{\mu}(x_i)+\bw_i'\widehat{\bgamma})^2\Big].
$$
It only relies on known or estimable quantities such as the derivative of the loss function $\psi(\cdot)$, the derivative of the inverse link function $\eta^{(1)}(\cdot)$, the residual $\widehat{\epsilon}_i$ and the binscatter estimates $\widehat{\mu}(\cdot)$ and $\widehat\bgamma$. Thus, $\widehat{\bSigma}$ and other types of heteroskedasticity-robust ``meat'' matrix estimators can be easily constructed using the data. Then, it remains to obtain an estimator $\widehat{\bQ}$ of $\bar\bQ$ (or $\bQ_0$), which in general relies on an estimator $\widehat{\Psi}_1(\cdot)$ of $\Psi_1(\cdot)$ and can be constructed on a case-by-case basis. In the following we discuss several examples.
\medskip

\textbf{Example 1} (Least Squares Regression).
For least squares regression, the loss function
$\rho(y; \eta)=\frac{1}{2}(y-\eta)^2$ and the (inverse) link function $\eta(\theta)=\theta$. Therefore, $\psi(y_i,\eta_i)=-\epsilon_i$ and $\eta_{i,1}=1$. Thus, the formula for $\widehat{\bQ}$ given in Section \ref{SA section: general loss fn} reduces to $\E_n[\widehat\bb_{p,s}(x_i)\widehat\bb_{p,s}(x_i)']$, which is immediately feasible in practice.
\medskip

\textbf{Example 2} (Logistic Regression).
For logistic regression, the loss function is given by the corresponding likelihood function, i.e., $-\rho(y;\eta)=y\log \eta+(1-y)\log (1-\eta)$, and the inverse link is given by the logistic function $\eta(\theta)=\frac{e^\theta}{1+e^\theta}$. Accordingly, an estimator of $\bQ_0$ is given by
\[
\widehat{\bQ}=\E_n\Big[\widehat{\bb}_{p,s}(x_i)\widehat{\bb}_{p,s}(x_i)'\widehat{\eta}_i(1-\widehat{\eta}_i)\Big], \quad
\widehat{\eta}_i=\eta(\widehat{\mu}(x_i)+\bw_i'\widehat{\bgamma}).
\]

\textbf{Example 3} (Quantile Regression).
For quantile regression, $\rho(y;\eta)=(q-\I(y<\eta))(y-\eta)$ for some $q\in(0,1)$ and $\eta(\theta)=\theta$. Accordingly, $\psi(y_i,\eta_i)=\I(\epsilon_i<0)-q$, and one needs to estimate
\[
\bQ_0=\E\Big[\bb_{p,s}(x_i)\bb_{p,s}(x_i)'f_{Y|XW}(\mu_0(x_i)+\bw_i'\bgamma_0|x_i,\bw_i)\Big].
\]
The key is to estimate the conditional density $f_{Y|XW}(\cdot|x_i,\bw_i)$ evaluated at the conditional quantile of interest $(\mu_0(x_i)+\bw_i'\bgamma_0)$, whose reciprocal is termed ``sparsity function'' in the literature. Many different methods have been proposed. For example, the sparsity function is simply the derivative of the conditional quantile function with respect to the quantile, which can be estimated by using the difference quotient of the estimated conditional quantile function. Alternatively, $\bQ_0$ can be viewed as a matrix-weighted density function, and one can construct a corresponding estimator based on kernel density estimation ideas. In addition, one can use bootstrapping methods to estimate the variance, avoiding the technical difficulty of estimating the sparsity function. See Section 3.4 and Section 3.9 of \cite{Koenker_2005_book} for more discussion of variance estimation for quantile regression.

\subsection{Number of Bins Selector}\label{SA section: select J}
We discuss the implementation details for data-driven selection of the number of bins, based on the approximate integrated mean squared error expansion in Theorem \ref{SA thm: IMSE, GL}.

We offer two procedures for estimating the bias and variance constants, and once these estimates ($\widehat{\mathscr{B}}_n(p,s,v)$ and $\widehat{\mathscr{V}}_n(p,s,v)$) are available, the estimated optimal $J$ is
\[
\widehat{J}_{\mathtt{IMSE}}=\bigg\lceil
\bigg(\frac{2(p-v+1)\widehat{\mathscr{B}}_n(p,s,v)}
{(1+2v)\widehat{\mathscr{V}}_n(p,s,v)}\bigg)^{\frac{1}{2p+3}}
n^{\frac{1}{2p+3}}
\bigg\rceil.
\]
We always let $\omega(x)=f_X(x)$ as weighting function for concreteness.

\subsubsection{Rule-of-thumb Selector}
For quantile-spaced partitions, a rule-of-thumb choice of $J$ can be obtained based on Corollary SA-3.2 in \cite{Cattaneo-Crump-Farrell-Feng_2024_AER}, which gives an explicit characterization of the variance and bias constants for least squares binscatter using piecewise polynomials ($s=0$). For other partitions covered by Theorem \ref{SA thm: IMSE, GL}, the partition-specific constants can instead be estimated by the direct plug-in method below.

Specifically,
the variance constant $\mathscr{V}(p,0,v)$ is estimated by
\[
\widehat{\mathscr{V}}(p,0,v)=\tr\Big\{\Big(\int_{0}^{1}\bm{\varphi}(z)\bm{\varphi}(z)'\d{z}\Big)^{-1}\int_0^1\bm{\varphi}^{(v)}(z)\bm{\varphi}^{(v)}(z)'\d{z}\Big\}\times
\frac{1}{n}\sum_{i=1}^{n}\widehat{\sigma}^2(x_i,\bw_i)\widehat{f}_X(x_i)^{2v}
\]
where $\bm{\varphi}(z)=(1, z, \ldots, z^p)'$, $\widehat{\sigma}^2(x_i,\bw_i)$ is some estimate of the conditional variance $\V[y_i|x_i,\bw_i]$ and $\widehat{f}_X(x_i)$ is some estimate of the density $f_X(x_i)$.
On the other hand, the bias constant $\mathscr{B}(p,0,v)$ is estimated by
\[
\widehat{\mathscr{B}}(p,0,v)=\frac{\int_{0}^{1}[\mathscr{B}_{p+1-v}(z)]^2\d{z}}{((p+1-v)!)^2}\times
\frac{1}{n}\sum_{i=1}^{n}\frac{[\widehat{\mu}^{(p+1)}(x_i)]^2}{\widehat{f}_X(x_i)^{2p+2-2v}}.
\]
where $\mathscr{B}_p(z)=(-1)^p\sum_{k=0}^p\binom{p}{k}\binom{p+k}{k}(-z)^k/\binom{2p}{p}$ for each $p\in\mathbb{Z}_+$ and $\widehat\mu^{(p+1)}(x_i)$ is some preliminary estimate of $\mu_0^{(p+1)}(x_i)$. The details about getting the estimates $\widehat\sigma^2(x_i,\bw_i)$, $\widehat{f}_X(x_i)$ and $\widehat\mu^{(p+1)}(x_i)$ can be found in Section SA-4.1 in \cite{Cattaneo-Crump-Farrell-Feng_2024_AER}.

This procedure still yields a choice of $J$ with the correct rate, though the constant approximations are inconsistent for general loss.

\subsubsection{Direct-plug-in Selector}

The direct-plug-in selector is implemented based on nonlinear binscatter estimators, which applies to any user-specified $p$, $s$ and $v$. It requires a preliminary choice of partitioning denoted by $J_{\mathtt{pre}}$.

More generally, suppose that a preliminary choice $J_{\mathtt{pre}}$ is given, and then a binscatter basis $\widehat{\bb}_{p,s}(x)$ (of order $p$) can be constructed immediately on the preliminary partition. Implementing a nonlinear binscatter estimation using this basis and partitioning, we can obtain the variance constant estimate using the variance matrix estimators
discussed in Section \ref{SA sec: standard error}.

Regarding the bias constant, the key unknown in the expression of the leading approximation error $r_{0,v}^\star(x)$ in Theorem \ref{SA thm: IMSE, GL} is $\mu_0^{(p+1)}(x)$, which can be estimated by implementing a nonlinear binscatter estimation of order $p+1$. The local width $h_{x_i}$ can be replaced directly by $\hat h_{x_i}$, the length of the interval in the preliminary partition containing $x_i$. Thus, this construction does not require replacing $h_{x_i}$ by the quantile-specific approximation $1/(Jf_X(x_i))$.
All other quantities in the expression of $\mathscr{B}(p,s,v)$ can be replaced by their sample analogues. Then, a bias constant estimate is available. Under the conditions of Theorem \ref{SA thm: IMSE, GL}, and provided that $J_{\mathtt{pre}}\to\infty$ and $J_{\mathtt{pre}}^{2p+3}/n=o(1)$, this construction employs consistent constant approximations for any nonlinear binscatter with any choice of $p$, $s$ and $v$.

\subsection{Fixed \texorpdfstring{$J$}{J} and choice of polynomial order}

Our theory treats $J$ as diverging with the sample size. This reflects the standard approach wherein a researcher selects $p$ and $s$ in advance (often $s=p=0$ or $s=p=3$) and then selects $J$ given the data, and further, is interested in recovery of the true nonparametric function $\mu_0(x)$. The partition must get finer to remove the nonparametric smoothing bias in estimating the function $\mu_0(x)$ (and along with it, $\vartheta_0(x,\evalw)$ or $\zeta_0(x,\evalw)$). Correct recovery (either for estimation or visualization) of these functions is the primary use of binscatter. However, researchers sometimes prefer to pre-specify a fixed $J = {\tt J}$, and we also discuss implementation and interpretation of binscatter in this case. In this subsection, when considering fixed $J = {\tt J}$, we also treat $d_\gamma$ as fixed, as this maintains focus on the relationship of $y_i$ and $x_i$. If $J = {\tt J}$ is fixed and $d_\gamma$ diverges this is high-dimensional linear model and is no longer in the spirit of binscatter, both in terms of applications and econometric theory.

A researcher may want to use a fixed $J = {\tt J}$, often small and/or round, instead of the optimal choice, to make the estimate more visually appealing or because the results can be directly interpreted.
In this case, instead of recovering the functions $\mu_0^{(v)}(x)$, $\vartheta_0(x,\evalw)$, and $\zeta_0(x,\evalw)$, the binscatter is interpreted as estimating their coarsened versions: the distribution of $y_i$ conditional on $x_i$ lying in a (fixed) bin, rather than at a single point.
For some ${\tt J}$, this remains interpretable and can be viewed as a finite-dimensional estimation problem.
For example, in our application we can take ${\tt J} = 10$ and study the distribution of uninsured rate for each decile of income.
The confidence bands then become pointwise confidence intervals that are simultaneously valid.
This could be used to examine inequality in health care access by asking if median uninsured rates are statistically different between the top and bottom decile.
All our inference results still apply to this case, with the caveat that
if control variables $\bw_i$ are added, the estimation error in $\widehat{\bgamma}$ cannot be ignored.

A fixed ${\tt J}$ is also interpretable, and applicable, if $x_i$ is discrete. Then each mass point is given its own bin and the results apply to the conditional distribution of $y_i$ given $x_i = x$. Again, our theoretical results apply directly to this case and one obtains simultaneous inference over the set of points. \cite{Cattaneo-Crump-Farrell-Feng_2024_AER} give further discussion and examples.

As a practical compromise between the visual appeal and interpretation of a small, fixed ${\tt J}$ and the demand for consistent estimation, we propose a novel, albeit ad-hoc, adjustment to the estimator aimed at addressing the smoothing bias left by fixing ${\tt J}$ by adjusting the choice of polynomial order $p$. The standard approach fixes $p$ in advance and selects $J$ based on the data, but we can invert this and search for the value of $p$ for which the fixed ${\tt J}$ would be IMSE-optimal. That is, we solve for
\begin{align}
    \label{eq:p IMSE}
    \widehat p_{\mathtt{IMSE}}({\tt J},s,v) = \argmin_{p\in\mathcal{P}} \Big| \widehat J_{\mathtt{IMSE}}(p,s,v) - \mathtt{J} \Big|,
\end{align}
where in principle the set $\mathcal{P}$ contains all integers $p\geq \max\{s,v\}$ (and $p\geq 1$ for $\zeta_0$), but in practice $\mathcal{P} = \{p_\mathtt{min},p_\mathtt{min}+1,\dots,p_\mathtt{max}-1,p_\mathtt{max}\}$, for some integers $\max\{s,v\} \leq p_\mathtt{min} \leq p_\mathtt{max}$ (and $p_\mathtt{min}\geq 1$ for $\zeta_0$). The (in)flexibility of fixed $J = {\tt J}$ is offset by changing the polynomial approximation. This may have some practical appeal, but our theoretical results in the next section continue in the standard case of fixed $p$ and diverging $J$.

To implement the data-driven choice $\widehat p_{\mathtt{IMSE}}({\tt J},s,v)$, users needs to specify the desired (often small) number of bins $\mathtt{J}$, the smoothness order $s$, the derivative order $v$ of interest, and a (finite) set $\mathcal{P}$ of acceptable polynomial orders. The size of $\mathcal{P}$ is usually small since in practice $p=3$ or $4$ often suffices to yield a small IMSE-optimal number of bins. Then, for each value of $p$ in $\mathcal{P}$, we can implement the rule-of-thumb or direct plug-in procedure as described in Section \ref{SA section: select J}  to obtain $\widehat J_{\mathtt{IMSE}}(p,s,v)$. The ``optimal'' choice $\widehat p_{\mathtt{IMSE}}({\tt J},s,v)$ is the value of $p$ with the resulting $\widehat J_{\mathtt{IMSE}}(p,s,v)$ closest to $\mathtt{J}$.

\newpage
\section{Proofs} \label{SA section: proof}

We begin with a subsection collecting some technical lemmas used in the proofs of our main results. We then collect all the proofs of the results presented in this supplemental appendix, which are in several cases more general than those discussed in the main text. Some of our technical results may be of broader independent interest in the nonlinear series estimation literature.

\subsection{Technical Lemmas}

We first give several simple facts about $\widehat\Delta$ in the following lemma, which are immediate from Assumption \ref{SA Assumption RP}(ii).

\begin{lem}[Quasi-Uniformity]\label{SA lem: quasi-uniform partition}
    Suppose that Assumption \ref{SA Assumption RP}(ii) holds. 
    Then, 
    (i) $J^{-1}\lesssim\min_{1\leq j\leq J}h_j\leq \max_{1\leq j\leq J}h_j\lesssim J^{-1}$,
    (ii) $\max_{0\leq j\leq J}|\hat{\tau}_j-\tau_j|\lesssim_\P \mathfrak{r}_{\tt RP}$, and
    (iii) $\widehat\Delta\in\Pi_{3c_{\tt QU}}$ w.p.a. $1$.
\end{lem}
\begin{proof}
    By Assumption \ref{SA Assumption RP}(ii), 
    $\mathrm{len}(\mathcal{X})=\sum_{j=1}^Jh_j\geq J\min_{1\leq j\leq J}h_j\geq 
    c_{\tt QU}^{-1}J\max_{1\leq j\leq J} h_j$ where $\mathrm{len}(\mathcal{X})$ denotes the length of $\mathcal{X}$ (which is a fixed number). 
    On the other hand,
    $\mathrm{len}(\mathcal{X})\leq J\max_{1\leq j\leq J}h_j\\
    \leq c_{\tt QU}J\min_{1\leq j\leq J} h_j$.
    Therefore, $c_{\tt QU}^{-1}J^{-1}\mathrm{len}(\mathcal{X})\leq \min_{1\leq j\leq J}h_j
    \leq \max_{1\leq j\leq J}h_j\leq c_{\tt QU}J^{-1}\mathrm{len}(\mathcal{X})$.

    Next, by Assumption \ref{SA Assumption RP}(ii),
    \[
    \max_{0\leq j\leq J}|\hat{\tau}_j-\tau_j|
    \leq |\hat{\tau}_0-\tau_0|
    +J\max_{1\leq l\leq J}|\hat{h}_l-h_l|
    \lesssim_\P \mathfrak{r}_{\tt RP}.
    \]
    In addition,
    $\max_{1\leq j\leq J}|\hat{h}_j-h_j|\leq \frac{1}{2}c_{\tt QU}^{-1}J^{-1}\mathrm{len}(\mathcal{X})
    \leq \frac{1}{2}\min_{1\leq j\leq J}h_j$ w.p.a. $1$, and thus
    \[
    \frac{\max_{1\leq j\leq J}\hat{h}_j}{\min_{1\leq j\leq J}\hat{h}_j}=
    \frac{\max_{1\leq j\leq J}h_j+\max_{1\leq j\leq J}|\hat{h}_j-h_j|}{\min_{1\leq j\leq J}h_j-\max_{1\leq j\leq J}|\hat{h}_j-h_j|}\leq 3c_{\tt QU}, \quad \text{w.p.a.} 1.
    \]
    Then, the proof is complete.
\end{proof}

The next lemma then verifies Assumption \ref{SA Assumption RP}(ii) for the special case of quantile-spaced partitions. The proof is available in the supplemental appendix of \cite{Cattaneo-Crump-Farrell-Feng_2024_AER} (see Section SA-3.1 therein) and thus omitted here.

\begin{lem}[Quasi-Uniformity of Quantile-Spaced Partitions]
	\label{SA lem: quantile partition}
	Suppose that Assumptions \ref{SA Assumption DGP}(i) and \ref{SA Assumption DGP}(ii) hold and $\widehat\Delta$ is the quantile-based partition defined in Section \ref{SA section: Setup}, i.e., $\hat\tau_0=\inf\mathcal X$, $\hat\tau_J=\sup\mathcal X$, and $\hat{\tau}_j=\hat{F}_X^{-1}(j/J)$ for $j=1,\ldots,J-1$. If
	$\frac{J\log J}{n}=o(1)$ and $\frac{\log n}{J}=o(1)$, then
    Assumption \ref{SA Assumption RP}(ii) holds with $\tau_0=\inf\mathcal X$, $\tau_J=\sup\mathcal X$, $\tau_j=F_X^{-1}(j/J)$ for $j=1,\ldots,J-1$, and $\mathfrak{r}_{\tt RP}=\Big(\frac{J\log J}{n}\Big)^{1/2}$.
\end{lem}

The next three lemmas \ref{SA lem: spline transform}--\ref{SA lem: uniform approx rate} 
concern the properties of binscatter basis functions. Their proofs are the same as those for quantile-based partitions that are available in the supplemental appendix of \cite{Cattaneo-Crump-Farrell-Feng_2024_AER} (see Section SA-3.1 therein) and are omitted here to conserve space.

\begin{lem}[Transformation Matrix]\label{SA lem: spline transform}
	Suppose that Assumption \ref{SA Assumption RP}(i) holds. Then
	$\widehat{\bb}_{p,s}(x)=\widehat{\mathbf{T}}_s\widehat{\bb}_{p,0}(x)$ with
	$\|\widehat{\mathbf{T}}_s\|_\infty\lesssim_\P 1$ 
        and $\|\widehat{\mathbf{T}}_s\|\lesssim_\P  1$. 
        If, in addition, Assumption \ref{SA Assumption RP}(ii) holds, then 
	$\|\widehat{\mathbf{T}}_s-\mathbf{T}_s\|_\infty\lesssim_\P\mathfrak{r}_{\tt RP}$ 
        and $\|\widehat{\mathbf{T}}_s-\mathbf{T}_s\|\lesssim_\P\mathfrak{r}_{\tt RP}$.
\end{lem}

\begin{lem}[Local Basis] \label{SA lem: local basis}
Suppose that Assumption \ref{SA Assumption RP}(i) holds. Then 
$\sup_{x\in\mathcal{X}}\|\widehat{\bb}_{p,s}^{(v)}(x)\|_0\leq (p+1)^2$ and
$\sup_{x\in\mathcal{X}}\|\widehat{\bb}_{p,s}^{(v)}(x)\|\lesssim_\P J^{\frac{1}{2}+v}$.	
\end{lem}

The following lemma provides a particular way to define $\bbeta_0(\Delta)$ and $\widehat\bbeta_0$ so that the required approximation rate is achieved. We define
$$
\bbeta_0^{\mathtt{LS}}(\Delta):=\argmin_{\bbeta\in\mathbb{R}^{K_{p,s}}}\;
\E[(\mu_0(x_i)-\bb_{p,s}(x_i;\Delta)'\bbeta)^2], \quad 
\widehat\bbeta_0^{\mathtt{LS}}=\bbeta_0^{\mathtt{LS}}(\widehat\Delta).
$$
\begin{lem}[Approximation Error] \label{SA lem: uniform approx rate}
	Suppose that Assumptions \ref{SA Assumption DGP}(i)(ii), \ref{SA Assumption GL}(v) and \ref{SA Assumption RP}(i) hold. Then 
 \begin{align*}
      \sup_{\Delta\in\Pi_C}\sup_{x\in\mathcal{X}}|\bb_{p,s}^{(v)}(x;\Delta)'\bbeta_0^{\mathtt{LS}}(\Delta)-\mu_0^{(v)}(x)|
      \lesssim J^{-p-1+v}\quad \text{and}\quad
      \sup_{x\in\mathcal{X}}|\widehat{\bb}_{p,s}^{(v)}(x)'\widehat{\bbeta}_0^{\mathtt{LS}}-\mu_0^{(v)}(x)|\lesssim_\P J^{-p-1+v}.
 \end{align*} 
\end{lem}

Next, the following maximal inequality is useful in our analysis and was established as Lemma SA24 in \citet{Cattaneo-Feng-Underwood_2024_jasa}.
Its proof is available therein and thus omitted here.

\begin{lem}[Maximal Inequality]\label{SA lem: maximal inequality}
Let $Z_1, \cdots, Z_n$ be independent but not necessarily identically distributed random variables taking values in a measurable space $(\mathcal{S}; \mathscr{S})$. Denote the joint distribution of $Z_1, \cdots, Z_n$ by $\P$ and the marginal distribution of $Z_i$ by $\P_i$, and let $\bar{\P}=\frac{1}{n}\sum_{i=1}^n\P_i$. Let $\mathcal{F}$ be a class of Borel measurable functions from $\mathcal{S}$ to $\mathbb{R}$ which is pointwise measurable. Let $\bar{F}$ be a measurable envelope function for $\mathcal{F}$. Suppose that $\|\bar{F}\|_{L_2(\bar{\P})}<\infty$. Let $\bar{\sigma}>0$ satisfy $\sup_{f\in\mathcal{F}}\|f\|_{L_2(\bar{\P})}\leq\bar{\sigma}\leq \|\bar{F}\|_{L_2(\bar{\P})}$ and define $\bar{\bar{F}}=\max_{1\leq i\leq n}\bar{F}(Z_i)$. Then, with $\delta=\bar{\sigma}/\|\bar{F}\|_{L_2(\bar{\P})}$,
\[
\E\Big[\sup_{f\in\mathcal{F}}\Big|\frac{1}{\sqrt{n}}\sum_{i=1}^{n}\Big(f(Z_i)-\E[f(Z_i)]\Big)\Big|\Big]\lesssim\|\bar{F}\|_{L_2(\bar{\P})}J(\delta, \mathcal{F}, \bar{F})+\frac{\|\bar{\bar{F}}\|_{L_2(\P)}J(\delta, \mathcal{F}, \bar{F})^2}{\delta^2\sqrt{n}},
\]
where
\[
J(\delta, \mathcal{F}, \bar{F})=\int_0^{\delta}\sqrt{1+\sup_{\mathbb{Q}}\log N(\mathcal{F}, L_2(\mathbb{Q}), \varepsilon\|\bar{F}\|_{L_2(\mathbb{Q})})}\d{\varepsilon}.
\]	
\end{lem}


\subsection{Proof of Lemma \ref{SA lem: Gram, GL}}

\begin{proof}	
    We write $\Psi_{i,1}:=\Psi_1(x_i,\bw_i;\eta_i)$.

    (i) We first prove a convergence result for $\bar{\bQ}$.  
        In view of Lemma \ref{SA lem: spline transform}, it suffices to show the convergence for $s=0$. Let $\mathcal{A}_n$ denote the event on which $\widehat{\Delta}\in\Pi_C$. By Assumption \ref{SA Assumption RP}(i), $\P(\mathcal{A}_n^c)=o(1)$. On $\mathcal{A}_n$,
	\[
	\begin{split}
		&\;\Big\|\E_n[\widehat{\bb}_{p,0}(x_i)\widehat{\bb}_{p,0}(x_i)'\Psi_{i,1}\eta_{i,1}^2]-
		\E_{\widehat{\Delta}}[\widehat{\bb}_{p,0}(x_i)\widehat{\bb}_{p,0}(x_i)'\Psi_{i,1}\eta_{i,1}^2]\Big\|\\
		\leq &  \sup_{\Delta\in\Pi_C}\|\E_n[\bb_{p,0}(x_i;\Delta)\bb_{p,0}(x_i;\Delta)'\Psi_{i,1}\eta_{i,1}^2]-
		\E[\bb_{p,0}(x_i;\Delta)\bb_{p,0}(x_i;\Delta)'\Psi_{i,1}\eta_{i,1}^2]\|_\infty.
	\end{split}
	\]
 
	Let $a_{kl}$ be a generic $(k,l)$th entry of the matrix inside the norm, i.e.,
	\[
		|a_{kl}|=\Big|\E_n[b_{p,0,k}(x_{i}; \Delta)b_{p,0,l}(x_{i};\Delta)\Psi_{i,1}\eta_{i,1}^2]-
		\E[b_{p,0,k}(x_{i};\Delta)b_{p,0,l}(x_{i};\Delta)\Psi_{i,1}\eta_{i,1}^2]\Big|.
	\] 
	Clearly, if $b_{p,0,k}(\cdot \,;\Delta)$ and $b_{p,0,l}(\cdot \,;\Delta)$ are basis functions with different supports, $a_{kl}$ is zero. Now define the following function class
		\[\mathcal{G}=\Big\{(x_1, \bw_1)\mapsto b_{p,0,k}(x_1;\Delta)b_{p,0,l}(x_1;\Delta)\Psi_1(x_1,\bw_1;\eta_1)\eta_{1,1}^2:
	1\leq k, l\leq J(p+1), \Delta\in\Pi_C\Big\}.
	\] 
	We have 
	$
	\sup_{g\in\mathcal{G}}|g|_\infty\lesssim J$ and
	$
	\sup_{g\in\mathcal{G}}\V[g]\leq 
	\sup_{g\in\mathcal{G}}\E[g^2]\lesssim J,
	$
	by Assumption \ref{SA Assumption GL}. 
        Also, by Proposition 3.6.12 of \cite{Gine-Nickl_2016_book}, 
       the collection $\mathcal{G}$ is of VC type with a bounded index. 
       Then, by Lemma \ref{SA lem: maximal inequality}, 
	$$\sup_{g\in\mathcal{G}}\Big|\frac{1}{n}\sum_{i=1}^{n}g(x_i)-\E[g(x_i)]\Big|
	\lesssim_\P \sqrt{J\log J/n},$$
	which implies $\|\E_n[\widehat{\bb}_{p,0}(x_i)\widehat{\bb}_{p,0}(x_i)'\Psi_{i,1}\eta_{i,1}^2]-
	\E_{\widehat{\Delta}}[\widehat{\bb}_{p,0}(x_i)\widehat{\bb}_{p,0}(x_i)'\Psi_{i,1}\eta_{i,1}^2]\|
	\lesssim_\P \sqrt{J\log J/n}$.	
 
   Then, the lower bound on the minimum eigenvalue of $\bar\bQ$ follows by Theorem 4.42 of \cite{Schumaker_2007_book} and Assumption \ref{SA Assumption RP}(i). The upper bound immediately follows by Assumption \ref{SA Assumption RP}(i) and Lemmas \ref{SA lem: spline transform} and \ref{SA lem: local basis}. 

    Given the above fact, it follows that $\|\bar{\bQ}^{-1}\|\lesssim_\P 1$. Notice that $\bar{\bQ}$ is a banded matrix with a finite band width. Then, the bounds on the elements of $\bar{\bQ}^{-1}$ and $\|\bar{\bQ}^{-1}\|_\infty$  hold by Theorem 2.2 of \cite{Demko_1977_SIAM}.
    
     (ii) By Assumption \ref{SA Assumption DGP} and \ref{SA Assumption GL}, $\Psi_{i,1}\eta_{i,1}^2$ is bounded and bounded away from zero uniformly over $1\leq i\leq n$. Then, $\E[\bb_{p,s}(x_i)\bb_{p,s}(x_i)']\lesssim \bQ_0\lesssim\E[\bb_{p,s}(x_i)\bb_{p,s}(x_i)']$. The desired bounds on the minimum and maximum eigenvalues of $\bQ_0$ follow from Lemma SA-3.5 of \cite{Cattaneo-Crump-Farrell-Feng_2024_AER}.
     
	Next, we show the convergence of $\bar\bQ$ to $\bQ_0$. 
        Let $\alpha_{kl}$ be a generic $(k,l)$th entry of
	$$
	\E_{\widehat{\Delta}}[\widehat{\bb}_{p,0}(x_{i})\widehat{\bb}_{p,0}(x_{i})'\Psi_{i,1}\eta_{i,1}^2]/J - \E[\bb_{p,0}(x_{i})\bb_{p,0}(x_{i})'\Psi_{i,1}\eta_{i,1}^2]/J.
	$$ 
        By definition, it is either equal to zero or
	\begin{align*}
		\alpha_{kl}
		=&\int_{\widehat{\mathcal{B}}_j}\Big(\frac{x-\hat{\tau}_{j-1}}{\hat{h}_j}\Big)^\ell
		\varphi(x) f_X(x)\d{x}
		-\int_{\mathcal{B}_j}\Big(\frac{x-\tau_{j-1}}{h_j}\Big)^\ell \varphi(x) f_X(x)\d{x} \\
		=&\hat{h}_j\int_0^1z^\ell \varphi(z\hat{h}_j+\hat{\tau}_{j-1})
		f_X(z\hat{h}_j+\hat{\tau}_{j-1})\d{z}
		-h_j\int_0^1 z^\ell \varphi(zh_j+\tau_{j-1})
		f_X(zh_j+\tau_{j-1})\d{z} \nonumber\\
		=&(\hat{h}_j-h_j)\int_0^1z^\ell \varphi(z\hat{h}_j+\hat{\tau}_{j-1})
		f_X(z\hat{h}_j+\hat{\tau}_{j-1})\d{z}\\
		&+h_j\int_0^1z^\ell\Big(\varphi(z\hat{h}_j+\hat{\tau}_{j-1})f_X(z\hat{h}_j+\hat{\tau}_{j-1})-
		\varphi(zh_j+\tau_{j-1})
		f_X(zh_j+\tau_{j-1})\Big)\d{z}
	\end{align*} 
	for some $1\leq j\leq J$ and $0\leq\ell\leq 2p$ and $\varphi(x)=\E[\varkappa(x,\bw_i)\mid x_i=x]$.
	Because $\varphi f_X$ is bounded and Lipschitz, Lemma \ref{SA lem: quasi-uniform partition} and Assumption \ref{SA Assumption RP}(ii) imply $\max_{k,l}|\alpha_{kl}|\lesssim_\P J^{-1}\mathfrak r_{\tt RP}$. Since the matrices have uniformly bounded bandwidth, rescaling by $J$ therefore gives  $$\|\E_{\widehat{\Delta}}[\widehat{\bb}_{p,0}(x_i)\widehat{\bb}_{p,0}(x_i)'\Psi_{i,1}\eta_{i,1}^2]-\bQ_0\|\lesssim_\P 
    \mathfrak{r}_{\tt RP}.$$ 
	
	Since $\bar{\bQ}$ and $\bQ_0$ are banded matrices with finite band widths. Then, the bound $\|\bar{\bQ}^{-1}-\bQ_0^{-1}\|_\infty$ hold by Theorem 2.2 of \cite{Demko_1977_SIAM}. This completes the proof.
\end{proof}

\subsection{Proof of Lemma \ref{SA lem: asymp variance, GL}}

\begin{proof}
	Since $\E[\psi(y_i, \eta_i)^2|x_i=x, \bw_i=\bw]$ and $(\eta^{(1)}(\mu_0(x)+\bw'\bgamma_0))^2$ is bounded and bounded away from zero uniformly over $x\in\mathcal{X}$ and $\bw\in\mathcal{W}$,
	$\E_n[\widehat{\bb}_{p,s}(x_i)\widehat{\bb}_{p,s}(x_i)']\lesssim \bar{\bSigma}\lesssim\E_n[\widehat{\bb}_{p,s}(x_i)\widehat{\bb}_{p,s}(x_i)']$. By the same argument in the proof of Lemma \ref{SA lem: Gram, GL} (we can simply drop the additional term $\Psi_{i,1}\eta_{i,1}^2$ in $\bar\bQ$), the eigenvalues of $\E_n[\widehat\bb_{p,s}(x_i)\widehat\bb_{p,s}(x_i)']$ and thus $\bar\bSigma$ are bounded and bounded away from zero. Then, the desired results follow from Lemma \ref{SA lem: Gram, GL} and the fact that $\inf_{x\in\mathcal{X}}\|\widehat\bb_{p,s}^{(v)}(x)\|\gtrsim J^{1/2+v}$ w.p.a. $1$ (it was shown in the proof of Lemma SA-3.6 of \cite{Cattaneo-Crump-Farrell-Feng_2024_AER}).
	Under Assumption \ref{SA Assumption RP}(ii), Lemmas \ref{SA lem: spline transform} and \ref{SA lem: local basis} also give
	$\sup_{x\in\mathcal X}\|(\widehat\bT_s-\bT_s)\widehat\bb_{p,0}^{(v)}(x)\|\lesssim_\P \mathfrak r_{\tt RP}J^{1/2+v}=o_\P(J^{1/2+v})$, so $\bT_s\widehat\bb_{p,0}^{(v)}(x)$ obeys the same uniform upper and lower bounds as $\widehat\bb_{p,s}^{(v)}(x)$. This proves the claims for the population variances under their revised definitions.
\end{proof}

\subsection{Proof of Lemma \ref{SA lem: uniform converge var part, GL}}
\begin{proof}
	By Lemmas \ref{SA lem: spline transform}, \ref{SA lem: local basis} and \ref{SA lem: Gram, GL},  $\sup_{x\in\mathcal{X}}\|\widehat{\bb}_{p,s}^{(v)}(x)\|_1\lesssim_\P J^{1/2+v}$, $\|\bar{\bQ}^{-1}\|_\infty\lesssim_\P 1$ and $\|\widehat{\bT}_s\|_\infty\lesssim_\P 1$. 
    Recall that by Assumption \ref{SA Assumption GL}, 
    $\psi(y_i,\eta_i)=\psi^\dagger(y_i-\eta_i)\psi^\ddagger(\eta_i)=\psi^\dagger(\epsilon_i)\psi^\ddagger(\eta_i)$.
    Define the following function class
	\[
	\mathcal{G}=\Big\{(x_1, \bw_1, \epsilon_1)\mapsto b_{p,0,l}(x_1;\Delta)\eta^{(1)}      (\mu_0(x_1)+\bw_1'\bgamma_0)\psi^\dagger(\epsilon_1)\psi^\ddagger(\eta_1):1\leq l\leq J(p+1), \Delta\in\Pi_C
	\Big\}.
	\]
 Then, $\sup_{g\in\mathcal{G}}|g|\lesssim \sqrt{J}|\psi^\dagger(\epsilon_1)|$, and hence take an envelop $\bar{G}=C\sqrt{J}|\psi^\dagger(\epsilon_1)|$ for some $C$ large enough. Moreover, $\sup_{g\in\mathcal{G}}\V[g]\lesssim 1$ and $\mathcal{G}$ is of VC type with a bounded index. 
 By Proposition 6.1 of \citet{Belloni-Chernozhukov-Chetverikov-Kato_2015_JoE},
	\[
 \sup_{g\in\mathcal{G}}\Big|\frac{1}{n}\sum_{i=1}^{n}g(x_i, \epsilon_i)\Big|\lesssim_\P \sqrt{\frac{\log J}{n}}+\frac{J^{\frac{\nu}{2(\nu-2)}}\log J}{n}\lesssim\sqrt{\frac{\log J}{n}},
	\]
	and the desired result follows.
\end{proof}

\subsection{Proof of Lemma \ref{SA lem: proj approx error, GL}}

\begin{proof} 
	Let $\hat{z}_i=\widehat{\bb}_{p,s}(x_i)'\widehat{\bbeta}_0+\bw_i'\bgamma_0$ and 
       $\mathfrak{r}(x_i,\bw_i, y_i):=\mathfrak{r}(x_i, \bw_i, y_i;\widehat{\Delta}):=
	\eta_{i,1}\psi(y_i,\eta_i)-
     \eta^{(1)}(\hat{z}_i)
     \psi(y_i, \eta(\hat{z}_i))\\
     =A_1(x_i,  \bw_i, y_i)+A_2(x_i, \bw_i, y_i)$ where 
	\begin{align*}
	&A_1(x_i, \bw_i, y_i):=A_1(x_i,\bw_i,  y_i;\widehat{\Delta}):=
 [\eta_{i,1}\psi^\ddagger(\eta_i)-\eta^{(1)}(\hat{z}_i)\psi^\ddagger(\eta(\hat{z}_i))]
 \psi^\dagger(y_i-\eta_i) \;\;\text{and}\\
	&A_2(x_i, \bw_i, y_i):=A_2(x_i,\bw_i,  y_i;\widehat{\Delta}):=
 \eta^{(1)}(\hat{z}_i)\psi^\ddagger(\eta(\hat{z}_i))[\psi^\dagger(y_i-\eta_i)-\psi^\dagger(y_i-\eta(\hat{z}_i))].
	\end{align*}
 
	First, by Assumption \ref{SA Assumption GL} and Lemma \ref{SA lem: uniform approx rate}, $\sup_{x\in\mathcal{X}, \bw\in\mathcal{W}}|\eta_{i,1}\psi^\ddagger(\eta_i)-\eta^{(1)}(\hat{z}_i)\psi^\ddagger(\eta(\hat{z}_i))|\lesssim J^{-p-1}$ w.p.a. 1. Also, for every $1\leq l\leq K_{p,s}$ and $\Delta\in\Pi_C$,
	\[\begin{split}
	&b_{p,s,l}(x;\Delta)\Big(\eta_{i,1}\psi^\ddagger(\eta_i)-
	\eta^{(1)}(\bb_{p,s}(x;\Delta)'\bbeta_0(\Delta)+\bw'\bgamma_0)\psi^\ddagger\!\left(\eta\!\left(\bb_{p,s}(x;\Delta)'\bbeta_0(\Delta)+\bw'\bgamma_0\right)\right)\Big)\\
	=&\;b_{p,s,l}(x;\Delta)\eta_{i,1}\psi^\ddagger(\eta_i)-\\
	&\;b_{p,s,l}(x;\Delta)\eta^{(1)}\bigg(\sum_{k=\underline{k}_l}^{\underline{k}_l+p}b_{p,s,k}(x;\Delta)\beta_{0,k}(\Delta)+\bw'\bgamma_0\bigg)\psi^\ddagger\!\left(\eta\!\left(\sum_{k=\underline{k}_l}^{\underline{k}_l+p}b_{p,s,k}(x;\Delta)\beta_{0,k}(\Delta)+\bw'\bgamma_0\right)\right)
	\end{split}\]
	for some integer $\underline{k}_l\in[1, K_{p,s}]$ where $\beta_{0,k}(\Delta)$ denotes the $k$th element in  $\bbeta_0(\Delta)$. Then, the function class $\mathcal{G}=\{(x,\bw, y)\mapsto b_{p,s,l}(x;\Delta)A_1(x,\bw, y;\Delta): 1\leq l\leq K_{p,s}, \Delta\in\Pi_C\}$ is of VC type with a bounded index. By the same argument given in the proof of Lemma \ref{SA lem: uniform converge var part, GL},
	\[
	\|\E_n[\widehat{\bb}_{p,s}(x_i)A_1(x_i,\bw_i, y_i)]\|_\infty\lesssim_\P J^{-p-1} \Big(\frac{\log J}{n}\Big)^{1/2}.
	\]
	
	Next, let $\mathscr{F}_{XW\Delta}$ be the $\sigma$-field generated by $\{(x_i, \bw_i)\}_{i=1}^n$ and $\widehat\Delta$. Note that
	\[
	\begin{split}
	\E_n[\widehat{\bb}_{p,s}(x_i)A_2(x_i, \bw_i, y_i)] =\;&\E_n[\E[\widehat{\bb}_{p,s}(x_i)A_2(x_i, \bw_i , y_i)|\mathscr{F}_{XW\Delta}]]+\\
	&\E_n\Big[\widehat{\bb}_{p,s}(x_i)A_2(x_i, \bw_i, y_i)-\E[\widehat{\bb}_{p,s}(x_i)A_2(x_i, \bw_i,  y_i)|\mathscr{F}_{XW\Delta}]\Big].
	\end{split}
	\] 
	By Assumption \ref{SA Assumption GL}(iii) and Lemma \ref{SA lem: uniform approx rate},
	\begin{align*}
        &\max_{1\leq i\leq n}
        |\E[A_2(x_i, \bw_i, y_i)|\mathscr{F}_{XW\Delta}]|\\
        =&\max_{1\leq i\leq n}|-\eta^{(1)}(\widehat{\bb}_{p,s}(x_i)'\widehat{\bbeta}_0+\bw_i'\bgamma_0)
	\Psi(x_i, \bw_i;\eta(\hat{z}_i))|
	\lesssim_\P J^{-p-1}.
	\end{align*} 
        Then, 
	$\|\E_n[\E[\widehat{\bb}_{p,s}(x_i)A_2(x_i, \bw_i, y_i)|\mathscr{F}_{XW\Delta}]]\|_\infty
	\lesssim_\P J^{-p-1-1/2}$ by the same argument in the proof of Lemma \ref{SA lem: Gram, GL}. 
	On the other hand, define the following function class
	\[
	\mathcal{G}:=\Big\{(x,\bw, y)\mapsto b_{p,s,l}(x;\Delta)A_2(x, \bw, y;\Delta):1\leq l\leq K_{p,s}, \Delta\in\Pi_C \Big\}.
	\]
	By Assumption \ref{SA Assumption GL}, 
	$\sup_{g\in\mathcal{G}}\|g\|_\infty\lesssim J^{1/2}$, and $\sup_{g\in\mathcal{G}}\V[g(x_i,\bw_i,y_i)]\lesssim J^{-p-1}$. By a similar argument given before, this function class is of VC type with a bounded index. Then, by Lemma \ref{SA lem: maximal inequality},
	\[
	\sup_{g\in\mathcal{G}}\Big|\frac{1}{n}\sum_{i=1}^{n}(g(x_i, \bw_i, y_i)-\E[g(x_i,\bw_i,y_i)])\Big|\lesssim_\P
	J^{-\frac{p+1}{2}}\sqrt{\frac{\log J}{n}}+\frac{J^{1/2}\log J}{n}.	
	\]
	Collecting these results, we conclude that
	\[
	\sup_{x\in\mathcal X}\left|\widehat{\bb}_{p,s}^{(v)}(x)'\bar{\bQ}^{-1}\E_n[\widehat{\bb}_{p,s}(x_i)\mathfrak{r}(x_i, \bw_i, y_i)]\right|
	\lesssim_\P J^{-p-1+v}+J^{\frac{2v-p-1}{2}}\Big(\frac{J\log J}{n}\Big)^{1/2}+\frac{J^{1+v}\log J}{n}.
	\]
	The proof is complete.
\end{proof}

\subsection{Proof of Lemma \ref{SA lem: uniform consistency, GL}}
\begin{proof}
	By convexity of $\rho(y;\eta(\cdot))$, we only need to consider $\bbeta=\widehat{\bbeta}_0+\varepsilon\bm{\alpha}/\sqrt{J}$ for any sufficiently small fixed $\varepsilon>0$ and $\bm{\alpha}\in\mathbb{R}^{K_{p,s}}$ such that $\|\bm{\alpha}\|=1$. For notational simplicity, let $\widehat{\bb}_i:=\widehat{\bb}_{p,s}(x_i)$. 
	For this choice of $\bbeta$ and $\bgamma\in\mathbb{R}^{d_\gamma}$,
	\[
        \begin{split}
	\delta_i(\bbeta, \bgamma)&=\rho(y_i;\eta(\widehat{\bb}_i'\bbeta+\bw_i'\bgamma))-       \rho(y_i;\eta(\widehat{\bb}_i'\widehat{\bbeta}_0+\bw_i'\bgamma))\\
	&=\int_0^{\varepsilon\widehat{\bb}_i'\bm{\alpha}/\sqrt{J}}\psi\Big(y_i,\eta(\widehat{\bb}_i'\widehat{\bbeta}_0+\bw_i'\bgamma+t)\Big)\eta^{(1)}(\widehat{\bb}_i'\widehat{\bbeta}_0+\bw_i'\bgamma+t)\d{t}.
	\end{split}\] 
	Let $\mathscr{F}_{XW\Delta}$ be the $\sigma$-field generated by $\{(x_i,\bw_i)\}_{i=1}^n$ and $\widehat\Delta$.  We have
	\[
	\E_n[\delta_i(\bbeta, \widehat{\bgamma})]
	=\frac{1}{\sqrt{n}}\G_n[\delta_i(\bbeta, \widehat{\bgamma})]+
	\E_n\Big[\E[\delta_i(\bbeta, \widehat{\bgamma})|\mathscr{F}_{XW\Delta}]\Big],
	\]
	where $\G_n[\cdot]$ denotes $\sqrt{n}(\E_n[\cdot]-\E[\cdot|\mathscr{F}_{XW\Delta}])$, and $\E[\delta_i(\bbeta, \widehat{\bgamma})|\mathscr{F}_{XW\Delta}]:=\E[\delta_i(\bbeta, \bgamma)|\mathscr{F}_{XW\Delta}]|_{\bgamma=\widehat{\bgamma}}$, i.e., the conditional expectation with $\widehat{\bgamma}$ viewed as fixed.
	By Assumption \ref{SA Assumption GL},
	\begin{align*}
	&\E[\delta_i(\bbeta, \widehat{\bgamma})|\mathscr{F}_{XW\Delta}]=	\int_0^{\varepsilon\widehat{\bb}_i'\bm{\alpha}/\sqrt{J}}\Psi\Big(x_i,\bw_i;\eta(\widehat{\bb}_i'\widehat{\bbeta}_0+\bw_i'\widehat{\bgamma}+t)\Big)\eta^{(1)}(\widehat{\bb}_i'\widehat{\bbeta}_0+\bw_i'\widehat{\bgamma}+t)\d{t}\\
	=\;&\int_0^{\varepsilon\widehat{\bb}_i'\bm{\alpha}/\sqrt{J}}
	\Psi_1(x_i,\bw_i;\xi_{i,t})(\eta(\widehat{\bb}_i'\widehat{\bbeta}_0+\bw_i'\widehat{\bgamma}+t)-\eta_{i})\eta^{(1)}(\widehat{\bb}_i'\widehat{\bbeta}_0+\bw_i'\widehat{\bgamma}+t)\d{t},
     \end{align*}
    where $\xi_{i,t}$ is between $\eta(\widehat{\bb}_i'\widehat{\bbeta}_0+\bw_i'\widehat{\bgamma}+t)$ and $\eta(\mu_0(x_i)+\bw_i'\bgamma_0)$ and we use the fact that $\Psi(x,\bw_i;\eta_i)=0$. 
	By Lemma \ref{SA lem: uniform approx rate} 
	and the imposed conditions that $\eta(\cdot)$ is strictly monotonic,
	$\widehat{\bgamma}-\bgamma_0\lesssim_\P \mathfrak{r}_\gamma$, 
	{
	$\sup_i|\bw_i'(\widehat\bgamma-\bgamma_0)|\lesssim_\P\sqrt{d_\gamma}\mathfrak r_\gamma=o_\P(\sqrt{J/n}+J^{-p-1})$ and
	$\sqrt J\big(J^{-p-1}+\sqrt{d_\gamma}\mathfrak r_\gamma\big)=o(1)$,}
	we have 
	$\E_n[\E[\delta_i(\bbeta, \widehat{\bgamma})|\mathscr{F}_{XW\Delta}]]
	\gtrsim_\P\varepsilon^2\bm{\alpha}'\E_n[\widehat{\bb}_i\widehat{\bb}_i']\bm{\alpha}/J
	\gtrsim_\P J^{-1}\varepsilon^2$. 
	
	On the other hand, let $\mathcal{H}:=\{\bgamma:\|\bgamma-\bgamma_0\|\leq C\mathfrak{r}_\gamma\}$ and define the following function class
	\[
	\mathcal{G}:=\Big\{(x_i,\bw_i,y_i)\mapsto \delta_i(\bbeta, \bgamma): \bm{\alpha}\in\mathcal{S}^{K_{p,s}}, \bgamma\in\mathcal{H}\Big\}.
	\]
	Note that
	\begin{align*}
	\delta_i(\bbeta, \bgamma)=  &\int_0^{\varepsilon\widehat{\bb}_i'\bm{\alpha}/\sqrt{J}}\Big(\psi(y_i,\eta(\widehat{\bb}_i'\widehat{\bbeta}_0+\bw_i'\bgamma+t))-\psi(y_i,\eta_{i})\Big)\eta^{(1)}(\widehat{\bb}_i'\widehat{\bbeta}_0+\bw_i'\bgamma+t)\d{t}\;+\\
	&\int_0^{\varepsilon\widehat{\bb}_i'\bm{\alpha}/\sqrt{J}}\psi(y_i,\eta_i)\eta^{(1)}(\widehat{\bb}_i'\widehat{\bbeta}_0+\bw_i'\bgamma+t)\d{t}.
	\end{align*}
	By Assumption \ref{SA Assumption GL}, we have $\sup_{g\in\mathcal{G}}|g|\lesssim \varepsilon(1+|\psi(y_i,\eta_i)|)$,
	$\|\max_{1\leq i\leq n}|\psi(y_i,\eta_i)|\|_{L_2(\P)}\lesssim n^{1/\nu}$, 
	 $\sup_{g\in\mathcal{G}}\E_n[\E[g^2|\mathscr{F}_{XW\Delta}]]
	 \lesssim_\P J^{-1}\varepsilon^2$, and {local support and the piecewise-Lipschitz score condition give the covering bound
	\[ \sup_{\mathbb Q}\log N\big(\mathcal G,L_2(\mathbb Q),t\|\varepsilon(1+|\psi|)\|_{\mathbb Q,2}\big)\lesssim(K_{p,s}+d_\gamma)\log(C/t). \]}
        Therefore, by Lemma  \ref{SA lem: maximal inequality} and the rate restriction, 
	\[
	\sup_{g\in\mathcal{G}}
	\Big|\frac{1}{\sqrt{n}}\G_n[\delta_i(\bbeta, \bgamma)]\Big|
	{\lesssim_\P J^{-1}\Big(\frac{J(J+d_\gamma)\log J}{n}\Big)^{1/2}\varepsilon+J^{-1}\frac{J(J+d_\gamma)\log J}{n^{1-\frac{1}{\nu}}}\varepsilon=o(\varepsilon/J).}
	\]
	Thus, for any fixed (sufficiently small) $\varepsilon>0$, $\E_n[\delta_i(\bbeta, \widehat{\bgamma})]>0$ when $n$ is  sufficiently large. Thus, $\|\widehat{\bbeta}-\widehat{\bbeta}_0\|=o_\P(J^{-1/2})$, implying  $\|\widehat{\bbeta}-\widehat{\bbeta}_0\|_\infty=o_\P(J^{-1/2})$ immediately.
\end{proof}

\subsection{Proof of Theorem \ref{SA thm: Bahadur representation, GL}}
\begin{proof} 
	The proof is long. We divide it into several steps.
	
	\textbf{Step 0:}
	We first prepare some notation and useful facts. To simplify the presentation, in this proof we drop the scaling factor $\sqrt{J}$ in the basis by defining 
	$$\breve{\bb}_i:=\widehat{\bb}_{p,s}(x_i)/\sqrt{J}=(\widehat{b}_{p,s,1}(x_i), \cdots, \widehat{b}_{p,s,K_{p,s}}(x_i))'/\sqrt{J}\quad 
	\text{and}\quad \breve{\bbeta}:=\sqrt{J}\widehat{\bbeta},\quad
	\breve{\bbeta}_0=\sqrt{J}\widehat{\bbeta}_0.
	$$                     
	The Hessian in this rescaled parameterization is
	\[
	\breve{\bQ}:=\frac{\bar{\bQ}}{J},
	\qquad
	\breve{\bQ}^{-1}=J\bar{\bQ}^{-1}.
	\]
	Throughout the proof, $C, c, C_1, c_1, C_2, c_2, \cdots$ denote (strictly positive) absolute constants,  $\mathscr{F}_{XW\Delta}$ denotes the $\sigma$-field generated by $\{(x_i,\bw_i)\}_{i=1}^n$ and $\widehat\Delta$, and $\supp(g(\cdot))$ denotes the support of a generic function $g(\cdot)$. 
	Moreover, define 
	\[
	\begin{split}
		&\mathcal{V}=\{(v_1, \cdots, v_{K_{p,s}})':  \exists k\in\{1,\cdots, K_{p,s}\},  |v_{\ell}|\leq\varrho^{|k-\ell|}\varepsilon_n \text{ for }|\ell-k|\leq M_n \text{ and } v_\ell=0 \text{ otherwise} \},\\
		&\mathcal{H}_l=\{\bv\in\mathbb{R}^{K_{p,s}}:\|\bv\|_\infty\leq r_{l,n}\}\;\;
		\text{for}\; l=1,2,\quad \text{and}\quad
		\mathcal{H}_3=\{\bv\in\mathbb{R}^{d_\gamma}: \|\bv\|\leq r_{3,n}\},
	\end{split}
	\]
	where $\varrho\in(0,1)$ is the constant given in Lemma \ref{SA lem: Gram, GL}, 
	$r_{1,n}=C_1[(J\log n/n)^{1/2}+J^{-p-1}]$, 
	$r_{2,n}=\mathfrak{z}\mathfrak{r}_{2,n}$ for $\mathfrak{z}>0$, 
	$\varepsilon_n=\mathfrak{z}'\mathfrak{r}_{2,n}$ for $\mathfrak{z}'>0$,
	{$\mathfrak{r}_{2,n}=\mathfrak r_{\mathrm{BR},n}$},
	$r_{3,n}=C\mathfrak{r}_\gamma$, and $M_n=c_1\log n$. In the last step of the proof, we will consider $\mathfrak{z}=2^\ell$, $\ell=L, L+1, \cdots, \bar{L}$ where $\bar{L}$ is the smallest number such that $2^{\bar{L}}\mathfrak r_{2,n}\geq c$ for some sufficiently small constant $c>0$, and $\varepsilon_n$ is a quantity that we can choose. By {Assumption \ref{SA Assumption HLE}(i)}, $\widehat{\bgamma}-\bgamma_0\in\mathcal{H}_3$ with probability approaching one for $C$ large enough, and by Lemma \ref{SA lem: uniform consistency, GL}, $\|\breve{\bbeta}-\breve{\bbeta}_0\|\leq c$ with probability approaching one.

	For any $\bbeta_1\in\mathcal{H}_1, \bbeta_2\in\mathcal{H}_2$, $\bups\in\mathcal{V}$ and $\bgamma:=\bgamma_0+\bgamma_1$ with $\bgamma_1\in\mathcal{H}_3$, define
	\[
	\begin{split}
		\delta_i(\bbeta_1, \bbeta_2, \bups,\bgamma)
		=\;&\rho\Big(y_i; \eta(\breve{\bb}_i'(\breve{\bbeta}_0+\bbeta_1+\bbeta_2)+\bw_i'\bgamma)\Big)-
		\rho\Big(y_i;\eta(\breve{\bb}_i'(\breve{\bbeta}_0+\bbeta_1+\bbeta_2-\bups)+\bw_i'\bgamma)\Big)\\
		&-\Big[\eta(\breve{\bb}_i'(\breve{\bbeta}_0+\bbeta_1+\bbeta_2)+\bw_i'\bgamma)-
		\eta(\breve{\bb}_i'(\breve{\bbeta}_0+\bbeta_1+\bbeta_2-\bups)+\bw_i'\bgamma)\Big]\\
		&\hspace{21em}\times\psi(y_i,\eta(\breve{\bb}_i'\breve{\bbeta}_0+\bw_i'\bgamma_0))\\
		=\;&\int_{-\breve{\bb}_i'\bups}^{0}
		\Big[\psi\Big(y_i, \eta(\breve{\bb}_i'(\breve{\bbeta}_0+\bbeta_1+\bbeta_2)+\bw_i'\bgamma+t)\Big)
		-\psi\Big(y_i,\eta(\breve{\bb}_i'\breve{\bbeta}_0+\bw_i'\bgamma_0)\Big)\Big]\\
		&\hspace{15em}\times\eta^{(1)}\Big(\breve{\bb}_i'(\breve{\bbeta}_0+\bbeta_1+\bbeta_2)+\bw_i'\bgamma+t\Big)\d{t}.
	\end{split}
	\]
	Note that $\delta_i(\bbeta_1, \bbeta_2, \bups, \bgamma)\neq 0$ only if $\breve{\bb}_i'\bups\neq 0$. For each $\bups\in\mathcal{V}$, let $\mathcal{J}_{\bups}=\{j: \upsilon_j\neq 0\}$. By construction, the cardinality of $\mathcal{J}_{\bups}$ is bounded by $2M_n+1$. We have $\delta_i(\bbeta_1, \bbeta_2, \bups, \bgamma)\neq 0$ only if $\breve{b}_{j}(x_i)\neq 0$ for some $j\in\mathcal{J}_\upsilon$, which happens only when $x_i\in\supp(\breve{b}_j(\cdot))$ for some $j\in\mathcal{J}_{\bups}$. Let $\mathcal{I}_{\bups}=\cup_{j\in\mathcal{J}_{\bups}}\supp(\breve{b}_j(\cdot))$.  Since the basis functions are locally supported, $\mathcal{I}_{\bups}$ includes at most $c_2M_n$ (connected) intervals for all $\bups\in\mathcal{V}$. Moreover, at most $c_3M_n$ basis functions in $\breve{\bb}(\cdot)$ have supports overlapping with $\mathcal{I}_{\bups}$. Denote the set of indices for such basis functions by $\bar{\mathcal{J}}_{\bups}$.
	Let $\breve{\beta}_{0,j}$, $\beta_{1,j}$ and $\beta_{2,j}$ be the $j$th entries of $\breve{\bbeta}_0$, $\bbeta_1$, and $\bbeta_2$ respectively, and $\upsilon_j$ be the $j$th entry of $\bups$. Based on the above observations, we have $\delta_i(\bbeta_1, \bbeta_2, \bups, \bgamma)\equiv\delta_i(\bbeta_{1,\bar{\mathcal{J}}_{\bups}}, \bbeta_{2,\bar{\mathcal{J}}_{\bups}}, \bups, \bgamma)$ where
	\[
	\begin{split}
		\delta_i(\bbeta_{1,\bar{\mathcal{J}}_{\bups}}, \bbeta_{2,\bar{\mathcal{J}}_{\bups}}, \bups, \bgamma)&:=
		\int_{-\underset{j\in\mathcal{J}_{\bups}}{\sum}\breve{b}_{i,j}\upsilon_j}^{0}\Big[
		\psi\Big(y_i,\eta\Big(\sum_{l\in\bar{\mathcal{J}}_{\bups}}
		\breve{b}_{i,l}(\breve{\beta}_{0,l}+\beta_{1,l}+\beta_{2,l})+\bw_i'\bgamma+t\Big)\Big)
		\\
		-\psi\Big(y_i,\,&\eta\Big(\sum_{l\in\bar{\mathcal{J}}_{\bups}}
		\breve{b}_{i,l}\breve{\beta}_{0,l}+\bw_i'\bgamma_0\Big)\Big)\Big]
		\times\eta^{(1)}\Big(\sum_{l\in\bar{\mathcal{J}}_{\bups}}\breve{b}_{i,l}(\breve{\beta}_{0,l}+\beta_{1,l}+\beta_{2,l})+\bw_i'\bgamma+t\Big)\d{t}\,\I_{i,\bups},
	\end{split}
	\]
	$\I_{i,\bups}=\I(x_i\in\mathcal{I}_{\bups})$, and $\bbeta_{1,\bar{\mathcal{J}}_{\bups}}$ and $\bbeta_{2,\bar{\mathcal{J}}_{\bups}}$ respectively denote the subvectors of $\bbeta_1$ and $\bbeta_2$ whose indices belong to $\bar{\mathcal{J}}_{\bups}$.
	Accordingly, define the following function class
	\begin{align*}
	\mathcal{G}=\Big\{(x_i, \bw_i, y_i)\mapsto \delta_i(\widetilde{\bbeta}_1, 
	\widetilde{\bbeta}_2, \bups, \bgamma): \bups\in\mathcal{V}, \,
	&\widetilde{\bbeta}_1\in\mathbb{R}^{c_3M_n}, 
	\widetilde{\bbeta}_2\in\mathbb{R}^{c_3M_n},\\
	&\|\widetilde{\bbeta}_1\|_\infty\leq r_{1,n}, 
	\|\widetilde{\bbeta}_2\|_\infty\leq r_{2,n}, 
	\bgamma-\bgamma_0\in\mathcal{H}_3
	\Big\}.
	\end{align*}
	
	\textbf{Step 1:}
	We bound $\sup_{g\in\mathcal{G}}|\E_n[g(x_i,\bw_i,y_i)]-\E[g(x_i,\bw_i,y_i)|\mathscr{F}_{XW\Delta}]|$ in this step.
	Let $a_i(t):=\eta(\sum_{l\in\bar{\mathcal{J}}_{\bups}}\breve{b}_{i,l}\breve{\beta}_{0,l}+\bw_i'\bgamma_0+t)$. Define
	\[\begin{split}
	&
	\underline{a}_i=\min\Big\{a_i(0), a_i\Big(\sum_{l\in\bar{\mathcal{J}}_{\bups}}\breve{b}_{i,l}(\beta_{1,l}+\beta_{2,l})+\bw_i'\bgamma_1\Big), a_i\Big(\sum_{l\in\bar{\mathcal{J}}_{\bups}}\breve{b}_{i,l}(\beta_{1,l}+\beta_{2,l})+\bw_i'\bgamma_1+\sum_{j\in\mathcal{J}_{\bups}}\breve{b}_{i,j}\upsilon_j\Big)\Big\} \text{ and} \\
	&\bar{a}_i=\max\Big\{a_i(0), a_i\Big(\sum_{l\in\bar{\mathcal{J}}_{\bups}}\breve{b}_{i,l}(\beta_{1,l}+\beta_{2,l})+\bw_i'\bgamma_1\Big), a_i\Big(\sum_{l\in\bar{\mathcal{J}}_{\bups}}\breve{b}_{i,l}(\beta_{1,l}+\beta_{2,l})+\bw_i'\bgamma_1+\sum_{j\in\mathcal{J}_{\bups}}\breve{b}_{i,j}\upsilon_j\Big)\Big\}.
	\end{split}
	\]
	Consider the following two cases. 
	
	First, 
	suppose that $(y_i-\bar{a}_i, y_i-\underline{a}_i)$ does not contain any discontinuity points. By Assumption \ref{SA Assumption GL}, for all $t$ in the interval of integration
	$[-\sum_{j\in\mathcal{J}_{\bups}}\breve{b}_{i,j}\upsilon_j, 0]$ (or 
	$[0, -\sum_{j\in\mathcal{J}_{\bups}}\breve{b}_{i,j}\upsilon_j])$,
		{$$\Big|\psi\Big(y_i, a_i\Big(\sum_{l\in\bar{\mathcal{J}}_{\bups}}\breve{b}_{i,l}(\beta_{1,l}+\beta_{2,l})+\bw_i'\bgamma_1+t\Big)\Big)-\psi(y_i, a_i(0))\Big|\lesssim (1+|\psi(y_i,\eta_i)|)(r_{1, n}+r_{2,n}+\varepsilon_n+\sqrt{d_\gamma}r_{3,n}).$$}
	Second, if $(y_i-\bar{a}_i, y_i-\underline{a}_i)$  contains at least one discontinuity point, say $\jmath$.   
	For any $t$ in the interval of integration, by Assumption \ref{SA Assumption GL}, 
	{\[
	\begin{aligned}
	&\Big|\psi\Big(y_i, a_i\Big(\sum_{l\in\bar{\mathcal{J}}_{\bups}}\breve{b}_{i,l}(\beta_{1,l}+\beta_{2,l})+\bw_i'\bgamma_1+t\Big)\Big)-\psi(y_i, a_i(0))\Big|\\
      &\qquad\lesssim 1+\sqrt{d_\gamma}r_{3,n}+(1+|\psi(y_i,\eta_i)|)(r_{1,n}+r_{2,n}+\varepsilon_n+\sqrt{d_\gamma}r_{3,n}).
	\end{aligned}
	\]}
	for any $(x_i,\bw_i,y_i)$, and in this case $y_i\in(\jmath+\underline{a}_i, \jmath+\bar{a}_i)$. {By Assumption \ref{SA Assumption GL}, $$|\bar{a}_i-\underline{a}_i|\lesssim (r_{1,n}+r_{2,n}+\sqrt{d_\gamma}r_{3,n}+\varepsilon_n)(|\eta_{i,1}|+r_{1,n}+r_{2,n}+\sqrt{d_\gamma}r_{3,n}+\varepsilon_n).$$}
	By construction, for each $\bups\in\mathcal{V}$, there exists some $k_{\bups}$ such that $|\upsilon_\ell|\leq \varrho^{|\ell-k_{\bups}|}\varepsilon_n$ for $|\ell-k_{\bups}|\leq M_n$. Therefore, we can further write  $\I_{i,\bups}=\sum_{j:\widehat{\mathcal{B}}_j\subset\mathcal{I}_{\bups}}\I_{i,\bups,j}$ where each $\I_{i,\bups,j}$ is an indicator of the subinterval involved in $\mathcal{I}_{\bups}$, and 
	the above facts imply that for any $x_i\in\widehat{\mathcal{B}}_{l}$ for some $\widehat{\mathcal{B}}_{l}\subset\mathcal{I}_{\bups}$,
	\[
	\begin{aligned}
		\V[\delta_i(\bbeta_1, \bbeta_2, \bups,\bgamma)|\mathscr{F}_{XW\Delta}]\lesssim\;& \varrho^{2|(p-s+1)l-k_{\bups}|}\varepsilon_n^2\\
		&\times{(r_{1,n}+r_{2,n}+\varepsilon_n+\sqrt{d_\gamma}r_{3,n})(|\eta_{i,1}|+r_{1,n}+r_{2,n}+\varepsilon_n+\sqrt{d_\gamma}r_{3,n}).}
	\end{aligned}
	\]
	In addition, since	
	$\delta_i(\bbeta_1, \bbeta_2, \bups, \bgamma)\neq 0$ only if $x_i\in\mathcal{I}_{\bups}$, 
	for all $g\in\mathcal{G}$ (each corresponds to a particular $\bups$),
	\[
	\E_n[\V[g(x_i,\bw_i,y_i)|\mathscr{F}_{XW\Delta}]]\lesssim
	{\varepsilon_n^2(r_{1,n}+r_{2,n}+\varepsilon_n+\sqrt{d_\gamma}r_{3,n})}\sum_{l:\widehat{\mathcal{B}}_l\subset\mathcal{I}_{\bups}}\E_n[\I_{i,\bups,l}]\varrho^{2|(p-s+1)l-k_{\bups}|}.
	\]
	This inequality holds for any event in $\mathscr{F}_{XW\Delta}$.
	Define an event $\mathcal{A}_1$ on which 
	$\sup_{1\leq j\leq J}\E_n[\I_{i,j}]\leq C_2J^{-1}$ 
	for some large enough $C_2>0$ where $\I_{i,j}=\I(x_i\in\widehat{\mathcal{B}}_j)$. By the argument in Lemma \ref{SA lem: Gram, GL},  $\P(\mathcal{A}_1^c)\to 0$. On $\mathcal{A}_1$, 
	$$
	{\bar{\sigma}^2:=\sup_{g\in\mathcal{G}}\;\E_n[\V[g(x_i,\bw_i,y_i)|\mathscr{F}_{XW\Delta}]]\lesssim \varepsilon_n^2J^{-1}(r_{1,n}+r_{2,n}+\varepsilon_n+\sqrt{d_\gamma}r_{3,n}).}$$
	
	{On the other hand,}
	{\[
	\begin{aligned}
	\bar{G}:=\sup_{g\in\mathcal{G}}|g(x_i,\bw_i,y_i)|\lesssim\;&\ \varepsilon_n\{1+\sqrt{d_\gamma}r_{3,n}+|\psi(y_i,\eta_i)|(r_{1,n}+r_{2,n}+\varepsilon_n+\sqrt{d_\gamma}r_{3,n})\}\\
	&\times(|\eta_{i,1}|+r_{1,n}+r_{2,n}+\varepsilon_n+\sqrt{d_\gamma}r_{3,n}).
	\end{aligned}
	\]}
	Also, for any $g,\tilde{g}\in\mathcal{G}$, denote the corresponding parameters defining $g$ and $\tilde{g}$ by $(\bbeta_1, \bbeta_2, \bups, \bgamma)$ and $(\tilde{\bbeta}_1, \tilde{\bbeta}_2, \tilde{\bups}, \tilde{\bgamma})$. We have
	\begin{align*}
			\big|\tilde{g}(x_i,\bw_i,y_i)-g(x_i,\bw_i,y_i)\big|=&
			\Bigg|\int_{0}^{\varLambda_1}\Big[\psi(y_i,\eta(\breve{\bb}_i'(\breve{\bbeta}_0+\bbeta_1+\bbeta_2)+\bw_i'\bgamma+t))\\
			&\hspace{1.5em}-\psi(y_i,\eta(\breve{\bb}_i'\breve{\bbeta}_0+\bw_i'\bgamma_0))\Big]
			\times\eta^{(1)}(\breve{\bb}_i'(\breve{\bbeta}_0+\bbeta_1+\bbeta_2)+\bw_i'\bgamma+t)\d{t}\\
			&-\int_{0}^{\varLambda_2}\Big[\psi(y_i,\eta(\breve{\bb}_i'(\breve{\bbeta}_0+\bbeta_1+\bbeta_2-\bups)+\bw_i'\bgamma+t))\\
			&-\psi(y_i,\eta(\breve{\bb}_i'\breve{\bbeta}_0+\bw_i'\bgamma_0))\Big]
			\times\eta^{(1)}(\breve{\bb}_i'(\breve{\bbeta}_0+\bbeta_1+\bbeta_2-\bups)+\bw_i'\bgamma+t)\d{t}\Bigg|\\
			{\lesssim}&\; {\{1+\sqrt{d_\gamma}r_{3,n}+|\psi(y_i,\eta_i)|(r_{1,n}+r_{2,n}+\varepsilon_n+\sqrt{d_\gamma}r_{3,n})\}}\\
			&\times{\Big(|\eta_{i,1}|+r_{1,n}+r_{2,n}+\varepsilon_n+\sqrt{d_\gamma}r_{3,n}}
			{+|\varLambda_1|+|\varLambda_2|\Big)}\\
			&\times{\Big(\|\tilde{\bbeta}_1-\bbeta_1\|_\infty+\|\tilde{\bbeta}_2-\bbeta_2\|_\infty}
			{+\|\tilde{\bups}-\bups\|_\infty+\sqrt{d_\gamma}\|\tilde{\bgamma}-\bgamma\|\Big)},
	\end{align*}
	where $\varLambda_1=\breve{\bb}_i'(\tilde{\bbeta}_1+\tilde{\bbeta}_2-\bbeta_1-\bbeta_2)+\bw_i'(\tilde{\bgamma}-\bgamma)$ and
	$\varLambda_2=\varLambda_1-\breve{\bb}_i'(\tilde{\bups}-\bups)$.
	Based on these observations, 
	{\[
	\begin{aligned}
	&\|\bar{G}\|_{\bar{\P},2}\int_0^{\frac{\bar\sigma}{\|\bar{G}\|_{\bar{\P},2}}}\sqrt{1+\sup_{\mathbb{Q}}\log N(\mathcal{G}, L_2(\mathbb{Q}), t\|\bar{G}\|_{\mathbb{Q},2})}\d{t}\\
	&\qquad\lesssim\bar{\sigma}\Big(\sqrt{\log J}+\sqrt{(d_\gamma+\log n)\log \frac{e\|\bar G\|_{\bar\P,2}}{\bar{\sigma}}}\Big)\lesssim\bar{\sigma}\ell_{\gamma,n}.
	\end{aligned}
	\]}
	where the supremum is taken over all finite discrete probability measures $\mathbb{Q}$.
	Then, by Lemma \ref{SA lem: maximal inequality},
	\[
	\E\bigg[\sup_{g\in\mathcal{G}}
	\Big|\G_n[g(x_i, \bw_i, y_i)]\Big|\bigg|\mathscr{F}_{XW\Delta}\bigg]
	{\lesssim \bar{\sigma}\ell_{\gamma,n}+\frac{\sqrt{\E[\bar{\bar{G}}^2]}\ell_{\gamma,n}^2}{\sqrt{n}},}
	\]
	where $\bar{\bar{G}}=\max_{1\leq i\leq n}\bar{G}(x_i,\bw_i,y_i)$. Note that
	{$(\E[\bar{\bar{G}}^2])^{1/2}\lesssim\varepsilon_n (1+n^{1/\nu}(r_{1,n}+r_{2,n}+\sqrt{d_\gamma}r_{3,n}+\varepsilon_n))$.}
	
	Therefore, on $\mathcal{A}_1$ (whose probability approaches one),
	\[
	\begin{split}
		&\sup_{\bbeta_1\in\mathcal{H}_1, \bbeta_2\in\mathcal{H}_2, \bups\in\mathcal{V}, \bgamma_1\in\mathcal{H}_3}\;
		\Big|\E_n\Big[\delta_i(\bbeta_1, \bbeta_2, \bups, \bgamma)\Big]-
		\E_n\Big[\E[\delta_i(\bbeta_1, \bbeta_2, \bups, \bgamma)|\mathscr{F}_{XW\Delta}]\Big]\Big|\\
		\lesssim &\;\bigg(J^{-1}\varepsilon_n\sqrt{\mathfrak{L}_n}
		{\sqrt{\frac{J}{n}}\ell_{\gamma,n}+\frac{\varepsilon_n(1+n^{1/\nu}\mathfrak{L}_n)\ell_{\gamma,n}^2}{n}}\bigg)
	\end{split}
	\]
	{for $\mathfrak{L}_n=r_{1,n}+r_{2,n}+\sqrt{d_\gamma}r_{3,n}+\varepsilon_n$.}
	
	\textbf{Step 2:}
	For $\widetilde{\bQ}:=
	\E_n[\breve{\bb}_i\breve{\bb}_i'
	\Psi_1(x_i,\bw_i;\eta(\breve{\bb}_i'\breve{\bbeta}_0+\bw_i'\bgamma_0))
	(\eta^{(1)}(\breve{\bb}_i'\breve{\bbeta}_0+\bw_i'\bgamma_0))^2]$, by Assumption \ref{SA Assumption GL} and the same argument in the proof of Lemma \ref{SA lem: Gram, GL},
	\[
	\|\breve{\bQ}-\widetilde{\bQ}\|_\infty
	\vee\|\breve{\bQ}-\widetilde{\bQ}\|
	\lesssim J^{-p-2}.
	\]
	Therefore,
	$$
	\sup_{\bbeta_1\in\mathcal{H}_1, \bbeta_2\in\mathcal{H}_2, \bups\in\mathcal{V}}|\bups'(\widetilde{\bQ}-\breve{\bQ})(\bbeta_1+\bbeta_2)|\lesssim J^{-p-2}\varepsilon_n(r_{1,n}+r_{2,n}).
	$$
	In addition, by Lemmas \ref{SA lem: uniform converge var part, GL} and \ref{SA lem: proj approx error, GL}, 
	$\|\bar{\bbeta}\|_\infty\leq r_{1,n}$ with probability approaching one for $C_1$ large enough, where
	\[
	\bar{\bbeta}:=-\breve{\bQ}^{-1}\E_n\Big[\breve{\bb}_i\eta^{(1)}(\breve{\bb}_i'\breve{\bbeta}_0+\bw_i'\bgamma_0)\psi\Big(y_i,\eta(\breve{\bb}_i'\breve{\bbeta}_0+\bw_i'\bgamma_0)\Big)\Big].
	\]

	\textbf{Step 3:}
	Let $\widetilde{\varkappa}_i=\Psi_1(x_i,\bw_i;\eta(\breve{\bb}_i'\breve{\bbeta}_0+\bw_i'\bgamma_0))
	(\eta^{(1)}(\breve{\bb}_i'\breve{\bbeta}_0+\bw_i'\bgamma_0))^2$.
	By Taylor expansion, we have
	\begin{align*}
		&\;\E_n\Big[\E[\delta_i(\bbeta_1, \bbeta_2, \bups, \bgamma)|\mathscr{F}_{XW\Delta}]\Big]\\
		=&\;\E_n\bigg[\int_{-\breve{\bb}_i'\bups}^{0}
		\Big\{\Psi(x_i, \bw_i; \eta(\breve{\bb}_i'(\breve{\bbeta}_0+\bbeta_1+\bbeta_2)+\bw_i'\bgamma+t))\\
		&\hspace{5.5em}-\Psi(x_i, \bw_i;\eta(\breve{\bb}_i'\breve{\bbeta}_0+\bw_i'\bgamma_0))\Big\}
		\times\eta^{(1)}\Big(\breve{\bb}_i'(\breve{\bbeta}_0+\bbeta_1+\bbeta_2)+\bw_i'\bgamma+t\Big)\d{t}\bigg]\\
		=&\;\E_n\bigg[\int_{-\breve{\bb}_i'\bups}^{0}\Big\{
		\Psi_1(x_i, \bw_i;\eta(\breve{\bb}_i'\breve{\bbeta}_0+\bw_i'\bgamma_0))
		\Big(\eta^{(1)}(\breve{\bb}_i'\breve{\bbeta}_0+\bw_i'\bgamma_0)
		(\breve{\bb}_i'(\bbeta_1+\bbeta_2)+\bw_i'\bgamma_1+t)\\
		&\hspace{18em}+\frac{1}{2}\eta^{(2)}(\xi_{i,t})(\breve{\bb}_i'(\bbeta_1+\bbeta_2)+\bw_i'\bgamma_1+t)^2\Big)\\
		&\hspace{6em}+\frac{1}{2}\Psi_2(x_i, \bw_i; \tilde{\xi}_{i,t})\Big(\eta(\breve{\bb}_i'(\breve{\bbeta}_0+\bbeta_1+\bbeta_2)+\bw_i'\bgamma+t)-
		\eta(\breve{\bb}_i'\breve{\bbeta}_0+\bw_i'\bgamma_0)\Big)^2\Big\}\\
		&\hspace{9em}\times \Big(\eta^{(1)}(\breve{\bb}_i'\breve{\bbeta}_0+\bw_i'\bgamma_0)+
		\eta^{(2)}(\check{\xi}_{i,t})(\breve{\bb}_i'(\bbeta_1+\bbeta_2)+\bw_i'\bgamma_1+t)\Big)\d{t}\bigg]\\
		=&\;\bups'\widetilde{\bQ}(\bbeta_1+\bbeta_2)+\bups'\E_n[\breve{\bb}_i\widetilde{\varkappa}_i\bw_i']\bgamma_1-\frac{1}{2}\bups'\widetilde{\bQ}\bups+\mathrm{I}+\mathrm{II}+\mathrm{III},
	\end{align*}
	where $\xi_{i,t}$ and $\check{\xi}_{i,t}$ are between $\breve{\bb}_i'\breve{\bbeta}_0+\bw_i'\bgamma_0$ and $\breve{\bb}_i'(\breve{\bbeta}_0+\bbeta_1+\bbeta_2)+\bw_i'\bgamma+t$, 
	$\tilde{\xi}_{i,t}$ is between $\eta(\breve{\bb}_i'\breve{\bbeta}_0+\bw_i'\bgamma_0)$ and $\eta(\breve{\bb}_i'(\breve{\bbeta}_0+\bbeta_1+\bbeta_2)+\bw_i'\bgamma+t)$, 
	{$\sup_{1\leq i\leq n}|\bw_i'\bgamma_1|\lesssim\sqrt{d_\gamma}r_{3,n}$, $\E_n[|\bups'\breve\bb_i|]\lesssim\varepsilon_n/J$, and hence $|\bups'\E_n[\breve{\bb}_i\widetilde{\varkappa}_i\bw_i']\bgamma_1|\lesssim\varepsilon_n\sqrt{d_\gamma}r_{3,n}/J$},
	$|\bups'\widetilde{\bQ}\bups|\lesssim \varepsilon_n^2/J$, and $\mathrm{I}, \mathrm{II}$, and $\mathrm{III}$ are defined and bounded as follows:
	\begin{align*}
		\mathrm{I}&=\E_n\bigg[\int_{-\breve{\bb}_i'\bups}^{0}
		\Psi_1(x_i,\bw_i;\eta(\breve{\bb}_i'\breve{\bbeta}_0+\bw_i'\bgamma_0))
		\eta^{(1)}(\breve{\bb}_i'\breve{\bbeta}_0+\bw_i'\bgamma_0)\\
		&\hspace{5em}\times\eta^{(2)}(\check{\xi}_{i,t})(\breve{\bb}_i'(\bbeta_1+\bbeta_2)+\bw_i'\bgamma_1+t)^2\d{t}\,\I_{i,\bups}
		\bigg]
		{\lesssim\varepsilon_nJ^{-1}(r_{1,n}+r_{2,n}+\varepsilon_n+\sqrt{d_\gamma}r_{3,n})^2},\\
		\mathrm{II}&=\E_n\bigg[\int_{-\breve{\bb}_i'\bups}^{0}
		\Psi_1(x_i,\bw_i;\eta(\breve{\bb}_i'\breve{\bbeta}_0+\bw_i'\bgamma_0))\times
		\frac{1}{2}\eta^{(2)}(\xi_{i,t})(\breve{\bb}_i'(\bbeta_1+\bbeta_2)+\bw_i'\bgamma_1+t)^2\\
		&\hspace{5em}\times\eta^{(1)}\Big(\breve{\bb}_i'(\breve{\bbeta}_0+\bbeta_1+\bbeta_2)+\bw_i'\bgamma+t\Big)\d{t}\,\I_{i,\bups}\bigg]
		{\lesssim\varepsilon_nJ^{-1}(r_{1,n}+r_{2,n}+\varepsilon_n+\sqrt{d_\gamma}r_{3,n})^2},\\
		\mathrm{III}&=\E_n\bigg[\int_{-\breve{\bb}_i'\bups}^{0}
		\frac{1}{2}\Psi_2(x_i,\bw_i;\tilde{\xi}_{i,t})\Big(\eta(\breve{\bb}_i'(\breve{\bbeta}_0+\bbeta_1+\bbeta_2)+\bw_i'\bgamma+t)-\eta(\breve{\bb}_i'\breve{\bbeta}_0+\bw_i'\bgamma_0)\Big)^2\\
		&\hspace{5em}\times\eta^{(1)}\Big(\breve{\bb}_i'(\breve{\bbeta}_0+\bbeta_1+\bbeta_2)+\bw_i'\bgamma+t\Big)\d{t}\,\I_{i,\bups}\bigg]
		{\lesssim\varepsilon_nJ^{-1}(r_{1,n}+r_{2,n}+\varepsilon_n+\sqrt{d_\gamma}r_{3,n})^2}.
	\end{align*}
	These bounds hold uniformly for $\bups\in\mathcal{V}$, $\bbeta_1\in\mathcal{H}_1$, $\bbeta_2\in\mathcal{H}_2$ and $\bgamma_1\in\mathcal{H}_3$ (that is, uniformly over the function class $\mathcal{G}$), and on an event $\mathcal{A}_1\cap\mathcal{A}_2$ where $\mathcal{A}_2=\{\lambda_{\max}(\widetilde{\bQ})\leq c_4J^{-1}\}$ for some large enough $c_4>0$. Note that $\P(\mathcal{A}_1\cap\mathcal{A}_2)\to 1$ by Lemma \ref{SA lem: Gram, GL}.

	\textbf{Step 4:}	
    By Assumption \ref{SA Assumption GL} and {the integral form of} Taylor's expansion,
	\begin{align*}
		\mathrm{IV}&=\E_n\bigg[\Big(\eta(\breve{\bb}_i'(\breve{\bbeta}_0+\bbeta_1+\bbeta_2)+\bw_i'\bgamma)-
		\eta(\breve{\bb}_i'(\breve{\bbeta}_0+\bbeta_1+\bbeta_2-\bups)+\bw_i'\bgamma)\Big)
		\psi(y_i,\eta(\breve{\bb}_i'\breve{\bbeta}_0+\bw_i'\bgamma_0))\bigg]\\
		&\qquad-\E_n\Big[\bups'\breve{\bb}_i\psi(y_i, \eta(\breve{\bb}_i'\breve{\bbeta}_0+\bw_i'\bgamma_0))
		\eta^{(1)}(\breve{\bb}_i'\breve{\bbeta}_0+\bw_i'\bgamma_0)\Big]\\
		&={\sum_j\upsilon_j\E_n\Bigg[\breve b_{i,j}\psi(y_i, \eta(\breve{\bb}_i'\breve{\bbeta}_0+\bw_i'\bgamma_0))
		\int_0^1\int_0^{\breve{\bb}_i'(\bbeta_1+\bbeta_2)+\bw_i'\bgamma_1-t\breve{\bb}_i'\bups}
		\eta^{(2)}(\breve{\bb}_i'\breve{\bbeta}_0+\bw_i'\bgamma_0+s)\d{s}\,\d{t}\Bigg]}\\
		&{\lesssim J^{-1}\left\{\sqrt{\frac{J(d_\gamma+\log J)}{n}}+\frac{J(d_\gamma+\log J)}{n^{1-1/\nu}}+J^{-p-1}\right\}(\varepsilon_n+r_{1,n}+r_{2,n}+\sqrt{d_\gamma}r_{3,n})\varepsilon_n,}
	\end{align*}
    {The last line holds uniformly over $\bbeta_1\in\mathcal{H}_1$, $\bbeta_2\in\mathcal{H}_2$, $\bups\in\mathcal{V}$, and $\bgamma_1\in\mathcal{H}_3$ on $\mathcal A_1\cap\mathcal A_3$, where $\P(\mathcal A_3)\to1$. To verify it, fix $j$ in the preceding sum and, recalling $\mathfrak L_n$ from Step 1, write
    \[
    \begin{aligned}
    R_{i,j}:={}&\breve b_{i,j}\psi(y_i,\eta(\breve{\bb}_i'\breve{\bbeta}_0+\bw_i'\bgamma_0))
    \int_0^1\int_0^{\breve{\bb}_i'(\bbeta_1+\bbeta_2)+\bw_i'\bgamma_1-t\breve{\bb}_i'\bups}
    \eta^{(2)}(\breve{\bb}_i'\breve{\bbeta}_0+\bw_i'\bgamma_0+s)\d{s}\,\d{t},\\
    |R_{i,j}|\lesssim{}&
    |\breve b_{i,j}\psi(y_i,\eta(\breve{\bb}_i'\breve{\bbeta}_0+\bw_i'\bgamma_0))|\mathfrak L_n,
    \qquad
    \E_n[\I(\breve b_{i,j}\neq0)]\lesssim J^{-1}\quad\text{on }\mathcal A_1.
    \end{aligned}
    \]
    On the support of $\breve b_{i,j}$, local support implies that only a fixed number of coordinates of $\bbeta_1$, $\bbeta_2$, and $\bups$ enter $R_{i,j}$. 
	Its smoothness and the variation in $\bgamma_1$ therefore give covering entropy bounded by $Cd_\gamma\log(C/t)$ relative to the displayed envelope. 
	Taking the union over $O(J)$ possible values of $j$ adds $\log J$. 
	Consequently, with each supremum below taken over $1\leq j\leq K_{p,s}$ and the parameter sets above, Lemma \ref{SA lem: maximal inequality} and the conditional mean bound give
    \[
    \begin{aligned}
    \sup_{j,\bbeta_1,\bbeta_2,\bups,\bgamma_1}
    \left|\E_n\left[R_{i,j}-\E[R_{i,j}\mid\mathscr F_{XW\Delta}]\right]\right|
    &\lesssim J^{-1}\left\{\sqrt{\frac{J(d_\gamma+\log J)}{n}}
    +\frac{J(d_\gamma+\log J)}{n^{1-1/\nu}}\right\}\mathfrak L_n,\\
    \sup_{j,\bbeta_1,\bbeta_2,\bups,\bgamma_1}
    \left|\E_n\left[\E[R_{i,j}\mid\mathscr F_{XW\Delta}]\right]\right|
    &\lesssim J^{-1}J^{-p-1}\mathfrak L_n,\\
    |\mathrm{IV}|
    \leq\sum_j|\upsilon_j|
    \sup_{j,\bbeta_1,\bbeta_2,\bups,\bgamma_1}|\E_n[R_{i,j}]|
    &\lesssim J^{-1}\left\{\sqrt{\frac{J(d_\gamma+\log J)}{n}}
    +\frac{J(d_\gamma+\log J)}{n^{1-1/\nu}}+J^{-p-1}\right\}\mathfrak L_n\varepsilon_n.
    \end{aligned}
    \]
    Here the second line uses $\E[\psi(y_i,\eta(\breve{\bb}_i'\breve{\bbeta}_0+\bw_i'\bgamma_0))\mid\mathscr F_{XW\Delta}]\lesssim J^{-p-1}$, 
	while the last line uses the geometric bound $\sum_j|\upsilon_j|\lesssim\varepsilon_n$. This also makes clear why no factor proportional to $M_n$ appears.}
	
	\textbf{Step 5:}
	Let
	\[
	\bar\bups=c_5\varepsilon_nJ^{-1}\operatorname{sgn}(\beta_{2,k})
	[\breve{\bQ}^{-1}]_{\cdot k}
	\]
	for some $k$ such that $|\beta_{2,k}|=\|\bbeta_2\|_\infty$ and some $c_5>0$, where $[\breve{\bQ}^{-1}]_{\cdot k}$ denotes the $k$th column of $\breve{\bQ}^{-1}$. Since $J^{-1}\breve{\bQ}^{-1}=\bar{\bQ}^{-1}$,
	\[
	\bar\bups'\breve{\bQ}\bbeta_2
	=c_5\varepsilon_nJ^{-1}|\beta_{2,k}|.
	\]
	Take
	$\bups=(\upsilon_1, \cdots, \upsilon_{K_{p,s}})$ where $\upsilon_j=\bar{\upsilon}_j$ for $|j-k|\leq M_n$ and zero otherwise. Clearly, $\bups\in\mathcal{V}$ on an event $\mathcal{A}_4$ with $\P(\mathcal{A}_4)\to 1$. On $\mathcal{A}_2\cap\mathcal{A}_4$,
	\[
	|(\bups-\bar{\bups})'\breve{\bQ}\bbeta_2|\lesssim \varepsilon_nJ^{-1}r_{2,n}n^{-c_6}
	\]
	for some large $c_6>0$ if we let $c_1$ be sufficiently large.
	
	\textbf{Step 6:}
	Partition the parameter space into shells
	$\mathcal{O}=\cup_{\ell=-\infty}^{\bar{L}}\mathcal{O}_{\ell}$, where
	\[
	\mathcal{O}_{\ell}=\left\{\bbeta\in\mathbb{R}^{K_{p,s}}:
	2^{\ell-1}\mathfrak{r}_{2,n}\leq
	\|\bbeta-\breve{\bbeta}_0-\bar{\bbeta}\|_\infty
	\leq2^\ell\mathfrak{r}_{2,n}\right\},
	\]
	and let $\bar L$ be the smallest integer such that
	$2^{\bar L}\mathfrak r_{2,n}\geq c$. Also, define
	\[
	\mathcal H_{2,\ell}=\left\{\bbeta_2\in\mathbb R^{K_{p,s}}:
	2^{\ell-1}\mathfrak r_{2,n}\leq\|\bbeta_2\|_\infty
	\leq2^\ell\mathfrak r_{2,n}\right\}.
	\]
	By construction,
	\[
	\breve{\bQ}\bar{\bbeta}
	=-\E_n\Big[\breve{\bb}_i
	\eta^{(1)}(\breve{\bb}_i'\breve{\bbeta}_0+\bw_i'\bgamma_0)
	\psi\big(y_i,\eta(\breve{\bb}_i'\breve{\bbeta}_0+\bw_i'\bgamma_0)\big)\Big].
	\]

	For generic $(\bbeta_1,\bbeta_2,\bups,\bgamma)$ in the sets defined in
	Step 0, let
	\[
	\begin{split}
	A_i(\bbeta_1,\bbeta_2,\bups,\bgamma)
	=\;&\eta\big(\breve{\bb}_i'(\breve{\bbeta}_0+\bbeta_1+\bbeta_2)
	+\bw_i'\bgamma\big)\\
	&-\eta\big(\breve{\bb}_i'(\breve{\bbeta}_0+\bbeta_1+\bbeta_2-\bups)
	+\bw_i'\bgamma\big),\\
	s_i^0=\;&\psi\big(y_i,
	\eta(\breve{\bb}_i'\breve{\bbeta}_0+\bw_i'\bgamma_0)\big).
	\end{split}
	\]
	The definition of $\delta_i$ in Step 0 gives the exact identity
	\[
	\begin{split}
	&\E_n\Big[\rho\big(y_i;\eta(\breve{\bb}_i'(\breve{\bbeta}_0+\bbeta_1+\bbeta_2)
	+\bw_i'\bgamma)\big)
	-\rho\big(y_i;\eta(\breve{\bb}_i'(\breve{\bbeta}_0+\bbeta_1+\bbeta_2-\bups)
	+\bw_i'\bgamma)\big)\Big]\\
	&\qquad=
	\E_n\Big[\E\big[\delta_i(\bbeta_1,\bbeta_2,\bups,\bgamma)
	\mid\mathscr F_{XW\Delta}\big]\Big]
	+\E_n\big[A_i(\bbeta_1,\bbeta_2,\bups,\bgamma)s_i^0\big]
	+\frac{1}{\sqrt n}\G_n[\delta_i(\bbeta_1,\bbeta_2,\bups,\bgamma)],
	\end{split}
	\]
	where, throughout this step,
	$\G_n[f_i]=\sqrt n\,\E_n\big[f_i-
	\E[f_i\mid\mathscr F_{XW\Delta}]\big]$.
	Conditional expectations involving random parameter values are understood as
	the generic-parameter conditional expectations evaluated at those values;
	the bounds in Steps 1--4 are uniform over the corresponding parameter sets.

	Let $\mathcal A=\cap_{j=1}^4\mathcal A_j$. For
	$\bbeta_1=\bar{\bbeta}$, $\bbeta_2\in\mathcal H_{2,\ell}$, and
	$\bgamma_1=\widehat{\bgamma}-\bgamma_0$, Steps 2--5 and the signed direction
	constructed in Step 5 give, 
	the uniform lower bound
	\[
	\begin{split}
	&\E_n\Big[\E\big[\delta_i(\bbeta_1,\bbeta_2,\bups,\bgamma)
	\mid\mathscr F_{XW\Delta}\big]\Big]
	+\E_n\big[A_i(\bbeta_1,\bbeta_2,\bups,\bgamma)s_i^0\big]\\
	&\quad\geq
	c_5J^{-1}\varepsilon_n2^{\ell-1}\mathfrak r_{2,n}
	-CJ^{-1}\varepsilon_n\Big\{
	{\sqrt{d_\gamma}r_{3,n}}+\varepsilon_n+\mathfrak L_{n,\ell}^2
	+q_n\mathfrak L_{n,\ell}
	+J^{-p-1}\mathfrak L_{n,\ell}
	+2^\ell\mathfrak r_{2,n}n^{-c_6}
	\Big\},
	\end{split}
	\]
	with
	{\[
	\begin{aligned}
	\mathfrak L_{n,\ell}&=r_{1,n}+2^\ell\mathfrak r_{2,n}+\sqrt{d_\gamma}r_{3,n}+\varepsilon_n,\\
	q_n&=\left(\frac{J\log n}{n}\right)^{1/2}+\sqrt{\frac{J(d_\gamma+\log J)}{n}}+\frac{J(d_\gamma+\log J)}{n^{1-1/\nu}}+J^{-p-1}.
	\end{aligned}
	\]}
	{To verify the required absorption explicitly, the definition of $\mathfrak r_{2,n}$ and Young's inequality give
	\begin{equation}\label{SA eq: shell absorption root}
	\sqrt{\frac Jn}\ell_{\gamma,n}\sqrt{r_{1,n}+\sqrt{d_\gamma}r_{3,n}}
	\lesssim\left(\frac{J\log n}{n}\right)^{3/4}\ell_{\gamma,n}
	+J^{-\frac{p+1}{2}}\sqrt{\frac Jn}\ell_{\gamma,n}
	+\frac{J\ell_{\gamma,n}^2}{n}+\sqrt{d_\gamma}r_{3,n}.
	\end{equation}
	\begin{equation}\label{SA eq: shell absorption square}
	(r_{1,n}+\sqrt{d_\gamma}r_{3,n})^2
	\lesssim\left(\frac{J\log n}{n}\right)^{3/4}\ell_{\gamma,n}
	+J^{-p-1}+\sqrt{d_\gamma}r_{3,n}.
	\end{equation}
	The second rate condition in the theorem further gives
	\begin{equation}\label{SA eq: shell absorption heavy tail}
	\frac{J\ell_{\gamma,n}^2}{n^{1-1/\nu}}
	(r_{1,n}+\sqrt{d_\gamma}r_{3,n})
	\lesssim(r_{1,n}+\sqrt{d_\gamma}r_{3,n})^2,
	\end{equation}
	while $J(d_\gamma+\log J)/n\lesssim J\ell_{\gamma,n}^2/n$ and $J(d_\gamma+\log J)/n^{1-1/\nu}\lesssim r_{1,n}$ show that the Step 4 base-radius terms are absorbed by \eqref{SA eq: shell absorption root}--\eqref{SA eq: shell absorption heavy tail}. Set $\varepsilon_n=c_7 2^L\mathfrak r_{2,n}$ for a sufficiently small fixed $c_7>0$. The outer-shell terms must instead be compared with the shell radius. Uniformly for $L\leq\ell\leq\bar L$,
	\[
	\mathfrak L_{n,\ell}^2\lesssim\mathfrak r_{2,n}+Cc\,2^\ell\mathfrak r_{2,n},\qquad q_n\mathfrak L_{n,\ell}\lesssim\mathfrak r_{2,n}+o(1)2^\ell\mathfrak r_{2,n},
	\]
	and the same second bound holds with $q_n$ replaced by $J^{-p-1}$. Indeed, $2^\ell\mathfrak r_{2,n}\lesssim c$, $\varepsilon_n\leq c_7 2^\ell\mathfrak r_{2,n}$, the base-radius bounds in \eqref{SA eq: shell absorption root} and \eqref{SA eq: shell absorption square} are $O(\mathfrak r_{2,n})$, and $q_n=o(1)$. Consequently, the terms independent of the shell radius are absorbed by taking $L$ large, while the terms proportional to $2^\ell\mathfrak r_{2,n}$ are absorbed by taking $c$ and $c_7$ small.} Hence, by first taking the localization constant $c$ and $c_7$
	sufficiently small and then $L$ sufficiently large, the preceding display
	implies, uniformly on $\mathcal A$,
	\[
	\E_n\Big[\E\big[\delta_i(\bbeta_1,\bbeta_2,\bups,\bgamma)
	\mid\mathscr F_{XW\Delta}\big]\Big]
	+\E_n\big[A_i(\bbeta_1,\bbeta_2,\bups,\bgamma)s_i^0\big]
	\geq C_4J^{-1}2^\ell\mathfrak r_{2,n}\varepsilon_n
	\]
	for some $C_4>0$. Exact optimality of
	$\breve{\bbeta}$ yields a non-strict objective comparison. Therefore,
	by Lemma \ref{SA lem: uniform consistency, GL}, the preceding display, and
	conditional Markov's inequality,
	\begin{align*}
	&\P\Big(\|\breve{\bbeta}-\breve{\bbeta}_0-\bar{\bbeta}\|_\infty
	\geq2^L\mathfrak r_{2,n}\mid\mathscr F_{XW\Delta}\Big)\\
	&\quad\leq
	\P\Bigg(\bigcup_{\ell=L}^{\bar L}
	\Bigg\{
	\sup_{\bbeta_1\in\mathcal H_1}
	\sup_{\bbeta_2\in\mathcal H_{2,\ell}}
	\sup_{\bgamma_1\in\mathcal H_3}
	\sup_{\bups\in\mathcal V}
	\frac{1}{\sqrt n}
	\big|\G_n[\delta_i(\bbeta_1,\bbeta_2,\bups,\bgamma)]\big|
	\geq C_4J^{-1}2^\ell\mathfrak r_{2,n}\varepsilon_n
	\Bigg\}\cap\mathcal A
	\,\Bigg|\,\mathscr F_{XW\Delta}\Bigg)+o_\P(1)\\
	&\quad\leq
	\sum_{\ell=L}^{\bar L}
	(C_6J^{-1}2^\ell\mathfrak r_{2,n}\varepsilon_n)^{-1}
	\I(\mathcal A_1)
	\E\Bigg[
	\sup_{\bbeta_1\in\mathcal H_1}
	\sup_{\bbeta_2\in\mathcal H_{2,\ell}}
	\sup_{\bgamma_1\in\mathcal H_3}
	\sup_{\bups\in\mathcal V}
	\frac{1}{\sqrt n}
	\big|\G_n[\delta_i(\bbeta_1,\bbeta_2,\bups,\bgamma)]\big|
	\,\Bigg|\,\mathscr F_{XW\Delta}\Bigg]+o_\P(1).
	\end{align*}
	{Using Step 1, dividing its bound by the shell threshold in the preceding display, and summing over the shells gives
	\[
	\begin{split}
	&\sum_{\ell=L}^{\bar L}\left\{\frac{\sqrt{J/n}\,\ell_{\gamma,n}\sqrt{r_{1,n}+\sqrt{d_\gamma}r_{3,n}+2^\ell\mathfrak r_{2,n}+\varepsilon_n}}{2^\ell\mathfrak r_{2,n}}+\frac{J\ell_{\gamma,n}^2}{n2^\ell\mathfrak r_{2,n}}\right.\\
	&\hspace{11em}\left.+\frac{J\ell_{\gamma,n}^2(r_{1,n}+\sqrt{d_\gamma}r_{3,n}+2^\ell\mathfrak r_{2,n}+\varepsilon_n)}{n^{1-1/\nu}2^\ell\mathfrak r_{2,n}}\right\}\\
		&\qquad\lesssim
		\underbrace{2^{-L/2}}_{\text{first summand}}
		+\underbrace{2^{-L}}_{\text{second summand}}
		+\underbrace{\left(2^{-L}+\frac{J\ell_{\gamma,n}^2}{n^{1-1/\nu}}\bar L\right)}_{\text{third summand}}.
		\end{split}
		\]
		The first group follows from \eqref{SA eq: shell absorption root}, the definition of $\mathfrak r_{2,n}$, and $\varepsilon_n=c_7 2^L\mathfrak r_{2,n}$. The second follows from $J\ell_{\gamma,n}^2/n\leq\mathfrak r_{2,n}$. For the third group, \eqref{SA eq: shell absorption heavy tail} followed by \eqref{SA eq: shell absorption square} gives the $2^{-L}$ term, while the second rate condition in the theorem gives
		\[
		\frac{J\ell_{\gamma,n}^2}{n^{1-1/\nu}}\bar L
		\lesssim r_{1,n}\log J=o(1),
		\]
		because $\bar L=O(\log(1/\mathfrak r_{2,n}))=O(\log J)$. Thus the displayed shell sum can be made arbitrarily small by first choosing $L$ sufficiently large and then letting $n\to\infty$.} Hence
	\[
	\|\breve{\bbeta}-\breve{\bbeta}_0-\bar{\bbeta}\|_\infty
	=O_\P(\mathfrak r_{2,n}).
	\]
	Moreover,
	\[
	\frac{\bar{\bbeta}}{\sqrt J}
	=-\bar{\bQ}^{-1}\E_n\Big[\widehat{\bb}_{p,s}(x_i)
	\eta^{(1)}(\widehat{\bb}_{p,s}(x_i)'\widehat{\bbeta}_0+\bw_i'\bgamma_0)
	\psi\big(y_i,\eta(\widehat{\bb}_{p,s}(x_i)'\widehat{\bbeta}_0
	+\bw_i'\bgamma_0)\big)\Big].
	\]
	Therefore, Lemma \ref{SA lem: proj approx error, GL} gives
	\[
	\begin{split}
	&\sup_{x\in\mathcal X}\left|
	\widehat{\bb}_{p,s}^{(v)}(x)'
	\left\{\frac{\bar{\bbeta}}{\sqrt J}
	+\bar{\bQ}^{-1}\E_n[\widehat{\bb}_{p,s}(x_i)
	\eta_{i,1}\psi(y_i,\eta_i)]\right\}\right|\\
	&\qquad\lesssim_\P
	J^v\left\{J^{-p-1}
	+J^{-(p+1)/2}\left(\frac{J\log J}{n}\right)^{1/2}
	+\frac{J\log J}{n}\right\}.
	\end{split}
	\]
	The last two terms are absorbed by the first two stochastic terms in
	$\mathfrak r_{2,n}$ under the theorem's rate conditions. Combining these
	displays with the approximation bound
	$\sup_x|\widehat r_{0,v}(x)|\lesssim_\P J^{-p-1+v}$ and converting back
	through $\breve{\bbeta}=\sqrt J\widehat{\bbeta}$ proves the leading term
	{and the remainder stated in part (i).}

    \textbf{Step 7:} To show part (ii) and part (iii), by Taylor expansion and the result in part (i),
	\begin{align*}
		&\eta(\widehat{\mu}(x)+\widehat\evalw'\widehat{\bgamma})-\eta(\mu_0(x)+\evalw'\bgamma_0)\\
		=\;&\eta^{(1)}(\mu_0(x)+\evalw'\bgamma_0)
		\Big(\widehat{\bb}_{p,s}(x)'\widehat{\bbeta}-\mu_0(x)\Big)\\
		&\hspace{5em}+{O_\P\Big(\|\widehat\evalw-\evalw\|+\sqrt{d_\gamma}\mathfrak r_\gamma+\frac{J\log n}{n}+J^{-2p-2}+\mathfrak r_{\mathrm{BR},n}^2\Big)}\\
		=\;&-\eta^{(1)}(\mu_0(x)+\evalw'\bgamma_0)\widehat{\bb}_{p,s}(x)'\bar{\bQ}^{-1}\E_n[\widehat{\bb}_{p,s}(x_i)\eta_{i,1}\psi(y_i,\eta_i)]\\
		&\hspace{5em}+{O_\P\Big(\mathfrak r_{\mathrm{BR},n}+\|\widehat\evalw-\evalw\|\Big)},
	\end{align*}
	and
	\begin{align*}
		&\eta^{(1)}(\widehat{\mu}(x)+\widehat\evalw'\widehat{\bgamma})\widehat{\mu}^{(1)}(x)-
		\eta^{(1)}(\mu_0(x)+\evalw'\bgamma_0)\mu_0^{(1)}(x)\\
		=&\;\eta^{(1)}(\mu_0(x)+\evalw'\bgamma_0)
		\Big(\widehat{\mu}^{(1)}(x)-\mu_0^{(1)}(x)\Big)\\
		&+{O_\P\Big(\Big(\frac{J\log n}{n}\Big)^{1/2}+J^{-p-1}+\|\widehat\evalw-\evalw\|+\mathfrak r_{\mathrm{BR},n}\Big)
  O_\P\Big(1+J\Big(\Big(\frac{J\log n}{n}\Big)^{1/2}+J^{-p-1}+\mathfrak r_{\mathrm{BR},n}\Big)\Big)}\\
		=&\;-\eta^{(1)}(\mu_0(x)+\evalw'\bgamma_0)\widehat{\bb}_{p,s}^{(1)}(x)'\bar{\bQ}^{-1}\E_n[\widehat{\bb}_{p,s}(x_i)\eta_{i,1}\psi(y_i,\eta_i)]+\\
		&{O_\P\Big(\Big(\frac{J\log n}{n}\Big)^{1/2}+J^{-p}+J\mathfrak r_{\mathrm{BR},n}
  +	\|\widehat{\evalw}-\evalw\|\Big(1+\Big(\frac{J^3\log n}{n}\Big)^{1/2}\Big)\Big)}.
	\end{align*}
    In the above derivation the probability bound holds uniformly over $x\in\mathcal{X}$ as well. Then the proof is complete.
\end{proof}

\subsection{Proof of Theorem \ref{SA thm: meat matrix, GL}}
\begin{proof}
	Since $\widehat{\epsilon}_{i}:=\epsilon_{i}+\eta_i-\widehat{\eta}_i=:\epsilon_{i}+u_{i}$, we can write
	\begin{align*}
		&\E_n[\widehat{\bb}_{p,s}(x_{i})\widehat{\bb}_{p,s}(x_{i})'\widehat{\eta}_{i,1}^2\psi^\ddagger(\widehat{\eta}_i)^2\psi^\dagger(\widehat{\epsilon}_{i})^2]
		-\E[\bb_{p,s}(x_{i})\bb_{p,s}(x_{i})'\eta_{i,1}^2\sigma^2(x_i, \bw_i)]\\
		=\;&\E_n\Big[\widehat{\bb}_{p,s}(x_{i})\widehat{\bb}_{p,s}(x_{i})'
		\widehat{\eta}_{i,1}^2\psi^\ddagger(\widehat{\eta}_i)^2\Big(\psi^\dagger(\epsilon_i+u_i)^2-\psi^\dagger(\epsilon_i)^2\Big)\Big]\\
		&+\E_n\Big[\widehat{\bb}_{p,s}(x_{i})
		\widehat{\bb}_{p,s}(x_{i})'\Big(\widehat{\eta}_{i,1}^2\psi^\ddagger(\widehat{\eta}_i)^2-\eta_{i,1}^2\psi^\ddagger(\eta_i)^2\Big)\psi^\dagger(\epsilon_i)^2\Big]\\
		&+\E_n[\widehat{\bb}_{p,s}(x_{i})\widehat{\bb}_{p,s}(x_{i})'
		\eta_{i,1}^2(\psi(y_i, \eta_i)^2-\sigma^2(x_i, \bw_i))]\\
		&+\Big(\E_n[\widehat{\bb}_{p,s}(x_{i})\widehat{\bb}_{p,s}(x_{i})'\eta_{i,1}^2\sigma^2(x_i, \bw_i)]-
		\E[\bb_{p,s}(x_{i})\bb_{p,s}(x_{i})'\eta_{i,1}^2\sigma^2(x_i, \bw_i)]\Big)\\
		=:&\bV_1+\bV_2+\bV_3+\bV_4.
	\end{align*}
	We bound each term in the following. The first part of the theorem only concerns $\bV_1+\bV_2+\bV_3$, and the second part needs a bound on $\bV_4$ as well where the additional Assumption \ref{SA Assumption RP}(ii) is used.
	
	\textbf{Step 1:} For $\bV_1$, we further write $\bV_1=\bV_{11}+\bV_{12}$ where
	\[
	\begin{split}
		&\bV_{11}:=\E_n\Big[\widehat{\bb}_{p,s}(x_{i})\widehat{\bb}_{p,s}(x_{i})'
		\eta_{i,1}^2\psi^\ddagger(\eta_i)^2\Big(\psi^\dagger(\epsilon_i+u_i)^2-\psi^\dagger(\epsilon_i)^2\Big)\Big],\\
		&\bV_{12}:=\E_n\Big[\widehat{\bb}_{p,s}(x_{i})\widehat{\bb}_{p,s}(x_{i})'
		\Big(\widehat{\eta}_{i,1}^2\psi^\ddagger(\widehat{\eta}_i)^2-\eta_{i,1}^2\psi^\ddagger(\eta_i)^2\Big)\Big(\psi^\dagger(\epsilon_i+u_i)^2-\psi^\dagger(\epsilon_i)^2\Big)\Big].
	\end{split}
	\]
	Let $r_{1,n}=C_1(J\log n/n)^{1/2}+J^{-p-1}$ for a constant $C_1>0$. By Assumption \ref{SA Assumption GL} and Corollary \ref{SA coro: uniform convergence, GL}, $\max_{1\leq i\leq n}|u_i|\leq r_{1,n}$ with arbitrarily large probability for $C_1$ sufficiently large. 
	For $\bV_{11}$, 
	let $\mathcal{J}$ be the set of all discontinuity points of $\psi(\cdot)$. Define $\I_{i,\mathcal{D}}:=\I(\epsilon_i\in\mathcal{D})$ and $\I_{i,\mathcal{D}^c}:=(1-\I_{i,\mathcal{D}})$ where $\mathcal{D}:=\{a: |a-\jmath|\leq r_{1,n} \text{ for some } \jmath\in\mathcal{J} \}$. Define
	\begin{align*}
		&\bV_{111}:=\E_n\Big[\widehat{\bb}_{p,s}(x_{i})\widehat{\bb}_{p,s}(x_{i})'
		\eta_{i,1}^2\psi^\ddagger(\eta_i)^2\Big(\psi^\dagger(\epsilon_i+u_i)^2-\psi^\dagger(\epsilon_i)^2\Big)\I_{i,\mathcal{D}}\Big],\\
		&\bV_{112}:=\E_n\Big[\widehat{\bb}_{p,s}(x_{i})\widehat{\bb}_{p,s}(x_{i})'
		\eta_{i,1}^2\psi^\ddagger(\eta_i)^2\Big(\psi^\dagger(\epsilon_i+u_i)^2-\psi^\dagger(\epsilon_i)^2\Big)\I_{i,\mathcal{D}^c}\Big].
	\end{align*}
	By definition of $\mathcal{D}$ and Assumption \ref{SA Assumption GL}, 
	\[
    \|\bV_{111}\|
	\lesssim\|\E_n[\widehat{\bb}_{p,s}(x_{i})\widehat{\bb}_{p,s}(x_{i})'
	\E[\I_{i,\mathcal{D}}|\mathscr{F}_{XW\Delta}]]\|+
	\|\E_n[\widehat{\bb}_{p,s}(x_{i})\widehat{\bb}_{p,s}(x_{i})'
	(\I_{i,\mathcal{D}}-\E[\I_{i,\mathcal{D}}|\mathscr{F}_{XW\Delta}])]\|.
	\]
	By Assumption \ref{SA Assumption GL} and Lemma SA-3.5 of \cite{Cattaneo-Crump-Farrell-Feng_2024_AER}, the first term on the right-hand side is $O_\P(r_{1,n})$. For the second term, conditional on $\mathscr{F}_{XW\Delta}$, it is an independent sequence with mean zero. Thus, we can apply the argument given in Step 3 below and conclude that the second term is $O_\P(\sqrt{r_{1,n}J\log J/n}+J\log J/n)$. In this case, the indicator $\I_{i,\mathcal{D}}$ is trivially bounded uniformly.
	
	On the other hand, by Assumption \ref{SA Assumption GL}, 
	$$\|\bV_{112}\|\lesssim r_{1,n}
	\|\E_n[\widehat{\bb}_{p,s}(x_{i})\widehat{\bb}_{p,s}(x_{i})'
	\eta_{i,1}^2\psi^\ddagger(\eta_i)^2|\psi^\dagger(\epsilon_i+u_i)+\psi^\dagger(\epsilon_i)|]\|.
	$$
	Since $|c|\leq \frac{1}{2}(1+c^2)$ for any scalar $c$, we have 
	\[
	\E_n\Big[\widehat{\bb}_{p,s}(x_{i})\widehat{\bb}_{p,s}(x_{i})'
	\eta_{i,1}^2\psi^\ddagger(\eta_i)^2|\psi^\dagger(\epsilon_i)|\Big]\leq
	\frac{1}{2}\E_n\Big[\widehat{\bb}_{p,s}(x_{i})\widehat{\bb}_{p,s}(x_{i})'
	\eta_{i,1}^2\psi^\ddagger(\eta_i)^2(1+\psi^\dagger(\epsilon_i)^2)\Big]\lesssim_\P 1,
	\]
	by Lemma \ref{SA lem: Gram, GL} and the result in Step 3. In addition,	we further write
	\begin{align*}
       &\E_n\Big[\widehat{\bb}_{p,s}(x_{i})\widehat{\bb}_{p,s}(x_{i})'
	\eta_{i,1}^2\psi^\ddagger(\eta_i)^2|\psi^\dagger(\epsilon_{i}+u_i)|\Big]\\
       =&\,\E_n\Big[\widehat{\bb}_{p,s}(x_{i})\widehat{\bb}_{p,s}(x_{i})'
	\eta_{i,1}^2\psi^\ddagger(\eta_i)^2
	|\psi^\dagger(\epsilon_i)+(\psi^\dagger(\epsilon_i+u_i)-\psi^\dagger(\epsilon_i))|\Big].
	\end{align*}
    Repeat the previous argument to bound this term.
    We conclude that $\|\bV_{11}\|\lesssim_\P r_{1,n}$.
	
	$\bV_{12}$ can be treated using the previous argument combined with the argument given in Step 2 and the result in Step 3. It leads to $\|\bV_{12}\|\lesssim_\P r_{1,n}$.
	
	\textbf{Step 2:} For $\bV_2$,
	by Assumption \ref{SA Assumption GL}, Corollary \ref{SA coro: uniform convergence, GL} and the argument given later in Step 3, we have   
	$$\|\bV_2\|\leq \max_{1\leq i\leq n}|\widehat{\eta}_{i,1}^2\psi^\ddagger(\widehat\eta_i)^2-\eta_{i,1}^2\psi^\ddagger(\eta_i)^2|\|\E_n[\widehat{\bb}_{p,s}(x_i)\widehat{\bb}_{p,s}(x_i)'\psi^\dagger(\epsilon_i)^2]\|
	\lesssim_\P
	(J\log n/n)^{1/2}+J^{-p-1}.$$
	
	\textbf{Step 3:}
	For $\bV_3$, Assumption \ref{SA Assumption RP}(i) implies that $\widehat\Delta\in\Pi_C$ w.p.a. $1$ and that $\E[\psi(y_i,\eta_i)^2-\sigma^2(x_i,\bw_i)\mid\mathscr F_{XW\Delta}]=0$. Thus, in view of Lemma \ref{SA lem: spline transform}, it suffices to show that
	\[\sup_{\Delta\in\Pi_C}\Big\|\E_n[\bb_{p,0}(x_{i};\Delta)
	\bb_{p,0}(x_{i};\Delta)'\eta_{i,1}^2(\psi(y_i, \eta_i)^2-\sigma^2(x_i, \bw_i))]\Big\|\lesssim_\P 
	\Big(\frac{J\log J}{n^{\frac{\nu-2}{\nu}}}\Big)^{1/2}.
	\]
	For notational simplicity, we write $\varphi_{i}=\psi(y_i, \eta_i)^2-\sigma^2(x_i, \bw_i)$, 
	$\varphi_{i}^-=\varphi_{i}\I(|\varphi_{i}|\leq M)-\E[\varphi_{i}\I(|\varphi_{i}|\leq M)|x_i,\bw_i]$,
	$\varphi_{i}^+=\varphi_{i}\I(|\varphi_{i}|> M)-\E[\varphi_{i}\I(|\varphi_{i}|> M)|x_i,\bw_i]$ for some $M>0$ to be specified later. Since $\E[\varphi_{i}|x_i, \bw_i]=0$, $\varphi_{i}=\varphi_{i}^-+\varphi_{i}^+$. Then, define a function class
	\[\mathcal{G}=\Big\{(x_{1}, \bw_1, \varphi_{1})\mapsto b_{p,0,l}(x_{1};\Delta)b_{p,0,k}(x_{1};\Delta)\eta_{1,1}^2\varphi_{1}:1\leq l\leq J(p+1), 1\leq k\leq J(p+1), \Delta\in\Pi_C\Big\}.
	\]
	For $g\in\mathcal{G}$, $\sum_{i=1}^{n}g(x_i, \bw_i, \varphi_i)
	=\sum_{i=1}^{n}g(x_i, \bw_i, \varphi_i^+)+\sum_{i=1}^{n}g(x_i, \bw_i, \varphi_i^-)$.
	
	For the truncated piece, we have
	$\sup_{g\in\mathcal{G}}|g(x_i, \bw_i, \varphi_i^-)|\lesssim JM$,
	and
	\begin{align*}
		\sup_{g\in\mathcal{G}}\V[g(x_{i}, \bw_i, \varphi_{i}^-)]
		&\lesssim 
		\sup_{x\in\mathcal{X}, \bw\in\mathcal{W}}\E[(\varphi_{i}^-)^2|x_i=x, \bw_i=\bw]\sup_{\Delta\in\Pi_C}\sup_{1\leq l,k\leq J(p+1)}\E[b_{p,0,l}^2(x_i;\Delta)b_{p,0,k}^2(x_i;\Delta)\eta_{i,1}^4] \\
		&\lesssim  JM \sup_{x\in\mathcal{X}, \bw\in\mathcal{W}}
		\E\Big[|\varphi_i|\Big|x_i=x,\bw_i=\bw\Big]\lesssim JM.
	\end{align*}
	The VC condition holds by the same argument given in the proof of Lemma \ref{SA lem: Gram, GL}. Then, by Lemma \ref{SA lem: maximal inequality},
	\[
	\E\Big[\sup_{g\in\mathcal{G}}\Big|\E_n[g(x_{i}, \bw_i, \varphi_{i}^-)]\Big|\Big]
	\lesssim \sqrt{\frac{JM\log (JM)}{n}}+\frac{JM\log(JM)}{n}.
	\]
	
	Regarding the tail, we apply Theorem 2.14.1 of \citet*{vandevarrt-Wellner_1996_book} and obtain
	\begin{align*}
		\E\Big[\sup_{g\in\mathcal{G}}\Big|\E_n[g(x_{i}, \bw_i, \varphi_{i}^+)]\Big|\Big]
		&\lesssim \frac{1}{\sqrt{n}}J
		\E\Big[\sqrt{\E_n[|\varphi_{i}^+|^2]}\Big]\\
		&\leq \frac{1}{\sqrt{n}}J
		(\E[\max_{1\leq i\leq n}|\varphi_{i}^+|])^{1/2}(\E[\E_n[|\varphi_{i}^+|]])^{1/2}\\
		&\lesssim \frac{J}{\sqrt{n}}\cdot \frac{n^{\frac{1}{\nu}}}{M^{(\nu-2)/4}},
	\end{align*}
	where the second line follows from Cauchy-Schwarz inequality and the third line uses the fact that
	\[
		\E[\max_{1\leq i\leq n}|\varphi_{i}^+|]\lesssim
		\E[\max_{1\leq i\leq n}\psi(y_i, \eta_i)^2]\lesssim
		n^{2/\nu} \quad \text{and}\quad
		\E[\E_n[|\varphi_{i}^+|]]\leq \E[|\varphi_{1}^+|]\lesssim \frac{\E[|\psi(y_1,\eta_1)|^{\nu}]}{M^{(\nu-2)/2}}.
	\]
	Then the desired result follows simply by setting $M=J^{\frac{2}{\nu-2}}$ and the sparsity of the basis.
	
	\textbf{Step 4:} For $\bV_4$, since by Assumption \ref{SA Assumption GL},
	$\sup_{x\in\mathcal{X}, \bw\in\mathcal{W}}\E[\psi(y_i, \eta_i)^2|x_i=x]\lesssim 1$. Then, by the same argument given in the proof of Lemma \ref{SA lem: Gram, GL},  
	\begin{align*}
		&\sup_{\Delta\in\Pi_C}\Big\|\frac{1}{\sqrt{n}}\G_n[\bb_{p,s}(x_{i};\Delta)\bb_{p,s}(x_{i};\Delta)'
		\eta_{i,1}^2\sigma^2(x_i, \bw_i)]\Big\|
		\lesssim_\P\sqrt{J\log J/n} \quad\text{and}\\
		&\Big\|\E_{\widehat{\Delta}}\Big[\widehat{\bb}_{p,s}(x_{i})\widehat{\bb}_{p,s}(x_{i})'
		\eta_{i,1}^2\psi(y_i, \eta_i)^2\Big]-
		\E\Big[\bb_{p,s}(x_{i})\bb_{p,s}(x_{i})'\eta_{i,1}^2\psi(y_i, \eta_i)^2\Big]\Big\|\lesssim_\P \sqrt{J\log J/n}+\mathfrak{r}_{\tt RP}.
	\end{align*}
    The proof for the first conclusion is complete.

    \textbf{Step 5:}
    The results about $\widehat\Omega_{\mu^{(v)}}(x)$, $\widehat\Omega_\vartheta(x)$ and $\widehat\Omega_\zeta(x)$ follow by Assumption \ref{SA Assumption GL}, {Assumption \ref{SA Assumption HLE}(ii)}, Lemmas \ref{SA lem: local basis} and \ref{SA lem: Gram, GL}, and Corollary \ref{SA coro: uniform convergence, GL}.
    In particular,
    \begin{align*}
    \widehat\Omega_\vartheta(x)&=\widehat\eta_{0,1}(x,\widehat\evalw)^2\widehat\Omega_{\mu^{(0)}}(x), &
    \bar\Omega_\vartheta(x)&=\eta_{0,1}(x,\evalw)^2\bar\Omega_{\mu^{(0)}}(x),\\
    \widehat\Omega_\zeta(x)&=\widehat\eta_{0,1}(x,\widehat\evalw)^2\widehat\Omega_{\mu^{(1)}}(x), &
    \bar\Omega_\zeta(x)&=\eta_{0,1}(x,\evalw)^2\bar\Omega_{\mu^{(1)}}(x).
    \end{align*}
    Moreover, a first-order expansion and Corollary \ref{SA coro: uniform convergence, GL} give
    \[
    \sup_{x\in\mathcal X}\big|\widehat\eta_{0,1}(x,\widehat\evalw)^2-\eta_{0,1}(x,\evalw)^2\big|
    \lesssim_\P \Big(\frac{J\log n}{n}\Big)^{1/2}+J^{-p-1}+\sqrt{d_\gamma}\mathfrak r_\gamma+\|\widehat\evalw-\evalw\|.
    \]
    Assumption \ref{SA Assumption HLE}(i) and the theorem's rate conditions absorb the first three terms into the stated rates, yielding the transformed-parameter conclusions.
    For the conclusions involving the population variances, Lemmas \ref{SA lem: spline transform}, \ref{SA lem: local basis} and \ref{SA lem: Gram, GL} show that replacing $\widehat\bT_s$ by $\bT_s$ contributes at most $O_\P(J^{1+2v}\mathfrak r_{\tt RP})$, $O_\P(J\mathfrak r_{\tt RP})$ and $O_\P(J^3\mathfrak r_{\tt RP})$ to the three respective variance differences, which are absorbed by the stated rates.
    The proof is complete.

\end{proof}

\subsection{Proof of Theorem \ref{SA thm: pointwise normality, GL}}
\begin{proof}
	We first show that for each fixed $x\in\mathcal{X}$,
	\[
	\bar{\Omega}_{\mu^{(v)}}(x)^{-1/2}\widehat{\bb}_{p,s}^{(v)}(x)'\bar{\bQ}^{-1}\G_n[\widehat{\bb}_{p,s}(x_i)\eta_{i,1}\psi(y_i,\eta_i)]=:\G_n[a_{i}\psi(y_i, \eta_i)]
	\] 
	is asymptotically normal.	
	Conditional on $\mathscr{F}_{XW\Delta}$, the $\sigma$-field generated by $\{(x_i,\bw_i)\}_{i=1}^n$ and $\widehat\Delta$, it is an independent mean-zero sequence over $i$ with variance equal to $1$.
	Then by Berry-Esseen inequality,
	\begin{align*}
		\sup_{u\in\mathbb{R}}
		\Big|{\P(\G_n[a_i\psi(y_i,\eta_i)]\leq u\mid\mathscr F_{XW\Delta})}-\Phi(u)\Big|
		\leq \min\bigg(1, \;\frac{\sum_{i=1}^{n}\E[|a_{i}\psi(y_i, \eta_i)|^3|\mathscr{F}_{XW\Delta}]}{n^{3/2}}\bigg).
	\end{align*}
	By Lemmas \ref{SA lem: local basis}, \ref{SA lem: Gram, GL} and \ref{SA lem: asymp variance, GL},
	\begin{align*}
		&\quad\;\frac{1}{n^{3/2}}\sum_{i=1}^{n}\E\Big[|a_{i}\psi(y_i, \eta_i)|^3\Big|\mathscr{F}_{XW\Delta}\Big]\\
		&\lesssim \bar{\Omega}_{\mu^{(v)}}(x)^{-3/2}\frac{1}{n^{3/2}}\sum_{i=1}^{n}\E\Big[
		|\widehat{\bb}_{p,s}^{(v)}(x)'\bar{\bQ}^{-1}\widehat{\bb}_{p,s}(x_{i})\eta_{i,1}\psi(y_i, \eta_i)|^3\Big|\mathscr{F}_{XW\Delta}\Big]\\
		&\lesssim \bar{\Omega}_{\mu^{(v)}}(x)^{-3/2}\frac{1}{n^{3/2}}\sum_{i=1}^n|\widehat{\bb}_{p,s}^{(v)}(x)'\bar{\bQ}^{-1}\widehat{\bb}_{p,s}(x_{i})|^3\\
		&\leq\bar{\Omega}_{\mu^{(v)}}(x)^{-3/2}\frac{\sup_{x\in\mathcal{X}}\sup_{z\in\mathcal{X}}
			|\widehat{\bb}_{p,s}^{(v)}(x)'\bar{\bQ}^{-1}\widehat{\bb}_{p,s}(z)|}{n^{3/2}}
		\sum_{i=1}^{n}
		|\widehat{\bb}_{p,s}^{(v)}(x)'\bar{\bQ}^{-1}\widehat{\bb}_{p,s}(x_{i})|^2\\
		&\lesssim_\P\frac{1}{J^{3/2+3v}}\cdot\frac{J^{1+v}}{\sqrt{n}}\cdot J^{1+2v}
		\to 0
	\end{align*}
	since $J/n=o(1)$.
	By Theorem \ref{SA thm: meat matrix, GL}, the above weak convergence still holds if $\bar{\Omega}_{\mu^{(v)}}(x)$ is replaced by $\widehat{\Omega}_{\mu^{(v)}}(x)$.
	Also, Theorem \ref{SA thm: meat matrix, GL} and Lemma \ref{SA lem: asymp variance, GL} imply, under the respective evaluation-point conditions in parts (ii) and (iii),
	\begin{align*}
	\sup_{x\in\mathcal X}\left|\frac{\widehat\Omega_\vartheta(x)}{\bar\Omega_\vartheta(x)}-1\right|
	&\lesssim_\P J^{-p-1}+\Big(\frac{J\log n}{n^{1-2/\nu}}\Big)^{1/2}+\|\widehat\evalw-\evalw\|=o_\P(1),\\
	\sup_{x\in\mathcal X}\left|\frac{\widehat\Omega_\zeta(x)}{\bar\Omega_\zeta(x)}-1\right|
	&\lesssim_\P J^{-p-1}+\Big(\frac{J\log n}{n^{1-2/\nu}}\Big)^{1/2}+\|\widehat\evalw-\evalw\|=o_\P(1).
	\end{align*}
	Thus, $\widehat\Omega_\vartheta(x)\asymp_\P J$ and $\widehat\Omega_\zeta(x)\asymp_\P J^3$.
For $\widehat\vartheta(x,\widehat\evalw)$, the evaluation-point contribution after Studentization is
\[
\frac{\|\widehat\evalw-\evalw\|}{\sqrt{J/n}}=o_\P(1).
\]
For $\widehat\zeta(x,\widehat\evalw)$, Theorem \ref{SA thm: Bahadur representation, GL}(iii) and $\widehat\Omega_\zeta(x)\asymp_\P J^3$ imply that the corresponding contribution is bounded in probability by a constant times
\[
\frac{\|\widehat\evalw-\evalw\|\{1+\sqrt{J^3\log n/n}\}}{\sqrt{J^3/n}}=\|\widehat\evalw-\evalw\|\left\{\sqrt{n/J^3}+\sqrt{\log n}\right\}=o_\P(1),
\]
where the last equality follows from the rate condition in part (iii). The remaining Bahadur remainders are negligible under the maintained rate conditions, and hence the three desired conclusions follow.
\end{proof}

\subsection{Proof of Theorem \ref{SA thm: IMSE, GL}}
\begin{proof}
 We let $\widehat\bbeta_0$ and $\widehat{r}_{0,v}$ be defined as in Lemma \ref{SA lem: uniform approx rate}. {Let $\check{\eta}_i=\eta(\widehat{\bb}_{p,s}(x_i)'\widehat\bbeta_0+\bw_i'\bgamma_0)$. To retain the leading approximation error explicitly, first note the exact decomposition
	\[
	\begin{split}
	&\E_n\!\left[\widehat\bb_{p,s}(x_i)\eta^{(1)}\!\left(\widehat\bb_{p,s}(x_i)'\widehat\bbeta_0+\bw_i'\bgamma_0\right)\psi(y_i,\check\eta_i)\right]\\
	=&\;\E_n[\widehat\bb_{p,s}(x_i)\eta_{i,1}\psi(y_i,\eta_i)]+\E_n[\widehat\bb_{p,s}(x_i)\eta_{i,1}\Psi(x_i,\bw_i;\check\eta_i)]\\
	&+\E_n\!\left[\widehat\bb_{p,s}(x_i)\left\{\eta^{(1)}\!\left(\widehat\bb_{p,s}(x_i)'\widehat\bbeta_0+\bw_i'\bgamma_0\right)\psi(y_i,\check\eta_i)-\eta_{i,1}\psi(y_i,\eta_i)\right.\right.\\
	&\hspace{12em}\left.\left.-\E\!\left[\eta^{(1)}\!\left(\widehat\bb_{p,s}(x_i)'\widehat\bbeta_0+\bw_i'\bgamma_0\right)\psi(y_i,\check\eta_i)-\eta_{i,1}\psi(y_i,\eta_i)\mid\mathscr F_{XW\Delta}\right]\right\}\right]\\
	&+\E_n\!\left[\widehat\bb_{p,s}(x_i)\left\{\eta^{(1)}\!\left(\widehat\bb_{p,s}(x_i)'\widehat\bbeta_0+\bw_i'\bgamma_0\right)-\eta_{i,1}\right\}\Psi(x_i,\bw_i;\check\eta_i)\right].
	\end{split}
	\]
	The last line is uniformly $O_\P(J^{-2p-2})$ before projection. The same conditional maximal-inequality argument used in the proof of Lemma \ref{SA lem: proj approx error, GL} gives
	\[
	\begin{split}
	&\sup_{x\in\mathcal X}\left|\widehat\bb_{p,s}^{(v)}(x)'\bQ_0^{-1}\E_n\!\left[\widehat\bb_{p,s}(x_i)\left\{\eta^{(1)}\!\left(\widehat\bb_{p,s}(x_i)'\widehat\bbeta_0+\bw_i'\bgamma_0\right)\psi(y_i,\check\eta_i)\right.\right.\right.\\
	&\hspace{15em}\left.\left.\left.-\eta_{i,1}\psi(y_i,\eta_i)-\E[\,\cdot\mid\mathscr F_{XW\Delta}]\right\}\right]\right|\\
	&\hspace{5em}\lesssim_\P J^v\left\{J^{-\frac{p+1}{2}}\sqrt{\frac{J\log J}{n}}+\frac{J\log J}{n}\right\}.
	\end{split}
	\]
	where $\E[\,\cdot\mid\mathscr F_{XW\Delta}]$ denotes the conditional expectation of the two preceding score terms inside braces.} By Lemmas \ref{SA lem: uniform approx rate} and \ref{SA lem: Gram, GL}, Theorem \ref{SA thm: Bahadur representation, GL} and the results given in the proof of Lemma \ref{SA lem: proj approx error, GL}, {the deterministic $J^{-p-1}$ component is therefore represented by the explicit conditional-mean and approximation-error terms below}, and
	\[
	\begin{split}
		\widehat{\mu}^{(v)}(x)-\mu_0^{(v)}(x)
		=&\,{\widehat{\bb}_{p,s}^{(v)}(x)'(\widehat{\bbeta}-\widehat{\bbeta}_0)}-\widehat{r}_{0,v}(x)\\
		=&-\widehat{\bb}_{p,s}^{(v)}(x)'\bQ_0^{-1}\E_n[\widehat{\bb}_{p,s}(x_i)\eta_{i,1}\psi(y_i,\eta_i)]
		-\widehat{\bb}_{p,s}^{(v)}(x)'\bQ_0^{-1}\E_n[\widehat{\bb}_{p,s}(x_i)\eta_{i,1}
		\Psi(x_i, \bw_i;\check{\eta}_i)]\\
		&\hspace{2em}-\widehat{r}_{0,v}(x)\\
		&\hspace{2em}+{O_\P\left(J^v\left\{\left(\frac{J\log n}{n}\right)^{3/4}\ell_{\gamma,n}+J^{-\frac{p+1}{2}}\sqrt{\frac Jn}\ell_{\gamma,n}+\frac{J\ell_{\gamma,n}^2}{n}+\sqrt{d_\gamma}\mathfrak r_\gamma\right\}\right)},
	\end{split}
	\]
	{The $O_\P(\cdot)$ in the last line holds uniformly over $x\in\mathcal{X}$ and is $o_\P\big(J^v(\sqrt{J/n}+J^{-p-1})\big)$ under the maintained rate conditions. Thus, the integral of its square is $o_\P(J^{1+2v}/n+J^{-2(p+1-v)})$.} Then,
	\begin{align*}
	\mathtt{AISE}_{\mu^{(v)}}=\int_{\mathcal{X}}\Big(&\widehat{\bb}_{p,s}^{(v)}(x)'\bQ_0^{-1}\E_n[\widehat{\bb}_{p,s}(x_i)\eta_{i,1}\psi(y_i,\eta_i)]\\
    &+\widehat{\bb}_{p,s}^{(v)}(x)'\bQ_0^{-1}\E_n[\widehat{\bb}_{p,s}(x_i)\eta_{i,1}
	\Psi(x_i, \bw_i;\check{\eta}_i)]
	+\widehat{r}_{0,v}(x)\Big)^2\omega(x)\d{x}.
	\end{align*}
	Next, taking conditional expectation given $\bX$, $\bW$ and $\widehat\Delta$ and using the argument in the proof of Lemma \ref{SA lem: Gram, GL} again, we have
	\begin{align*}
		\E[\mathtt{AISE}_{\mu^{(v)}}|\bX, \bW, \widehat\Delta]=&\;\frac{1}{n}\tr\Big(\bQ_0^{-1}\bSigma_0\bQ_0^{-1}\int_{\mathcal{X}}\bb_{p,s}^{(v)}(x)\bb_{p,s}^{(v)}(x)'\omega(x)\d{x}\Big)+o_\P(J^{2v+1}/n)\\
		&+\int_{\mathcal{X}}\Big(\widehat{\bb}_{p,s}^{(v)}(x)'\widehat{\bbeta}_0-\mu_0^{(v)}(x)\Big)^2\omega(x)\d{x}
		\\
		&+\int_{\mathcal{X}}\Big(\widehat{\bb}_{p,s}^{(v)}(x)'\bQ_0^{-1}\E_n[\widehat{\bb}_{p,s}(x_i)\eta_{i,1}
		\Psi(x_i, \bw_i;\check{\eta}_i)]\Big)^2\omega(x)\d{x}\\
		&+2\int_{\mathcal{X}}\widehat{\bb}_{p,s}^{(v)}(x)'\bQ_0^{-1}\E_n[\widehat{\bb}_{p,s}(x_i)\eta_{i,1}
		\Psi(x_i, \bw_i;\check{\eta}_i)]\widehat{r}_{0,v}(x)\omega(x)\d{x}.
	\end{align*}
	By Assumption \ref{SA Assumption GL}, $\Psi(x_i,\bw_i;\check{\eta}_i)=-\Psi_1(x_i,\bw_i;{\eta_i})\eta_{i,1}\widehat{r}_{0}(x_i)+O_\P(J^{-2p-2})$ where $O_\P(\cdot)$ holds uniformly over $i$. The terms in the last three lines correspond to the integrated squared bias. Also, using the same argument in the proof of Lemma \ref{SA lem: Gram, GL}, $\E_n[\cdot]$ in the last two lines can be safely replaced by $\E_{\widehat\Delta}[\cdot]$, which only introduces some additional approximation error of order $o_\P(J^{-2p-2+2v})$.

For any $\Delta\in\Pi_C$, let $\tau_x^{\mathtt L}(\Delta)$ and $h_x(\Delta)$ denote, respectively, the left endpoint and width of the interval containing $x$, and define
\[
q_{0,v}(x;\Delta)
:=\frac{\mu_0^{(p+1)}(x)}{(p+1-v)!}h_x(\Delta)^{p+1-v}
\mathscr E_{p+1-v}\left(\frac{x-\tau_x^{\mathtt L}(\Delta)}{h_x(\Delta)}\right).
\]
For $s=0$, apply Taylor's theorem on each bin around its left endpoint and decompose the order-$(p+1)$ term using $z^{p+1}=\mathscr E_{p+1}(z)+\{z^{p+1}-\mathscr E_{p+1}(z)\}$. The term in braces is a polynomial of degree $p$ and can therefore be absorbed separately in each bin into the piecewise-polynomial approximation. Differentiating $v$ times and using the uniform bound on $\mu_0^{(p+2)}$ gives, for $\Delta=\widehat\Delta$ and $\Delta=\Delta_0$,
\[
r_{0,v}(x;\Delta)=q_{0,v}(x;\Delta)+\bb_{p,0}^{(v)}(x;\Delta)'\mathbf a(\Delta)+o_\P(J^{-p-1+v}).
\]
For $s>0$, first note that the additional condition in the theorem and Assumption \ref{SA Assumption RP}(ii) imply
\[
\max_{2\leq j\leq J}|\hat h_j-\hat h_{j-1}|\leq\max_{2\leq j\leq J}|h_j-h_{j-1}|+2\max_{1\leq j\leq J}|\hat h_j-h_j|=o_\P(J^{-1}).
\]
Under this condition, the proof of Lemma SA-6.1 in \cite{Cattaneo-Farrell-Feng_2020_AoS} shows that
there exists a maximally smooth degree-$p$ spline whose $v$th-order approximation error is $q_{0,v}(x;\Delta)+o_\P(J^{-p-1+v})$
uniformly over $x\in\mathcal X$, for $\Delta=\widehat\Delta$ and $\Delta=\Delta_0$.
Indeed, the uniform bound on $\mu_0^{(p+2)}$ makes the Taylor remainder $O(J^{-p-2+v})=o(J^{-p-1+v})$, while the contribution of unequal widths is bounded by $J^{-p+v}\max_{2\leq j\leq J}|h_j-h_{j-1}|=o(J^{-p-1+v})$ (and analogously in probability for $\widehat\Delta$). Since this approximant also belongs to the spline space for $0<s<p$, the argument following Equation (SA-5.6) in the proof of Theorem SA-3.4 in \cite{Cattaneo-Crump-Farrell-Feng_2024_AER}, applied to the least-squares projection defining $\bbeta_0(\Delta)$, yields
\[
r_{0,v}(x;\Delta)
=q_{0,v}(x;\Delta)+\bb_{p,s}^{(v)}(x;\Delta)'\mathbf a(\Delta)
+o_\P(J^{-p-1+v})
\]
uniformly over $x\in\mathcal X$, where the approximation remainders for both $s=0$ and $s>0$ are nonrandom for $\Delta=\Delta_0$. The spline components in the two approximation displays cancel after weighted projection. Indeed, if
\[
\bQ(\Delta):=\E_{\Delta}\!\left[\bb_{p,s}(x_i;\Delta)\bb_{p,s}(x_i;\Delta)'
\varkappa(x_i,\bw_i)\right],
\]
where $\E_{\Delta}[\cdot]$ treats $\Delta$ as fixed. Then
\begin{align*}
&r_{0,v}(x;\Delta)-\bb_{p,s}^{(v)}(x;\Delta)'\bQ(\Delta)^{-1}
\E_{\Delta}\!\left[\bb_{p,s}(x_i;\Delta)\varkappa(x_i,\bw_i)r_{0,0}(x_i;\Delta)\right]\\
={}&q_{0,v}(x;\Delta)-\bb_{p,s}^{(v)}(x;\Delta)'\bQ(\Delta)^{-1}
\E_{\Delta}\!\left[\bb_{p,s}(x_i;\Delta)\varkappa(x_i,\bw_i)q_{0,0}(x_i;\Delta)\right]
+o_\P(J^{-p-1+v}).
\end{align*}
Combining this cancellation with the preceding Taylor expansion, the conditional Gram-matrix argument, and Assumption \ref{SA Assumption RP}(ii), the integrated squared bias is
\begin{align*}
\int_{\mathcal X}\Big(&q_{0,v}(x;\Delta_0)
-\bb_{p,s}^{(v)}(x)'\bQ_0^{-1}
\E[\bb_{p,s}(x_i)\varkappa(x_i,\bw_i)q_{0,0}(x_i;\Delta_0)]\Big)^2
\omega(x)\d{x}
+o_\P(J^{-2p-2+2v}).
\end{align*}
Since $q_{0,v}(x;\Delta_0)=r_{0,v}^\star(x)$ by definition, the expression of $\mathscr B_n(p,s,v)$ given in the theorem follows.

Finally, consider the contribution of estimating the evaluation point. For part (ii), the stated condition gives $\|\widehat\evalw-\evalw\|=o_\P(\sqrt{J/n}+J^{-p-1})$, which makes its squared contribution and its cross-products negligible relative to $J/n+J^{-2p-2}$. For part (iii), let
\[
L_{\zeta,n}:=\sqrt{J^3/n+J^{-2p}}
\quad\text{and}\quad
C_{\zeta,n}:=1+\sqrt{J^3\log n/n}.
\]
Theorem \ref{SA thm: Bahadur representation, GL}(iii) bounds the evaluation-point contribution uniformly over $x$ by
\[
O_\P\big(\|\widehat\evalw-\evalw\|C_{\zeta,n}\big)=o_\P(L_{\zeta,n}).
\]
Its integrated square is therefore $o_\P(J^3/n+J^{-2p})$, and its cross-products with the leading stochastic and bias terms have the same order by Cauchy--Schwarz. The remaining terms are controlled as in part (i), proving parts (ii) and (iii).
\end{proof}

\subsection{Proof of Theorem \ref{SA thm: strong approximation, GL}}
\begin{proof}
	The proof is divided into several steps.
	
	\textbf{Step 1:}	
	Note that
	\begin{align*}
		&\sup_{x\in\mathcal{X}}\bigg|\frac{\widehat{\mu}^{(v)}(x)-\mu_0^{(v)}(x)}{\sqrt{\widehat{\Omega}_{\mu^{(v)}}(x)/n}}-\frac{\widehat{\mu}^{(v)}(x)-\mu_0^{(v)}(x)}{\sqrt{\bar\Omega_{\mu^{(v)}}(x)/n}}\bigg|\\
		\leq& \sup_{x\in\mathcal{X}}\bigg|\frac{\widehat{\mu}^{(v)}(x)-\mu_0^{(v)}(x)}{\sqrt{\bar\Omega_{\mu^{(v)}}(x)/n}}\bigg|
		\sup_{x\in\mathcal{X}}\bigg|\frac{\widehat{\Omega}_{\mu^{(v)}}(x)^{1/2}-\bar\Omega_{\mu^{(v)}}(x)^{1/2}}{\widehat{\Omega}_{\mu^{(v)}}(x)^{1/2}}\bigg|\\
		\lesssim&_\P \Big(\sqrt{\log n}+\sqrt{n}J^{-p-1-1/2}\Big)\Big(J^{-p-1}+\sqrt{\frac{J\log n}{n^{1-\frac{2}{\nu}}}}\Big)
	\end{align*}
	where the last step uses Lemma \ref{SA lem: asymp variance, GL} and Corollary \ref{SA coro: uniform convergence, GL}.
	Then, in view of Lemmas \ref{SA lem: uniform approx rate}, \ref{SA lem: proj approx error, GL}, Theorems \ref{SA thm: Bahadur representation, GL}, \ref{SA thm: meat matrix, GL} and the rate restriction given in the lemma, we have
	\[
	\sup_{x\in\mathcal{X}}\bigg|\frac{\widehat{\mu}^{(v)}(x)-\mu_0^{(v)}(x)}
	{\sqrt{\widehat{\Omega}_{\mu^{(v)}}(x)/n}} +
	\frac{\widehat{\bb}_{p,s}^{(v)}(x)'\bar{\bQ}^{-1}}{\sqrt{\bar\Omega_{\mu^{(v)}}(x)}}\G_n[\widehat{\bb}_{p,s}(x_i)\eta_{i,1}\psi(y_i,\eta_i)]\bigg|=o_\P(a_n^{-1}).
	\]

	\textbf{Step 2:}
	Let us write 
	$\mathscr{K}(x,x_i)=\bar\Omega_{\mu^{(v)}}(x)^{-1/2}\widehat{\bb}_{p,s}^{(v)}(x)'\bar{\bQ}^{-1}\widehat{\bb}_{p,s}(x_i)$ (the dependence of $\widehat{\bb}_{p,s}^{(v)}(x)$, $\bar\bQ$ and $\bar\Omega_{\mu^{(v)}}(x)$ on $\bX$, $\bW$ and $\widehat\Delta$ is omitted for simplicity), and $\tilde\sigma^2(x_i,\bw_i)=\E[\psi^\dagger(\epsilon_i)^2|x_i,\bw_i]$.
	Now we
	rearrange $\{x_i\}_{i=1}^n$ as a sequence of order statistics $\{x_{(i)}\}_{i=1}^n$, i.e., $x_{(1)}\leq\cdots\leq x_{(n)}$.
	Accordingly, $\{\epsilon_i\}_{i=1}^n$, $\{\bw_i\}_{i=1}^n$ and $\{\tilde\sigma^2(x_i, \bw_i)\}_{i=1}^n$ are ordered as concomitants $\{\epsilon_{[i]}\}_{i=1}^n$, $\{\bw_{[i]}\}$ and 
    $\{\tilde{\sigma}^2_{[i]}\}_{i=1}^n$ where 
     $\tilde\sigma^2_{[i]}=\tilde\sigma^2(x_{(i)},\bw_{[i]})$. Clearly, conditional on $\mathscr{F}_{XW\Delta}$ (the $\sigma$-field generated by $\{(x_i,\bw_i)\}$ and $\widehat\Delta$), $\{\psi^\dagger(\epsilon_{[i]})\}_{i=1}^n$ is still an independent mean-zero sequence. Then by Assumptions \ref{SA Assumption DGP}, \ref{SA Assumption GL} and the result of  \cite{Sakhanenko_1991_SAM}, there exists a sequence of i.i.d. standard normal random variables $\{\zeta_{[i]}\}_{i=1}^n$ such that
	\begin{align*}
	\max_{1\leq \ell\leq n}|S_{\ell}|:=
	\max_{1\leq \ell \leq n}&\bigg|\sum_{i=1}^\ell\eta^{(1)}(\mu_0(x_{(i)})+\bw_{[i]}'\bgamma_0)\psi^\ddagger(\eta(\mu_0(x_{(i)})+\bw_{[i]}'\bgamma_0))\psi^\dagger(\epsilon_{[i]}) \\
 &- \sum_{i=1}^\ell\eta^{(1)}(\mu_0(x_{(i)})+\bw_{[i]}'\bgamma_0)\psi^\ddagger(\eta(\mu_0(x_{(i)})+\bw_{[i]}'\bgamma_0))
     \tilde\sigma_{[i]}\zeta_{[i]}\Big|\lesssim_\P n^{\frac{1}{\nu}}.
	\end{align*}
	Then, using summation by parts,
	\begin{align*}
		&\,\sup_{x\in\mathcal{X}}\left|\sum_{i=1}^{n}\mathscr{K}(x, x_{(i)})\eta^{(1)}(\mu_0(x_{(i)})+\bw_{[i]}'\bgamma_0)\psi^\ddagger(\eta(\mu_0(x_{(i)})+\bw_{[i]}'\bgamma_0))(\psi^\dagger(\epsilon_{[i]})-\tilde\sigma_{[i]}\zeta_{[i]})\right|\\
		=&\,\sup_{x\in\mathcal{X}}\left|\mathscr{K}(x, x_{(n)})S_{n}
		-\sum_{i=1}^{n-1}S_{i}\left(\mathscr{K}(x,x_{(i+1)})
		-\mathscr{K}(x,x_{(i)})\right)\right|\\
		\leq&\,\sup_{x\in\mathcal{X}}\max_{1\leq i\leq n}|\mathscr{K}(x,x_i)||S_{n}|
		+\sup_{x\in\mathcal{X}}\left|
		\frac{\widehat{\bb}_{p,s}^{(v)}(x)'\bar{\bQ}^{-1}}{\sqrt{\bar\Omega_{\mu^{(v)}}(x)}}
		\sum_{i=1}^{n-1}S_{i}\Big(\widehat{\bb}_{p,s}(x_{(i+1)}) -
		\widehat{\bb}_{p,s}(x_{(i)})\Big)\right|\\
		\leq&\,\sup_{x\in\mathcal{X}}\max_{1\leq i\leq n}|\mathscr{K}(x,x_i)||S_{n}|
		+\sup_{x\in\mathcal{X}}
		\Bigg\|\frac{\bar{\bQ}^{-1}\widehat{\bb}_{p,s}^{(v)}(x)}{\sqrt{\bar\Omega_{\mu^{(v)}}(x)}}\Bigg\|_1
		\left\|\sum_{i=1}^{n-1}S_{i}\Big(\widehat{\bb}_{p,s}(x_{(i+1)})-
		\widehat{\bb}_{p,s}(x_{(i)})\Big)\right\|_\infty.
	\end{align*}
	
	By Lemmas \ref{SA lem: local basis}, \ref{SA lem: Gram, GL} and \ref{SA lem: asymp variance, GL},
	$\sup_{x\in\mathcal{X}}\sup_{x_i\in\mathcal{X}}|\mathscr{K}(x,x_i)|\lesssim_\P \sqrt{J}$, and $$\sup_{x\in\mathcal{X}}\Bigg\|\frac{\bar{\bQ}^{-1}\widehat{\bb}_{p,s}^{(v)}(x)}{\sqrt{\bar\Omega_{\mu^{(v)}}(x)}}\Bigg\|_1\lesssim_\P 1.$$
	Then, notice that
	\begin{equation*}
		\max_{1\leq l \leq K_{p,s}}
		\bigg|\sum_{i=1}^{n-1}\Big(\widehat{b}_{p,s,l}(x_{(i+1)})-\widehat{b}_{p,s,l}(x_{(i)})\Big)S_{i}\bigg|
		\leq \max_{1\leq l \leq K_{p,s}}
		\sum_{i=1}^{n-1}\Big|\widehat{b}_{p,s,l}(x_{(i+1)})-\widehat{b}_{p,s,l}(x_{(i)})\Big|
		\max_{1\leq \ell \leq n}\Big|S_{\ell}\Big|.
	\end{equation*}
	By construction of the ordering, $\max_{1\leq l \leq K_{p,s}}
	\sum_{i=1}^{n-1}\Big|\widehat{b}_{p,s,l}(x_{(i+1)})-\widehat{b}_{p,s,l}(x_{(i)})\Big|\lesssim \sqrt{J}$. Under the rate restriction in the theorem, this suffices to show that for any $\xi>0$, 
	$$
	\P\bigg(\sup_{x\in\mathcal{X}}\,
	\Big|\G_n[\mathscr{K}(x,x_i)\eta^{(1)}(\mu_0(x_i)+\bw_i'\bgamma_0)(\psi(y_i,\eta_i)-\sigma(x_i,\bw_i)\zeta_i)]\Big|
	>\xi a_n^{-1}\Big|\mathscr{F}_{XW\Delta}\bigg)=o_\P(1),
	$$ 
	where we recover the original ordering.
	Here, $=_{d|\mathscr{F}_{XW\Delta}}$ denotes equality in distribution conditional on $\mathscr{F}_{XW\Delta}$.
	Since $\G_n[\widehat{\bb}_{p,s}(x_i)\zeta_i\sigma(x_i,\bw_i)\eta_{i,1}]=_{d|\mathscr{F}_{XW\Delta}}\mathbf{N}(0, \bar{\bSigma})$, the above steps construct the following approximating process:
	\[ \bar{Z}_{\mu^{(v)},p}(x):=\frac{\widehat{\bb}_{p,s}^{(v)}(x)'\bar{\bQ}^{-1}}{\sqrt{\bar\Omega_{\mu^{(v)}}(x)}}\bar{\bSigma}^{1/2}\bN_{K_{p,s}}.
	\]
	
	\textbf{Step 3:}	
	Suppose that Assumption \ref{SA Assumption RP}(ii) also holds. Note that
	\[
	\begin{split}
		&\sup_{x\in\mathcal{X}}|\bar{Z}_{\mu^{(v)},p}(x)-Z_{\mu^{(v)},p}(x)|\\
		\leq&
		\sup_{x\in\mathcal{X}}\bigg|\frac{\widehat{\bb}^{(v)}(x)'(\bar{\bQ}^{-1}-\bQ_0^{-1})}{\sqrt{\Omega_{\mu^{(v)}}(x)}}\bar{\bSigma}^{1/2}\bN_{K_{p,s}}\bigg|\\
		&+\sup_{x\in\mathcal{X}}\bigg|\frac{\widehat{\bb}^{(v)}(x)'\bQ_0^{-1}}{\sqrt{\Omega_{\mu^{(v)}}(x)}}\Big(\bar{\bSigma}^{1/2}-\bSigma_0^{1/2}\Big)\bN_{K_{p,s}}\bigg|\\
		&+\sup_{x\in\mathcal{X}}\bigg|\frac{\widehat{\bb}_{p,0}^{(v)}(x)'
			(\widehat{\bT}_s-\bT_s)'\bQ_0^{-1}}{\sqrt{\Omega_{\mu^{(v)}}(x)}}\bSigma_0^{1/2}\bN_{K_{p,s}}\bigg|\\
       &+\sup_{x\in\mathcal{X}}\bigg|\Big(
       \frac{1}{\sqrt{\bar\Omega_{\mu^{(v)}}(x)}}-
       \frac{1}{\sqrt{\Omega_{\mu^{(v)}}(x)}}
       \Big)
       \widehat{\bb}_{p,0}^{(v)}(x)'
			\widehat{\bT}_s\bar\bQ^{-1}\bar\bSigma^{1/2}\bN_{K_{p,s}}\bigg|\\
   =&\; I+II+III+IV,
	\end{split}
	\]
	where each term on the right-hand side is a mean-zero Gaussian process conditional on $\mathscr{F}_{XW\Delta}$. By Theorem \ref{SA thm: meat matrix, GL} (see Steps 4 and 5 of its proof),
    $\sup_{x\in\mathcal{X}}|\bar\Omega_{\mu^{(v)}}(x)-\Omega_{\mu^{(v)}}(x)|\lesssim_\P J^{1+2v}(\sqrt{J\log n/n}+\mathfrak{r}_{\tt RP})$. By a similar calculation given in Step 1 and the rate condition imposed, the last term is $o_\P(a_n^{-1})$. 	
    By Lemmas \ref{SA lem: spline transform} and \ref{SA lem: Gram, GL}, and the argument in the proof of Lemma \ref{SA lem: local basis} together with Theorem X.3.8 of \citet*{Bhatia_2013_book},
    \begin{align*}
    \|\bar{\bQ}^{-1}-\bQ_0^{-1}\|&\lesssim_\P\sqrt{J\log J/n}+\mathfrak r_{\tt RP},&
    \|\widehat{\bT}_s-\bT_s\|&\lesssim_\P\mathfrak r_{\tt RP},\\
    \|\bar{\bSigma}^{1/2}-\bSigma_0^{1/2}\|&\lesssim_\P\sqrt{J\log J/n}+\mathfrak r_{\tt RP}.
    \end{align*}
	By Gaussian Maximal Inequality \citep*[Corollary 2.2.8]{vandevarrt-Wellner_1996_book},
	\begin{align*}
	\E\Big[I+II+III\Big|\mathscr{F}_{XW\Delta} \Big]
	&\lesssim_\P \sqrt{\log J}
	\Big(\|\bar{\bSigma}^{1/2}-\bSigma_0^{1/2}\|+\|\bar{\bQ}^{-1}-\bQ_0^{-1}\|+\|\widehat{\bT}_s-\bT_s\|\Big)\\
	&\lesssim_\P\sqrt{\log J}\left\{\sqrt{\frac{J\log J}{n}}+\mathfrak r_{\tt RP}\right\}=o_\P(a_n^{-1}).
	\end{align*}
	where the last line follows from the imposed rate restriction.
	Then the proof for part (i) is complete.

For part (ii), the evaluation-point remainder after Studentization is
\[
\frac{\|\widehat\evalw-\evalw\|}{\sqrt{J/n}}=o_\P(a_n^{-1}).
\]
For part (iii), Theorem \ref{SA thm: Bahadur representation, GL}(iii) and $\widehat\Omega_\zeta(x)\asymp_\P J^3$ give
\[
\sup_{x\in\mathcal X}\frac{\|\widehat\evalw-\evalw\|\{1+\sqrt{J^3\log n/n}\}}{\sqrt{\widehat\Omega_\zeta(x)/n}}\lesssim_\P\|\widehat\evalw-\evalw\|\left\{\sqrt{n/J^3}+\sqrt{\log n}\right\}=o_\P(a_n^{-1}).
\]
{The definition of $\widehat\Omega_\zeta(x)$, a first-order expansion of the transformation, and the proof of Theorem \ref{SA thm: meat matrix, GL} give
\[
\sup_{x\in\mathcal X}\left|\frac{\widehat\Omega_\zeta(x)}{\bar\Omega_\zeta(x)}-1\right|\lesssim_\P J^{-p-1}+\left(\frac{J\log n}{n^{1-2/\nu}}\right)^{1/2}+\sqrt{d_\gamma}\mathfrak r_\gamma+\|\widehat\evalw-\evalw\|.
\]
The right-hand side multiplied by $\sqrt{\log n}$ is $o_\P(a_n^{-1})$ under the rate conditions in the theorem and the evaluation-point condition in part (iii). Since the supremum of the leading Studentized process is $O_\P(\sqrt{\log n})$, replacing $\bar\Omega_\zeta(x)$ by $\widehat\Omega_\zeta(x)$ therefore contributes $o_\P(a_n^{-1})$ uniformly in $x$.} All other remainder terms are controlled by the common rate conditions. The leading variance term for $\widehat\vartheta(x,\widehat\evalw)$ or $\widehat\zeta(x,\widehat\evalw)$ differs from that for $\widehat\mu(x)$ or $\widehat\mu^{(1)}(x)$ only by the stated transformation, so the preceding coupling argument completes parts (ii) and (iii).
\end{proof}

\subsection{Proof of Theorem \ref{SA thm: plug-in approx, GL}}
\begin{proof}
	This conclusion follows from Lemmas \ref{SA lem: local basis}, \ref{SA lem: Gram, GL}, Theorem \ref{SA thm: meat matrix, GL} and Gaussian Maximal Inequality as applied in Step 3 in the proof of Theorem \ref{SA thm: strong approximation, GL}.
\end{proof}


\subsection{Proof of Theorem \ref{SA thm: sup approx}}
\begin{proof}
	We first show that
	\[
	\sup_{u\in\mathbb{R}}\Big|\P\Big(\sup_{x\in\mathcal{X}}
	|T_{\mu^{(v)},p}(x)|\leq u\Big) -
	\P\Big(\sup_{x\in\mathcal{X}}|Z_{\mu^{(v)},p}(x)|\leq u\Big)\Big|=o(1).
	\]
	By Theorem \ref{SA thm: strong approximation, GL}, there exists a sequence of constants $\xi_n$ such that $\xi_n=o(1)$ and
	\[ \P\Big(\Big|\sup_{x\in\mathcal{X}}|T_{\mu^{(v)},p}(x)|-\sup_{x\in\mathcal{X}}|Z_{\mu^{(v)},p}(x)|\Big|>\xi_n/a_n\Big)=o(1).
	\]
	Then,
	\begin{align*}
	\P\Big(\sup_{x\in\mathcal{X}}|T_{\mu^{(v)},p}(x)|\leq u\Big)&\leq
	\P\Big(\Big\{\sup_{x\in\mathcal{X}}|T_{\mu^{(v)},p}(x)|\leq u\Big\}\cap\Big\{\Big|\sup_{x\in\mathcal{X}}|T_{\mu^{(v)},p}(x)|-
	\sup_{x\in\mathcal{X}}|Z_{\mu^{(v)},p}(x)|\Big|\leq\xi_n/a_n\Big\}\Big)+o(1)\\
	&\leq \P\Big(\sup_{x\in\mathcal{X}}|Z_{\mu^{(v)},p}(x)|\leq u+\xi_n/a_n\Big)
	+o(1)\\
	&\leq \P\Big(\sup_{x\in\mathcal{X}}|Z_{\mu^{(v)},p}(x)|\leq u\Big)+
	\sup_{u\in\mathbb{R}}\E\Big[\P\Big(\Big|\sup_{x\in\mathcal{X}}|Z_{\mu^{(v)},p}(x)|-u\Big|\leq \xi_n/a_n\Big|\widehat{\Delta}\Big)\Big]\\
	&\leq \P\Big(\sup_{x\in\mathcal{X}}|Z_{\mu^{(v)},p}(x)|\leq u\Big)+
	\E\Big[\sup_{u\in\mathbb{R}}\P\Big(\Big|\sup_{x\in\mathcal{X}}
	|Z_{\mu^{(v)},p}(x)|-u\Big|\leq \xi_n/a_n\Big|\widehat{\Delta}\Big)\Big]
	+o(1).
	\end{align*}
	Apply the Anti-Concentration Inequality conditional on $\widehat\Delta$ \citep{Chernozhukov-Chetverikov-Kato_2014b_AoS} to the second term:
	\begin{align*}
	\sup_{u\in\mathbb{R}}\;
	\P\Big(\Big|\sup_{x\in\mathcal{X}}|Z_{\mu^{(v)},p}(x)|-u\Big|\leq \xi_n/a_n\Big|\widehat{\Delta}\Big)&\leq 4\xi_na_n^{-1}\E\Big[\sup_{x\in\mathcal{X}}|Z_{\mu^{(v)},p}(x)|\Big|\widehat\Delta\Big]+o(1)\\
	&\lesssim_\P \xi_n a_n^{-1}\sqrt{\log J}+o(1)\to 0
	\end{align*}
	where the last step uses Gaussian Maximal Inequality
	\citep*[see][Corollary 2.2.8]{vandevarrt-Wellner_1996_book}.
	By Dominated Convergence Theorem,
	\[\E\Big[
	\sup_{u\in\mathbb{R}}\P\Big(\Big|\sup_{x\in\mathcal{X}}|Z_{\mu^{(v)},p}(x)|-u\Big|\leq \xi_n/a_n\Big|\widehat{\Delta}\Big)\Big]=o(1).
	\]
	The other side of the inequality follows similarly.
	
	By similar argument, using Theorem \ref{SA thm: plug-in approx, GL}, we have
	\[
	\sup_{u\in\mathbb{R}}\Big|\P\Big(\sup_{x\in\mathcal{X}}|\widehat{Z}_{\mu^{(v)},p}(x)|\leq u\Big|\bD,\widehat{\Delta}\Big) -
	\P\Big(\sup_{x\in\mathcal{X}}|Z_{\mu^{(v)},p}(x)|\leq u\Big|\widehat{\Delta}\Big)\Big|=o_\P(1).
	\]
	Then, it remains to show that
	\begin{equation} \label{eq: sup ks distance 1}
	\sup_{u\in\mathbb{R}}\Big|\P\Big(\sup_{x\in\mathcal{X}}|Z_{\mu^{(v)},p}(x)|\leq u\Big)-
	\P\Big(\sup_{x\in\mathcal{X}}|Z_{\mu^{(v)},p}(x)|\leq u|\widehat{\Delta}\Big)\Big|=o_\P(1).
	\end{equation}
	We can write
	\[ Z_{\mu^{(v)},p}(x)=\frac{\widehat{\bb}_{p,0}^{(v)}(x)'}{\sqrt{\widehat{\bb}_{p,0}^{(v)}(x)'\bV_0\widehat{\bb}_{p,0}^{(v)}(x)}}\breve{\bN}_{K_{p,0}}
	\]
	where $\bV_0=\bT_s'\bQ_0^{-1}\bSigma_0\bQ_0^{-1}\bT_s$ and $\breve{\bN}_{K_{p,0}}:=\bT_s'\bQ_0^{-1}\bSigma_0^{1/2}\bN_{K_{p,s}}$ is a $K_{p,0}$-dimensional Gaussian random vector. Importantly, by this construction, $\breve{\bN}_{K_{p,0}}$ and $\bV_0$ do not depend on $\widehat{\Delta}$ and $x$, and they are only determined by the deterministic partition $\Delta_0$.
	
	First consider $v=0$. For any two partitions $\Delta_1, \Delta_2\in\Pi_C$, for any $x\in\mathcal{X}$, there exists $\check{x}\in\mathcal{X}$ such that
	\[\bb_{p,0}^{(0)}(x;\Delta_1)=\bb_{p,0}^{(0)}(\check{x};\Delta_2),\]
	and vice versa.
	Therefore, the following two events are equivalent: $\{\omega:\sup_{x\in\mathcal{X}}|Z_p(x;\Delta_1)|\leq u\}=
	\{\omega:\sup_{x\in\mathcal{X}}|Z_p(x;\Delta_2)|\leq u\}$ for any $u$.
	Thus, 
	$$\E\Big[\P\Big(\sup_{x\in\mathcal{X}}|Z_{\mu^{(v)},p}(x)|\leq u
	\Big|\widehat\Delta\Big)\Big]=\P\Big(\sup_{x\in\mathcal{X}}|Z_{\mu^{(v)},p}(x)|\leq u
	\Big|\widehat\Delta\Big)+o_\P(1).$$
	Then for $v=0$, the desired result follows.
	
	For $v>0$, simply notice that 
	$\widehat{\bb}_{p,0}^{(v)}(x)=\widehat{\mathfrak{T}}_v\widehat{\bb}_{p,0}(x)$ for some transformation matrix $\widehat{\mathfrak{T}}_v$. Clearly, $\widehat{\mathfrak{T}}_v$ takes a similar structure as $\widehat{\bT}_s$: each row and each column only have a finite number of nonzeros. Each nonzero element is simply $\hat{h}_j^{-v}$ up to some constants. 
    Under Assumption \ref{SA Assumption RP}(ii), let $\mathfrak{T}_v$ be the fixed analogue of $\widehat{\mathfrak{T}}_v$ obtained by replacing $\hat h_j$ with $h_j$. Since $\hat h_j\asymp_\P h_j\asymp J^{-1}$, the mean value theorem and the finite number of nonzero elements in each row and column give
    \[
    \|\widehat{\mathfrak{T}}_v-\mathfrak{T}_v\|
    \lesssim_\P J^v\mathfrak r_{\tt RP},
    \qquad
    \sup_{x\in\mathcal X}\|(\widehat{\mathfrak{T}}_v-\mathfrak{T}_v)
    \widehat{\bb}_{p,0}(x)\|
    \lesssim_\P J^{v+1/2}\mathfrak r_{\tt RP}.
    \]
    Write $Z_{\mu^{(v)},p}(x;\mathfrak{T})$ for the displayed representation of $Z_{\mu^{(v)},p}(x)$ above with $\widehat{\bb}_{p,0}^{(v)}(x)$ replaced by $\mathfrak{T}\widehat{\bb}_{p,0}(x)$ in both the numerator and denominator, so $Z_{\mu^{(v)},p}(x)=Z_{\mu^{(v)},p}(x;\widehat{\mathfrak T}_v)$. Lemma \ref{SA lem: asymp variance, GL}, a first-order expansion of the denominator, and Gaussian Maximal Inequality yield
    \[
    \E\left[\sup_{x\in\mathcal X}
    \left|Z_{\mu^{(v)},p}(x;\widehat{\mathfrak{T}}_v)
    -Z_{\mu^{(v)},p}(x;\mathfrak{T}_v)\right|
    \,\middle|\,\widehat\Delta\right]
    \lesssim_\P \mathfrak r_{\tt RP}\sqrt{\log J}
    =o_\P((\log J)^{-1/2}),
    \]
    where the last equality follows from the rate restriction in the theorem. Thus, $\widehat{\mathfrak{T}}_v$ can be replaced by $\mathfrak{T}_v$ without affecting the approximation rate, and the desired result follows by repeating the argument for $v=0$.

    Finally, the result for $T_{\vartheta,p}(x)$ ($T_{\zeta,p}(x)$) follows by the fact that $Z_{\vartheta,p}(x)$ and $\widehat{Z}_{\vartheta,p}(x)$ ($Z_{\zeta,p}(x)$ and $\widehat{Z}_{\zeta,p}(x)$) differ from $Z_{\mu^{(v)},p}(x)$ and $\widehat{Z}_{\mu^{(v)},p}(x)$ up to a sign only. 
\end{proof}

\subsection{Proof of Theorem \ref{SA thm: CB}}
\begin{proof}
      We only consider $\widehat{I}_{\mu^{(v)},p}(x)$. The results in part (ii) and part (iii) follow similarly. 
      
	 Let $\xi_{1,n}=o(1)$, $\xi_{2,n}=o(1)$ and $\xi_{3,n}=o(1)$. Then,
	\begin{align*}
	\P\left[\sup_{x\in\mathcal{X}}|T_{\mu^{(v)},p}(x)|\leq \cval_{\mu^{(v)}}
	\right]&\leq
		\P\left[\sup_{x\in\mathcal{X}}|\bar{Z}_{\mu^{(v)},p}(x)|\leq \cval_{\mu^{(v)}}+\xi_{1,n}/a_n
	\right]+o(1)\\
	&\leq 
	\P\left[\sup_{x\in\mathcal{X}}|\bar{Z}_{\mu^{(v)},p}(x)|\leq c^0(1-\alpha+\xi_{3,n})+(\xi_{1,n}+\xi_{2,n})/a_n
	\right]+o(1)\\
	&\leq \P\left[\sup_{x\in\mathcal{X}}|\bar{Z}_{\mu^{(v)},p}(x)|\leq c^0(1-\alpha+\xi_{3,n})\right]+o(1)\to 1-\alpha,
	\end{align*}
	where $c^0(1-\alpha+\xi_{3,n})$ denotes the $(1-\alpha+\xi_{3,n})$-quantile of $\sup_{x\in\mathcal{X}}|\bar{Z}_{\mu^{(v)},p}(x)|$ conditional on $\mathscr{F}_{XW\Delta}$ (the $\sigma$-field generated by $\bX$, $\bW$ and the partition $\widehat\Delta$), the first inequality holds by Theorem \ref{SA thm: strong approximation, GL}, the second by Lemma A.1 of \cite{Belloni-Chernozhukov-Chetverikov-Kato_2015_JoE}, and the third by Anti-Concentration Inequality in \cite{Chernozhukov-Chetverikov-Kato_2014b_AoS}. The other side of the bound follows similarly.
\end{proof}

\subsection{Proof of Theorem \ref{SA thm: testing specification}}
\begin{proof}
	       We only consider the proof for part (i). The results in part (ii) and part (iii) follow similarly.
	       {For those parts, the displayed function-level convergence conditions absorb the triangular-array dependence of $\bar{\bgamma}$, while the imposed $\liminf$ conditions ensure that the fixed-alternative signal used in the power argument remains bounded away from zero uniformly in $n$.}

        Throughout this proof, we let $\xi_{1,n}=o(1)$, $\xi_{2,n}=o(1)$ and
	$\xi_{3,n}=o(1)$ be sequences of vanishing constants. Moreover, let $A_n$ be a sequence of diverging constants such that $\sqrt{\log J}A_n\lesssim \sqrt{\frac{n}{J^{1+2v}}}$. 	
	Note that
	\[ \sup_{x\in\mathcal{X}}
	|\dot{T}_{\mu^{(v)},p}(x)|\leq \sup_{x\in\mathcal{X}}
		\bigg|\frac{\widehat{\mu}^{(v)}(x)-
		\mu_0^{(v)}(x)}
	{\sqrt{\widehat\Omega_{\mu^{(v)}}(x)/n}}\bigg|+
	\sup_{x\in\mathcal{X}}\bigg|
	\frac{\mu_0^{(v)}(x)-
		m^{(v)}(x;\widetilde{\btheta})}
	{\sqrt{\widehat{\Omega}_{\mu^{(v)}}(x)/n}}\bigg|.
	\]
	Therefore, under $\dot{\mathsf{H}}_0^{\mu^{(v)}}$,
	\begin{align*}
	\P\Big[\sup_{x\in\mathcal{X}}|\dot{T}_{\mu^{(v)},p}(x)|>\cval_{\mu^{(v)}}\Big]&
	\leq
	\P\bigg[\sup_{x\in\mathcal{X}}|T_{\mu^{(v)},p}(x)|>\cval_{\mu^{(v)}}-
	\sup_{x\in\mathcal{X}}\bigg|\frac{\mu_0^{(v)}(x)-m^{(v)}(x;\widetilde{\btheta})}{\sqrt{\widehat{\Omega}_{\mu^{(v)}}(x)/n}}\bigg|\bigg] \\
	&\leq \P\bigg[\sup_{x\in\mathcal{X}}|\bar{Z}_{\mu^{(v)},p}(x)|>\cval_{\mu^{(v)}}-\xi_{1,n}/a_n - 
    \sup_{x\in\mathcal{X}}\bigg|\frac{\mu_0^{(v)}(x)-m^{(v)}(x;\widetilde{\btheta})}{\sqrt{\widehat{\Omega}_{\mu^{(v)}}(x)/n}}\bigg|\bigg] + o(1)\\
	&\leq \P\bigg[\sup_{x\in\mathcal{X}}|\bar{Z}_{\mu^{(v)},p}(x)|>c^0(1-\alpha-\xi_{3,n})-(\xi_{1,n}+\xi_{2,n})/a_n - \\
	&\qquad \sup_{x\in\mathcal{X}}\bigg|\frac{\mu_0^{(v)}(x)-m^{(v)}(x;\widetilde{\btheta})}{\sqrt{\widehat{\Omega}_{\mu^{(v)}}(x)/n}}\bigg|\bigg] + o(1)\\
	&\leq \P\Big[\sup_{x\in\mathcal{X}}|\bar{Z}_{\mu^{(v)},p}(x)|>c^0(1-\alpha-\xi_{3,n})\Big]+o(1)\\
	&=\alpha+o(1)
	\end{align*}
	where $c^0(1-\alpha-\xi_{3,n})$ denotes the $(1-\alpha-\xi_{3,n})$-quantile of $\sup_{x\in\mathcal{X}}|\bar{Z}_{\mu^{(v)},p}(x)|$ conditional on $\mathscr{F}_{XW\Delta}$ (the $\sigma$-field generated by $\bX$, $\bW$ and $\widehat\Delta$), the second inequality holds by Theorem \ref{SA thm: strong approximation, GL}, the third by Lemma A.1 of \cite{Belloni-Chernozhukov-Chetverikov-Kato_2015_JoE}, the fourth by the fact that $\sup_{x\in\mathcal{X}}\big|\frac{\mu_0^{(v)}(x)-m^{(v)}(x;\widetilde{\btheta})}{\sqrt{\widehat{\Omega}_{\mu^{(v)}}(x)/n}}\big|=o_\P(\frac{1}{\sqrt{\log J}})$ and Anti-Concentration Inequality in \cite{Chernozhukov-Chetverikov-Kato_2014b_AoS}. The other side of the bound follows similarly.
	
	On the other hand, under $\dot{\mathsf{H}}_\text{A}^{\mu^{(v)}}$,
	\begin{align*}
	&\P\Big[\sup_{x\in\mathcal{X}}|\dot{T}_{\mu^{(v)},p}(x)|>\cval_{\mu^{(v)}}\Big] \\
	=\,&\P\Big[\sup_{x\in\mathcal{X}}\Big|T_{\mu^{(v)},p}(x)+
	\frac{\mu_0^{(v)}(x)-m^{(v)}(x;\bar{\btheta})}{\sqrt{\widehat{\Omega}_{\mu^{(v)}}(x)/n}}+
	\frac{m^{(v)}(x;\bar{\btheta})-m^{(v)}(x;\widetilde{\btheta})}{\sqrt{\widehat{\Omega}_{\mu^{(v)}}(x)/n}}\Big|>\cval_{\mu^{(v)}}\Big]\\
	\geq\,& \P\bigg[\sup_{x\in\mathcal{X}}|T_{\mu^{(v)},p}(x)|< \sup_{x\in\mathcal{X}}\bigg|\frac{\mu_0^{(v)}(x)-m^{(v)}(x;\bar{\btheta})}
	{\sqrt{\widehat{\Omega}_{\mu^{(v)}}(x)/n}} +
	\frac{m^{(v)}(x;\bar{\btheta})-m^{(v)}(x;\widetilde{\btheta})}{\sqrt{\widehat{\Omega}_{\mu^{(v)}}(x)/n}}\bigg| 
	-\cval_{\mu^{(v)}}\bigg] \\
	\geq\,& \P\Big[\sup_{x\in\mathcal{X}} |\bar{Z}_{\mu^{(v)},p}(x)|\leq \sqrt{\log J}A_n-\xi_{1,n}/a_n\Big]-o(1)\\
	\geq\,& 1-o(1).
	\end{align*}
	where the fourth line holds by Lemma \ref{SA lem: asymp variance, GL}, 
	        Theorem \ref{SA thm: meat matrix, GL}, Theorem \ref{SA thm: strong approximation, GL}, the fixed-separation condition, the condition that $J^v\sqrt{J\log J/n}=o(1)$ and the definition of $A_n$, and the last by the Talagrand-Samorodnitsky Concentration Inequality \cite*[Proposition A.2.7]{vandevarrt-Wellner_1996_book}.
\end{proof}

\subsection{Proof of Theorem \ref{SA thm: testing shape restriction}}
\begin{proof}
	        We only consider the proof for part (i). The results in part (ii) and part (iii) follow similarly.
	        {For those parts, the displayed function-level convergence conditions absorb the triangular-array dependence of $\bar{\bgamma}$, while the imposed $\liminf$ conditions ensure that the one-sided signal used in the power argument remains bounded away from zero uniformly in $n$.}

	Throughout this proof, the definitions of $A_n$, $\xi_{1,n}, \xi_{2,n}$ and $\xi_{3,n}$ are the same as in the proof of Theorem \ref{SA thm: testing specification}. 
	Under $\ddot{\mathsf{H}}_0^{\mu^{(v)}}$,
	\[
	\sup_{x\in\mathcal{X}}\ddot{T}_{\mu^{(v)},p}(x)
	\leq\sup_{x\in\mathcal{X}}T_{\mu^{(v)},p}(x)+
	\sup_{x\in\mathcal{X}}\frac{|m^{(v)}(x;\bar{\btheta})-
		m^{(v)}(x;\widetilde{\btheta})|}
	{\sqrt{\widehat{\Omega}_{\mu^{(v)}}(x)/n}}.
	\]
	Then,
	\begin{align*}
	\P\Big[ \sup_{x\in\mathcal{X}} \ddot{T}_{\mu^{(v)},p}(x)> \cval_{\mu^{(v)}}\Big]
	&\leq \P\bigg[ \sup_{x\in\mathcal{X}} T_{\mu^{(v)},p}(x)> \cval_{\mu^{(v)}}
	- \sup_{x\in\mathcal{X}}\frac{|m^{(v)}(x;
		\bar{\btheta})-
		m^{(v)}(x;\widetilde{\btheta})|}
	{\sqrt{\widehat{\Omega}_{\mu^{(v)}}(x)/n}}
	\bigg] \\
	&\leq \P\Big[ \sup_{x\in\mathcal{X}} \bar{Z}_{\mu^{(v)},p}(x)> \cval_{\mu^{(v)}}-\xi_{1,n}/a_n\Big] + o(1)\\
	&\leq \P\Big[ \sup_{x\in\mathcal{X}} \bar{Z}_{\mu^{(v)},p}(x)> c^0(1-\alpha-\xi_{3,n})-(\xi_{1,n}+\xi_{2,n})/a_n\Big] + o(1)\\ 
	&\leq \P\Big[ \sup_{x\in\mathcal{X}} \bar{Z}_{\mu^{(v)},p}(x)> c^0(1-\alpha-\xi_{3,n})\Big] + o(1)\\
	&=\alpha+o(1)
	\end{align*}
	where $c^0(1-\alpha-\xi_{3,n})$ denotes the $(1-\alpha-\xi_{3,n})$-quantile of $\sup_{x\in\mathcal{X}}\bar{Z}_{\mu^{(v)},p}(x)$ conditional on $\mathscr{F}_{XW\Delta}$ (the $\sigma$-field generated by $\bX$, $\bW$ and $\widehat\Delta$), the second line holds by Theorem \ref{SA thm: strong approximation, GL}, the third by Lemma A.1 of \cite{Belloni-Chernozhukov-Chetverikov-Kato_2015_JoE}, the fourth by Anti-Concentration Inequality in \cite{Chernozhukov-Chetverikov-Kato_2014b_AoS}.
	
	On the other hand, under $\ddot{\mathsf{H}}_\text{A}^{\mu^{(v)}}$,
	\begin{align*}
	\P\Big[ \sup_{x\in\mathcal{X}} \ddot{T}_{\mu^{(v)},p}(x)> \cval_{\mu^{(v)}}\Big]
	&=\P\bigg[ \sup_{x\in\mathcal{X}}\Big( T_{\mu^{(v)},p}(x)+
	\frac{\mu_0^{(v)}(x)-
		m^{(v)}(x;\widetilde{\btheta})}
	{\sqrt{\widehat{\Omega}_{\mu^{(v)}}(x)/n}}-\cval_{\mu^{(v)}}\Big)>0\bigg]\\
	&\geq
	\P\Big[ \sup_{x\in\mathcal{X}} |T_{\mu^{(v)},p}(x)|< \sup_{x\in\mathcal{X}}\frac{
		\mu_0^{(v)}(x)-
		m^{(v)}(x;\widetilde{\btheta})}
	{\sqrt{\widehat{\Omega}_{\mu^{(v)}}(x)/n}}-\cval_{\mu^{(v)}}, \\
	&\hspace{2.5em}\sup_{x\in\mathcal{X}}\frac{\mu_0^{(v)}(x)-
		m^{(v)}(x;\widetilde{\btheta})}
	{\sqrt{\widehat{\Omega}_{\mu^{(v)}}(x)/n}}>\cval_{\mu^{(v)}}\Big]\\
	&\geq \P\Big[ \sup_{x\in\mathcal{X}} |T_{\mu^{(v)},p}(x)|< \sup_{x\in\mathcal{X}}\frac{\mu_0^{(v)}(x)-
		m^{(v)}(x;\widetilde{\btheta})}{\sqrt{\widehat{\Omega}_{\mu^{(v)}}(x)/n}}
	-\cval_{\mu^{(v)}}\Big]-o(1)\\
	&\geq \P\Big[ \sup_{x\in\mathcal{X}} |T_{\mu^{(v)},p}(x)|< \sqrt{\log J}A_n\Big]-o(1)\\
	&\geq \P\Big[ \sup_{x\in\mathcal{X}} |\bar{Z}_{\mu^{(v)},p}(x)|<\sqrt{\log J}A_n-\xi_{1,n}/a_n\Big] - o(1) \\
	&\geq 1 - o(1) 
	\end{align*}
	where the third line holds by  Lemma \ref{SA lem: asymp variance, GL},  Theorem \ref{SA thm: meat matrix, GL}, Lemma A.1 of \cite{Belloni-Chernozhukov-Chetverikov-Kato_2015_JoE}, the fixed-separation condition, the assumption that $\sup_{x\in\mathcal{X}}|m^{(v)}(x;\widetilde{\btheta})-
	m^{(v)}(x;\bar{\btheta})|=o_\P(1)$ and $J^v\sqrt{J\log J/n}=o(1)$, the fourth by definition of $A_n$, and the fifth by Theorem \ref{SA thm: strong approximation, GL}, and the last by Proposition A.2.7 in \cite{vandevarrt-Wellner_1996_book}.
	
\end{proof}

\subsection{Proof of Lemma \ref{SA lem: growing d gamma rate}}
\begin{proof}
\textbf{Step 1: Reparametrization and key facts.}
By construction,
\begin{equation}\label{SA eq: growing d orthogonality}
	\E_n[\varkappa_i\widehat{\bb}_i\widetilde{\bw}_i']=0.
\end{equation}
The one-to-one reparametrization
\[
	\btheta=\bbeta+\bar{\bQ}^{-1}
	\E_n[\widehat{\bb}_i\bw_i'\varkappa_i]\bgamma
\]
satisfies
\[
	\widehat{\bb}_i'\bbeta+\bw_i'\bgamma
	=\widehat{\bb}_i'\btheta+\widetilde{\bw}_i'\bgamma.
\]
Thus, if $\widehat{\btheta}$ denotes the transform of
$(\widehat{\bbeta},\widehat{\bgamma})$, then
$(\widehat{\btheta},\widehat{\bgamma})$ minimizes
\[
	Q_n^*(\btheta,\bgamma)
	=\E_n\big[\rho(y_i;\eta(\widehat{\bb}_i'\btheta+\widetilde{\bw}_i'\bgamma))\big].
\]

By Lemma \ref{SA lem: uniform approx rate},
\[
	\sup_{x\in\mathcal X}|\widehat r_{0,0}(x)|
	\lesssim_\P J^{-p-1}.
\]
A coordinatewise approximation error of order $c_J$ implies
$\mathfrak a_{\pi,J}\lesssim\sqrt{d_\gamma}\,c_J$, while
$\mathfrak a_{\pi,J}\lesssim c_J$ requires a uniform directional, or vector-valued, smoothness condition.

The weights $\varkappa_i$ are bounded and bounded away from zero. Hence,
the sparsity of the binscatter basis and the bounded absolute row-sum norm
of the inverse weighted Gram matrix give, coordinatewise,
\[
	\max_{1\leq i\leq n}|[\bpi_w(x_i)]_k|
	\lesssim_\P\max_{1\leq i\leq n}|w_{i,k}|\lesssim 1,
	\qquad 1\leq k\leq d_\gamma.
\]
Compact support and
$\|\widetilde{\bw}_i\|\leq\sqrt{d_\gamma}
\|\widetilde{\bw}_i\|_\infty$
then give
\begin{equation}\label{SA eq: growing d residualized covariate bound}
	\max_{1\leq i\leq n}\|\widetilde{\bw}_i\|_\infty
	\lesssim_\P 1,
	\qquad
	\max_{1\leq i\leq n}\|\widetilde{\bw}_i\|
	\lesssim_\P\sqrt{d_\gamma}.
\end{equation}

By condition (i), w.p.a. $1$,
\[
	0<\underline{\lambda}_{w,n}
	\leq\lambda_{\min}(\widehat{\bG}_w)
	\leq\lambda_{\max}(\widehat{\bG}_w)
	\leq\overline{\lambda}_{w,n},
	\qquad
	\widehat{\bG}_w=\E_n[\varkappa_i\widetilde{\bw}_i\widetilde{\bw}_i'],
\]
so that
\[
	\E_n[\varkappa_i\|\widetilde{\bw}_i\|^2]
	=\tr(\widehat{\bG}_w)\leq d_\gamma\overline{\lambda}_{w,n}.
\]
Condition (ii) gives
\[
	\left\|\E_n[\varkappa_i\widetilde{\bw}_i\widehat r_{0,0}(x_i)]-\E_{\widehat\Delta}\left[\varkappa_i\{\bw_i-\bpi_{w,J}(x_i)\}\widehat r_{0,0}(x_i)\right]\right\|\lesssim_\P J^{-p-1}\sqrt{\overline{\lambda}_{w,n}d_\gamma/n}.
\]
By iterated expectations, Cauchy--Schwarz, the definition of
$\mathfrak a_{\pi,J}$, and Euclidean duality,
\begin{align*}
	&\left\|\E_{\widehat\Delta}\left[
	\varkappa_i\{\bw_i-\bpi_{w,J}(x_i)\}\widehat r_{0,0}(x_i)
	\right]\right\|\\
	&\qquad=\sup_{\|\balpha\|=1}\left|
	\E_{\widehat\Delta}\left[
	\E[\varkappa_i\mid x_i]
	\balpha'\{\bpi_{w,0}(x_i)-\bpi_{w,J}(x_i)\}
	\widehat r_{0,0}(x_i)
	\right]\right|
	\lesssim_\P J^{-p-1}\mathfrak a_{\pi,J}.
\end{align*}
Consequently,
\begin{equation}\label{SA eq: growing d product bias}
	\big\|\E_n[\varkappa_i\widetilde{\bw}_i\widehat r_{0,0}(x_i)]\big\|
	\lesssim_\P
	J^{-p-1}\mathfrak a_{\pi,J}
	+J^{-p-1}\sqrt{\overline{\lambda}_{w,n}d_\gamma/n}.
\end{equation}
Since $J^{-p-1}=o(1)$, the second term in
\eqref{SA eq: growing d product bias} is
$o_\P(\sqrt{\overline{\lambda}_{w,n}d_\gamma/n})$.

Equation \eqref{SA eq: growing d orthogonality} implies
\begin{equation}\label{SA eq: growing d block curvature}
	\E_n[\varkappa_i\{u_i(\btheta)+\widetilde{\bw}_i'\bh\}^2]
	=\E_n[\varkappa_i u_i(\btheta)^2]+\bh'\widehat{\bG}_w\bh.
\end{equation}
It remains to establish a preliminary rate for the full index and to prove
$\|\widehat{\bgamma}-\bgamma_0\|=O_\P(\mathfrak r_{\gamma,n})$.

\textbf{Step 2: Preliminary rate.}
Let $\mathscr F_{XW\Delta}$ be the $\sigma$-field generated by
$\{(x_i,\bw_i)\}_{i=1}^n$ and $\widehat\Delta$, and set
\[
	\G_n[f_i]=\sqrt n\,\E_n\big[f_i-\E[f_i\mid\mathscr F_{XW\Delta}]\big].
\]
For this step, write
\[
	q_i=u_i(\btheta)+\widetilde{\bw}_i'\bh.
\]
Since
$\widehat{\bb}_i'\btheta_0+\widetilde{\bw}_i'\bgamma_0
=\varpi_i-\widehat r_{0,0}(x_i)$, absolute continuity of the loss gives
\begin{align*}
	&\E\left[
	\rho\big(y_i;\eta(\widehat{\bb}_i'\btheta+\widetilde{\bw}_i'(\bgamma_0+\bh))\big)
	-\rho\big(y_i;\eta(\widehat{\bb}_i'\btheta_0+\widetilde{\bw}_i'\bgamma_0)\big)
	\mid\mathscr F_{XW\Delta}\right]\\
	&\qquad=\int_0^{q_i}
	\Psi\big(x_i,\bw_i;
	\eta(\varpi_i-\widehat r_{0,0}(x_i)+t)\big)
	\eta^{(1)}(\varpi_i-\widehat r_{0,0}(x_i)+t)\d{t}.
\end{align*}
Because $\Psi(x_i,\bw_i;\eta_i)=0$, the derivative of the integrand with respect to $t$, evaluated at
$\widehat r_{0,0}(x_i)=t=0$, is $\varkappa_i$. Assumption \ref{SA Assumption GL} therefore gives, uniformly on a sufficiently small neighborhood,
\[
\begin{split}
	&\Psi\big(x_i,\bw_i;
	\eta(\varpi_i-\widehat r_{0,0}(x_i)+t)\big)
	\eta^{(1)}(\varpi_i-\widehat r_{0,0}(x_i)+t)\\
	&\qquad=\varkappa_i\{t-\widehat r_{0,0}(x_i)\}+R_{i,t},
	\qquad
	|R_{i,t}|\lesssim\{|t|+|\widehat r_{0,0}(x_i)|\}^2.
\end{split}
\]
Integrating and averaging yields
\begin{align*}
	&\E_n\left[\E\left[
	\rho\big(y_i;\eta(\widehat{\bb}_i'\btheta+\widetilde{\bw}_i'(\bgamma_0+\bh))\big)
	-\rho\big(y_i;\eta(\widehat{\bb}_i'\btheta_0+\widetilde{\bw}_i'\bgamma_0)\big)
	\mid\mathscr F_{XW\Delta}\right]\right]\\
	&\qquad=
	\frac{1}{2}\E_n[\varkappa_iq_i^2]
	-\E_n[\varkappa_i\widehat r_{0,0}(x_i)q_i]
	+\E_n\left[\int_0^{q_i}R_{i,t}\d{t}\right].
\end{align*}
By Cauchy--Schwarz and Lemma \ref{SA lem: uniform approx rate},
\[
	\left|\E_n[\varkappa_i\widehat r_{0,0}(x_i)q_i]\right|
	\lesssim_\P
	J^{-p-1}\E_n[\varkappa_iq_i^2]^{1/2}.
\]
Moreover,
\[
	\left|\E_n\left[\int_0^{q_i}R_{i,t}\d{t}\right]\right|
	\lesssim_\P
	\{J^{-p-1}+\max_i|q_i|\}\E_n[q_i^2]
	+J^{-2p-2}\E_n[|q_i|].
\]
Since $\varkappa_i$ is bounded and bounded away from zero, the first term on the right is absorbed by the quadratic term when the neighborhood is sufficiently small, while the second is bounded by
$C J^{-p-1}\E_n[\varkappa_iq_i^2]^{1/2}$.
Finally, \eqref{SA eq: growing d block curvature} gives
\[
	\E_n[\varkappa_iq_i^2]
	=\E_n[\varkappa_i u_i(\btheta)^2]+\bh'\widehat{\bG}_w\bh.
\]
Combining these bounds,
\begin{align*}
	&\E_n\left[\E\left[
	\rho\big(y_i;\eta(\widehat{\bb}_i'\btheta+\widetilde{\bw}_i'(\bgamma_0+\bh))\big)
	-\rho\big(y_i;\eta(\widehat{\bb}_i'\btheta_0+\widetilde{\bw}_i'\bgamma_0)\big)
	\mid\mathscr F_{XW\Delta}\right]\right]\\
	&\qquad\geq
	c\left\{\E_n[\varkappa_i u_i(\btheta)^2]+\bh'\widehat{\bG}_w\bh\right\}
	-C J^{-p-1}
	\left\{\E_n[\varkappa_i u_i(\btheta)^2]+\bh'\widehat{\bG}_w\bh\right\}^{1/2}.
\end{align*}
Denote the criterion-increment class in the preceding display by $\mathcal G$.
Since $d_\gamma=o(J)$, it is indexed by $K_{p,s}+d_\gamma\asymp J$
parameters. On the ball
\[
	\E_n[\varkappa_i u_i(\btheta)^2]+\bh'\widehat{\bG}_w\bh
	\leq M^2\mathfrak r_{\theta,n}^2,
\]
its conditional $L_2$ radius is
$\bar\sigma_G\lesssim M\mathfrak r_{\theta,n}$, and an envelope satisfies
\[
	\bar G_i\lesssim M\mathfrak r_{\theta,n}
	\left(\sqrt J+
	\sqrt{\frac{d_\gamma}{\underline{\lambda}_{w,n}}}\right)
	(1+|\psi(y_i,\eta_i)|).
\]
With $\delta_G=\bar\sigma_G/\|\bar G\|_{L_2(\P)}$, the localized entropy bound is
\[
	\sup_{\mathbb Q}\log N(\mathcal G,L_2(\mathbb Q),
	\varepsilon\|\bar G\|_{\mathbb Q,2})
	\lesssim J\log(C\delta_G/\varepsilon),
	\qquad 0<\varepsilon\leq\delta_G.
\]
Assumption \ref{SA Assumption GL}(iii) and condition (iv) give
\[
	\frac{J}{n}
	\left\|\max_{1\leq i\leq n}\bar G_i\right\|_{L_2(\P)}
	=o_\P(\mathfrak r_{\theta,n}^2).
\]
Hence, Lemma \ref{SA lem: maximal inequality} yields
\begin{align*}
	&\sup_{\substack{
	\E_n[\varkappa_i u_i(\btheta)^2]+\bh'\widehat{\bG}_w\bh
	\leq M^2\mathfrak r_{\theta,n}^2}}
	\left|\frac{1}{\sqrt n}\G_n\left[
	\rho\big(y_i;\eta(\widehat{\bb}_i'\btheta+\widetilde{\bw}_i'(\bgamma_0+\bh))\big)
	-\rho\big(y_i;\eta(\widehat{\bb}_i'\btheta_0+\widetilde{\bw}_i'\bgamma_0)\big)
	\right]\right|\\
	&\qquad\lesssim_\P
	M\mathfrak r_{\theta,n}\sqrt{J/n}
	+o_\P(\mathfrak r_{\theta,n}^2).
\end{align*}
Since $\mathfrak r_{\theta,n}=\sqrt{J/n}+J^{-p-1}$, convexity and the preceding two displays imply
\begin{equation}\label{SA eq: growing d preliminary rate}
	\E_n[\varkappa_i u_i(\widehat{\btheta})^2]
	+(\widehat{\bgamma}-\bgamma_0)'\widehat{\bG}_w
	(\widehat{\bgamma}-\bgamma_0)
	=O_\P(\mathfrak r_{\theta,n}^2).
\end{equation}
In particular,
$\E_n[u_i(\widehat{\btheta})^2]^{1/2}=O_\P(\mathfrak r_{\theta,n})$.
The local support and normalization of $\widehat{\bb}_i$, together with the Gram bound, also give
\[
	\max_{1\leq i\leq n}|u_i(\widehat{\btheta})|
	\lesssim_\P\sqrt J\,\mathfrak r_{\theta,n}=o_\P(1),
\]
where the last equality follows from condition (iii).
{Consequently, for every $\varepsilon>0$ there exists a
sufficiently large fixed $C_T=C_T(\varepsilon)<\infty$ such that}
\[
{
	\mathcal T_n(C_T):=
	\left\{\btheta:
	\max_i|u_i(\btheta)|\leq C_T\sqrt J\mathfrak r_{\theta,n},\quad
	\E_n[u_i(\btheta)^2]^{1/2}\leq C_T\mathfrak r_{\theta,n}
	\right\}
}
\]
{and}
\[
{
	\liminf_{n\to\infty}\P[\widehat{\btheta}\in\mathcal T_n(C_T)]
	\geq1-\varepsilon.
}
\]

\textbf{Step 3: Conditional mean.}
For $\btheta\in\mathcal T_n(C_T)$ and $\bh=\bgamma-\bgamma_0$, define
\[
	\delta_i(\btheta,\bh)
	=\;\rho\big(y_i;\eta(\widehat{\bb}_i'\btheta+
	\widetilde{\bw}_i'(\bgamma_0+\bh))\big)
	-\rho\big(y_i;\eta(\widehat{\bb}_i'\btheta+
	\widetilde{\bw}_i'\bgamma_0)\big).
\]
Because $\widehat{\bb}_i'\btheta_0+\widetilde{\bw}_i'\bgamma_0
=\varpi_i-\widehat r_{0,0}(x_i)$,
\begin{align*}
	\E[\delta_i(\btheta,\bh)\mid\mathscr F_{XW\Delta}]
	=\int_0^{\widetilde{\bw}_i'\bh}
	&\Psi\big(x_i,\bw_i;
	\eta(\varpi_i-\widehat r_{0,0}(x_i)+u_i(\btheta)+t)\big)\\
	&\times\eta^{(1)}(\varpi_i-\widehat r_{0,0}(x_i)+u_i(\btheta)+t)\d{t}.
\end{align*}
Expanding the integrand around $\varpi_i$ gives the leading terms
\[
	\frac{1}{2}\bh'\widehat{\bG}_w\bh
	-\bh'\E_n[\varkappa_i\widetilde{\bw}_i\widehat r_{0,0}(x_i)]
	+\E_n[\varkappa_i u_i(\btheta)\widetilde{\bw}_i'\bh].
\]
The last term is zero by \eqref{SA eq: growing d orthogonality}. By condition
(iii), whose product-moment bound follows under the primitive condition in
Remark \ref{SA rem: growing d primitive covariance}, uniformly over
$\btheta\in\mathcal T_n(C_T)$ and the local values of $\bh$ used below, the remaining terms are bounded by
\[
	C\sqrt{\overline{\lambda}_{w,n}}\|\bh\|\mathfrak r_{\theta,n}^2
	+o_\P(\underline{\lambda}_{w,n}\|\bh\|^2).
\]
Therefore, by \eqref{SA eq: growing d product bias},
\begin{equation}\label{SA eq: growing d mean bound}
\begin{split}
	\E_n\big[\E[\delta_i(\btheta,\bh)\mid\mathscr F_{XW\Delta}]\big]
	\geq\;&\frac{\underline{\lambda}_{w,n}}{4}\|\bh\|^2\\
	&-C\|\bh\|\left(
		J^{-p-1}\mathfrak a_{\pi,J}
		+J^{-p-1}\sqrt{\overline{\lambda}_{w,n}d_\gamma/n}
		+\sqrt{\overline{\lambda}_{w,n}}\mathfrak r_{\theta,n}^2
	\right).
\end{split}
\end{equation}

\textbf{Step 4: Centered empirical process.}
Decompose the objective increment as
\[
	\delta_i(\btheta,\bh)=\delta_i^0(\bh)+\delta_i^u(\btheta,\bh),
\]
where
\[
\begin{split}
	\delta_i^0(\bh)
	=\;&\rho\big(y_i;\eta(\widehat{\bb}_i'\btheta_0+
	\widetilde{\bw}_i'(\bgamma_0+\bh))\big)
	-\rho\big(y_i;\eta(\widehat{\bb}_i'\btheta_0+
	\widetilde{\bw}_i'\bgamma_0)\big),\\
	\delta_i^u(\btheta,\bh)
	=\;&\delta_i(\btheta,\bh)-\delta_i^0(\bh).
\end{split}
\]
Denote the first class by $\mathcal G_0$ and the second class by $\mathcal G_u$. 

For $\mathcal{G}_0$, it is indexed only by $\bh\in\mathbb R^{d_\gamma}$ and has envelope
\[
	\bar F_0\lesssim M\mathfrak r_{\gamma,n}\|\widetilde{\bw}_i\|
	(1+|\psi(y_i,\eta_i)|).
\]
For every $\|\balpha\|=1$, Assumption \ref{SA Assumption GL} and the eigenvalue bound give
\[
	\E_n\left[\E\left[
	(\balpha'\widetilde{\bw}_i)^2
	(1+|\psi(y_i,\eta_i)|)^2
	\mid\mathscr F_{XW\Delta}\right]\right]
	\lesssim \balpha'\widehat{\bG}_w\balpha
	\leq\overline{\lambda}_{w,n}.
\]
Thus, a conditional $L_2$ radius for $\mathcal G_0$ is
$\bar\sigma_0\lesssim
	M\mathfrak r_{\gamma,n}\sqrt{\overline{\lambda}_{w,n}}$.
Moreover, the $L_2$ norm of the pointwise envelope satisfies
\begin{align*}
	\|\bar F_0\|_{L_2(\P)}^2
	&\lesssim M^2\mathfrak r_{\gamma,n}^2
	\E_n\left[\E\left[
	\|\widetilde{\bw}_i\|^2(1+|\psi(y_i,\eta_i)|)^2
	\mid\mathscr F_{XW\Delta}\right]\right]\\
	&\lesssim M^2\mathfrak r_{\gamma,n}^2
	\tr(\widehat{\bG}_w)
	\leq M^2\mathfrak r_{\gamma,n}^2d_\gamma\overline{\lambda}_{w,n}.
\end{align*}
Let $\delta_0=\bar\sigma_0/\|\bar F_0\|_{L_2(\P)}$. The localized entropy bound is
\[
	\sup_{\mathbb Q}\log N(\mathcal G_0,L_2(\mathbb Q),
	\varepsilon\|\bar F_0\|_{\mathbb Q,2})
	\lesssim d_\gamma\log(C\delta_0/\varepsilon),
	\qquad 0<\varepsilon\leq\delta_0,
\]
and hence
$J(\delta_0,\mathcal G_0,\bar F_0)\lesssim\delta_0\sqrt{d_\gamma}$.
Lemma \ref{SA lem: maximal inequality} gives
\begin{equation}\label{SA eq: growing d gamma process}
	\sup_{\|\bh\|\leq M\mathfrak r_{\gamma,n}}\left|\frac{1}{\sqrt n}\G_n[\delta_i^0(\bh)]\right|\lesssim_\P M\mathfrak r_{\gamma,n}\sqrt{\overline{\lambda}_{w,n}d_\gamma/n}+\frac{d_\gamma}{n}\left\|\max_{1\leq i\leq n}\bar F_0(x_i,\bw_i,y_i)\right\|_{L_2(\P)}.
\end{equation}
Assumption \ref{SA Assumption GL}(iii) implies
$\|\max_{1\leq i\leq n}(1+|\psi(y_i,\eta_i)|)\|_{L_2(\P)}
\lesssim n^{1/\nu}$. Hence,
\[
	\frac{d_\gamma}{n}
	\left\|\max_{1\leq i\leq n}\bar F_0(x_i,\bw_i,y_i)\right\|_{L_2(\P)}
	\lesssim_\P
	M\mathfrak r_{\gamma,n}\frac{d_\gamma^{3/2}}{n^{1-1/\nu}}.
\]
Condition (iv) gives
\begin{equation}\label{SA eq: growing d gamma envelope condition}
	\frac{d_\gamma^{3/2}}{n^{1-1/\nu}}
	=o_\P(\underline{\lambda}_{w,n}\mathfrak r_{\gamma,n}).
\end{equation}

For the second class $\mathcal G_u$, decompose it according to the
globally Lipschitz and step-function components of the piecewise-Lipschitz
score. The globally Lipschitz component gives the pointwise bound
\[
	|u_i(\btheta)|\,|\widetilde{\bw}_i'\bh|(1+|\psi(y_i,\eta_i)|)
	+|u_i(\btheta)|\,|\widetilde{\bw}_i'\bh|
	\{|u_i(\btheta)|+|\widetilde{\bw}_i'\bh|\}.
\]
The localized entropy argument for this component gives a centered-process
bound of order
\[
	M\mathfrak r_{\gamma,n}\sqrt{\overline{\lambda}_{w,n}}
	\mathfrak r_{\theta,n}\sqrt{J/n}
	+o_\P(\underline{\lambda}_{w,n}\mathfrak r_{\gamma,n}^2).
\]

For each step-function component, the crossing contribution is bounded by
$\min\{|u_i(\btheta)|,|\widetilde{\bw}_i'\bh|\}$ and can be nonzero only if
$y_i$ lies in an interval of length proportional to
$|u_i(\btheta)|+|\widetilde{\bw}_i'\bh|$ around the corresponding
discontinuity threshold. Assumption \ref{SA Assumption DGP}(iii) therefore
bounds its conditional second moment by a constant multiple of
$|u_i(\btheta)|\,|\widetilde{\bw}_i'\bh|
\min\{|u_i(\btheta)|,|\widetilde{\bw}_i'\bh|\}$. For the sum of the finitely
many crossing components, an envelope may be chosen as
\[
	\bar F_u\lesssim M\mathfrak r_{\gamma,n}\|\widetilde{\bw}_i\|.
\]
The first bound in condition (iii), Cauchy--Schwarz, and the eigenvalue bound
give
\[
	\E_n[|u_i(\btheta)|(\widetilde{\bw}_i'\bh)^2]\lesssim_\P
	M^2\overline{\lambda}_{w,n}\mathfrak r_{\gamma,n}^2\mathfrak r_{\theta,n},
\]
so we may take
\[
	\bar\sigma_u=\min\left\{\|\bar F_u\|_{L_2(\P)},
	CM\mathfrak r_{\gamma,n}
	\sqrt{\overline{\lambda}_{w,n}\mathfrak r_{\theta,n}}\right\},
	\qquad
	\delta_u=\frac{\bar\sigma_u}{\|\bar F_u\|_{L_2(\P)}},
\]
as the conditional $L_2$ radius and its ratio to the envelope norm. The
threshold-crossing components are VC-type with index proportional to
$K_{p,s}+d_\gamma\asymp J$, and hence the conservative entropy bound is
\[
	\sup_{\mathbb Q}\log N(\mathcal G_u,L_2(\mathbb Q),
	\varepsilon\|\bar F_u\|_{\mathbb Q,2})
	\lesssim J\log(C/\varepsilon),
	\qquad 0<\varepsilon\leq\delta_u.
\]
The envelope and eigenvalue bounds give
$\delta_u\gtrsim_\P\min\{1,\sqrt{\mathfrak r_{\theta,n}/d_\gamma}\}$.
Together with $d_\gamma=o(J)$ and
$\mathfrak r_{\theta,n}\geq\sqrt{J/n}$, this yields
$\log(C/\delta_u)\lesssim_\P\log n$. Therefore,
\[
	J(\delta_u,\mathcal G_u,\bar F_u)
	\lesssim_\P\delta_u\sqrt{J\log n},
	\qquad
	\frac{J\log n}{n}
	\left\|\max_{1\leq i\leq n}\bar F_u(x_i,\bw_i,y_i)\right\|_{L_2(\P)}
	\lesssim_\P M\mathfrak r_{\gamma,n}
	\frac{J\sqrt{d_\gamma}\log n}{n}.
\]
Applying Lemma \ref{SA lem: maximal inequality} separately to the smooth and
crossing components gives
\begin{equation}\label{SA eq: growing d nuisance process}
\begin{split}
	&\sup_{\substack{\btheta\in\mathcal T_n(C_T)\\
	\|\bh\|\leq M\mathfrak r_{\gamma,n}}}
	\left|\frac{1}{\sqrt n}\G_n[\delta_i^u(\btheta,\bh)]\right|\\
	&\qquad\lesssim_\P
	M\mathfrak r_{\gamma,n}
	\sqrt{\frac{\overline{\lambda}_{w,n}J\mathfrak r_{\theta,n}\log n}{n}}
	+M\mathfrak r_{\gamma,n}\frac{J\sqrt{d_\gamma}\log n}{n}
	+o_\P(\underline{\lambda}_{w,n}\mathfrak r_{\gamma,n}^2).
\end{split}
\end{equation}

\textbf{Step 5: Conclusion.}
Since $J^{-p-1}=o(1)$, its product with
$\sqrt{\overline{\lambda}_{w,n}d_\gamma/n}$ is absorbed by the leading stochastic term.
Combining \eqref{SA eq: growing d mean bound}--\eqref{SA eq: growing d nuisance process}, uniformly over $\btheta\in\mathcal T_n(C_T)$ and $\|\bh\|=M\mathfrak r_{\gamma,n}$,
\[
\begin{split}
	\E_n[\delta_i(\btheta,\bh)]
	\geq\;&\frac{\underline{\lambda}_{w,n}}{8}\|\bh\|^2\\
	&-C\|\bh\|\left(
	\sqrt{\overline{\lambda}_{w,n}d_\gamma/n}
	+J^{-p-1}\mathfrak a_{\pi,J}
	+\sqrt{\overline{\lambda}_{w,n}}\mathfrak r_{\theta,n}^2\right)\\
	&-C\|\bh\|\left(
	{\sqrt{\overline{\lambda}_{w,n}J\mathfrak r_{\theta,n}\log n/n}
	+J\sqrt{d_\gamma}\log n/n}\right)
	+o_\P(\underline{\lambda}_{w,n}\|\bh\|^2).
\end{split}
\]
{Fix $\varepsilon>0$ and choose $C_T=C_T(\varepsilon)$ as in
Step 2. By the definition of $\mathfrak r_{\gamma,n}$, there exists a
sufficiently large fixed $M=M(\varepsilon,C_T)$ such that the probability
that the right-hand side is positive uniformly over
$\btheta\in\mathcal T_n(C_T)$ has limit inferior at least
$1-\varepsilon$. On this event, convexity gives}
\[
	\argmin_{\bgamma}Q_n^*(\btheta,\bgamma)
	\subseteq\{\bgamma:\|\bgamma-\bgamma_0\|\leq M\mathfrak r_{\gamma,n}\}
\]
{uniformly over $\btheta\in\mathcal T_n(C_T)$. Combining this
event with Step 2 and then letting $\varepsilon$ be arbitrary gives}
\[
	\|\widehat{\bgamma}-\bgamma_0\|
	=O_\P\!\left(
	\frac{\sqrt{\overline{\lambda}_{w,n}d_\gamma/n}
		+J^{-p-1}\mathfrak a_{\pi,J}
		+\sqrt{\overline{\lambda}_{w,n}}\mathfrak r_{\theta,n}^2
		{{}+\sqrt{\overline{\lambda}_{w,n}J\mathfrak r_{\theta,n}\log n/n}
		+J\sqrt{d_\gamma}\log n/n}}
	{\underline{\lambda}_{w,n}}
	\right).
\]
{Consequently, the rate condition on $\widehat{\bgamma}$ in Assumption \ref{SA Assumption HLE}(i) holds if there exists a deterministic sequence $\mathfrak r_\gamma$ such that
\[ \mathfrak r_{\gamma,n}=O_\P(\mathfrak r_\gamma),\qquad \sqrt{d_\gamma}\mathfrak r_\gamma=o\!\left(\sqrt{J/n}+J^{-p-1}\right). \]}
\end{proof}


\section*{References} 
\begingroup
\renewcommand{\section}[2]{}	
\bibliography{CCFF_2026_RESTAT--bib}
\bibliographystyle{jasa}
\endgroup